\documentclass{article}
\usepackage{amssymb}
\usepackage{graphicx}
\usepackage{amsmath}
\usepackage{geometry}
\usepackage{appendix}

\let\dprod\prod
\let\dsum\sum

\newtheorem{theorem}{Theorem}

\newtheorem{notation}[theorem]{Notation}

\DeclareMathOperator\erf{erf}

\begin{document}

\title{A Path Integral Approach to Business Cycle Models with Large Number
of Agents}
\author{Pierre Gosselin\thanks{%
Pierre Gosselin : Universit\'{e} Grenoble Alpes, Institut Fourier, UMR 5582
CNRS-UJF, BP 74, 38402 Saint Martin d'H\`{e}res, France.\ E-Mail:
pierre.gosselin@univ-grenoble-alpes.fr} \and A\"{\i}leen Lotz\thanks{%
A\"{\i}leen Lotz: Cerca Trova, BP 114, 38001 Grenoble Cedex 1, France.\
E-mail: a.lotz@erc-cercatrova.eu} \and Marc Wambst\thanks{%
Marc Wambst : IRMA, UMR 7501 CNRS, Universit\'{e} de Strasbourg, France.\
E-Mail: wambst@math.unistra.fr}}
\date{October 2018}
\maketitle

\begin{abstract}
This paper presents an analytical treatment of economic systems with an
arbitrary number of agents that keeps track of the systems' interactions and
agents' complexity. This formalism does not seek to aggregate agents.\ It
rather replaces the standard optimization approach by a probabilistic
description of both the entire system and agents' behaviors. This is done in
two distinct steps.

A first step considers an interacting system involving an arbitrary number
of agents, where each agent's utility function is subject to unpredictable
shocks. In such a setting, individual optimization problems need not be
resolved. Each agent is described by a time-dependent probability
distribution centered around his utility optimum. The entire system of
agents is thus defined by a composite probability depending on time, agents'
interactions and forward-looking behaviors. This dynamic system is described
by a path integral formalism in an abstract space -- the space of the
agents' actions -- and is very similar to a statistical physics or quantum
mechanics system. We show that this description, applied to the space of
agents' actions, reduces to the usual optimization results in simple cases.

Compared to a standard optimization, such a description markedly eases the
treatment of systems with small number of agents. It becomes however useless
for a large number of agents. In a second step therefore, we show that for a
large number of agents, the previous description is equivalent to a more
compact description in terms of field theory. This yields an analytical
though approximate treatment of the system. This field theory does not model
the aggregation of a microeconomic system in the usual sense.\ It rather
describes an environment of a large number of interacting agents. From this
description, various phases or equilibria may be retrieved, along with
individual agents' behaviors and their interactions with the environment.

For illustrative purposes, this paper studies a Business Cycle model with a
large number of agents.

\textbf{Key words:} path integrals, statistical field theory, business
cycle, budget constraint, multi-agent model, interacting agents.

\textbf{JEL Classification:} C02,C60, E00, E1.
\end{abstract}

\pagebreak

\section*{Introduction}

In many instances, representative agent models have proven unrealistic,
lacking both the collective and emerging effects stemming from agents'
interactions. Complex systems, Networks, Agent Based Systems or Econophysics
are among the various paths that have been explored to remedy these
pitfalls. However Agent Based and Networks Models rely on numerical
simulations and may lack microeconomic foundations. Econophysics builds on
statistical facts and empirical aggregate rules to derive macroeconomic
laws.\ These laws are prone, like ad-hoc macroeconomics, to the Lucas
critique (see Lucas 1976). The gap between microeconomic foundations and
multi-agent systems remains.

The present paper attempts to fill this gap by adapting statistical physics
methods to describe multiple interacting agents. It is an introduction to
the method developped in (Gosselin, Lotz, Wambst 2017), illustrated by a
basic economic application to a Business Cycle model. Our setup models
individual, i.e. micro interactions in a context of statistical uncertainty,
to recover a global, macroeconomic description of the system. This approach
allows an analytical treatment of a broad class of economic models with an
arbitrary number of agents, while keeping track of the system's interactions
and complexity at the individual level. It is at the crossroads of
statistical physics and economics: it preserves the microeconomic concepts
of standard economic models to describe fully or partly rational agents,
while enabling the study of the transition from individual to collective
scale given by statistical physics.

Are microeconomic concepts still relevant at the statistical system - macro
- level? Some microscopic features are known to fade away at large scales,
whereas others become predominant. The relevance - or irrelevance in the
physical sense - of some micro interactions when moving to a macro scale
could indirectly shed some light on the change of scale in economics.\textbf{%
\ }Our work is a first attempt to address these questions.

Translating standard economic models into statistical ones requires a
statistical field formalism. Such a formalism partly differs from those used
for physical systems. The field formalism presented in this paper keeps
track of the individual behaviors shaping the field theoretic description,
as well as the results at the macro scale. The field description, in turn,
describes the impact of the macroeconomic environment on individual
behaviors.

The statistical approach of economic systems presented here is a two-step
process. In a first step, the usual model of optimizing agents is replaced
by a probabilistic description of the system. In an interacting system
involving an arbitrary number of agents, each agent is described by an
intertemporal utility function depending on an arbitrary number of
variables. However each agent's utility function is subject to unpredictable
shocks. In such a setting, individual optimization problems are discarded.
Each agent is described by a time-dependent probability distribution
centered around this agent's utility optimum. Unpredictable shocks deviate
each agent from his optimal action, depending on each individual shock
variance. When these variances are null, standard optimization results are
recovered. This so to speak blurred behavior can be justified by the
inherent complexity of agents: each period, their goals and behavior can be
modified by some internal, unobservable and individual shocks.

This setup is a path integral formalism in the abstract space of agents'
actions.\ It is actually very similar to the statistical physics or quantum
mechanics systems. This description is a good approximation of standard
descriptions and allows to solve otherwise intractable problems. Compared to
standard optimization techniques, such a description markedly eases the
treatment of systems with a small number of agents. Working with a
probability distribution is often easier than solving optimization
equations. This approach is thus consistent and useful in itself.\ It
provides an alternative to the standard modeling in the case of a small
number of interacting agents. The average dynamics recovered is close and at
times identical to the standard approach. It also allows to study the set of
agents' dynamics and its fluctuations under some external shocks.

This formalism, useful for small sets, becomes intractable for a large
number of agents.\ It is nonetheless conveniently and classicaly modified
using methods of statistical field theory (Kleinert 1989), into another and
more efficient description directly grounded on our initial path integral
formalism. In a second step, therefore, the individual agents' description
is replaced by a model of field theory that replicates the properties of the
system when $N$, the number of agents, is large. This modeling, although
approximate, is compact enough to allow an analytical treatment of the
system. A double transformation is thus performed with respect to the usual
optimization models. The optimization problem is first replaced by a
statistical system of $N$ agents, that is then itself replaced by a specific
field theory with a large number of degrees of freedom.

This field theory does not represent an aggregation of microeconomic systems
in the usual sense. It rather describes an environment of an infinite number
of interacting agents, from which various phases or equilibria may be
retrieved, as well as the behaviors of the agents, and the way they are
influenced by, or interact, with their environment. This is the so-called
"phase transition" of field theory: the configuration of the ground state
represents an equilibrium for the whole set of agents, and shapes
interactions and individual dynamics. Depending on the parameters of the
system, the form of the ground state may change drastically the description
at the individual level. It is thus possible to compare the particular
features of the macro state of a system and those of the individual level.
As such, it may confirm or invalidate some aspects of the representative
agent models.

By several aspects, our work is related to the Multi-Agents System economic
literature, notably Agent Based Models (see Gaffard Napoletano 2012) and
Economic Networks (Jackson 2010). Both rely on numerical simulation of
Multi-Agents System, but are often concerned with different types of model.
Agent Based Models deal with general macroeconomics models, whereas Network
Models rather deal with lower scale models, such as Contract Theory,
Behavior Diffusion, Information Sharing or Learning. In both type of
settings, agents are typically defined by, and follow, various set of rules.
These rules allow for equilibria and dynamics that would otherwise remain
inaccessible to the representative agent setup.

The Agent-Based approach is similar to ours in that it does not seek to
aggregate all agents, but considers the interacting system in itself. It is
however highly numerical, model-dependent, and relies on microeconomic
relations, such as ad-hoc reaction functions, that may be too simplistic. On
the contrary, Statistical Field Theory accounts for the transition between
scales. Macroeconomic patterns do not emerge from the sole dynamics of a
large set of agents, but are grounded on particular behaviors and
interactions structures. Describing these structures in terms of Field
Theory allows the study the emergence of a particular phase at the macro
scale, and in turn its impact at the individual level.

Econophysics is closer to our approach (for a review, see Chakraborti, Muni
Toke, Patriarca and Abergel (2011a) and (2011b) and references therein). It
often considers the set of agents as a statistical system.\ Moreover,
Kleinert (2009) has already used path integrals to model the stock prices'
dynamics. However, Econophysics does not fully apply the potentiality of
Field Theory to economic systems and rather focuses on empirical laws. But
the absence of microfoundations casts some doubts on the robustness of these
observed empirical laws. Our approach, in contrast, keeps track of usual
microeconomics concepts such as utility functions, expectations, forward
looking behaviors. It includes these behaviors in the analytical treatment
of Multi-Agents Systems by translating the main characteristics of a system
of optimizing agents in terms of a statistical system.

To sum up, the advantages of statistical field theories are threefold. They
allow, at least approximatively, to deal analytically with systems with
large degrees of freedom, without reducing them to mere aggregates.\ They
reveal features otherwise hidden in an aggregate context. Actually, they
allow switching from micro to macro description, and vice-versa, and to
interpret one scale in the light of the other. Moreover, and relevantly for
economic systems, these model may exhibit phase transition. Depending on the
parameters of the model, the system may experience structural changes in
behaviors, at the individual and collective scale. In that, they allow to
consider the question of multiple equilibria in economics.

The first section presents a probabilistic formalism for a system with $N$
identical economic agents, interacting through mutual constraints. Section
two introduces and discusses the associated field formalism for a large
number of agents.\ In section three, we present an application of this
formalism to a business cycle model. Section four concludes.

\section{A probabilistic description of economic agents in interaction}

This section presents a probabilistic formalism for a system with $N$
identical economic agents in interaction. Agents are described by
intertemporal utility functions, but do not optimize these utilities.\
Instead, each agent chooses for his action a path randomly distributed
around the optimal path. The agent's behavior can be described as a weight
that is an exponential of the intertemporal utility, that concentrates the
probability around the optimal path. This feature models some internal
uncertainty as well as non-measurable shocks. Gathering all agents, it
yields a probabilistic description of the system in terms of a probabilistic
weight. This weight includes utility functions and internalizes
forward-looking behaviors, such as intertemporal budget constraint and
interactions among agents. These interactions may for instance arise through
constraints, since revenue flows depend on other agents demand. The
probabilist description then allows to compute the transition functions of
the system, and in turn compute the probability for a system to evolve from
an initial state to a final state within a given time span. They have the
form of Euclidean path integrals.

For the sake of clarity, the description in terms of probabilistic
representation is first explained discarding constraints.\ This
modelization, however suitable for simple models, would be a limitation for
most economic models in which constraints are relevant. So that these
constraints will be considered in a second step. In a third step, we will
show that the interactions between agents are best described in terms of
mutual constraints. We end the section by discussing the transition
functions associated with a system for a large number of economic agents,
and the transition to the field formalism.

\subsection{Principles}

To keep track of the agents' main microeconomic features, several conditions
must be satisfied. First, optimization equations should, at least in some
basic cases and in average, be recovered. Second, this probabilistic
description should account for the agents' individual characteristics, such
as constraints, interactions and forward-looking behavior.

The probabilistic description presented here involves a probability density
for the state of the system at each period $t$. In a system composed of $N$\
agents, each defined by a vector of action $X_{i}\left( t\right) $.

\begin{notation}
We denote $X\left( t\right) $ the concatenation $\left( X_{1}\left( t\right)
,...,X_{N}(t)\right) $\ of the action vectors.
\end{notation}

\begin{notation}
We denote$\ X=\left( X\left( 0\right) ,...,X\left( T\right) \right) $\ the
concatenation of these vectors over the entire timespan, where $T$\ is the
time horizon.
\end{notation}

We define a probability density $P\left( X\left( t\right) \right) $\ for the
set of actions $X\left( t\right) $\ that describes the state of the system
at time $t$. Consider first the intertemporal utility of an agent $i$:%
\begin{equation*}
U_{t}^{\left( i\right) }=\sum_{n\geqslant 0}\beta ^{n}u_{t+n}^{\left(
i\right) }\left( X_{i}\left( t+n\right) ,X\left( t+n-1\right) \right)
\end{equation*}%
where $u_{t+n}^{\left( i\right) }$ is the instantaneous utility at time $t+n$%
. In the optimization setup, $X_{i}\left( t+n\right) $ is the agent $i$
control variable, the variables $\left( X_{j}\left( t+n-1\right) \right)
_{j\neq i}$ are the actions of the other agents, and $\left( X_{j}\left(
t+n-1\right) \right) $\ are the actions of the set of all agents.

The above utility can encompass any quantity optimized, such as the
production or utility functions of consumer/producer models.\ It may also
describe the interaction of several substructures within an individual agent
(see Gosselin Lotz Wambst 2018, and previous formulations in 2013, 2015), or
the motion mechanisms (decision and control) in the neurosciences literature.

We assume that agent $i$\ has no information about others (see Gosselin Lotz
Wambst 2018). Their actions are perceived as random shocks by agent $i$.
Rather than optimizing $U_{t}^{\left( i\right) }$ on $X_{i}\left( t\right) $%
, we postulate that agent $i$ will choose an action $X_{i}\left( t\right) $
and a plan, updated every period, $X_{i}\left( t+n\right) $, $n>0$, for its
future actions. This plan follows a conditional probabilistic law
proportional to:

\begin{equation}
\exp \left( U_{t}^{\left( i\right) }\right) =\exp \left( \sum_{n\geqslant 0}%
\frac{\beta ^{n}}{\sigma _{i}^{2}}u_{t+n}^{\left( i\right) }\left(
X_{i}\left( t+n\right) ,X\left( t+n-1\right) \right) \right)
\label{ntrtmpwg}
\end{equation}%
This is a probabilistic law for $X_{i}\left( t\right) $\ and the plan $%
X_{i}\left( t+n\right) $ when $n>0$.\ It is conditional on the action
variables $X\left( t+n-1\right) $, perceived as exogenous by agent $i$. The
uncertainty about agent $i$\ behavior, or the variability of agents actions,
is denoted $\sigma _{i}^{2}$.

Remark that, for a usual convex utility with a maximum, the closest the
choices of the $X_{i}\left( t+n\right) $\ to $U_{t}^{\left( i\right) }$
optimum, the higher the probability associated to $X_{i}\left( t+n\right) $.
When $\sigma _{i}^{2}\rightarrow 0$, the agent's action is optimal. Our
choice of utility is therefore coherent with a probability peaked around the
optimization optimum. It is thus different from the usual description in
terms of optimal path of actions, but encompasses this approach in average.

This probabilistic description is simplified for non-strategic agents with
no information about others. The agent considers the variables $X_{j}\left(
t+n\right) $ as random noises and integrate them out.\ The probability for $%
X_{i}\left( t\right) $\ and $X_{i}\left( t+n\right) $, $n>0$\ will then be:%
\begin{equation*}
\int \exp \left( U_{t}^{\left( i\right) }\right) \exp \left( -\frac{%
X_{j}^{2}\left( s\right) }{\sigma _{j}^{2}}\right) \prod_{j\neq
i}\prod_{s\geqslant t}dX_{j}\left( s\right)
\end{equation*}%
Here, $\exp \left( -\frac{X_{j}^{2}\left( s\right) }{\sigma _{j}^{2}}\right) 
$ is the subjective weight attributed by $i$ to the $X_{j}\left( s\right) $.
In general if no information is available to agent $i$, we can assume that $%
\sigma _{j}^{2}\rightarrow \infty $ and $\exp \left( -\frac{X_{j}^{2}\left(
s\right) }{\sigma _{j}^{2}}\right) \rightarrow \delta \left( X_{j}\left(
s\right) \right) $, where $\delta \left( X_{j}\left( s\right) \right) $ is
the Dirac delta function, i.e. a function that is peaked on $0$, and null
everywhere else. As a consequence, as long as no further information is
available, other agents may be considered as random perturbations : agent $i$
set their future actions to $0$, discarding them from his planning.

When there are no constraint and no inertia in $u_{t}^{\left( i\right) }$,
i.e. when $u_{t}^{\left( i\right) }$\ solely depends on $X_{i}\left(
t\right) $\ and other agents' previous actions $\left( X_{j}\left(
t-1\right) \right) _{j\neq i}$, periods are independent. Actually, action $%
X_{i}\left( s\right) $ at times $s>t$ are independent of $X_{i}\left(
t\right) $. Consequently, $\exp \left( U_{t}^{\left( i\right) }\right) $\ is
a product of independent terms of the kind $\exp \left( \beta
^{n}u_{t+n}^{\left( i\right) }\left( X_{i}\left( t+n\right) ,X\left(
t+n-1\right) \right) \right) $. The $X_{i}\left( s\right) $ can thus be
integrated out, and the probability associated to the action $X_{i}\left(
t\right) $ is then: 
\begin{equation*}
\int \!\!\left( \int \exp \left( U_{t}^{\left( i\right) }\right) \exp \left( -%
\frac{X_{j}^{2}\left( s\right) }{\sigma _{j}^{2}}\right) \prod_{j\neq
i}\prod_{s\geqslant t}dX_{j}\left( s\right) \right) \prod_{s>t}dX_{i}\left(
s\right) \propto \exp \left( \frac{u_{t}^{\left( i\right) }\left(
X_{i}\left( t\right) ,X\left( t-1\right) \right) }{\sigma _{i}^{2}}\right)
\end{equation*}%
or in term of conditional probabilities:%
\begin{equation}
P\left( X_{i}\left( t\right) \mid \left( X\left( t-1\right) \right) \right) =%
\frac{\exp \left( \frac{u_{t}^{\left( i\right) }\left( X_{i}\left( t\right)
,X\left( t-1\right) \right) }{\sigma _{i}^{2}}\right) }{\mathcal{N}_{i}}
\label{ffct}
\end{equation}%
where the normalization factor is defined by:%
\begin{equation}
\mathcal{N}_{i}\mathcal{=}\int \exp \left( u_{t}^{\left( i\right) }\left(
X_{i}\left( t\right) ,X\left( t-1\right) \right) \right) dX_{i}\left(
t\right)  \label{Nrmlzt}
\end{equation}

From now on, the normalization factor will be skipped, and reintroduced if
needed. Formula (\ref{ffct}) shows that each agent is described by his
instantaneous utility.\ The lack of information induces a short sighted
behavior: in absence of any period overlap, i.e. without any constraint, the
behavior of agent $i$\ is described by a random distribution peaked around
the optimum of $u_{t}^{\left( i\right) }\left( X_{i}\left( t\right) ,X\left(
t-1\right) \right) $ which models exactly the optimal behavior of an agent
influenced by individual random shocks.

As a consequence, gathering all $N$\ agents, the full system $X\left(
t\right) $, is described by a probability weight at each time $t$:%
\begin{equation*}
\prod_{i}\exp \left( \frac{u_{t}^{\left( i\right) }\left( X_{i}\left(
t\right) ,X\left( t-1\right) \right) }{\sigma _{i}^{2}}\right)
\end{equation*}%
for any $\left( X_{k}\left( t-1\right) \right) $.\ Assuming that $\sigma
_{i}^{2}=\sigma ^{2}$ for each agent, a\ particular path for the whole
system $X$, is~defined by the probabililty, up to the normalization factors:%
\begin{equation}
P\left( X\right) =\prod_{t}\prod_{i}\exp \left( \frac{u_{t}^{\left( i\right)
}\left( X_{i}\left( t\right) ,X\left( t-1\right) \right) }{\sigma ^{2}}%
\right)  \label{Prbgb}
\end{equation}

\subsection{Introducing constraints}

Let us now consider the introduction of constraints, in an exact way for
simple cases, or as first approximation in the general case. To do so, we
have two distinguish two types of constraints.

\subsubsection{ Instantaneous constraints}

We define an instantaneous constraint as a dynamic identity between the
control variables of the system. A~standard example is the dynamic for
capital accumulation of a single producer/consumer:%
\begin{equation}
K_{i}\left( t+1\right) -\left( 1-\delta \right) K_{i}\left( t\right)
=Y_{i}\left( t\right) -C_{i}\left( t\right) +\epsilon _{i}\left( t\right)
\label{nstntcst}
\end{equation}%
where $C_{t}$ is the consumption, $\delta $ the depreciation rate, $\epsilon
_{i}\left( t\right) $ a gaussian random shock centered around $0$ and of
variance $\eta ^{2}$, and $Y_{t}=F\left( K_{t}\right) $ the revenue. The
function $F$ may depend on other variables, such as technology, that
themselves depend on the environment provided by other agents. We can
generalize equation (\ref{nstntcst}) for an arbitrary action variable vector 
$X_{i}\left( t\right) $ :%
\begin{equation*}
X_{i}\left( t+1\right) -X_{i}\left( t\right) -H\left( X_{i}\left( t\right)
,X\left( t-1\right) \right) =\epsilon _{i}\left( t\right)
\end{equation*}%
for some function $H$. The inclusion of this constraint in our probabilistic
description is straightforward. If~we assume that $\epsilon _{i}\left(
t\right) $ is independent from any of the variables, the density of
probability for the system (\ref{ffct}) is modified by the adjunction of a
gaussian term: 
\begin{equation}
P\left( X_{i}\left( t\right) \mid \left( X\left( t-1\right) \right) \right)
=\exp \left( \frac{u_{t}^{\left( i\right) }\left( X_{i}\left( t\right)
,X\left( t-1\right) \right) }{\sigma ^{2}}\right) \exp \left( \frac{-\left(
X_{i}\left( t+1\right) -X_{i}\left( t\right) -H\left( X_{i}\left( t\right)
,X\left( t-1\right) \right) \right) ^{2}}{\eta ^{2}}\right)  \label{ffctbb}
\end{equation}%
Summing over agents and periods yields a statistical weight for a path $X$
of the system:%
\begin{equation}
P\left( X\right) =\exp \left( \sum_{i,t}\frac{u_{t}^{\left( i\right) }\left(
X_{i}\left( t\right) ,X\left( t-1\right) \right) }{\sigma ^{2}}\right) \exp
\left( -\sum_{i,t}\frac{\left( X_{i}\left( t+1\right) -X_{i}\left( t\right)
-H\left( X_{i}\left( t\right) ,X\left( t-1\right) \right) \right) ^{2}}{\eta
^{2}}\right)  \label{ffcttt}
\end{equation}%
In continuous time, an integral replaces the sum over $t$ and formula (\ref%
{glblwght2tx}) becomes:%
\begin{equation}
P\left( X\right) =\exp \left( \sum_{i}\int \frac{u_{t}^{\left( i\right)
}\left( X_{i}\left( t\right) ,X\left( t-1\right) \right) }{\sigma ^{2}}%
dt\right) \exp \left( -\frac{1}{\eta ^{2}}\sum_{i}\int_{0}^{T}\left( \frac{d%
}{dt}X_{i}\left( t\right) -H\left( X_{i}\left( t\right) \right) \right)
^{2}dt\right)  \label{ffctxx}
\end{equation}

\subsubsection{Intertemporal constraints}

We will first consider an economic agent optimizing a quadratic utility
under some budget constraint. We~will then extend the result to $N$\ agents
with quadratic utilities under linear arbitrary constraints. Finally, we
will consider the general case of arbitrary utility.

Consider the quadratic utility of an agent whose action vector $X_{i}\left(
t\right) $\ is his sole consumption. His utility reduces to:%
\begin{equation}
u_{t}^{\left( i\right) }\left( X_{i}\left( t\right) ,X\left( t-1\right)
\right) =u\left( C_{i}\left( t\right) \right)  \label{intertemputil}
\end{equation}%
His current account intertemporal constraint is of the form:

\begin{equation}
C_{i}\left( t\right) =B_{i}\left( t\right) +Y_{i}\left( t\right)
-B_{i}\left( t+1\right)  \label{BC}
\end{equation}%
where $Y_{i}\left( t\right) $\ is first considered as an exogenous random
variable, such as revenue in standard optimization models. The state
variable $B_{i}\left( t\right) $\ represents the usual Treasury Bond. Both
the interest rate $r$ and the discount factor $\beta $\ are discarded here
for the sake of simplicity. They can be reintroduced when needed (see
Gosselin Lotz Wambst 2017). We do this explicitly for the interest rate in
our example in section $3$.

If we were to keep the state variable $B_{i}\left( t\right) $ in our
description, we could consider (\ref{BC}) as an instantaneous constraint
similar to (\ref{nstntcst}). However, we will rather replace the state
variable\ $B_{i}\left( t\right) $ and describe the system in terms of the
usual control variable $C_{i}\left( t\right) $. This is in line with the
usual models with intertemporal constraint such as:%
\begin{equation*}
\sum_{t\geqslant 0}Y_{i}\left( t\right) -\sum_{t\geqslant 0}C_{i}\left(
t\right) =0
\end{equation*}%
The usual consumption smoothing is imposed through the Euler equation.
However, in our formalism, both these elements will appear in a
probabilistic form.\textbf{\ }

Successive periods are interconnected through the constraint.\ When $%
C_{i}\left( t\right) $\ is replaced by the state variable $B_{i}\left(
t\right) $, the intertemporal probability weight (\ref{ntrtmpwg}) becomes:%
\begin{eqnarray}
&&\exp \left( u\left( C_{i}\left( t\right) \right) +\sum_{n>0}u\left(
C_{i}\left( t+n\right) \right) \right)   \label{sttcs} \\
&=&\exp \left( u\left( B_{i}\left( t\right) +Y_{i}\left( t\right)
-B_{i}\left( t+1\right) \right) +\sum_{n>0}u\left( B_{i}\left( t+n\right)
+Y_{i}\left( t+n\right) -B_{i}\left( t+n+1\right) \right) \right)   \notag
\end{eqnarray}%
This measures the probability for a choice $C_{i}\left( t\right) $\ and $%
C_{i}\left( t+n\right) $, $n=1...T$. Alternately it is the probability for
the state variable $B_{i}$ to follow a path $B_{i}\left( t+n\right) $, $%
n\geqslant 0$\ starting from $B\left( t\right) _{i}$. The time horizon $T$\
represents the expected remaining duration at time $t$ of the interaction
process'. It should depend decreasingly on $t$, but, for the sake of
simplicity, it is assumed to follow a random poisson process. As a
consequence, the mean expected duration will be a constant written $T$,
irrespective of $t$. Integrating over the $B_{i}\left( t+n\right) $ with $%
n\geqslant 2$,\ yields a transition probability between $B_{i}\left(
t\right) $\ and $B_{i}\left( t+1\right) $\ written $P\left( B_{i}\left(
t\right) ,B_{i}\left( t+1\right) \right) $, the probability to reach $%
B_{i}\left( t+1\right) $\ given $B_{i}\left( t\right) $. It is equal to: 
{\small
\begin{equation*} 
P\left( B_{i}\left( t\right) ,B_{i}\left( t+1\right) \right) =  
\int
\dprod\limits_{i=2}^{T}dB_{t+i}\exp\!\!\left( u\left( B_{i}\left( t\right)
+Y_{i}\left( t\right) -C_{i}\left( t+1\right) \right) +\sum_{n>0}u\left(
B_{i}\left( t+n\right) +Y_{i}\left( t+n\right) -C_{i}\left( t+n+1\right)
\right) \right) 
\end{equation*}
{\LARGE \ }Computing $P\left( B_{i}\left( t\right) ,B_{i}\left( t+1\right)
\right) $\ rather than the transition function for $C_{i}\left( t\right) $
does not change our approach.\ It merely requires that it be applied to the
state variable $B_{i}\left( t\right) $\ rather than to the control variable $%
C_{i}\left( t\right) $. In this case, due to the overlapping nature of state
variables, the probability transition $P\left( B_{i}\left( t\right)
,B_{i}\left( t+1\right) \right) $\ now measures a probability involving two
successive periods, so that the probability for the path $C_{i}\left(
t+n\right) $, $n\geqslant 0$\ has to be rebuilt from the data $P\left(
B_{i}\left( t\right) ,B_{i}\left( t+1\right) \right) $.

Consider a quadratic utility function of the form $u\left( C_{i}\left(
t\right) \right) =-\alpha \left( C_{i}\left( t\right) -\bar{C}\right) ^{2}$
with objective $\bar{C}$ or, should it be non quadratic, its second order
approximation. Rescale it for the sake of simplicity as $-\alpha \left(
C_{i}\left( t\right) -\bar{C}\right) ^{2}\rightarrow -C_{i}^{2}\left(
t\right) $. The constant $\bar{C}$\ can be reintroduced at the end of the
computation. We assume for now that $\sigma ^{2}=1$\textbf{.} The transition
probability between two consecutive state variables thus becomes:%
\begin{eqnarray}
P\left( B_{i}\left( t\right) ,B_{i}\left( t+1\right) \right) &=&\int
\dprod\limits_{i=2}^{T}dB_{i}\left( t+n\right) \exp \left( u\left(
C_{i}\left( t\right) \right) +\sum_{n>0}u\left( C_{i}\left( t+n\right)
\right) \right)  \notag \\
&=&\int \dprod\limits_{i=2}^{T}dB_{i}\left( t+n\right) \exp \left( -\left(
C_{i}\left( t\right) -\bar{C}\right) ^{2}-\sum_{i>0}\left( C_{i}\left(
t+n\right) -\bar{C}\right) ^{2}\right)  \notag
\end{eqnarray}%
The successive integrals can be performed using the budget constraint (\ref%
{BC}).\ We find: 
\begin{equation}
P\left( B_{i}\left( t\right) ,B_{i}\left( t+1\right) \right) =\exp \left(
-\left( B_{i}\left( t\right) +Y_{i}\left( t\right) -B_{i}\left( t+1\right) -%
\bar{C}\right) ^{2}-\frac{1}{T}\left( B_{i}\left( t+1\right)
+\sum_{n>0}\left( Y_{i}\left( t+n\right) -\bar{C}\right) \right) ^{2}\right)
\notag
\end{equation}%
where the transversality condition $B_{i}\left( t\right) \rightarrow 0$ as $%
t\rightarrow T$ has been imposed. Recall that the number of periods $T$ is
the expected mean process duration. Appendix 1 shows how to recover the
transition probability for $C_{i}\left( t\right) $ by integrating over $%
B_{i}\left( t\right) $ and $B_{i}\left( t+1\right) $, under the condition
that the revenue $Y_{i}\left( t+n\right) $ is centered on $\bar{Y}$ with
variance $\theta ^{2}$. Defining the centered variables:%
\begin{equation*}
\hat{C}_{i}\left( t\right) =C_{i}\left( t\right) -\bar{Y}
\end{equation*}%
and:%
\begin{equation*}
\hat{Y}_{i}\left( t\right) =C_{i}\left( t\right) -\bar{Y}
\end{equation*}%
the statistical weight for the consumption path $C_{i}$ becomes in first
approximation for $T$ large:%
\begin{equation}
P\left( \hat{C}_{i}\left( t\right) \right) =\exp \left(
-\sum_{t=1}^{T}\left( \hat{C}_{i}\left( t\right) -\hat{C}_{i}\left(
t+1\right) \right) ^{2}-\frac{\left( \sum_{t=1}^{T}\hat{Y}_{i}\left(
t\right) -\sum_{t=1}^{T}\hat{C}_{i}\left( t\right) \right) ^{2}}{\theta ^{2}}%
\right)  \label{glblwghttxx}
\end{equation}%
Equation (\ref{glblwghttxx}) modifies (\ref{Prbgb}) when agents are facing
some constraints.

We can generalize (\ref{glblwghttxx}) for any system with a constraint
similar to (\ref{BC}), by coming back to the notation $X_{i}\left( t\right) $
for a general action variable and by assuming that individual utilities have
a quadratic approximation around some reference value $X_{0}$: 
\begin{equation*}
u_{t}^{\left( i\right) }\left( X_{i}\left( t\right) ,X\left( t-1\right)
\right) \simeq \text{Constant}+\left( u_{t}^{\left( i\right) }\right)
^{\prime \prime }\left( X_{0}\right)\!\!\left( X_{i}\left( t\right)
-X_{0}\right) ^{2}
\end{equation*}%
Assuming an intertemporal constraint of the form:%
\begin{equation}
\sum_{0\leqslant t\leqslant T}\hat{Y}_{i}\left( t\right) =\sum_{0\leqslant
t\leqslant T}X_{i}\left( t\right)  \label{ntrtmp}
\end{equation}%
for some exogenous flow variable $\hat{Y}_{T+i}$, the individual weight for
an individual path $X_{i}$ becomes (after normalizing $\left( u_{t}^{\left(
i\right) }\right) ^{\prime \prime }\left( X_{0}\right) =-1$, and
reintroducing the variance $\sigma ^{2}$) :%
\begin{equation}
P\left( X_{i}\right) =\exp \left( -\sum_{0\leqslant t\leqslant T}\frac{%
\left( X_{i}\left( t\right) -X_{i}\left( t+1\right) \right) ^{2}}{\sigma ^{2}%
}-\frac{\left( \sum_{0\leqslant t\leqslant T}\hat{Y}_{i}\left( t\right)
-\sum_{0\leqslant t\leqslant T}X_{i}\left( t\right) \right) ^{2}}{\theta ^{2}%
}\right)  \label{wght2tx}
\end{equation}%
This yields the global weight for a path $X$ of the system when an
intertemporal constraint is considered:%
\begin{equation}
P\left( X\right) =\exp \left( -\sum_{i,0\leqslant t\leqslant T}\frac{\left(
X_{i}\left( t\right) -X_{i}\left( t+1\right) \right) ^{2}}{\sigma ^{2}}%
-\sum_{i,0\leqslant t\leqslant T}\frac{\left( \sum_{0\leqslant t\leqslant T}%
\hat{Y}_{i}\left( t\right) -\sum_{0\leqslant t\leqslant T}X_{i}\left(
t\right) \right) ^{2}}{\theta ^{2}}\right)  \label{glblwght2tx}
\end{equation}%
If we consider a continuous time, an integral replaces the sum over $t$ and
formula (\ref{glblwght2tx}) becomes:%
\begin{equation}
P\left( X\right) =\exp \left( -\sum_{i}\left( \frac{1}{\sigma ^{2}}%
\int_{0}^{T}\left( \frac{d}{dt}X_{i}\left( t\right) \right) ^{2}dt+\frac{%
\left( \int_{0}^{T}dt\hat{Y}_{i}\left( t\right) -\int_{0}^{T}dtX_{i}\left(
t\right) \right) ^{2}}{\theta ^{2}}\right) \right)  \label{glblwght3tx}
\end{equation}%
Equations (\ref{glblwght2tx}) and (\ref{glblwght3tx}) describe the
statistical weight associated to a path for a system with intertemporal
constraint. Gosselin, Lotz and Wambst (2018) shows that non-quadratic
corrections to the utility can be considered by adding terms of the form $%
V_{1}\left( X_{i}\left( t\right) \right) $\ to the weights (\ref{glblwght2tx}%
) and (\ref{glblwght3tx}). In continuous time, this yields:%
\begin{equation}
P\left( X\right) =\exp \left( -\sum_{i}\left( \int_{0}^{T}\left( \frac{1}{%
\sigma ^{2}}\int_{0}^{T}\left( \frac{d}{dt}X_{i}\left( t\right) \right)
^{2}+V_{1}\left( X_{i}\left( t\right) \right) \right) dt+\frac{\left(
\int_{0}^{T}dt\hat{Y}_{i}\left( t\right) -\int_{0}^{T}dtX_{i}\left( t\right)
\right) ^{2}}{\theta ^{2}}\right) \right)  \label{Crtnnqdr}
\end{equation}%
Ultimately, we can directly generalize (\ref{Crtnnqdr}) when the revenue $%
\hat{Y}_{i}\left( t\right) $ is itself a function of the variables of the
system. This will be the case in section 3, when considering a Business
Cycle model. Assuming the form $\hat{Y}_{i}\left( t\right) =F\left(
X_{i}\left( t\right) \right) $, yields:%
\begin{equation}
P\left( X\right) =\exp \left( -\sum_{i}\left( \int_{0}^{T}\left( \frac{1}{%
\sigma ^{2}}\int_{0}^{T}\left( \frac{d}{dt}X_{i}\left( t\right) \right)
^{2}+V_{1}\left( X_{i}\left( t\right) \right) \right) dt+\frac{\left(
\int_{0}^{T}dtF\left( X_{i}\left( t\right) \right)
-\int_{0}^{T}dtX_{i}\left( t\right) \right) ^{2}}{\theta ^{2}}\right) \right)
\label{Crtnnqdrbb}
\end{equation}

\subsubsection{Interdependent constraints}

The above computations were performed under the assumption that the
constraint included some exogenous, i.e. totally independent from other
agents, variable $Y_{i}\left( t\right) $\textbf{.} However for a system of $%
N $\ agents, constraints are more likely imposed on agents by the entire set
of interacting agents. For example the variable $Y_{i}\left( t\right) $\ in
the constraint (\ref{BC}) represented the agent's revenue.\ In a context of $%
N$\ interacting agents, this variable depends on others' activity.\ In our
simple model (\ref{intertemputil}), it is on their consumption. In a system
of consumer/producer, the others' consumption generates the flow of revenue $%
Y_{i}\left( t\right) $. In other word, agent $i$\ \ revenue $Y_{i}\left(
t\right) $\ depends on other agents' consumptions $C_{j}\left( t\right) $\ -
or possibly $C_{j}\left( t-1\right) $\ if we assume a lag between agents
actions and their effect. More generally, for a system with a large number
of agents, the revenue $Y_{i}\left( t\right) $, may depend on endogenous
variables that can still be considered as exogenous in agent $i^{\prime }$s
perspective. Thus, our benchmark hypothesis in this section will be that
agents are too numerous to be manipulated by a single agent.

The previous procedures developed for the constraint of a single agent
remain valid and can be generalized directly. Again, we impose a constraint
of the form (\ref{ntrtmp}) in continuous time:%
\begin{equation*}
\int_{0}^{T}Y_{i}\left( t\right) dt=\int_{0}^{T}X_{i}\left( t\right) dt
\end{equation*}%
for each agent. When the individual agent considers $Y_{i}\left( t\right) $
as exogenous, (\ref{glblwght3tx}) applies. But, if $Y_{i}\left( t\right) $
depends endogenously on other agents, (\ref{glblwght3tx}) must be modified
accordingly. Assume for example that $Y_{i}\left( t\right) =\sum_{j}\alpha
_{j}^{i}X_{j}\left( t\right) $ for the i-th agent. Appendix 1 shows that
under some assumptions about $\theta ^{2}$ and $\sigma ^{2}$, the last term
in (\ref{glblwght3tx}) for the i-th agent can be replaced by:%
\begin{equation}
\frac{\left( \int dtY_{i}\left( t\right) -\int dtX_{i}\left( t\right)
\right) ^{2}}{\theta ^{2}}=\frac{1}{\theta ^{2}}\int \int X_{i}\left(
s\right) X_{i}\left( t\right) dsdt+\frac{1}{\theta ^{2}}\sum_{i,j}\int \int
V_{2}\left( X_{i}\left( s\right) ,X_{j}\left( t\right) \right) dsdt
\label{ctnt}
\end{equation}%
for some constant $\nu $ depending on the system, and with:%
\begin{equation}
V_{2}\left( X_{i}\left( s\right) ,X_{j}\left( t\right) \right) =\left(
\sum_{k}\alpha _{j}^{k}\alpha _{i}^{k}-2\alpha _{j}^{i}\right) X_{i}\left(
s\right) X_{j}\left( t\right)  \label{ctntbb}
\end{equation}%
The term:%
\begin{equation}
\frac{1}{\theta ^{2}}\int \int X_{i}\left( s\right) X_{i}\left( t\right) dsdt
\label{qdtrm}
\end{equation}%
depends only on the individual agent $i$.\ It is irrelevant in modeling the
interactions between agents. Moreover, (Gosselin, Lotz, Wambst 2017) shows
that it can often be approximated by a term proportional to $\int
X_{i}^{2}\left( t\right) dsdt$. It can thus be included in the contribution $%
V_{1}\left( X_{i}\left( t\right) \right) $\ of (\ref{Crtnnqdr}). As a
result, equation (\ref{ctntbb}) transcribes the constraints in some
non-local interactions between agents.\ Each agent's constraint is shaped by
the environment other agents create. Equation (\ref{ctntbb})\ also accounts,
when necessary, for some non-linear constraints $V\left( X_{i}\left(
s\right) ,X_{j}\left( t\right) \right) $, where $V$ is an arbitrary function.

This discussion can be generalized straightforwardly to constraints
involving up to $k$\ agents. In that case, any interdependent intertemporal
constraint, or any interaction between $k$\ agents is modeled by:%
\begin{equation}
\sum_{k\geqslant 2}\sum_{i_{1},...,i_{k}}\int_{0}^{T}\int_{0}^{T}\frac{%
V_{k}\left( X_{i_{1}}\left( s_{1}\right) ,...,X_{i_{k}}\left( s_{1}\right)
\right) }{\theta ^{2}}ds_{1}...ds_{k}  \label{ntrtprl}
\end{equation}%
The functions $V_{k}$\ depend on the particular interactions to model.%
\textbf{\ }Section 3 details an example involving technology and capital.
Gathering the results of (\ref{Crtnnqdr}), (\ref{ffctxx}) and (\ref{ntrtprl}%
) leads to the global statistical weight for the set of agents in the
continuous time version, including both instantaneous constraint (\ref%
{ffctxx}), individual intertemporal constraint (\ref{Crtnnqdr}), and
intertemporal constraint (\ref{ntrtprl}):%
\begin{equation}
P\left( X\right) =\exp \left( -A_{1}-A_{2}\right)  \label{gblw3}
\end{equation}%
with:%
\begin{eqnarray*}
A_{1} &=&\sum_{i}\int_{0}^{T}\left( \frac{1}{\sigma ^{2}}\int_{0}^{T}\left( 
\frac{d}{dt}X_{i}\left( t\right) \right) ^{2}+V_{1}\left( X_{i}\left(
t\right) \right) \right) dt+\frac{1}{\eta ^{2}}\sum_{i}\int_{0}^{T}\left( 
\frac{d}{dt}X_{i}\left( t\right) -H\left( X_{i}\left( t\right) \right)
\right) ^{2}dt \\
A_{2} &=&\sum_{k\geqslant 2}\sum_{i_{1},...,i_{k}}\int_{0}^{T}\int_{0}^{T}%
\frac{V_{k}\left( X_{i_{1}}\left( s_{1}\right) ,...,X_{i_{k}}\left(
s_{1}\right) \right) }{\theta ^{2}}ds_{1}...ds_{k}
\end{eqnarray*}%
Contribution $A_{1}$\ is the individual part of the statistical weight.\ It
depends on individual agents. It includes a utility with possible individual
intertemporal constraint and instantaneous constraint. For the intertemporal
part, we have chosen (\ref{Crtnnqdr}). Contribution $A_{2}$\ models the
agents' interactions through a potential depending on several agents.

Remark that the term:%
\begin{equation*}
\frac{\left( \int_{0}^{T}ds\hat{Y}_{i}\left( s\right)
-\int_{0}^{T}dsX_{i}\left( s\right) \right) ^{2}}{\theta ^{2}}
\end{equation*}%
present in (\ref{Crtnnqdr})\textbf{, }has desappeared.\ It has been
decomposed into an individual part, included in $V_{1}\left( X_{s}^{\left(
i\right) }\right) $, and an interaction part, included in $A_{2}$.

Finally, a slight generalization of (\ref{gblw3}) will later prove useful.
Assuming the $N$ agents have different lifespan $T_{1}$,..., $T_{1}$, we
define $P_{T_{1},...,T_{1}}\left( X\right) $ the probability for a path with
variable individual lifespan by: 
\begin{equation}
P_{T_{1},...,T_{1}}\left( X\right) =\exp \left( -A_{1}^{\prime
}-A_{2}^{\prime }\right)  \label{gblw3pr}
\end{equation}%
with:%
\begin{eqnarray*}
A_{1}^{\prime } &=&\sum_{i}\int_{0}^{T_{i}}\left( \frac{1}{\sigma ^{2}}%
\int_{0}^{T_{i}}\left( \frac{d}{dt}X_{i}\left( t\right) \right)
^{2}+V_{1}\left( X_{i}\left( t\right) \right) \right) dt+\frac{1}{\eta ^{2}}%
\sum_{i}\int_{0}^{T_{i}}\left( \frac{d}{dt}X_{i}\left( t\right) -H\left(
X_{i}\left( t\right) \right) \right) ^{2}dt \\
A_{2}^{\prime } &=&\sum_{k\geqslant
2}\sum_{i_{1},...,i_{k}}\int_{0}^{T_{i}}\int_{0}^{T_{j}}\frac{V_{k}\left(
X_{i_{1}}\left( s_{1}\right) ,...,X_{i_{k}}\left( s_{1}\right) \right) }{%
\theta ^{2}}ds_{1}...ds_{k}
\end{eqnarray*}

\subsection{Probability transition functions}

Our formalism replaces the optimization problem with a probabilistic
approach.\ It thus allows to compute the probability transition functions
(or transition functions in short) for the system between an initial and a
final state. To do so, we first define the paths with initial state $%
\underline{X}$ and final state $\overline{X}$ as the set of paths $X$ such
that $X(0)=\underline{X}$ and $X(T)=\overline{X}$. In formula (\ref{gblw3}), 
$P\left( X\right) $ is the probability density for a given path $X$. We then
define the probability of transition between an initial state $\underline{X}$%
, and a final state $\overline{X}$ of the system, as a sum of (\ref{gblw3})
over all paths. This probability is computed as a multiple integral:%
\begin{equation}
P_{T}\left( \underline{X},\overline{X}\right) =\int ...\int P\left( X\right)
\prod_{i}\prod_{t}dX_{i}\left( t\right)  \label{trsntnfc}
\end{equation}%
\ The integrand $\prod_{i}\prod_{t}dX_{i}\left( t\right) $ can be understood
as the sum over the paths $X\left( t\right) $ between $\underline{X}=X\left(
0\right) $ and $\overline{X}=X\left( T\right) $. A compact notation for this
Path integral is $\prod_{i}\prod_{t}dX_{i}\left( t\right) \equiv \prod_{i}%
\mathcal{D}X_{i}$ (see Peskin, Schroeder, 1995). Similarly, using (\ref%
{gblw3pr}) we can define:%
\begin{equation}
P_{T_{1},...,T_{1}}\left( \underline{X},\overline{X}\right) =\int ...\int
P_{T_{1},...,T_{1}}\left( X\right) \prod_{i}\prod_{t}dX_{i}\left( t\right)
\label{trsntnfcpr}
\end{equation}%
for agents with variable lifespan.

However, the integrals in (\ref{trsntnfc}) and (\ref{trsntnfcpr}) are
difficult to compute, particularly when the number of agents is large.
Moreover, the inclusion of non-local terms in (\ref{gblw3}) is another
source of complexity. Techniques such as perturbation expansions of the
potential term in terms of Feynman graphs exist and may be used in some
case. Nevertheless, for a large number of agents another method exists,
based on Statistical Field Theory. This formalism will consider a set of an
infinite number of agents and compute (\ref{trsntnfc}) for any number $N$ of
agents among this set. More precisely, we will rather compute the Laplace
transform of (\ref{trsntnfcpr}), defined by:

\begin{equation}
G_{\alpha }\left( \underline{X},\overline{X}\right) =\int_{0}^{T}\exp \left(
-\alpha \left( T_{1}+...T_{N}\right) \right) P_{T_{1},...,T_{1}}\left( 
\underline{X},\overline{X}\right) dT_{1}...dT_{N}  \label{gblw3bb}
\end{equation}%
Once (\ref{gblw3bb}) computed, the function $P_{T}\left( \underline{X},%
\overline{X}\right) $ can be recovered, either analytically, through an
inverse Laplace transform, either numerically. However, the function $%
G_{\alpha }\left( \underline{X},\overline{X}\right) $ has an interest in
itself. It represents a transition probability for a variable lifespan
between agents: the timespans $T_{i}$\ are assumed to be a Poisson random
process with mean $1/\alpha $. As a consequence, $G_{\alpha }\left( 
\underline{X},\overline{X}\right) $\ is the transition of probability for a
system of agents with variable lifespan $T$.

\section{A field theoretic formulation for interactions between large number
agents}

When the present formalism is applied to a large number of agents,
transition functions can be computed as the so-called correlation functions
of a field theory (see Kleinert 1989) whose action is directly derived from
individual agents' statistical weight defined in section 1. Starting from
the expression (\ref{gblw3}) defining $P\left( X\right) $ for a system, a
functional of an abstract quantity, or "field", is built, that will both
keep the collective aspects of the system, and allow to compute the
transition functions of individual agents (\ref{gblw3bb}) defined in section
1. Field theory allows to inspect the phases of the system, phases that
describe the background in which individual agents evolve. Given the
parameters of the system, several phases may exist: the system may
experience phase transition, switching from one type of dynamic to another.

We can now explain how to associate a field representation to (\ref{gblw3}).
The idea is the following. For a large number of agents, the system
described by (\ref{gblw3}), involves a large number of variables $X\left(
t\right) $\ that are dificult to handle. We consider the space $H$\ of
complex functions defined on the space of a single agent's actions. The
space $H$ describes the collective behavior of the system. Each function $%
\Psi $ of $H$ encodes a particular state of the system. Then, to\ each
function $\Psi $ of $H$, we associate a statistical weight, i.e. a
probability describing the state encoded in $\Psi $. This probability is
written $\exp \left( -S\left( \Psi \right) \right) $, where $S\left( \Psi
\right) $ is a functional, i.e. a function of the function $\Psi $. The form
of $S\left( \Psi \right) $ is derived directly from the form of (\ref{gblw3}%
).

This description does not represent an aggregation to a single
representative agents. It keeps the information about individual agents
among the whole system, and will allow to compute the transition functions
for several agents among the system. Moreover, it makes possible to find
some collective features of the system as a whole.

The method presented here is an adaptation of tools in statistical field
theory described in (Kleinert 1989). It relies on building a field action,
starting from the probabilistic description (\ref{gblw3}) in the following
successive steps.

\paragraph{Replacing the action variable by a field}

Rather than describing agents by a set of action variables $X\left( t\right) 
$, we consider a (complex valued) function $\Psi \left( x\right) $ where the
vector $x$ belongs to the same space as the $X_{i}\left( t\right) $. For
agents described by their consumption $C_{i}\left( t\right) $ and a stock of
individual capital $K_{i}\left( t\right) $, the field will be a function $%
\Psi \left( x\right) =\Psi \left( c,k\right) $.

This function is an abstract encoding of the distribution of consumption and
capital among the whole set of agents. It is not a distribution of
probability for these values.\ Only a functional $S\left( \Psi \right) $ of
this field will give some information about this distribution.

\paragraph{Translating the individual part of $P\left( X\right) $ in terms
of field}

The individual part $A_{1}$ of (\ref{gblw3}) is the weight depending only on
individual agents, excluding their mutual interactions. Recall that:%
\begin{equation*}
A_{1}=\sum_{i}\int_{0}^{T}\left( \frac{1}{\sigma ^{2}}\int_{0}^{T}\left( 
\frac{d}{ds}X_{i}\left( s\right) \right) ^{2}+V_{1}\left( X_{s}^{\left(
i\right) }\right) \right) ds+\frac{1}{\eta ^{2}}\sum_{i}\int_{0}^{T}\left( 
\frac{d}{dt}X_{i}\left( t\right) -H\left( X_{i}\left( t\right) \right)
\right) ^{2}dt
\end{equation*}%
Under some conditions on $\sigma ^{2}$ (see Gosselin, Lotz, Wambst 2017), we
can associate to $A_{1}$ the following functional:%
\begin{equation*}
S_{0}\left( \Psi \right) =\int \!\!\left( \Psi ^{\dag }\left( x\right)\!\!\left(
-\sigma ^{2}\nabla ^{2}+V_{1}\left( x\right) +\alpha \right) \Psi \left(
x\right) \right) dx-\sum_{i}\int_{0}^{T}\Psi ^{\dag }\left( x\right)\!\!\left(
\eta ^{2}\nabla ^{2}+\nabla .H\left( x\right) \right) \Psi \left( x\right)
\end{equation*}%
where $\alpha $ is the parameter arising in the Laplace transform described
in the first section, formula (\ref{gblw3bb}), and where $\Psi ^{\dag
}\left( x\right) $ denotes the complex conjugate of $\Psi \left( x\right) $.
The operator $\nabla $ is the gradient operator, a vector whose $i$-th
coordinate is the first derivative $\frac{\partial }{\partial x_{i}}$: $%
\nabla =\left( \frac{\partial }{\partial x_{i}}\right) $. The operator $%
\nabla ^{2}$ denotes the Laplacian:%
\begin{equation*}
\nabla ^{2}=\sum_{i}\frac{\partial ^{2}}{\partial x_{i}^{2}}
\end{equation*}%
where the sum runs over the coordinates $x_{i}$ of the vector $x$. Applying
this to our previous example, where $\Psi \left( x\right) =\Psi \left(
c,k\right) $, we get $\nabla ^{2}=\frac{\partial ^{2}}{\partial c^{2}}+\frac{%
\partial ^{2}}{\partial k^{2}}$.

\paragraph{Adding the interaction terms of $P\left( X\right) $}

The last part of $P\left( X\right) $ describes specifically the interaction
between the different agents. We will call it $A_{2}$.%
\begin{equation*}
A_{2}=\sum_{k\geqslant 2}\sum_{i_{1},...,i_{k}}\int_{0}^{T}\int_{0}^{T}\frac{%
V_{k}\left( X_{s_{1}}^{\left( i_{1}\right) },...,X_{s_{k}}^{\left(
i_{k}\right) }\right) }{\theta ^{2}}ds_{1}...ds_{k}
\end{equation*}%
In terms of field, it is translated into a functional:%
\begin{equation*}
S_{I}\left( \Psi \right) =\frac{1}{\theta ^{2}}\sum_{k\geqslant 2}\int \Psi
\left( x_{1}\right) ...\Psi \left( x_{k}\right) V_{k}\left(
x_{1}...x_{k}\right) \Psi ^{\dag }\left( x_{1}\right) ...\Psi ^{\dag }\left(
x_{k}\right) dx_{1}...dx_{k}
\end{equation*}%
Adding $S_{I}\left( \Psi \right) $ to $S_{0}\left( \Psi \right) $ yields the
field action: 
\begin{eqnarray*}
S\left( \Psi \right) &=&S_{0}\left( \Psi \right) +\int \!\!\left( \Psi ^{\dag
}\left( x\right) \Psi ^{\dag }\left( y\right)\!\!\left( V_{2}\left( x,y\right)
\right) \Psi \left( x\right) \Psi \left( y\right) \right) dxdy \\
&=&\int \!\!\left( \Psi ^{\dag }\left( x\right)\!\!\left( -\sigma ^{2}\nabla
^{2}+V_{1}\left( x\right) +\alpha \right) \Psi \left( x\right) \right)
dx-\sum_{i}\int_{0}^{T}\Psi ^{\dag }\left( x\right)\!\!\left( \eta ^{2}\nabla
^{2}+\nabla .H\left( x\right) \right) \Psi \left( x\right) \\
&&+\frac{1}{\theta ^{2}}\sum_{k\geqslant 2}\sum_{k\geqslant 2}\int \Psi
\left( x_{1}\right) ...\Psi \left( x_{k}\right) V_{k}\left(
x_{1}...x_{k}\right) \Psi ^{\dag }\left( x_{1}\right) ...\Psi ^{\dag }\left(
x_{k}\right) dx_{1}...dx_{k}
\end{eqnarray*}

\paragraph{Adding source fields to the action}

The above functional $S\left( \Psi \right) $ gathers all the necessary
information. But to compute the transition functions associated to the
system described by (\ref{gblw3}), it has to be supplemented with so-called
source fields.

Consider a complex function $J\left( x\right) $ in $\mathcal{H}$, and add to 
$S\left( \Psi \right) $ the quadratic terms: 
\begin{equation*}
\int \!\!\left( J\left( x\right) \Psi ^{\dag }\left( x\right) +J^{\dag }\left(
x\right) \Psi \left( x\right) \right) dx
\end{equation*}%
The system can be described by an action with source:%
\begin{eqnarray*}
S\left( \Psi ,J\left( x\right) \right) &=&\int \!\!\left( \Psi ^{\dag }\left(
x\right)\!\!\left( -\sigma ^{2}\nabla ^{2}+V_{1}\left( x\right) +\alpha \right)
\Psi \left( x\right) \right) dx-\sum_{i}\int_{0}^{T}\Psi ^{\dag }\left(
x\right)\!\!\left( \eta ^{2}\nabla ^{2}+\nabla .H\left( x\right) \right) \Psi
\left( x\right) \\
&&+\frac{1}{\theta ^{2}}\sum_{k\geqslant 2}\sum_{i_{1},...i_{k}}\Psi \left(
x_{i_{1}}\right) ...\Psi \left( x_{i_{k}}\right) V_{k}\left(
x_{i_{1}}...x_{i_{k}}\right) \Psi ^{\dag }\left( x_{i_{1}}\right) ...\Psi
^{\dag }\left( x_{i_{k}}\right) dx_{i_{1}}...dx_{i_{k}} \\
&&+\int \!\!\left( J\left( x\right) \Psi ^{\dag }\left( x\right) +J^{\dag
}\left( x\right) \Psi \left( x\right) \right) dx
\end{eqnarray*}%
The successive derivatives of $S\left( \Psi ,J\left( x\right) \right) $ with
respect to $J\left( x\right) $ and $J^{\dag }\left( x\right) $ will allow to
recover the transition functions.

\paragraph{Computing the transition functions via path integrals over $\Psi
\left( x\right) $}

Once the action $S\left( \Psi ,J\left( x\right) \right) $ derived from (\ref%
{gblw3}), we can compute the transition functions defined in (\ref{gblw3bb}%
). To do so, we first need to introduce two notations.

\begin{notation}
We denote $X^{\left[ k\right] }(t)$ the vector $\left(
X_{1}(t),...,X_{k}(t)\right) $ of $k$ action vectors. Agents being
identical, any set of $k$ agents among the entire set of $N$ agents is
equivalent.
\end{notation}

\begin{notation}
We denote $\underline{X}^{\left[ k\right] }$ and $\overline{X}^{\left[ k%
\right] }$ any initial and final conditions for this set of agents.
\end{notation}

\bigskip

Then, a result of statistical field theory (see Kleinert 1989) is the
following:

The transition probability $G_{\alpha }\left( \underline{X}^{\left[ k\right]
},\overline{X}^{\left[ k\right] }\right) $ defined in (\ref{gblw3bb})\ for $%
k $\ agents between the initial state $\underline{X}^{\left[ k\right] }$\
and the final state $\overline{X}^{\left[ k\right] }$\ for the system
defined by the probability weight (\ref{gblw3}) is:

\begin{eqnarray}
&&G_{\alpha }\left( \underline{X}^{\left[ k\right] },\overline{X}^{\left[ k%
\right] }\right)  \label{trnsgrtx} \\
&=&\left[ \left( \frac{\delta }{\delta J\left( x_{i_{1}}\right) }\frac{%
\delta }{\delta J^{\dag }\left( y_{i_{1}}\right) }\right) ...\left( \frac{%
\delta }{\delta J\left( x_{i_{N}}\right) }\frac{\delta }{\delta J^{\dag
}\left( y_{i_{N}}\right) }\right) \exp \left( \int \!\!\left( \Psi ^{\dag
}\left( x\right)\!\!\left( -\sigma ^{2}\nabla ^{2}+V_{1}\left( x\right) +\alpha
\right) \Psi \left( x\right) \right) dx\right. \right.  \notag \\
&&-\sum_{i}\int_{0}^{T}\Psi ^{\dag }\left( x\right)\!\!\left( \eta ^{2}\nabla
^{2}+\nabla .H\left( x\right) \right) \Psi \left( x\right)  \notag \\
&&\left. \left. +\frac{1}{\theta ^{2}}\sum_{k\geqslant
2}\sum_{i_{1},...i_{k}}\Psi \left( x_{i_{1}}\right) ...\Psi \left(
x_{i_{k}}\right) V_{k}\left( x_{i_{1}}...x_{i_{k}}\right) \Psi ^{\dag
}\left( x_{i_{1}}\right) ...\Psi ^{\dag }\left( x_{i_{k}}\right) +J\left(
x\right) \Psi ^{\dag }\left( x\right) +J^{\dag }\left( x\right) \Psi \left(
x\right) \right) \mathcal{D}\Psi \mathcal{D}\Psi ^{\dag }\right] _{J=J^{\dag
}=0}  \notag
\end{eqnarray}%
As before, the notation $\mathcal{D}\Psi \mathcal{D}\Psi ^{\dag }$ denotes
an integration over the space of functions $\Psi \left( x\right) $ and $\Psi
^{\dag }\left( x\right) $, that is integral in an infinite dimensional
space. Actually, these integrals are formal, and solely computed in simple
cases. The form of $S\left( \Psi \right) $ is often sufficient to derive
good qualitative insights about the results. In terms of field theory,
formula (\ref{trnsgrtx}) means that the transition functions are the
correlation functions of the field theory with action $S\left( \Psi \right) $%
.

As announced, the formulation (\ref{trnsgrtx}) shows how the transition of
the agents, i.e. their dynamical and stochastic properties, take place in a
surrounding. We do not compute the dynamic of the whole system.\ We rather
derive agent's behaviors from the global properties of a substratum,\ i.e.
the global action for the field $\Psi \left( x\right) $.

Remark that this change in formulation has to be related to the introduction
of a variable number $k$\ of agents in (\ref{trnsgrtx}). In the previous
section, the system was described by a fixed number of agents. Here, the
focus being on the environment, we can compute the transition functions for
an arbitrary number of agents in this environment.

\subsection{Non trivial vacuum, phase transition and Green function}

In practice, it is often not necessary to compute $2k$\ derivatives in (\ref%
{trnsgrtx}) to compute the $k$\ agents transition functions. It is generally
enough to know the transition for one agent $G_{\alpha }\left( \underline{X}%
^{\left[ 1\right] },\overline{X}^{\left[ 1\right] }\right) $. From there,
the transition functions $G_{\alpha }\left( \underline{X}^{\left[ k\right] },%
\overline{X}^{\left[ k\right] }\right) $ for several agents can be deduced
by techniques such as Feynman graphs (see article complet for a detailed
treatment). The one agent transition functions are themselves computed
through graph expansion of the interaction term $S_{I}\left( \Psi \right) $.
However, some simplifications may arise and the graph expansion can be
avoided in first approximation.

We proceed in the following way.\ We first look for a field minimizing the
action $S\left( \Psi \right) $, that is a field $\Psi _{0}$ solution of $%
\frac{\delta }{\delta \Psi }S\left( \Psi \right) =0$. If such a non null
solution does exist, the system is said to have a non-trivial vacuum for $%
S\left( \Psi _{0}\right) $. Then, let $\Psi =\Psi _{0}+\delta \Psi $ and
expand $S\left( \Psi \right) $ to the second order in $\delta \Psi $. That
is:%
\begin{eqnarray*}
S\left( \Psi \right) &\simeq &S\left( \Psi _{0}\right) +\int \delta \Psi
^{\dag }\left( x\right)\!\!\left( -\sigma ^{2}\nabla ^{2}+V_{1}\left( x\right)
+\alpha \right) \delta \Psi \left( x\right) dx \\
&&+\int \delta \Psi ^{\dag }\left( x\right) V\left( \Psi _{0},x,y\right)
\delta \Psi \left( y\right) dxdy
\end{eqnarray*}%
with:%
\begin{equation*}
V\left( \Psi _{0},x\right) =\frac{\sigma ^{2}}{\theta ^{2}}\sum_{k\geqslant
2}\sum_{l_{1},l_{2}}\int \!\!\left[ \dprod\limits_{x_{i},i\neq l_{1}}\Psi
_{0}\left( x_{i}\right) \right] V_{k}\left( x_{i_{1}}...x_{i_{k}}\right) %
\left[ \dprod\limits_{x_{i},i\neq l_{2}}\Psi ^{\dag }\left( x_{i}\right) %
\right] \delta 
\left( x-x_{l_{1}}\right) \delta \left( y-x_{l_{2}}\right)
dx_{1}...dx_{k}
\end{equation*}%
and $\delta \left( x-x_{l_{1}}\right) $ is the Dirac function. It is then a
classical computation to show that in first approximation the one agent
transition function is determined by the quadratic part in $\delta \Psi $
and satisfies the differential equation:%
\begin{equation}
\left( -\sigma ^{2}\nabla ^{2}+\alpha +V_{1}\left( x\right) +V\left( \Psi
_{0},x,y\right) \right) G_{\alpha }\left( x,y\right) =\delta \left(
x-y\right)  \label{Trnstph}
\end{equation}%
in that case, the transition function describing the system can be computed
at least approximatively numerically. A consequence of this set up is the
notion of phase transition. For some values of the parameters, the only
vacuum of the theory may be $\Psi _{0}=0$. In that case, $V\left( \Psi
_{0},x,y\right) =0$, so that the transition function is in first
approximation given by the solution of: 
\begin{equation}
\left( -\sigma ^{2}\nabla ^{2}+\alpha +V_{1}\left( x\right) \right)
G_{\alpha }\left( x,y\right) =\delta \left( x-y\right)  \label{Trnstphbb}
\end{equation}%
On the other hand, if for another range of parameters $\Psi _{0}\neq 0$,
then the transition function is computed by (\ref{Trnstph}) and we say that
the system experiences a phase transition. The qualitative properties of the
system in the phase $\Psi _{0}\neq 0$ differs from those in the phase $\Psi
_{0}=0$. Probabilities of transition and average values of quantities may
differ from one phase to another.

\section{Application: revisiting a standard business cycle model}

In this section, we present an application of our formalism to a standard
Business Cycle model. The usual assumptions of the standard model are
maintained (see Romer 1996), but agents now interact through technology. In
such a model, we show that a non-trivial vacuum may appear. For some values
of the parameters, the equilibrium may experience a discontinuous shift. The
different phases of the system induce different individual behaviors. In the
following, we will present the model, compute the effect of the agents'
interactions on individual dynamics for each phase, and provide an
interpretation of the results.

\subsection{Description}

\subsubsection{The model}

We consider a system with a large number of identical consumer/producer
agents. Each agent $i$\ consumes at time $t$\ a quantity $C_{i}\left(
t\right) $, has a stock of capital $K_{i}\left( t\right) $ and a technology $%
A_{i}\left( t\right) $. The saving variable $B_{i}\left( t\right) $\ is
equal to the stock of capital $K_{i}\left( t\right) $\ used in\ the
production function, as usually assumed in standard Business Cycle models.

On the consumer side, we consider a utility function of the standard form
(Romer 1996, Obstfeld Rogoff 1996):%
\begin{equation*}
u\left( C_{i}\left( t\right) \right) =\frac{C_{i}\left( t\right) ^{1-\theta
}-1}{1-\theta }
\end{equation*}%
$\allowbreak $where the coefficient $\theta $\ measures the relative risk
aversion, i.e. the inverse of the elasticity of substitution between
consumption at different dates. A quadratic approximation of $u\left(
C_{i}\left( t\right) \right) $\ can be found by an expansion around some
minimal value $\hat{C}$\ for the consumption:

\begin{equation}
u\left( C_{i}\left( t\right) \right) =\left( C_{i}\left( t\right) -\hat{C}%
\right) -\frac{\theta }{2}\left( C_{i}\left( t\right) -\hat{C}\right) ^{2}
\label{BCutility}
\end{equation}%
We can thus rewrite:\textbf{\ }%
\begin{equation}
u\left( C_{i}\left( t\right) \right) =\left( C_{i}\left( t\right) -\hat{C}%
\right) -\frac{\theta }{2}\left( C_{i}\left( t\right) -\hat{C}\right) ^{2}=-%
\frac{\theta }{2}\left( C_{i}\left( t\right) -\tilde{C}\right) ^{2}+\frac{1}{%
2\theta }  \label{QdUtil}
\end{equation}%
with $\tilde{C}=\hat{C}+\frac{1}{\theta }$. The constant $\tilde{C}$. We
assume $\tilde{C}\gg 1$. As usual, this constant ensures decreasing marginal
utility. The quadratic approximation (\ref{QdUtil}) will be the utility used
in the sequel.

As consumers, agents each optimize their intertemporal utility function.
Written in continuous time:%
\begin{equation*}
U\left( C\right) =\int_{0}^{T}u\left( C_{i}\left( t\right) \right) dt
\end{equation*}%
Since the discount factor $\beta $ does not alter this section main
arguments, we set it to $1$. Under the usual budget constraint:%
\begin{equation}
C_{i}\left( t\right) =r_{i}\left( t\right) B_{i}\left( t\right) +Y_{i}\left(
t\right) -\frac{d}{dt}B_{i}\left( t\right)  \label{BCintertempBC}
\end{equation}%
where $r_{i}\left( t\right) $\ is the $i$-th\ agent or sector interest rate.
In continuous time, integrating (\ref{BCintertempBC}) over the entire
periods yields the overall budget constraint :%
\begin{equation*}
\int_{0}^{T}\left( Y_{i}\left( t\right) -C_{i}\left( t\right) \right) \exp
\left( -\int r_{i}\left( t\right) dt\right) dt=0
\end{equation*}

On the production side, assuming some uncertainty in the capital
accumulation process yields a dynamic equation for capital:%
\begin{equation}
\dot{K}_{i}\left( t\right) =Y_{i}\left( t\right) -C_{i}\left( t\right)
-\delta \left( K_{i}\left( t\right) \right) +\varepsilon \left( t\right)
\label{Kdnmcs}
\end{equation}%
where $\varepsilon \left( t\right) $\ is a random term of variance $\nu ^{2}$%
\ and $\delta \left( K_{i}\left( t\right) \right) $\ describes the
depreciation of $K_{i}\left( t\right) $.

We endogenize the production $Y_{i}\left( t\right) $\ and treat it as a
function of capital: $Y_{i}\left( t\right) =A_{i}\left( t\right) F_{i}\left(
K_{i}\left( t\right) \right) $. From this relation we can deduce the form of
the interest rate faced by each sector: 
\begin{equation*}
r_{i}\left( t\right) =A_{i}\left( t\right) F_{i}^{\prime }\left( K_{i}\left(
t\right) \right) +r_{c}
\end{equation*}%
That includes an exogenous (or minimal) interest rate $r_{c}$, plus some
individual determinants depending on the capital depreciation, rates of
return, environment and technology of each sector. That is, $r_{i}$\ is
defined by the marginal productivity in the sector plus some collective
effect $r_{c}$. Remark that usually, $r_{c}=-\delta $, but we assume that
other determinants allow to consider $r_{c}$\ as an independent variable. It
is always possible to set $r_{c}=-\delta $\ if needed.

To complete the model, the dynamics of technology should be modeled. We
assume that $A_{i}\left( t\right) $ is a stochatic process with specific
fetures.\ Its dynamics includes an intrinsic part that fluctuates around a
technology growth path. Besides, we assume that technology and capital stock
influence each other. Part of the technology random process will thus
describe technology's interaction with capital. Since the dynamics of $%
A_{i}\left( t\right) $ is probabilistic, we will provide its precise
description in the next section.

\subsubsection{\textbf{Probabilistic description}}

Let us now apply the method presented in section 1. Three variables describe
our model. The variables $K$ and $C$ are standard control variables. As
such, we must give them a statistical weight describing their dynamics. The
third variable, technology, does not qualify as a control variable.\ It
could be treated as an exogenous parameter. However since our formalism aims
at studying interactions between variables, and explore the consequences of
these interactions, we will treat technology as a variable of the system
interacting with capital.\ Consequently, we will give it a statistical
weight. So that the probability describing the system can be decomposed into
several statistical weights, respectively due to consumption, capital and
technology.

\paragraph{Statistical weight of consumption}

The first weight corresponds to the consumption behavior of agents with
utility (\ref{BCutility}) under an intertemporal budget constraint (\ref%
{BCintertempBC}). Appendix 3 shows that the exogenous interest rate in the
constraint modifies the statistical weight (\ref{glblwght3tx}) associated to
consumption under constraint in: 
\begin{equation*}
\exp \left( -\sum_{t}\frac{1}{\varpi ^{2}}\left( C_{i}\left( t\right) -\bar{C%
}-\frac{\left( C_{i}\left( t+1\right) -\bar{C}\right) }{\left( 1+r\right) }%
\right) ^{2}+\sum_{t}C_{0}\right) \exp \left( -\frac{\left(
\int_{0}^{T}\left( Y_{i}\left( t\right) -C_{i}\left( t\right) \right) \exp
\left( -\int r_{i}\left( t\right) dt\right) dt\right) ^{2}}{\theta ^{2}}%
\right)
\end{equation*}%
where $\varpi ^{2}$ $=\frac{2\sigma ^{2}}{\theta }$ measures the uncertainty
in consumption behavior among agents. For a $\varpi ^{2}\ll 1$, the weight is
peaked around the usual optimal Euler equation in continuous time. The
parameter $\sigma ^{2}$ is the uncertainty in consumption behavior used in (%
\ref{glblwght3tx}). Remark that $C_{0}\equiv \frac{1}{2\theta \sigma ^{2}}$,
and that the sum $\sum_{t}C_{0}=TC_{0}$ measures the agents' relative risk
aversion, cumulated over the entire timespan, to change consumption.

Appendix 5 shows that the part due to the overall intertemporal constraint:%
\begin{equation*}
-\frac{\left( \int_{0}^{T}\left( Y_{i}\left( t\right) -C_{i}\left( t\right)
\right) \exp \left( -\int r_{i}\left( t\right) dt\right) dt\right) ^{2}}{%
\theta ^{2}}
\end{equation*}%
can be neglected in first approximation. In continuous time, this leaves us
with:%
\begin{equation}
\exp \left( -\int dt\frac{\left( \dot{C}_{i}\left( t\right) -r\left(
C_{i}\left( t\right) -\bar{C}\right) \right) ^{2}}{\varpi ^{2}}+C_{0}\int
dt\right)  \label{wcmpt}
\end{equation}

\paragraph{Statistical weight of capital}

The second weight models the capital dynamics. Equation (\ref{Kdnmcs}) shows
that $\varepsilon _{i}\left( t\right) =\dot{K}_{i}\left( t\right) -\left(
Y_{i}\left( t\right) -C_{i}\left( t\right) -\delta \left( K_{i}\left(
t\right) \right) \right) $ is a gaussian variable with variance $\nu ^{2}$.
The associated statistical weight is thus gaussian and writes:%
\begin{equation}
\exp \left( -\int dt\frac{\left( \dot{K}_{i}\left( t\right) -\left(
AF_{i}\left( K_{i}\left( t\right) \right) -C_{i}\left( t\right) -\delta
\left( K_{i}\left( t\right) \right) \right) \right) ^{2}}{\nu ^{2}}\right)
\label{wkp}
\end{equation}%
Since $\varepsilon _{i}\left( t\right) $ is independent from consumption,
this weight will be multiplied by (\ref{wcmpt}).

\paragraph{Statistical weight of technology}

The third weight accounts for technology.\ Recall that this is a particular
variable in our setting: its dynamics can be seen as intrinsic, or resulting
from capital interaction. This reflects on its weight.

\subparagraph{Statistical weight of intrinsic technology}

We first consider the contribution inherent to the technology itself. We
denote $\left\langle A\right\rangle $ the system's average technology, to be
computed later, but phase dependent. Let us also denote $A_{0}$\ an
exogenous level of technology and $\bar{A}$\ is the optimal technology level
for the agent in the system, with $\bar{A}=\varkappa \left\langle
A\right\rangle +A_{0}$, and $\varkappa <1$. We choose the technology
contribution to be of the following form:

\begin{equation}
\exp \left( -\int dt\left( \frac{\left( \dot{A}_{i}\left( t\right)
-gA_{i}\left( t\right) \right) ^{2}}{\lambda ^{2}}+\left( A_{i}\left(
t\right) -\bar{A}\right) ^{2}\right) \right)  \label{wtc}
\end{equation}%
The first term $\frac{\left( \dot{A}_{t}-gA_{t}\right) ^{2}}{\lambda ^{2}}$\
models the agent's technology endowment as fluctuating around a technology
growth path $g$. Actually, the distribution is centered around the paths
solutions of $\left( \dot{A}_{i}\left( t\right) -gA_{i}\left( t\right)
\right) =0$. In the sequel, we will set $g=0$\ to simplify, so that $%
A_{i}\left( t\right) $\ can be seen as a detrended variable, but the growth
factor $g$\ can be reintroduced if needed. We consider $\lambda ^{2}\gg 1$,
which means that the level of technology can adapt relatively quickly to $%
\bar{A}$. The second term, $\left( A_{i}\left( t\right) -\bar{A}\right) ^{2}$%
, is the difference between the agent's technology and agents' potential
level of technology in the system. So that, in the absence any other forces,
the agent should be driven towards this optimal level of technology. As a
consequence (\ref{wtc}) models an individual technology that is both driven
by individual factors, and a collective level of technology.

\subparagraph{Statistical weight of capital-technology interaction}

We have shown in section 1 how interactions can be modeled by adding a
potential involving several agents (see (\ref{ntrtprl})). Here, we model the
impact of capital on technology by introducing an additional term in (\ref%
{wtc}) such as:%
\begin{equation}
\exp \left( -\gamma \int \int_{t_{j}<t_{i}}\sum_{j}A_{i}\left( t_{i}\right)
H\left( K_{i}\left( t_{i}\right) ,K_{j}\left( t_{j}\right) \right)
K_{j}\left( t_{j}\right) dt_{j}dt_{i}\right)  \label{wtcpr}
\end{equation}%
This term describes the value added accumulated in the different sectors by
capital stocks. The function $H\ $is any positive functions and represents
the impact of sector $i$\ on sector $j$. We assume reciprocal interactions
and assume that $H$\ is symmetric: $H\left( K_{i}\left( t_{i}\right)
,K_{j}\left( t_{j}\right) \right) =H\left( K_{j}\left( t_{j}\right)
,K_{i}\left( t\right) \right) $. The constant $\gamma $\ measures the
magnitude of these interactions. To check that (\ref{wtcpr}) represents the
impact of capital stock on technology, notice that the agent's technology $%
A_{i}\left( t_{i}\right) $\ is multiplied by a weighted sum of other agents'
past capital stocks. This weight thus models the interaction between
technology and the global capital stock.

This interaction weight would however be incomplete, since the various
sectors' technology may also, in turn, accelerate the dynamics of $%
K_{i}\left( t\right) $. Considering the interaction between capital and
technology is reciprocal is equivalent to adding the term:%
\begin{equation}
\exp \left( -\gamma \int \int_{t_{j}<t_{i}}\sum_{j}A_{i}\left( t_{i}\right)
H\left( K_{i}\left( t_{i}\right) ,K_{j}\left( t_{j}\right) \right)
K_{j}\left( t_{j}\right) dt_{i}dt_{j}\right)  \label{wtcprprime}
\end{equation}%
Here, inverting $t_{j}<t_{i}$\ accounts for the reversal of roles: it is the
past technology that impacts capital stock. Consequently, the statistical
weight for technology and its interaction with capital stock is:%
\begin{eqnarray}
&&\exp \left( -\int dt\left( \frac{\left( \dot{A}_{i}\left( t\right)
-gA_{i}\left( t\right) \right) ^{2}}{\lambda ^{2}}+\left( A_{i}\left(
t\right) -\bar{A}\right) ^{2}\right) \right.  \label{wta} \\
&&~\ \ \ \ \ \ \ \ \ \ \ \ \ \ \ \ \ \ \ \ \left. -\gamma \int \int
\sum_{j}A_{i}\left( t_{i}\right) H\left( K_{i}\left( t_{i}\right)
,K_{j}\left( t_{j}\right) \right) K_{j}\left( t_{j}\right)
dt_{j}dt_{i}\right)  \notag
\end{eqnarray}%
The second term of (\ref{wta}) can be better understood if we note that it
is an approximation of the quadratic term for $\left\vert \gamma \right\vert
\ll 1$:%
\begin{equation}
+\frac{\gamma }{\sqrt{\left\vert \gamma \right\vert }}\int \sum_{i}\left(
A_{i}\left( t_{i}\right) -\frac{\sqrt{\left\vert \gamma \right\vert }}{2}%
\int \sum_{j}H\left( K_{i}\left( t_{i}\right) ,K_{j}\left( t_{j}\right)
\right) K_{j}\left( t_{j}\right) dt_{j}\right) ^{2}dt_{i}  \label{wtpr}
\end{equation}%
Actually, the expansion of (\ref{wta}) yields the second term of (\ref{wta}%
), plus a quadratic term in $A_{i}\left( t_{i}\right) $\ and a quadratic
term in $K_{j}\left( t_{j}\right) $. The quadratic term in $K_{j}\left(
t_{j}\right) $\ is of magnitude $\left( \sqrt{\left\vert \gamma \right\vert }%
\right) ^{3}$\ and can be neglected. The quadratic term in $A_{i}\left(
t_{i}\right) $\ is of magnitude $\sqrt{\left\vert \gamma \right\vert }%
A_{i}^{2}\left( t_{i}\right) $\ which is negligible with respect to $\left(
A_{i}\left( t\right) -\bar{A}\right) ^{2}$.

Equation (\ref{wtpr}) shows for $\gamma <0$, the interaction is attractive:
the higher capital stock, the higher the technology. Interactions between
capital and technology increase the likelihood for paths satisfying:%
\begin{equation*}
A_{i}\left( t_{i}\right) -\frac{\sqrt{\left\vert \gamma \right\vert }}{2}%
\int \sum_{j}H\left( K_{i}\left( t_{i}\right) ,K_{j}\left( t_{j}\right)
\right) K_{j}\left( t_{j}\right) dt_{j}=0
\end{equation*}%
On the contrary, for $\gamma >0$\ the interaction is repulsive: interactions
between capital and technology increase the likelihood of paths satisfying:%
\begin{equation*}
A_{i}\left( t_{i}\right) -\frac{\sqrt{\left\vert \gamma \right\vert }}{2}%
\int \sum_{j}H\left( K_{i}\left( t_{i}\right) ,K_{j}\left( t_{j}\right)
\right) K_{j}\left( t_{j}\right) dt_{j}\rightarrow \infty
\end{equation*}%
Depending on society's stock of capital, we can define a certain threshold $%
\tilde{A}$ of required technology:%
\begin{equation*}
\tilde{A}=\frac{\sqrt{\left\vert \gamma \right\vert }}{2}\int
\sum_{j}H\left( K_{i}\left( t_{i}\right) ,K_{j}\left( t_{j}\right) \right)
K_{j}\left( t_{j}\right) dt_{j}
\end{equation*}
Agents with technology endowment higher than this threshold have an
advantage and are thus driven on a technology growth path. Agents below this
threshold $\tilde{A}$\ will be evicted. We will study both cases $\gamma <0$%
\ and $\gamma >0$\ later on.

\paragraph{Overall statistical weight}

We can now gather the three contributions (\ref{wcmpt}) (\ref{wkp}) and (\ref%
{wta}). The overall statistical weight writes:%
\begin{eqnarray}
&&\exp \left( -\int dt\left( \frac{\left( \dot{C}_{i}\left( t\right)
+r\left( C_{i}\left( t\right) -\bar{C}\right) \right) ^{2}}{\varpi ^{2}}%
\right) +C_{0}\int dt\right.  \label{ndsttwgth} \\
&&\hspace{3cm}\left. -\sum_{i}\int dt\left( \frac{\left( \dot{K}_{i}\left(
t\right) -\left( A_{i}\left( t\right) F_{i}\left( K_{i}\left( t\right)
\right) -C_{i}\left( t\right) -\delta \left( K_{i}\left( t\right) \right)
\right) \right) ^{2}+\varsigma ^{2}C_{i}^{2}\left( t\right) }{\nu ^{2}}%
\right) \right)  \notag \\
&&\times \exp \left( -\sum_{i}\int dt_{i}\left( \frac{\left( \dot{A}%
_{i}\left( t_{i}\right) \right) ^{2}}{\lambda ^{2}}+\left( A_{i}\left(
t_{i}\right) -\bar{A}\right) ^{2}\right) -\gamma \int \int
\sum_{i,j}A_{j}\left( t\right) H\left( K_{j}\left( t_{j}\right) ,K_{i}\left(
t_{i}\right) \right) K_{i}\left( t_{i}\right) dt_{i}dt_{j}\right)  \notag
\end{eqnarray}

\subsubsection{Field theoretic description}

Now that the model has been described in terms of probabilities, we can
transcribe it in terms of field, and apply the method presented in section
2. We introduce a field $\Psi \left( K,C,A\right) $ depending on the
relevant variables of the system, consumption, capital and technology.
Appendix 6 presents the field theoretic formulation of the system given the
above assumptions. Choosing the usual linear depreciation function $\delta
\left( K\right) =\delta K$ and a "distance function" of the form $H\left(
K_{2},K_{1}\right) =1$ yields the field formulation of the system:%
\begin{eqnarray}
&&S\left( \Psi \right) =\int \Psi ^{\dag }\left( K,C,A\right) \mathcal{O}%
\Psi \left( K,C,A\right)  \label{fldqn} \\
&&+\frac{\gamma }{2}\int \Psi ^{\dag }\left( K_{1},C_{1},A_{1}\right) \Psi
^{\dag }\left( K_{2},C_{2},A_{2}\right)\!\!\left( A_{2}K_{1}+A_{1}K_{2}\right)
\Psi \left( K_{1},C_{1},A_{1}\right) \Psi \left( K_{2},C_{2},A_{2}\right) 
\notag
\end{eqnarray}%
where the second order differential operator $\mathcal{O}$ is defined by:

\begin{eqnarray*}
\mathcal{O} &\mathcal{=}&-\varpi ^{2}\frac{\partial ^{2}}{\partial C^{2}}-%
\frac{1}{\lambda ^{2}}\frac{\partial ^{2}}{\partial A^{2}}-\nu ^{2}\frac{%
\partial ^{2}}{\partial K^{2}} \\
&&+\left( A-\bar{A}\right) ^{2}-2\left( C-AF\left( K\right) +\delta K\right) 
\frac{\partial }{\partial K}+2\left( AF^{\prime }\left( K\right)
+r_{c}\right)\!\!\left( C-\bar{C}\right) \frac{\partial }{\partial C}+\varsigma
^{2}\left( C-\bar{C}\right) ^{2} \\
&&+\left( \alpha +2AF^{\prime }\left( K\right) +\left( r_{c}-\delta \right)
-C_{0}\right)
\end{eqnarray*}%
The quadratic part of the action $S\left( \Psi \right) $, namely $\int \Psi
^{\dag }\left( K,C,A\right) \mathcal{O}\Psi \left( K,C,A\right) $, describes
the individual behavior of the agents. The quartic part of the action
represents the interaction between technology and capital stocks among
agents.

\subsection{Results}

We have described the model in terms of Field Theory. We can now search for
non-trivial phases in the system, and study their properties.\ These
emerging phases will then allow us to compute the transition functions first
without, then with interactions.\ 

\subsubsection{Phases of the system}

Once the field action $S\left( \Psi \right) $ found, the minima that define
the various phases of the system can be computed.\ These phases correspond
to the system different economic global equilibria. The trivial phase, i.e. $%
\Psi _{1}\left( K,C,A\right) =0$, is an analogous to a system linearized
around its equilibrium. On the contrary, non-trivial phases reveal the
emergence of other equilibria. In each of these phases, the variables'
average values and the agents' transition probability functions can be
computed and studied.

The inspection of the non-trivial phases associated to the field action (\ref%
{fldqn}) of our model and the computation of the variables' average values
in each phase are performed in Appendix $6$.\ The production function,
assumed to be Cobb Douglas, $F\left( K\right) =AK^{\varepsilon }$ with $%
\varepsilon <1$, can be approximated by a Taylor expansion above a minimal
stock of capital $\bar{K}$: 
\begin{equation*}
AK^{\varepsilon }\simeq A\bar{K}^{\varepsilon }\left( 1+\varepsilon \left( 
\frac{K-\bar{K}}{\bar{K}}\right) -\frac{\varepsilon \left( 1-\varepsilon
\right) }{2}\left( \frac{K-\bar{K}}{\bar{K}}\right) ^{2}\right)
\end{equation*}%
The results show that a non-trivial minimum $\Psi _{1}\left( K,C,A\right) $
for $S\left( \Psi \right) $ exists, provided some conditions on the
parameters are fulfilled. The minimum $\Psi _{1}\left( K,C,A\right) $ is a
product of several gaussian functions in the variables $K,C,A$ whose precise
form is not necessary to the discussion (see Appendix 6 for a details).

This non-trivial minimum of $S\left( \Psi \right) $ exists when:%
\begin{eqnarray}
\gamma &>&0  \label{cnd1} \\
A_{0} &>&\left( 1+\sqrt{2}\right) \bar{C}  \notag \\
\lambda &>&>1  \notag
\end{eqnarray}%
and:%
\begin{equation}
1\gg \frac{A_{0}\varepsilon }{\left( 1-\varkappa \right) \bar{K}%
^{1-\varepsilon }}-\delta >0  \label{cnd2}
\end{equation}%
Given that $\varepsilon <1$,\ these conditions are jointly satisfied for $%
A_{0}$ and $\bar{K}$ relatively large, and when the return of capital
exceeds the depreciation value (equation (\ref{cnd2})). An additional
condition on the relative risk aversion $C_{0}$ exists (see Appendix 6 for
details). Qualitatively, for an intermediate range of values for $C_{0}$,
the non-trivial phase is possible and stable. We give below some
interpretation for these conditions.

The equilibrium values of the variables in both phases and the global
patterns of the system are computed using quadratic expansions of $S\left(
\Psi \right) $\ around the minima $\Psi _{1}\left( K,C,A\right) =0$\ and $%
\Psi _{1}\left( K,C,A\right) \neq 0$ respectively.\ This expansion in turn
allows to find the average values for $K$, $C$\ and $A$\ in each phase (see
appendix 6). For $\Psi _{1}\left( K,C,A\right) =0$ and neglecting the
interaction terms, the average values for the relevant variables are: 
\begin{eqnarray*}
\left\langle A\right\rangle _{0} &=&\bar{A}=\frac{A_{0}}{\left( 1-\varkappa
\right) } \\
\left\langle C\right\rangle _{0} &=&\bar{C}+\sqrt{\frac{2}{\pi }}\varpi \\
\left\langle K\right\rangle _{0} &=&\frac{\frac{A_{0}}{1-\varkappa }\bar{K}%
^{\varepsilon }\left( 1-\varepsilon \right) -x}{Y}
\end{eqnarray*}%
with:%
\begin{eqnarray*}
Y &=&\delta -\frac{A_{0}}{\left( 1-\varkappa \right) }\varepsilon \bar{K}%
^{\varepsilon -1} \\
K_{1}^{\prime } &=&-\frac{2}{\pi }\allowbreak \left\vert \bar{C}+\sqrt{\frac{%
2}{\pi }}\varpi -A\bar{K}^{\varepsilon }\left( 1-\varepsilon \right)
\right\vert -\frac{\sqrt{2\left\vert \delta -\Gamma _{3}\varepsilon \bar{K}%
^{\varepsilon -1}\right\vert }}{\sqrt{\pi }}\nu \\
x &=&\bar{C}+\sqrt{\frac{2}{\pi }}\varpi -K_{1}^{\prime }
\end{eqnarray*}%
For the phase with $\Psi _{1}\left( K,C,A\right) \neq 0$, one finds:%
\begin{eqnarray}
\left\langle A\right\rangle _{1} &=&\frac{A_{0}}{\left( 1-\varkappa \right) }%
-\frac{1}{2}\frac{K^{\varepsilon }A_{0}\left( 1-\varepsilon \right) -\left( 
\bar{C}+\sqrt{\frac{2}{\pi }}\varpi -K_{1}^{\prime }\right) Y\left(
1-\varkappa \right) }{Y^{2}\left( 1-\varkappa \right) ^{3}}\gamma \eta
\label{prmtvr} \\
\left\langle C\right\rangle _{1} &=&\bar{C}+\sqrt{\frac{2}{\pi }}\varpi -%
\frac{\varpi ^{2}\left( \frac{A_{0}}{1-\chi }\right) }{2\left( \varsigma
^{2}\varpi ^{2}+\left( AF^{\prime }\left( K\right) +r_{c}\right) ^{2}\right)
\left\vert \delta -\Gamma _{3}\varepsilon \bar{K}^{\varepsilon
-1}\right\vert }\gamma \eta  \notag
\end{eqnarray}%
\begin{eqnarray}
\left\langle K\right\rangle _{1} &=&\frac{\frac{A_{0}}{1-\varkappa }\bar{K}%
^{\varepsilon }\left( 1-\varepsilon \right) -x}{Y}-\frac{1}{2}K^{\varepsilon
}\frac{\allowbreak \allowbreak \allowbreak \left( K^{\varepsilon
}A_{0}\left( 1-\varepsilon \right) -Y\left( 1-\varkappa \right) x\right)
\left( \varepsilon x-K\delta \left( 1-\varepsilon \right) \right) }{%
Y^{4}K\left( 1-\varkappa \right) ^{3}}\gamma \eta  \label{Kptl} \\
&&-\frac{\bar{K}^{\varepsilon }\left( 1-\varepsilon \right)\!\!\left(
1-Y\right) K_{1}^{\prime }}{2Y^{3}}\gamma \eta -\frac{\nu ^{2}\frac{A_{0}}{%
1-\chi }}{2Y^{2}}\gamma \eta -\frac{\varpi ^{2}\left( \frac{A_{0}}{1-\chi }%
\right) }{2\left( \varsigma ^{2}\varpi ^{2}+\left( AF^{\prime }\left(
K\right) +r_{c}\right) ^{2}\right) Y^{2}}\gamma \eta  \notag
\end{eqnarray}%
The parameter $\eta \gamma $ depends on the parameters of the system and can
be estimated as: 
\begin{equation*}
\eta \gamma \ll 1
\end{equation*}

In the non-trivial phase, both, the average level of consumption $%
\left\langle C\right\rangle _{1}$\ and the average level of technology $%
\left\langle A\right\rangle _{1}$\ are lower than in the trivial phase: $%
\left\langle C\right\rangle _{1}<\left\langle C\right\rangle _{0}$, $%
\left\langle A\right\rangle _{1}<\left\langle A\right\rangle _{0}$. Indeed,
the non-trivial phase emerges under high relative risk aversion, and hence
low level of consumptions. The effect on average capital stock $\left\langle
K\right\rangle _{1}$\ is mitigated: since the technology level has
decreased, a higher level of capital stock may be required to reach average
levels of consumption. Equation (\ref{Kptl}) shows that when capital or/and
consumption volatility, $\nu $\ and $\varpi $\ respectively, are high,
capital stock is lower in the non-trivial phase. Uncertainty hinders
accumulation. But when these volatilities are low, capital stock is higher
than in the trivial phase.

One can compute the average production level in both phases. This is done in
Appendix 6. It appears, that in our range of approximations:%
\begin{equation*}
\left\langle Y\right\rangle _{1}<\left\langle Y\right\rangle _{0}
\end{equation*}

Let us now discuss the conditions (\ref{cnd1}) and (\ref{cnd2}) in which
this non trivial phase should appear. When $\gamma <0$, capital and
technology are mutually enhanced.\ This prevents the non trivial phase to
appear. A high level of consumption and capital can be reached. But when $%
\gamma >0$,\ the interaction between capital and technology is selective.\
In such a setting, the initial endowment in technology is crucial. Agents
endowed with a level of technology above a certain threshold are favored. On
the contrary, for agents poorly endowed, this level acts as a ceiling. In
average, our results show that overall, the society experiences lower
technology, lower production and lower consumption.

Finally, the non trivial phase corresponds to a large minimal technology and
intermediate values of $C_{0}$. Actually, risk aversion has to be large
enough to reduce capital accumulation.\ Still, this reduced capital
accumulation must be sufficient to reach optimal consumption. To do so,
technology must be relatively high, and compensate for a lower stock of
capital. This equilibrium must also be sustainable. Thus, a large $C_{0}$ is
outside the limits of our model.

\subsubsection{Transition functions without interaction}

In the above, we had first translated the classical business model in terms
of probabilistic description, then translated it into statistical field. Now
that the various phases of the system of the statistical field have been
described, we can turn back to and examine the transition functions of the
system.\textbf{\ }In other terms, having found the collective levels, we can
now turn back to the individual level, that may include several agents and
their interactions. The transition functions will now allow us to describe
the individual agents' dynamics in each phase. We will then deduce the
equilibrium values of these dynamics, as well as the dynamic equations of
their average paths.

\paragraph{Transition functions}

In the following, we will denote the phase $\iota =0$ the phase with minimum 
$\Psi _{1}=0$, the trivial phase.\ The phase $\iota =1$ will correspond to $%
\Psi _{1}\neq 0$, the non-trivial phase.

Given a phase $\iota $ of the system and neglecting the expression last
term, i.e. the quartic interaction term in (\ref{fldqn}), the probability of
transition for an agent from a state $\left( C^{\prime },K^{\prime
},A^{\prime }\right) $ to a state $\left( C,K,A\right) $ during a timespan $%
t $, is equal to: 
\begin{eqnarray}
&&G\left( C,K,A,C^{\prime },K^{\prime },A^{\prime },t\right) =\frac{1}{\sqrt{%
\left( 2\pi \right) ^{3}\frac{\Omega ^{2}\varpi ^{2}}{\lambda ^{2}}t^{3}}}%
\exp \left( -\frac{\left( \left( C-\bar{C}_{\iota }\right) -\left( C^{\prime
}-\bar{C}_{\iota }\right)\!\!\left( 1+\beta t\right) \right) ^{2}}{2\varpi ^{2}t%
}\right)  \label{Trstngntx} \\
&&\times \exp \left( -\frac{\left( \left( K-\bar{K}+\frac{\delta \bar{K}+%
\bar{C}_{\iota }}{\alpha }\right) -\left( \left( K^{\prime }-\bar{K}+\frac{%
\delta \bar{K}+\bar{C}_{\iota }}{\alpha }\right)\!\!\left( 1-\alpha t\right)
-\left( C^{\prime }-\bar{C}_{\iota }\right) t+A^{\prime }\bar{K}%
^{\varepsilon }t\right) \right) ^{2}}{2\Omega ^{2}t}\right)  \notag \\
&&\times \exp \left( -\frac{\lambda ^{2}\left( A-A^{\prime }\right) ^{2}}{2t}%
-\frac{\left( \frac{A+A^{\prime }}{2}-\bar{A}_{i}\right) ^{2}}{2}t-m_{\iota
}t\right)  \notag
\end{eqnarray}%
where $\left( C^{\prime },K^{\prime },A^{\prime }\right) $ is the initial
state of the system and $\left( C,K,A\right) $ the final state for a process
of duration $t$.

\paragraph{Parameters}

Two sorts of parameters appear in equation (\ref{Trstngntx}). Some
parameters, such as $\Omega ^{2},\alpha $ and $\beta ,$ do not depend on the
phase of the system.\ They are: 
\begin{equation*}
\Omega ^{2}=\frac{\varpi ^{2}}{\lambda ^{2}}\left( \nu ^{2}+\frac{2\bar{K}%
^{2\varepsilon }}{\lambda ^{2}\alpha ^{2}}+\frac{3\varpi ^{2}}{2\left( \beta
^{2}-\alpha ^{2}\right) }\right)
\end{equation*}%
\begin{eqnarray}
\alpha &=&\delta -\frac{A+A^{\prime }}{2}F^{\prime }\left( \frac{K+K^{\prime
}}{2}\right)  \label{cftdntx} \\
\beta &=&\frac{A+A^{\prime }}{2}F^{\prime }\left( \frac{K+K^{\prime }}{2}%
\right) +r_{c}-\delta  \notag
\end{eqnarray}%
The parameter $\Omega ^{2}$ is a global variance of the system which mixes
the variances of the variables $C$, $K$ and $A$. This parameter is
independent from the phase, since the phases do not affect volatilities but
average values in this particular model. For $r_{c}=\delta $, the parameter $%
\beta $ is the average rate of return of capital for a process starting from 
$\left( C^{\prime },K^{\prime },A^{\prime }\right) $ and reaching $\left(
C,K,A\right) $.\ Finally, the parameter $\alpha $ measures the spread
between marginal productivity and capital depreciation. These two variables
do not depend on the phase, but merely on the producer capital and
technology levels.

\bigskip

Other parameters in equation (\ref{Trstngntx})\ are phase-dependent. These
parameters are the average values of technology and capital in a given
phase. They are, for the first phase: 
\begin{eqnarray*}
\bar{A}_{0} &=&\left\langle A\right\rangle _{0}=\frac{A_{0}}{1-\chi } \\
\bar{C}_{0} &=&\left\langle C\right\rangle _{0}=\bar{C}+2\varpi \\
m_{0} &=&0
\end{eqnarray*}%
and for the second phase:%
\begin{eqnarray*}
\bar{A}_{1} &=&A_{0}+\chi \left\langle A\right\rangle _{1},\bar{C}%
_{1}=\Gamma _{1} \\
m_{1} &=&\bar{A}_{1}^{2}-\left( \left( 1-\varkappa \right) \bar{A}%
_{1}+\varkappa \Gamma _{3}\right) ^{2}+\frac{\left( \left( \left(
1-\varkappa \right) \bar{A}_{1}+\varkappa \Gamma _{3}\right) ^{2}-2\bar{C}%
\left( \left( 1-\varkappa \right) \bar{A}_{1}+\varkappa \bar{A}_{1}\right) -%
\bar{C}_{1}^{2}\right) }{\left( 1-\varepsilon \right) \bar{K}^{\varepsilon }}
\end{eqnarray*}%
where the average values $\left\langle A\right\rangle _{\iota }$ \ have been
defined in (\ref{prmtvr}).

Because some parameters are phase-dependent, the agent's transition
probability depends on the phase of the system.

The parameter $m_{\iota }$\ is a measure of the system's inertia. It is null
in the trivial phase, but positive in the non-trivial phase. The fact that
it should appear as an exponential term $\exp \left( -m_{\iota }t\right) $\
in (\ref{Trstngntx}) is relevant. Transitions are quickly dampened through
the interaction process in the non-trivial phase, and transition
probabilities are reduced.\ The strong inertia of the system keeps agents
closer to their initial values than in the trivial phase.

Note also that multi-agents interactions do not appear directly in (\ref%
{Trstngntx}), but only through parameters.\textbf{\ }These parameters encode
the collective effects of the interactions and their impact at the
individual level.

\paragraph{Average paths and classical dynamics}

Formula (\ref{Trstngntx}) is valid for small $t$ (see Appendix 6 for larger $%
t$). However, it is sufficient to find the agent's average path. This is
straightforward: for a gaussian weight, the average path is found by setting
the exponent in (\ref{Trstngntx}) to $0$.\ This expresses the fact that we
select the most likely path, which is the average path. This yields the
relations between the initial and final points:%
\begin{eqnarray}
0 &=&\left( \left( C-\bar{C}_{\iota }\right) -\left( C^{\prime }-\bar{C}%
_{\iota }\right)\!\!\left( 1+\left( \alpha +\beta \right) t\right) \right)
\label{dmcvr} \\
0 &=&\left( \left( K-\bar{K}+\frac{\delta \bar{K}+\bar{C}_{\iota }}{\alpha }%
\right) -\left( \left( K^{\prime }-\bar{K}+\frac{\delta \bar{K}+\bar{C}%
_{\iota }}{\alpha }\right)\!\!\left( 1-\alpha t\right) -\left( C^{\prime }-\bar{%
C}_{\iota }\right) t+A^{\prime }\bar{K}^{\varepsilon }t\right) \right) 
\notag \\
0 &=&\frac{\lambda ^{2}\left( A-A^{\prime }\right) ^{2}}{2t}+\frac{\left( 
\frac{A+A^{\prime }}{2}-\bar{A}_{i}\right) ^{2}}{2}t  \notag
\end{eqnarray}%
The treatment of these equations is usual. The equilibrium values are first
found by setting both initial and final values equal to the equilibrium:%
\begin{equation*}
\left( C^{\prime },K^{\prime },A^{\prime }\right) =\left( C,K,A\right)
=\left( C_{e},K_{e},A_{e}\right)
\end{equation*}%
One finds directly:%
\begin{eqnarray*}
K_{e} &=&\frac{\left( 1-\varepsilon \right) \bar{A}_{i}\bar{K}^{\varepsilon
}-\bar{C}_{i}}{\delta -\bar{K}^{\varepsilon -1}\bar{A}_{i}\varepsilon } \\
C_{e} &=&\bar{C}_{i} \\
A_{e} &=&\bar{A}_{i}
\end{eqnarray*}%
Then, replacing these values in (\ref{dmcvr}) yields directly the relations
for the most likely path, or equivalently, the average path: 
\begin{eqnarray*}
\left( C-\bar{C}_{i}\right) &=&\left( C^{\prime }-\bar{C}_{i}\right) +\left(
C^{\prime }-\bar{C}_{i}\right)\!\!\left( \alpha +\beta \right) t \\
\left( K-K_{e}\right) &=&\left( K^{\prime }-K_{e}\right) -\alpha \left(
K^{\prime }-K_{e}\right) t-\left( C^{\prime }-\bar{C}_{i}\right) t \\
\lambda ^{2}\left( A-A^{\prime }\right) &=&-\frac{\frac{A+A^{\prime }}{2}-%
\bar{A}_{i}}{2}t
\end{eqnarray*}%
In the limit of small $t$, and using (\ref{cftdn}) and $\frac{K+K^{\prime }}{%
2}\rightarrow K$ leads to a differential equation for the average path:%
\begin{eqnarray}
\frac{d}{dt}\left( C\left( t\right) -\bar{C}_{e}\right) &=&\left( C\left(
t\right) -\bar{C}_{e}\right)\!\!\left( AF^{\prime }\left( K\left( t\right)
\right) +r_{c}-\delta \right)  \label{vrgpts} \\
\frac{d}{dt}\left( K\left( t\right) -K_{e}\right) &=&\left( AF^{\prime
}\left( K\left( t\right) \right) -\delta \right)\!\!\left( K\left( t\right)
-K_{e}\right) -\left( C\left( t\right) -\bar{C}_{i}\right)  \notag \\
\lambda ^{2}\frac{d\left( A-\bar{A}_{i}\right) }{dt} &=&-\frac{\left( A-\bar{%
A}_{i}\right) }{2}  \notag
\end{eqnarray}%
The above equations are in fact those of a simplified model of capital
accumulation: the standard approach is recovered in average. The first
equation is the usual Euler equation with interest rate. The second and
third equations describe the dynamics of capital and technology
respectively. The fixed point $\bar{A}_{i}$ depends both on the system and
the system's interactions, as seen in (\ref{gmm}). Linearizing the dynamics
around the fixed point leads to a usual saddle path dynamics, with two
eigenvalues: 
\begin{equation*}
\frac{1}{2}r_{c}-\delta \pm \frac{1}{2}\sqrt{r_{c}^{2}-4\bar{A}_{i}}+\bar{A}%
_{i}\varepsilon F^{\prime }K_{e}
\end{equation*}%
Usually, for $r_{c}=\delta $, the first eigenvalue is negative and the
second positive. When the system moves along the unstable equilibrium, the
linear approximation breaks down. Actually, for large values of $\left(
K\left( t\right) -K_{e}\right) $, marginal productivity falls below the
depreciation rate, and capital accumulation stops. We will not dwell on this
point since once the phases of the system have been found, and the average
dynamics equations have been written, interpretations are standard.\ 

\subsubsection{Corrections due to the interaction term}

We go on studying the individual dynamics in the background created by
system as a whole. To take into account the individual interactions and
their impact on (\ref{Trstngntx}), we have to turn back to the field
theoretic formulation and to find the modification of the transition
functions due to the quartic interaction term in (\ref{fldqn}): 
\begin{equation}
I=\frac{\gamma }{2}\int \Psi ^{\dag }\left( K_{1},C_{1},A_{1}\right) \Psi
^{\dag }\left( K_{2},C_{2},A_{2}\right)\!\!\left( A_{2}K_{1}+A_{1}K_{2}\right)
\Psi \left( K_{1},C_{1},A_{1}\right) \Psi \left( K_{2},C_{2},A_{2}\right)
\label{ntrtrm}
\end{equation}%
Appendix 6 computes this correction. We come back to the definition of the
probability transitions, and add the contribution of (\ref{ntrtrm}) that was
discarded while computing (\ref{Trstngntx}). We show that the Green function
is modified at the first order in $\gamma $ as:%
\begin{equation*}
\bar{G}\left( C,K,A,C^{\prime },K^{\prime },A^{\prime },t\right) =G\left(
C,K,A,C^{\prime },K^{\prime },A^{\prime },t\right) \times \exp \left(
-\gamma V\left( C,K,A,C^{\prime },K^{\prime },A^{\prime },t\right) \right)
\end{equation*}%
where the function $V\left( C,K,A,C^{\prime },K^{\prime },A^{\prime
},t\right) $ depends on the initial and final states:%
\begin{eqnarray*}
&&V\left( C,K,A,C^{\prime },K^{\prime },A^{\prime },t\right) =\left(
2t^{2}AK+\frac{t^{3}}{12}\left( A-A^{\prime }\right)\!\!\left( C-C^{\prime
}\right) +\frac{t^{3}}{2}\left( A-A^{\prime }\right)\!\!\left( K-K^{\prime
}\right) -\frac{1}{12}\bar{K}^{\varepsilon }t^{3}\left( A-A^{\prime }\right)
^{2}\right) \\
&&+A\left( \frac{1}{3}t^{3}\left( C-C^{\prime }\right) +t^{2}\left(
K-K^{\prime }\right) -\frac{1}{3}\bar{K}^{\varepsilon }t^{3}\left(
A-A^{\prime }\right) \right) +\gamma t^{2}\left( A-A^{\prime }\right) K
\end{eqnarray*}%
It can be shown that the trajectories correspond to an average path with
initial conditions $\left( C\left( 0\right) ,K\left( 0\right) ,A\left(
0\right) \right) $, and that for small interaction timepans, $\left( \dot{C}%
\left( 0\right) ,\dot{K}\left( 0\right) ,\dot{A}\left( 0\right) \right) $ is
modified at the first order in $\gamma $ (see Appendix 6). Defining $\delta
C\left( t\right) $, $\delta K\left( t\right) $\ and $\delta A\left( t\right) 
$\ the respective consumption, capital and technology deviations from their
average paths (\ref{vrgpts}) due to the interactions, we can write:

\begin{eqnarray}
\delta C\left( t\right) &=&0  \label{dvtns} \\
\delta K\left( t\right) &=&\gamma \left( \frac{7ct^{5}}{720\bar{A}^{2}}%
C\left( 0\right) +\frac{ct^{4}}{48\bar{A}^{2}}K\left( 0\right) +\left( \frac{%
bt^{3}}{6K^{\varepsilon }\bar{A}^{2}}+\frac{K^{\varepsilon }ct^{5}}{90}%
\right) A\left( 0\right) \right)  \notag \\
&&+\gamma \left( -\frac{7ct^{6}}{1440\bar{A}^{2}}\dot{C}\left( 0\right) +%
\frac{ct^{5}}{60\bar{A}^{2}}\dot{K}\left( 0\right) +\left( \frac{bt^{4}}{%
24K^{\varepsilon }\bar{A}^{2}}+\frac{3K^{\varepsilon }ct^{6}}{160\bar{A}^{2}}%
\right) \dot{A}\left( 0\right) \right)  \notag \\
\delta A\left( t\right) &=&\gamma \left( \frac{ct^{3}}{6K^{\varepsilon }\bar{%
A}^{2}}K\left( 0\right) \right) +\gamma \left( \frac{ct^{4}}{24}\dot{K}%
\left( 0\right) +t\dot{A}\left( 0\right) \right)  \notag
\end{eqnarray}%
where:%
\begin{eqnarray*}
b &=&2\left( \nu ^{2}+\frac{2\bar{K}^{2\varepsilon }}{\lambda ^{2}\alpha ^{2}%
}+\frac{3\varpi ^{2}}{2\left( 2\alpha +\beta \right) \beta }\right) \\
c &=&\frac{2}{\lambda ^{2}}
\end{eqnarray*}%
The interpretation of (\ref{dvtns}) is the following. From (\ref{dvtns}) we
find directly the impact of the agent's initial state on the deviations $%
\delta K\left( t\right) $\ and $\delta A\left( t\right) $:%
\begin{eqnarray}
\frac{\partial \left( \delta K\left( t\right) \right) }{\partial K\left(
0\right) } &=&\frac{ct^{4}}{48\bar{A}^{2}}>0  \label{sngy} \\
\text{ }\frac{\partial \left( \delta K\left( t\right) \right) }{\partial
A\left( 0\right) } &=&\frac{bt^{3}}{6K^{\varepsilon }\bar{A}^{2}}+\frac{%
K^{\varepsilon }ct^{5}}{90}>0  \notag \\
\frac{\partial \left( \delta A\left( t\right) \right) }{\partial K\left(
0\right) } &=&\frac{ct^{3}}{6K^{\varepsilon }\bar{A}^{2}}>0\text{ }  \notag
\end{eqnarray}%
the interaction between individual capital and technology stocks produces a
synergy effect, both stocks increase faster. This effect is proportional to
the initial values of the stocks. The higher these initial individual
values, the faster both stocks increase. Moreover, the polynomial time
dependency of the elasticities shows that the accumulation dynamics is
faster than a linear process. Remark that this synergy effect is not
contradictory with the eviction effect described in the non-trivial phase.
Actually, the results presented here are only valid at the individual level.
In other words, the individual dynamics in a given phase does not detect the
collective mechanisms of interactions. The latters are only detectable when
analyzing the phase, and are indeed hidden in apparently exogenous
parameters shaping the agents environment.

The accumulation dynamics is however dampened by fluctuations in technology
stocks, measured by $\bar{A}^{2}\lambda ^{2}$. The higher these
fluctuations, the slower the accumulation process. The initial direction of
the system, given by the terms proportional to $\dot{X}\left( 0\right) $
amplify this synergy effect. Actually, we can also compute from (\ref{dvtns}%
) the impact of the agent's initial momentum on the deviations 
$\delta
K\left( t\right) $\ and $\delta A\left( t\right) $:%
\begin{eqnarray*}
\frac{\partial \left( \delta K\left( t\right) \right) }{\partial \dot{K}%
\left( 0\right) } &=&\gamma \frac{ct^{5}}{60\bar{A}^{2}}>0 \\
\text{ }\frac{\partial \left( \delta K\left( t\right) \right) }{\partial 
\dot{A}\left( 0\right) } &=&\gamma \frac{bt^{4}}{24K^{\varepsilon }\bar{A}%
^{2}}+\frac{3K^{\varepsilon }ct^{6}}{160\bar{A}^{2}}>0 \\
\frac{\partial \left( \delta A\left( t\right) \right) }{\partial \dot{A}%
\left( 0\right) } &=&\gamma t>0 \\
\frac{\partial \left( \delta A\left( t\right) \right) }{\partial \dot{K}%
\left( 0\right) } &=&\gamma \frac{ct^{4}}{24}>0\text{ }
\end{eqnarray*}%
a system that had started initially to accumulate both capital and
technology stocks will accelerate faster compared to a system that was at
first in a constant equilibrium.

The effect of the consumption initial value is ambiguous. Technology
improvement, measured by the dynamics of $A$, increases productivity, and
rates of return. It is thus optimal for agents to increase their savings and
reduce their consumption. Capital stock is positively correlated to $C\left(
0\right) $,\ as shown by its elasticity with respect to $C\left( 0\right) $:%
\begin{equation*}
\frac{\partial \left( \delta K\left( t\right) \right) }{\partial C\left(
0\right) }=\gamma \frac{7ct^{5}}{720\bar{A}^{2}}>0
\end{equation*}%
In other word, a high level of initial consumption is an indicator of
wealth. The agents interaction induce an accumulation process that favors
capital expenditures. Since consumption elasticity with respect to
consumption initial value $\frac{\partial \left( \delta C\left( t\right)
\right) }{\partial C\left( 0\right) }$ is null, any increase of wealth is
spent on capital stock\textbf{.}

Besides, any initial increase in the consumption rate impairs capital
accumulation, since: 
\begin{equation*}
\frac{\partial \left( \delta K\left( t\right) \right) }{\partial \dot{C}%
\left( 0\right) }=-\gamma \frac{7ct^{6}}{1440\bar{A}^{2}}<0
\end{equation*}%
An initial increase in consumption will be smoothed over the entire
timespan, and will eventually dampen the accumulation process.

\subsubsection{Two agents transition functions and interaction}

The field formalism presented here also allows the study of interaction
between individual agents. Consider the simplest example of a two agents
dynamics. Discarding interactions, the probability of transition between an
initial state:%
\begin{equation*}
\left( \left( K_{1},C_{1},A_{1}\right) _{i},\left( K_{2},C_{2},A_{2}\right)
_{i}\right)
\end{equation*}%
and a final state:%
\begin{equation*}
\left( \left( K_{1},C_{1},A_{1}\right) _{f},\left( K_{2},C_{2},A_{2}\right)
_{f}\right)
\end{equation*}%
is simply the product of the transition probabilities (\ref{Trstngntx}) for
each agent: 
\begin{eqnarray*}
&&G\left( \left( K_{1},C_{1},A_{1}\right) _{i},\left(
K_{2},C_{2},A_{2}\right) _{i},\left( K_{1},C_{1},A_{1}\right) _{f},\left(
K_{2},C_{2},A_{2}\right) _{f}\right) \\
&\equiv &G\left( \left( K_{1},C_{1},A_{1}\right) _{i},\left(
K_{2},C_{2},A_{2}\right) _{i},t\right) G\left( \left(
K_{1},C_{1},A_{1}\right) _{f},\left( K_{2},C_{2},A_{2}\right) _{f},t\right)
\end{eqnarray*}%
since they are considered to be independent. The global interaction effect,
the impact of the entire system on each agent is included in the phase of
the system, through the parameters of the transition probabilities.

When the interaction term is included,\textbf{\ }the transition probability
has to be modified. In Appendix 6 compute the correction for the transition
probability for two agents. We consider an initial state $\left( \left(
K_{1},C_{1},A_{1}\right) _{i},\left( K_{2},C_{2},A_{2}\right) _{i}\right) $
and a final state $\left( \left( K_{1},C_{1},A_{1}\right) _{f},\left(
K_{2},C_{2},A_{2}\right) _{f}\right) $.

In presence of the interaction term, appendix shows that:%
\begin{equation*}
I=\frac{\gamma }{2}\int \Psi ^{\dag }\left( K_{1},C_{1},A_{1}\right) \Psi
^{\dag }\left( K_{2},C_{2},A_{2}\right)\!\!\left( A_{2}K_{1}+A_{1}K_{2}\right)
\Psi \left( K_{1},C_{1},A_{1}\right) \Psi \left( K_{2},C_{2},A_{2}\right)
\end{equation*}%
the transition probability is modified in the following way: 
\begin{eqnarray*}
&&G_{I}\left( \left( K_{1},C_{1},A_{1},t\right) _{i},\left(
K_{2},C_{2},A_{2},t\right) _{i},\left( K_{1},C_{1},A_{1},t\right)
_{f},\left( K_{2},C_{2},A_{2},t\right) _{f}\right) \\
&\equiv &G\left( \left( K_{1},C_{1},A_{1}\right) _{i},\left(
K_{2},C_{2},A_{2}\right) _{i},\left( K_{1},C_{1},A_{1}\right) _{f},\left(
K_{2},C_{2},A_{2}\right) _{f}\right) \exp \left( -V_{I}\right)
\end{eqnarray*}%
where:%
\begin{equation*}
V_{I}=\gamma t^{2}\left( \left\langle A_{1}\right\rangle \left\langle
K_{2}\right\rangle +\left\langle K_{1}\right\rangle \left\langle
A_{2}\right\rangle \right) +\frac{\gamma t^{3}}{24}\left( \left\langle
A_{1}\right\rangle \Delta C_{2}+\Delta C_{1}\left\langle A_{2}\right\rangle
-K^{\varepsilon }\left( \left\langle A_{1}\right\rangle \Delta A_{2}+\Delta
A_{1}\left\langle A_{2}\right\rangle \right) \right)
\end{equation*}%
with:%
\begin{eqnarray*}
\left\langle X_{j}\right\rangle &=&\frac{\left( X_{j}\right) _{i}+\left(
X_{j}\right) _{f}}{2} \\
\Delta X_{j} &=&\left( X_{j}\right) _{f}-\left( X_{j}\right) _{i}
\end{eqnarray*}%
for any variable $X=C$, $K$, or $A$ and agent $j=1$ or $2$. The quantity $%
\left\langle X_{j}\right\rangle $ computes the average value of $X$ for
agent $j$ along the path, and $\Delta X_{j}$, the variation of $X$ along
this path.

Due to the interaction, the two agents' transition probabilities are now
entangled. The average trajectory for one agent is modified by the other
agent's path (see Appendix 6 ). We write $\delta X_{2\rightarrow 1}\left(
t\right) $ the correction of agent $1$'s trajectory due to agent $2$ for $%
X=C $, $K$, or $A$, and $\delta X_{1\rightarrow 2}\left( t\right) $ the
correction of agent $2$'s trajectory due to agent $1$. Appendix 6 shows that:

\begin{eqnarray}
\delta K_{2\rightarrow 1}\left( t\right) &=&\gamma bt\left\langle
A_{2}\right\rangle  \label{ntrctnfrpt} \\
\delta A_{2\rightarrow 1}\left( t\right) &=&\gamma \left( \frac{1}{6}c\frac{%
\Delta C_{2}}{2}-\frac{1}{6}cK^{\varepsilon }\frac{\Delta A_{2}}{2}\right)
t^{3}+\gamma ct\left\langle K_{2}\right\rangle  \notag \\
\delta K_{1\rightarrow 2}\left( t\right) &=&\gamma bt\left\langle
A_{1}\right\rangle  \notag \\
\delta A_{1\rightarrow 2}\left( t\right) &=&\gamma \left( \frac{1}{6}c\frac{%
\Delta C_{1}}{2}-\frac{1}{6}cK^{\varepsilon }\frac{\Delta A_{1}}{2}\right)
t^{3}+\gamma ct\left\langle K_{1}\right\rangle  \notag
\end{eqnarray}%
with:%
\begin{eqnarray*}
\bar{X}_{i} &=&\frac{X_{i}\left( 0\right) +X_{i}\left( t\right) }{2} \\
\frac{\Delta X_{i}}{2} &=&\frac{X_{i}\left( t\right) -X_{i}\left( 0\right) }{%
2}
\end{eqnarray*}%
Formula (\ref{ntrctnfrpt}) allows to find the dependency of an agent
behavior on other agent's path. The elasticities are:%
\begin{equation*}
\frac{\partial \left( \delta K_{1}\left( t\right) \right) }{\partial \bar{A}%
_{2}\left( 0\right) }=\frac{\partial \left( \delta A_{1}\left( t\right)
\right) }{\partial \bar{K}_{2}\left( t\right) }=t>0\text{ (and other
elasticities with respect to }\bar{X}_{2}\text{ are null)}
\end{equation*}%
which means that the average technology of agent $2$ impacts positively the
accumulation of capital for agent $1$, and that the accumulated stock of
agent $2$ accelerates the technology improvement for agent $1$. Agent $2$
participates to the environment of agent $1$, and both its capital and
technology stocks influence the other agents.

These elasticities are proportional to the interaction timespan: the longer
agents interact, the higher the final stocks. The elasticities with respect
to the initial direction of agent's $2$ path may seem counterintuitive.%
\begin{eqnarray*}
\frac{\partial \left( \delta A_{1}\left( t\right) \right) }{\partial \Delta
A_{2}\left( 0\right) } &=&-\frac{1}{6}cK^{\varepsilon }t^{3}<0 \\
\frac{\partial \left( \delta A_{1}\left( t\right) \right) }{\partial
C_{2}\left( t\right) } &=&\frac{1}{6}ct^{3}>0
\end{eqnarray*}%
Technology stock is negatively correlated to other agents' accumulation
rate. This is the consequence of the acceleration of accumulation process
for both agents and our choice of representation of a path as function of
the average value of the path $\bar{X}_{2}$: since the dynamic follows an
accelerating pattern, its representative curve is below the average $\bar{X}%
_{2}$ most of the time. As a consequence, the accumulated stock are below
the linear approximation in $\bar{X}_{2}$. The term proportional to \ $\frac{%
\Delta X_{2}}{2}$ is thus a correction to this linear approximation.

\subsection{Synthesis and Discussion}

The application of our formalism to a basic Business Cycle model has shown
the implications of introducing multiple agents interacting through
technology and capital stocks. This has allowed us to inspect setups not
accessible to the usual representative agents models.\ It has also allowed
to detect collective effects due to large number of agents, such as the
appearance of multiple phases or macro equilibria. In turn, we have seen how
these phases impact the individual agents' dynamics. These individual
dynamics are formally identical to those used in representative agents
models.\ It is at this point however that some major major differences with
standard models appear:

In our formalism, the individual agents' dynamics are derived from a
collective background.\ They emerge from the model, but cannot be imposed as
a defining point of the model. This translates into several features.

First, the individual dynamics parameters are not exogenous. As mentioned
above, they depend on the global system. A change of phase in the entire
system induces a structural break that actually modifies the parameters of
the agent's dynamic equations.

Second, the fact that individual behaviors emerge from the system extends to
any subset of agents. Their dynamics and interactions, too, can and should
be deduced from the collective background (see section 3.2.4). This allows a
straightforward and detailed analysis of agents' interactions, while
preserving the agents' heterogeneity.

As a result, and third, individual agents cannot be considered as
representative agents. Section (3.2.3) demonstrates that individual features
do not aggregate to produce similar effects at the macro level. The synergy
effect in equation (\ref{sngy}) shows that an agent may experience a
virtuous circle between his capital and technology, even in the non-trivial
phase characterized at a macro level by an eviction effect and a lower
production. These two macro features are present at the individual level,
but only as a hidden externality that shapes the agent's environment through
seemingly exogenous parameters.

To conclude, the representative agent paradigm cannot detect some macro
features from the description of particular agents. Some conclusions at the
individual level do not agregate.

\section{Conclusion}

This paper has presented an analytical treatment of economic systems with an
arbitrary number of agents that keeps track of the systems' interactions and
agents' complexity. As significant results, we have shown that a field
theory formalism may reveal some emerging equilibria, and studied the
influence of these equilibria on the agent's individual dynamics. This
method can be applied to various economic models.

In this paper we have, for the sake of clarity, deliberately set aside some
matters developped in (Gosselin, Lotz, Wambst 2017), such as strategic
behaviors and heterogeneity among agents, in information, goals, or actions.
However, our formalism extends to such cases.\ Social interactions and
economic networks could also be included. These subjects are under current
research.

Ultimately, our formalism should shed some lights on the matter of
aggregation. Indeed, despite the fact that field theory does not deal with
aggregates, it allows to recover macroeconomic quantities through averages.
A natural question would be to find even partial relations between these
macroeconomic quantities.

\pagebreak
\section*{References}

\begin{description}

\item
 Abergel F, Chakraborti
A, Muni Toke I and Patriarca M (2011) Econophysics review: I. Empirical
facts, Quantitative Finance, Vol. 11, No. 7, 991-1012.

\item
Abergel F,
Chakraborti A, Muni Toke I and Patriarca M (2011) Econophysics review: II.
Agent-based models, Quantitative Finance, Vol. 11, No. 7, 1013-1041.

\item
 Lucas, Robert (1976) "Econometric Policy Evaluation:
A Critique". In Brunner, K.; Meltzer, A. The Phillips Curve and Labor
Markets. Carnegie-Rochester Conference Series on Public Policy. 1. New York:
American Elsevier. pp. 19--46. ISBN 0-444-11007-0.

\item
Lotz A (2011) An Economic Approach to the Self: the
Dual Agent, Preprint, 2011.\\
https://mpra.ub.uni-muenchen.de/50771/1/MPRA\_paper\_50771.pdf

\item
 Gosselin P and Lotz A (2012) A dynamic model
of interactions between conscious and unconscious, Preprint, 2012.
https://mpra.ub.uni-muenchen.de/36697/1/MPRA\_paper\_36697.pdf.

\item
Gosselin P, Lotz A, and Wambst M.
(2013) On Apparent Irrational Behavior : Interacting Structures and the
Mind, Preprint, 2013. https://hal.archives-ouvertes.fr/hal-00851309/document

\item
Gosselin P, Lotz A, and Wambst M
(2015) From Rationality to Irrationality : Dynamic Interacting Structures,
IF PREPUB. 2015. https://hal.archives-ouvertes.fr/hal-01122078/document

\item
Gosselin P, Lotz A and Wambst M (2017)
A Path Integral Approach to Interacting Economic Systems with Multiple
Heterogeneous Agents. IF\_PREPUB. 2017. hal-01549586v2

\item
Kleinert H (1989) Gauge fields in condensed
matter Vol. I , Superflow and vortex lines, Disorder Fields, Phase
Transitions, Vol. II, Stresses and defects, Differential Geometry, Crystal
Melting, World Scientific, Singapore 1989.

\item
 Gaffard J-L and Napoletano M Editors:
Agent-based models and economic policy. Ofce 2012.

\item
Jackson M (2010) Social and Economic Networks,
Princeton University Press 2010.

\item
Kleinert H (2009) Path Integrals in Quantum
Mechanics, Statistics, Polymer Physics, and Financial Markets 5th edition,
World Scientific, Singapore 2009.

\item
Hamilton J. D. (1994) Time series analysis.
Princeton University Press, 1994.

\item
 Zinn-Justin J (1993) Quantum Field Theory and
Critical Phenomena, 2nd edition, Oxford Science Publications, 1993.

\item
Peskin ME, Schroeder DV (1995), An
introduction to Quantum Field Theory. Addison-Wesley Publishing Company 1995.

\item
 Romer, David. Advanced Macroeconomics. New York:
McGraw-Hill Companies, 1996

\item
Maurice Obstfeld \& Kenneth S. Rogoff,
1996. "Foundations of International Macroeconomics," MIT Press Books, The
MIT Press.
\end{description}
\pagebreak

\section*{Appendix 1}

We show the derivation of (\ref{glblwghttxx}). We will remove the agent's
index $i$ since the argument deals with one individual agent.

If $Y\left( t+n\right) $ is{\LARGE \ }centered on $\bar{Y}$ with variance $%
\sigma ^{2}$, $\sum_{n>0}Y\left( t+n\right) $ will be centered on $\bar{Y}$
with variance $T\sigma ^{2}$, and the integration over $Y\left( t+n\right) $
yields:%
\begin{eqnarray*}
&&\int \dprod dY\left( t+n\right) \exp \left( -\frac{1}{T}\left( B\left(
t+1\right) +\sum_{n>0}\left( Y\left( t+n\right) -\bar{C}\right) \right) ^{2}-%
\frac{1}{\sigma ^{2}}\sum_{n=1}^{T}\left( Y\left( t+n\right) -\bar{Y}\right)
^{2}\right) \\
&=&\int \dprod dY^{\prime }\left( t+n\right) \exp \left( -\frac{1}{T}\left(
B\left( t+1\right) +\sum_{n=1}^{T}\left( Y^{\prime }\left( t+n\right)
-\left( \bar{C}-\bar{Y}\right) \right) \right) ^{2}-\frac{1}{\sigma ^{2}}%
\sum_{n=1}^{T}\left( Y^{\prime }\left( t+n\right) \right) ^{2}\right)
\end{eqnarray*}%
with $Y^{\prime }\left( t+n\right) =Y\left( t+n\right) -\bar{Y}$. The
exponential rewrites:%
\begin{eqnarray*}
&&\exp \left( -\frac{1}{T}\left( B\left( t+1\right) +\sum_{n=1}^{T}\left(
Y^{\prime }\left( t+n\right) -\left( \bar{C}-\bar{Y}\right) \right) \right)
^{2}-\frac{1}{\sigma ^{2}}\sum_{n=1}^{T}\left( Y^{\prime }\left( t+n\right)
\right) ^{2}\right) \\
&=&\exp \left( -\frac{\left( B\left( t+1\right) -T\left( \bar{C}-\bar{Y}%
\right) \right) ^{2}}{T}-\frac{2\left( B\left( t+1\right) -T\left( \bar{C}-%
\bar{Y}\right) \right) }{T}\sum_{n=1}^{T}Y^{\prime }\left( t+n\right)
-\left( \frac{1}{\sigma ^{2}}+\frac{1}{T}\right) \sum_{n=1}^{T}\left(
Y^{\prime }\left( t+n\right) \right) ^{2}\right)
\end{eqnarray*}%
and the integration over the $Y^{\prime }\left( t+n\right) $ leads to a
weight: 
\begin{eqnarray}
&&\exp \left( -\frac{\left( B\left( t+1\right) -T\left( \bar{C}-\bar{Y}%
\right) \right) ^{2}}{T}-\frac{2\left( B\left( t+1\right) -T\left( \bar{C}-%
\bar{Y}\right) \right) }{T}\sum_{n=1}^{T}Y^{\prime }\left( t+n\right)
-\left( \frac{1}{\sigma ^{2}}+\frac{1}{T}\right) \sum_{n=1}^{T}\left(
Y^{\prime }\left( t+n\right) \right) ^{2}\right)  \notag \\
&=&\exp \left( -\frac{\left( B\left( t+1\right) -T\left( \bar{C}-\bar{Y}%
\right) \right) ^{2}}{T}+\frac{\sigma ^{2}\left( B\left( t+1\right) -T\left( 
\bar{C}-\bar{Y}\right) \right) ^{2}}{T\left( \sigma ^{2}+T\right) }\right) 
\notag \\
&=&\exp \left( -\frac{1}{T+\sigma ^{2}}\left( B\left( t+1\right) -T\left( 
\bar{C}-\bar{Y}\right) \right) ^{2}\right)  \label{gblppdx}
\end{eqnarray}%
We can now write $B\left( t+1\right) $\ as a function of past variables: 
\begin{equation}
B\left( t+1\right) =\sum_{n\leqslant 0}Y\left( t+n\right) -\sum_{n\leqslant
0}C\left( t+n\right)  \label{pstcrst}
\end{equation}%
and, along with the expression{\LARGE \ }$B_{t}+Y\left( t\right) -B\left(
t+1\right) -\bar{C}=C\left( t\right) -\bar{C}$, write the global weight (\ref%
{gblppdx}) as:

\begin{eqnarray}
&&\exp \left( -\left( C\left( t\right) -\bar{C}\right) ^{2}-\frac{1}{%
T+\sigma ^{2}}\left( \sum_{n\leqslant 0}Y\left( t+n\right) -\sum_{n\leqslant
0}C\left( t+n\right) -T\left( \bar{C}-\bar{Y}\right) \right) ^{2}\right)
\label{wgtqd} \\
&\simeq &\exp \left( -\left( C\left( t\right) -\bar{C}\right) ^{2}-\frac{1}{T%
}\left( \sum_{n\leqslant 0}Y\left( t+n\right) -\sum_{n\leqslant 0}C\left(
t+n\right) -T\left( \bar{C}-\bar{Y}\right) \right) ^{2}\right)  \notag
\end{eqnarray}%
for a large enough time scale, so that $T\gg \sigma ^{2}$. Recall that terms
in the exponential depend only on past variables, and do not modify the
statistical weight. This statistical weigh can thus be written:%
\begin{eqnarray*}
&&\exp \left( -\left( \frac{T+1}{T}\right)\!\!\left( C\left( t\right) -\frac{T}{%
T+1}\bar{C}-\frac{1}{T+1}\left( \sum_{n\leqslant 0}Y\left( t+n\right)
-\sum_{n<0}C\left( t+n\right) -T\left( \bar{C}-\bar{Y}\right) \right)
\right) ^{2}\right) \\
&=&\exp \left( -\left( \frac{T+1}{T}\right)\!\!\left( C\left( t\right) -\frac{1%
}{T+1}\left( \sum_{n\leqslant 0}Y\left( t+n\right) -\sum_{n<0}C\left(
t+n\right) +T\bar{Y}\right) \right) ^{2}\right) \\
&=&\exp \left( -\left( \frac{T+1}{T}\right)\!\!\left( C\left( t\right) -\bar{Y}-%
\frac{1}{T+1}\left( \sum_{n\leqslant 0}\left( Y\left( t+n\right) -\bar{Y}%
\right) -\sum_{n<0}\left( C\left( t+n\right) -\bar{Y}\right) \right) \right)
^{2}\right)
\end{eqnarray*}%
For $T\gg 1$, the last expression reduces to: 
\begin{equation*}
\exp \left( -\left( C\left( t\right) -\bar{Y}-\frac{1}{T}\left(
\sum_{n\leqslant 0}\left( Y\left( t+n\right) -\bar{Y}\right)
-\sum_{n<0}\left( C\left( t+n\right) -\bar{Y}\right) \right) \right)
^{2}\right)
\end{equation*}%
and defining $\hat{C}\left( t\right) =C\left( t\right) -\bar{Y}$, we are
left with: 
\begin{equation}
\exp \left( -\left( \hat{C}\left( t\right) -\frac{1}{T}\left(
\sum_{n\leqslant 0}\hat{Y}\left( t+n\right) -\sum_{n<0}\hat{C}\left(
t+n\right) \right) \right) ^{2}\right)  \label{prbc}
\end{equation}%
To define the global statistical weight for the system over all periods, we
can now sum the exponentiated terms in (\ref{prbc}) to obtain the weight:%
\begin{equation*}
\exp \left( -\sum_{t=1}^{T}\left( \hat{C}\left( t\right) -\frac{1}{T}\left(
\sum_{n\leqslant 0}\hat{Y}\left( t+n\right) -\sum_{n<0}\hat{C}\left(
t+n\right) \right) \right) ^{2}\right)
\end{equation*}%
This weight describes the variables:%
\begin{equation*}
X\left( t\right) =\hat{C}\left( t\right) -\frac{1}{T}\left( \sum_{n\leqslant
0}\hat{Y}\left( t+n\right) -\sum_{n<0}\hat{C}\left( t+n\right) \right)
\end{equation*}%
as gaussian and independent. It can now be written differently by remarking
that:%
\begin{eqnarray*}
&&X\left( t\right) -X\left( t+1\right) \\
&=&\hat{C}\left( t\right) -\frac{1}{T}\left( \sum_{n\leqslant 0}\hat{Y}%
\left( t+n\right) -\sum_{n<0}\hat{C}\left( t+n\right) \right) -\left( \hat{C}%
\left( t+1\right) -\frac{1}{T}\left( \sum_{n\leqslant 0}\hat{Y}\left(
t+1+n\right) -\sum_{n<0}\hat{C}\left( t+1+n\right) \right) \right) \\
&=&\hat{C}\left( t\right) -\hat{C}\left( t+1\right) +\frac{1}{T}\left( \hat{Y%
}\left( t+1\right) -\hat{C}\left( t\right) \right) \\
&=&\frac{T-1}{T}\hat{C}\left( t\right) -\hat{C}\left( t+1\right) +\frac{1}{T}%
\hat{Y}\left( t+1\right)
\end{eqnarray*}%
It then allows computing the density probability of: 
\begin{equation*}
\frac{T-1}{T}\hat{C}\left( t\right) -\hat{C}\left( t+1\right) +\frac{1}{T}%
\hat{Y}\left( t+1\right)
\end{equation*}%
by writing:%
\begin{eqnarray*}
&&\int \exp \left( -X^{2}\left( t\right) -X^{2}\left( t+1\right) \right) \\
&&\hspace{3cm}\times \delta \left( X\left( t\right) -X\left( t+1\right)
-\left( \frac{T-1}{T}\hat{C}\left( t\right) -\hat{C}\left( t+1\right) +\frac{%
1}{T}\hat{Y}\left( t+1\right) \right) \right) dX\left( t\right) dX\left(
t+1\right) \\
&=&\int \exp \left( -X^{2}\left( t\right) -\left( X\left( t\right) -\left( 
\frac{T-1}{T}\hat{C}\left( t\right) -\hat{C}\left( t+1\right) +\frac{1}{T}%
\hat{Y}\left( t+1\right) \right) \right) ^{2}\right) dX\left( t\right) \\
&=&\exp \left( -\frac{\left( \frac{T-1}{T}\hat{C}\left( t\right) -\hat{C}%
\left( t+1\right) +\frac{1}{T}\hat{Y}\left( t+1\right) \right) ^{2}}{2}%
\right)
\end{eqnarray*}%
For $T$ large, $\frac{T-1}{T}\hat{C}\left( t\right) \simeq \hat{C}\left(
t\right) $. This describes a brownian type process for $\hat{C}\left(
t\right) $, with variance $2$. This brownian motion is constrained to $%
X_{T}=0$ through the constraint:%
\begin{equation*}
X_{T}=\hat{C}\left( T\right) -\left( \sum_{n\leqslant 0}\hat{Y}\left(
T+n\right) -\sum_{n<0}\hat{C}\left( T+n\right) \right) =\sum_{n\leqslant 0}%
\hat{Y}\left( T+n\right) -\sum_{n\leqslant 0}\hat{C}\left( T+n\right) =0
\end{equation*}%
so that the overall weight becomes, in the $\hat{C}\left( t\right) $
representation and in first approximation for $T$ large:%
\begin{eqnarray}
&&\exp \left( -\sum \left( \hat{C}\left( t\right) -\hat{C}\left( t+1\right) +%
\frac{1}{T}\hat{Y}\left( t+1\right) \right) ^{2}-\left( \sum_{n\leqslant 0}%
\hat{Y}\left( T+n\right) -\sum_{n\leqslant 0}\hat{C}\left( T+n\right)
\right) \right)  \notag \\
&&\times \delta \left( \sum_{n\leqslant 0}\hat{Y}\left( T+n\right)
-\sum_{n\leqslant 0}\hat{C}\left( T+n\right) \right)  \notag \\
&\simeq &\exp \left( -\sum^{2}\left( \hat{C}\left( t\right) -\hat{C}\left(
t+1\right) +\frac{1}{T}\hat{Y}\left( t+1\right) \right) -\frac{\left(
\sum_{n\leqslant 0}\hat{Y}\left( T+n\right) -\sum_{n\leqslant 0}\hat{C}%
\left( T+n\right) \right) ^{2}}{\sigma ^{2}}\right)  \notag \\
&\simeq &\exp \left( -\sum \left( \hat{C}\left( t\right) -\hat{C}\left(
t+1\right) \right) ^{2}-\frac{\left( \sum_{n\leqslant 0}\hat{Y}\left(
T+n\right) -\sum_{n\leqslant 0}\hat{C}\left( T+n\right) \right) ^{2}}{\sigma
^{2}}\right)  \label{glblwght}
\end{eqnarray}%
with $\sigma ^{2}\ll 1$. The second line allows for small deviations from the
overall constraint.

\bigskip Formula (\ref{glblwght}) is straightforward to generalize for
agents with varying horizon. Actually, if the time horizon at time $t$ is $%
T-t$, the statistical weight (\ref{prbc}) at time $t$ becomes: 
\begin{equation}
\exp \left( -\left( \hat{C}\left( t\right) -\frac{1}{T-t}\left(
\sum_{n\leqslant 0}\hat{Y}\left( t+n\right) -\sum_{n<0}\hat{C}\left(
t+n\right) \right) \right) ^{2}\right)  \label{prtwg}
\end{equation}%
As before, defining the variables 
\begin{equation*}
X\left( t\right) =\left( \hat{C}\left( t\right) -\frac{1}{T-t}\left(
\sum_{n\leqslant 0}\hat{Y}\left( t+n\right) -\sum_{n<0}\hat{C}\left(
t+n\right) \right) \right)
\end{equation*}%
The expression (\ref{prtwg}) shows that the $X\left( t\right) $ are gaussian
and independent. Computing $X\left( t\right) -\left( \frac{T-t-1}{T-t}%
\right) X\left( t+1\right) $ yields:%
\begin{eqnarray*}
&&X\left( t\right) -\left( \frac{T-t-1}{T-t}\right) X\left( t+1\right) \\
&=&\hat{C}\left( t\right) -\frac{1}{T-t}\left( \sum_{n\leqslant 0}\hat{Y}%
\left( t+n\right) -\sum_{n<0}\hat{C}\left( t+n\right) \right) \\
&&-\left( \frac{T-t-1}{T-t}\right)\!\!\left( \hat{C}\left( t+1\right) -\frac{1}{%
T-t-1}\left( \sum_{n\leqslant 0}\hat{Y}\left( t+1+n\right) -\sum_{n<0}\hat{C}%
\left( t+1+n\right) \right) \right) \\
&=&\hat{C}\left( t\right) -\left( \frac{T-t-1}{T-t}\right) \hat{C}\left(
t+1\right) +\frac{1}{T-t}\left( \hat{Y}\left( t+1\right) -\hat{C}\left(
t\right) \right) \\
&=&\left( \frac{T-t-1}{T-t}\right)\!\!\left( \hat{C}\left( t\right) -\hat{C}%
\left( t+1\right) +\frac{1}{T-t}\hat{Y}\left( t+1\right) \right)
\end{eqnarray*}%
which implies that the probability density for $\hat{C}\left( t\right) -\hat{%
C}\left( t+1\right) +\frac{1}{T-t}\hat{Y}\left( t+1\right) $ can be computed
by:%
\begin{eqnarray*}
&&\int \exp \left( -X^{2}\left( t\right) -X^{2}\left( t+1\right) \right) \\
&&\times \delta \left( \left( \frac{T-t}{T-t-1}\right) X_{t}-X\left(
t+1\right) -\left( \hat{C}\left( t\right) -\hat{C}\left( t+1\right) +\frac{1%
}{T-t}\hat{Y}\left( t+1\right) \right) \right) dX_{t}dX\left( t+1\right) \\
&=&\int \exp \left( -X_{t}^{2}-\left( \frac{T-t}{T-t-1}\right) ^{2}\left(
X_{t}-\left( \frac{T-t-1}{T-t}\right)\!\!\left( \hat{C}\left( t\right) -\hat{C}%
\left( t+1\right) +\frac{1}{T-t}\hat{Y}\left( t+1\right) \right) \right)
^{2}\right) dX_{t} \\
&=&\exp \left( -\frac{\left( \hat{C}\left( t\right) -\hat{C}\left(
t+1\right) +\frac{1}{T-t}\hat{Y}\left( t+1\right) \right) ^{2}}{\left( \frac{%
T-t}{T-t-1}\right) ^{2}\left( 1+\left( \frac{T-t}{T-t-1}\right) ^{2}\right) }%
\right)
\end{eqnarray*}

\bigskip This describes a brownian type process for $\hat{C}\left( t\right) $
of variance $\left( \frac{T-t}{T-t-1}\right) ^{2}\left( 1+\left( \frac{T-t}{%
T-t-1}\right) ^{2}\right) $. For $T$ large, we have $\left( \frac{T-t}{T-t-1}\right)
^{2}\left( 1+\left( \frac{T-t}{T-t-1}\right) ^{2}\right) \simeq 2$ and one
recovers the brownian motion, for $t\ll T$. This brownian motion is
constrained to $X_{T}=0$ through the constraint:%
\begin{equation*}
X\left( T\right) =\hat{C}\left( T\right) -\left( \sum_{n\leqslant 0}\hat{Y}%
\left( T+n\right) -\sum_{n<0}\hat{C}\left( T+n\right) \right)
=\sum_{n\leqslant 0}\hat{Y}\left( T+n\right) -\sum_{n\leqslant 0}\hat{C}%
\left( T+n\right) =0
\end{equation*}%
so that the overall weight in the $\hat{C}\left( t\right) $ representation
becomes in first approximation for $T$ large:%
\begin{eqnarray}
&&\exp \left( -\sum \left( \hat{C}\left( t\right) -\hat{C}\left( t+1\right) +%
\frac{\hat{Y}\left( t+1\right) }{T-t}\right) ^{2}-\left( \sum_{n\leqslant 0}%
\hat{Y}\left( T+n\right) -\sum_{n\leqslant 0}\hat{C}\left( T+n\right)
\right) \right)  \notag \\
&&\times \delta \left( \left( \sum_{n\leqslant 0}\hat{Y}\left( T+n\right)
-\sum_{n\leqslant 0}\hat{C}\left( T+n\right) \right) \right)  \notag \\
&\simeq &\exp \left( -\sum \left( \hat{C}\left( t\right) -\hat{C}\left(
t+1\right) +\frac{\hat{Y}\left( t+1\right) }{T-t}\right) ^{2}-\frac{\left(
\sum_{n\leqslant 0}\hat{Y}\left( T+n\right) -\sum_{n\leqslant 0}\hat{C}%
\left( T+n\right) \right) ^{2}}{\bar{\sigma}^{2}}\right)  \label{glblwght2}
\end{eqnarray}%
The second line allows for small deviations from the overall constraint.
Under a strictly binding constraint, $\bar{\sigma}^{2}\ll 1$.

In the continuous version, we can replace the sum over $t$ by an integral.
Formulas (\ref{glblwght}) and (\ref{glblwght2}) become:%
\begin{eqnarray*}
\exp \left( U^{eff}\right) &\equiv &\exp \left( \int dtU^{eff}\left( \hat{C}%
\left( t\right) \right) \right) \\
&=&\exp \left( -\int dt\left( \frac{d}{dt}\hat{C}\left( t\right) +\frac{1}{%
T-t}\hat{Y}_{t+1}\right) ^{2}-\frac{\left( \int dt\hat{Y}_{t}-\int dt\hat{C}%
\left( t\right) \right) ^{2}}{\bar{\sigma}^{2}}\right)
\end{eqnarray*}%
The first term means (discarding the constraint) that $\frac{d}{dt}\hat{C}%
\left( t\right) +\frac{1}{T-t}\hat{Y}\left( t+1\right) $ is gaussian of
variance $1$. Thus, if $\hat{Y}\left( t+1\right) $ is considered purely
random, one can consider that $\frac{d}{dt}\hat{C}\left( t\right) $ is
gaussian with variance $1+Var\left( \frac{1}{T-t}\hat{Y}\left( t+1\right)
\right) =\frac{1}{\beta }>1$ (for $T$ large, we can consider $\beta $ as
constant) and replace the first term of $U^{eff}$ by $\beta \left( \frac{d}{%
dt}\hat{C}\left( t\right) \right) ^{2}$ to obtain:%
\begin{equation}
\exp U^{eff}=\exp \left( -\beta \int dt\left( \frac{d}{dt}\hat{C}\left(
t\right) \right) ^{2}-\frac{\left( \int dt\hat{Y}\left( t\right) -\int dt%
\hat{C}\left( t\right) \right) ^{2}}{\bar{\sigma}^{2}}\right)
\label{glblwght3}
\end{equation}

As a consequence of (\ref{glblwght}), (\ref{glblwght2}) and (\ref{glblwght3}%
), the introduction of a constraint is equivalent to the introduction of
non-local interaction terms. The non-local terms may, in some cases, be
approximated by some terms in the derivatives of \ $C\left(t\right) $ (see
Gosselin, Lotz Wambst 2018).

\section*{Appendix 2}

We consider some constraints within the context of non quadratic utilities.\
To do so, we start with a simple example and consider the budgent constraint
(\ref{BC}) for a single agent:%
\begin{equation}
C\left( t\right) =B\left( t\right) +Y\left( t\right) -B\left( t+1\right)
\label{BCgg}
\end{equation}%
At time $t$, the agent's statistical weight has the general form (\ref{sttcs}%
):

\begin{equation}
\int \dprod\limits_{n>1}\exp \left( U\left( B\left( t\right) +Y\left(
t\right) -B\left( t+1\right) \right) +\sum_{n>0}U\left( B\left( t+n\right)
+Y\left( t+n\right) -B\left( t+n+1\right) \right) \right) dB\left( t+n\right)
\label{sttcsnn}
\end{equation}%
Performing the following change of variables for $n>1$: 
\begin{eqnarray*}
B\left( t+n\right) &\rightarrow &B\left( t+n\right) -\sum_{m\geqslant
n}Y\left( t+m\right) \\
B\left( t+n\right) +Y\left( t+n\right) -B\left( t+n\right) &\rightarrow
&B\left( t+n\right) -B\left( t+1+n\right)
\end{eqnarray*}%
the statistical weight (\ref{sttcsnn}) become:%
\begin{equation}
\int \dprod\limits_{n>1}\exp   ( U( B( t) +Y(
t) -B( t+1) ) +U ( B ( t+1) -B(
t+2) +\sum_{n\geqslant 1}Y ( t+n )  ) +\sum_{n>1}U(
B( t+n) -B ( t+1+m ) ) ) dB( t+n)
\label{cvlt}
\end{equation}%
Except for the case of a quadratic utility function, the successive integrals%
\begin{equation}
\int \dprod\limits_{n>1}\exp \left( U\left( B\left( t+1\right) -B\left(
t+2\right) +\sum_{n\geqslant 1}Y\left( t+n\right) \right) +\sum_{n>1}U\left(
B\left( t+n\right) -B\left( t+1+n\right) \right) \right) dB\left( t+n\right)
\label{sscssv}
\end{equation}%
arising in (\ref{cvlt}) cannot be computed exactly, . However, we can still
define a function $\check{U}\left( B\left( t+1\right) \right) $ resulting
from the convolution integrals (\ref{sscssv}):%
\begin{eqnarray}
&&\exp \left( \check{U}\left( B\left( t+1\right) +\sum_{n\geqslant 1}Y\left(
t+n\right) \right) \right)  \label{dfh} \\
&=&\int \exp \left( U\left( B\left( t+1\right) -B\left( t+2\right)
+\sum_{n\geqslant 1}Y\left( t+n\right) \right) +\sum_{n>1}U\left( B\left(
t+n\right) -B\left( t+n+1\right) \right) \right) \dprod\limits_{n>1}dB\left(
t+n\right)  \notag
\end{eqnarray}%
The function $\check{U}$ can be approximatively computed - we will comment
on that later in the paragraph - however its precise form is not needed
here. Instead, we use the general formula (\ref{dfh}) to write (\ref{cvlt})
as:%
\begin{equation}
\exp \left( U\left( B\left( t\right) +Y\left( t\right) -B\left( t+1\right)
\right) +\check{U}\left( B\left( t+1\right) +\sum_{n\geqslant 1}Y\left(
t+n\right) \right) \right)  \label{wghtnqdrt}
\end{equation}%
Here again (see the first paragraph of this section), we can get rid of the
variables $Y\left( t+n\right) $ by considering them to be gaussian random
variables centered on $\bar{Y}$ for $n\geqslant 1$. The transition
probability for $B\left( t\right) $ is obtained by integrating (\ref%
{wghtnqdrt}) over the variables $Y\left( t+n\right) $ :\ 
\begin{equation}
\int \dprod dY\left( t+n\right) \exp \left( U\left( B\left( t\right)
+Y\left( t\right) -B\left( t+1\right) \right) +\check{U}\left( B\left(
t+1\right) +\sum_{n\geqslant 1}Y\left( t+n\right) \right) -\frac{1}{\sigma
^{2}}\sum_{n>0}\left( Y\left( t+n\right) -\bar{Y}\right) ^{2}\right)
\label{wghtnqdrtgg}
\end{equation}%
This expression can be simplified. Actually, in the gaussian integrals: 
\begin{equation}
\int \dprod dY\left( t+n\right) \exp \left( \check{U}\left( B\left(
t+1\right) +\sum_{n\geqslant 1}Y\left( t+n\right) \right) -\frac{1}{\sigma
^{2}}\sum_{n>0}\left( Y\left( t+n\right) -\bar{Y}\right) ^{2}\right)
\label{gss}
\end{equation}%
the variable $\sum_{n\geqslant 1}Y\left( t+n\right) $ has mean $T\bar{Y}$
and variance $T\sigma ^{2}$. As a consequence, if we assume $T$ large enough
so that $\sqrt{T}\gg \sigma $, then%
\begin{equation*}
\sum_{n\geqslant 1}Y\left( t+n\right) \simeq T\bar{Y}\pm \sqrt{T}\sigma
\simeq T\bar{Y}
\end{equation*}%
in first approximation. This allows to simplify (\ref{gss}):%
\begin{eqnarray*}
&&\int \dprod dY\left( t+n\right) \exp \left( \check{U}\left( B\left(
t+1\right) +\sum_{n\geqslant 1}Y\left( t+n\right) \right) -\frac{1}{\sigma
^{2}}\sum_{n>0}\left( Y\left( t+n\right) -\bar{Y}\right) ^{2}\right) \\
&\simeq &\exp \left( \check{U}\left( B\left( t+1\right) +T\bar{Y}\right)
\right)
\end{eqnarray*}%
so that, using the constraint (\ref{BCgg}) to write $B\left( t+1\right) $ as
a function of the past variables: 
\begin{eqnarray*}
B\left( t+1\right) +\sum_{n\geqslant 1}Y\left( t+n\right)
&=&\sum_{n\leqslant 0}Y\left( t+n\right) -\sum_{n\leqslant 0}C\left(
t+n\right) +\sum_{n\geqslant 1}Y\left( t+n\right) \\
&\simeq &\sum_{n\leqslant 0}Y\left( t+n\right) -\sum_{n\leqslant 0}C\left(
t+n\right) +T\bar{Y}
\end{eqnarray*}%
the weight (\ref{wghtnqdrtgg}) results in:

\begin{eqnarray}
&&\hskip-30pt\int \dprod dY\left( t+n\right) \exp \left( U\left( B ( t )
+Y ( t ) -B ( t+1 ) \right) +\check{U}\!\!\left( B (
t+1 ) +\sum_{n\geqslant 1}Y ( t+ 1) \right) -\frac{%
\sum_{n>0}\left( Y ( t+n ) -\bar{Y}\right) ^{2}}{\sigma ^{2}}\right)
\label{wghtpr} \\
&\simeq &\exp \!\!\left( U\left( B\left( t\right) +Y\left( t\right) -B\left(
t+1\right) \right) +\check{U}\left( \sum_{n\leqslant 0}Y ( t+n )
-\sum_{n\leqslant 0}C\left( t+n\right) +T\bar{Y}\right) \right)  \notag
\end{eqnarray}%
We can consider that the term:%
\begin{equation*}
\sum_{n\leqslant 0}Y\left( t+n\right) -\sum_{n\leqslant 0}C\left( t+n\right)
+T\bar{Y}
\end{equation*}%
has relatively small fluctuations with respect to its average $T\bar{Y}$, we
can approximate $\check{U}$ by its second order expansion:%
\begin{equation*}
\check{U}\left( \sum_{n\leqslant 0}Y\left( t+n\right) -\sum_{n\leqslant
0}C\left( t+n\right) +T\bar{Y}\right) \simeq C-\gamma \left(
\sum_{n\leqslant 0}Y\left( t+n\right) -\sum_{n\leqslant 0}C\left( t+n\right)
+\bar{Y}\right) ^{2}
\end{equation*}%
the values of $C$ and $\gamma $\ depending on (\ref{dfh}). Then, up to the
irrelevant constant $C$, (\ref{wghtpr}) simplifies to the second order
approximation:{\LARGE \ }%
\begin{eqnarray}
&&\exp \left( U\left( B\left( t\right) +Y\left( t\right) -B\left( t+1\right)
\right) -\gamma \left( \sum_{n\leqslant 0}Y\left( t+n\right)
-\sum_{n\leqslant 0}C\left( t+n\right) +\bar{Y}\right) ^{2}\right)
\label{sttwg} \\
&=&\exp \left( U\left( C\left( t\right) \right) -\gamma \left(
\sum_{n\leqslant 0}Y\left( t+n\right) -\sum_{n\leqslant 0}C\left( t+n\right)
+\bar{Y}\right) ^{2}\right)  \notag
\end{eqnarray}%
a result similar to the first example of this section. The constraint can be
introduced as a quadratic and non local contribution to the utility $U\left(
C_{t}\right) $. This result is not surprising. The constraint being imposed
on the whole path of the system, the inclusion of its intertemporal
quadratic expansion enforces the constraint on average, as needed. The
result is similar to (\ref{wgtqd}), except that the quadratic utility has
been replaced by a more general function. In first approximation, one can
thus share $U\left( C\left( t\right) \right) $ in a qudratic approximation
plus some perturbative terms $V\left( C\left( t\right) \right) $. Then
proceeding with the quadratic term as we did in Appendix $1$ to define the
statistical weight, one recovers a formula similar to (\ref{glblwght2}) for
the consumption path:%
\begin{equation}
\exp \left( -\sum \left( \left( \hat{C}\left( t\right) -\hat{C}\left(
t+1\right) \right) ^{2}+V\left( C\left( t\right) \right) \right) -\frac{%
\left( \sum_{n\leqslant 0}\hat{Y}\left( T+n\right) -\sum_{n\leqslant 0}\hat{C%
}\left( T+n\right) \right) ^{2}}{\bar{\sigma}^{2}}\right)  \notag
\end{equation}%
where $\frac{\hat{Y}\left( t+1\right) }{T-t}$ has been neglected for $T\gg 1$.

Let us close this section by quickly discussing the form of the function $%
\check{U}$ defined by (\ref{dfh}): 
\begin{eqnarray}
&&\exp \left( \check{U}\left( B\left( t+1\right) +\sum_{n\geqslant 1}Y\left(
t+n\right) \right) \right)  \label{dfhgn} \\
&=&\int \exp \left( U\left( B\left( t+1\right) -B\left( t+2\right)
+\sum_{n\geqslant 1}Y\left( t+n\right) \right) +\sum_{n>1}U\left( B\left(
t+n\right) -B_{t+n+1}\right) \right) \dprod\limits_{n>1}dB\left( t+n\right) 
\notag
\end{eqnarray}

These integrals can be approximatively computed with the saddle path
approximation technique developed in the first and second sections. The
saddle path result is not exact for a non quadratic utility, but constitutes
a sufficient approximation for us. The saddle path (\ref{dfhgn}) for the
function inside the exponential can be written as a difference equation $%
B\left( t+n\right) $ with $n>1$:%
\begin{equation*}
U^{\prime }\left( B\left( t+n\right) -B\left( t+n+1\right) \right)
-U^{\prime }\left( B\left( t+n-1\right) -B\left( t+n\right) \right) =0\text{
for }n>2
\end{equation*}%
and:%
\begin{equation*}
U^{\prime }\left( B\left( t+1\right) -B\left( t+2\right) \right) -U\left(
B\left( t+1\right) -B\left( t+2\right) +\sum_{m\geqslant 1}Y\left(
t+m\right) \right) \text{ for }n=2
\end{equation*}%
Once the saddle path $\bar{B}\left( t+n\right) $ is found, it can be
introduced in (\ref{dfhgn}) to yield:%
\begin{equation*}
\exp \left( \check{U}\left( B\left( t+1\right) +\sum_{n\geqslant 1}Y\left(
t+n\right) \right) \right) =\exp \left( U\left( B\left( t+1\right) -\bar{B}%
_{t+2}+\sum_{n\geqslant 1}Y\left( t+n\right) \right) +\sum_{n>1}U\left( \bar{%
B}\left( t+n\right) -\bar{B}\left( t+1+n\right) \right) \right)
\end{equation*}%
and a first approximation for $\check{U}$ is thus:%
\begin{equation}
\check{U}\left( B\left( t+1\right) +\sum_{n\geqslant 1}Y\left( t+n\right)
\right) =U\left( B\left( t+1\right) -\bar{B}\left( t+2\right)
+\sum_{n\geqslant 1}Y\left( t+n\right) \right) +\sum_{n>1}U\left( \bar{B}%
\left( t+n\right) -\bar{B}\left( t+1+n\right) \right)  \label{xprssn}
\end{equation}%
Some corrections to the saddle path can be included if we expand the RHS to
the second order around the saddle point by letting 
\begin{equation*}
B\left( t+n\right) =\bar{B}\left( t+n+1\right) +\delta B\left( t+n\right)
\end{equation*}%
and then integrate over $\delta B\left( t+n\right) $:%
\begin{eqnarray}
&&\exp \left( \check{U}\left( B\left( t+1\right) +\sum_{n\geqslant 1}Y\left(
t+n\right) \right) \right)  \label{xpnsc} \\
&=&\exp \left( U\left( B\left( t+1\right) -\bar{B}\left( t+2\right)
+\sum_{n\geqslant 1}Y\left( t+n\right) \right) +\sum_{n>1}U\left( \bar{B}%
\left( t+n\right) -\bar{B}\left( t+n+1\right) \right) \right)  \notag \\
&&\times \int \exp \left( U^{\prime \prime }\left( B\left( t+1\right) -\bar{B%
}\left( t+2\right) +\sum_{n\geqslant 1}Y\left( t+n\right) \right)\!\!\left(
\delta B\left( t+2\right) \right) ^{2}\right.  \notag \\
&&\hspace{0.96in}\hspace{0.6in}\left. +\sum_{n>1}U^{\prime \prime }\left( 
\bar{B}\left( t+n\right) -\bar{B}\left( t+n+1\right) \right)\!\!\left( \delta
B\left( t+n\right) -\delta B\left( t+n\right) \right) ^{2}\right)
\dprod\limits_{n>1}d\delta B\left( t+n\right)  \notag
\end{eqnarray}%
The log of the integrals in (\ref{xpnsc}) will yield some corrections to (%
\ref{xprssn}), but we will not inspect further the precise form of these
corrections.

\section*{Appendix 3}

When some discount rate is introduced, we go back to the initial individual
agent formulation and modify it accordingly. \bigskip Recall that the
transition probabilities between two consecutive state variables of the
system are defined by (\ref{pathscndordr}) with a discount rate $\beta $
added:%
\begin{equation*}
P\left( B_{i}\left( t\right) ,B_{i}\left( t+1\right) \right) =\int
\dprod\limits_{n=2}^{T}dB\left( t+n\right) \exp \left( U\left( C\left(
t\right) \right) +\sum_{n>0}\beta ^{n}U\left( C\left( t+n\right) \right)
\right)
\end{equation*}%
but now, the constraint rewrites:%
\begin{equation*}
B\left( t+1\right) =\left( 1+r\right)\!\!\left( B\left( t\right) +Y\left(
t\right) -C\left( t\right) \right)
\end{equation*}%
or equivalently:%
\begin{equation*}
C\left( t\right) =B\left( t\right) +Y\left( t\right) -\frac{B\left(
t+1\right) }{\left( 1+r\right) }
\end{equation*}%
Then, the integral over the $B\left( t+n\right) $ is similar to the previous
one, since one can change the variables: $\frac{B\left( t+n\right) }{\left(
1+r\right) ^{n}}\rightarrow $ $B\left( t+n\right) $ for $n>1$. We assume
that the uncertainty about future periods increases with a factor $\left(
1+r\right) $.

\begin{eqnarray*}
&=&\int \dprod\limits_{n=2}^{T}dB\left( t+n\right) \exp \left( -\left(
C\left( t\right) -\bar{C}\right) ^{2}-\sum_{n>0}\left( C\left( t+n\right) -%
\bar{C}\right) ^{2}\right) \\
&=&\int \dprod\limits_{n=2}^{T}dB\left( t+n\right) \exp \left( -\left(
B\left( t\right) +Y\left( t\right) -\frac{B\left( t+1\right) }{\left(
1+r\right) }-\bar{C}\right) ^{2}\right. \\
&&\hspace{3cm}\left. -\sum_{n>0}\left( \frac{\beta }{\left( 1+r\right) }%
\right) ^{n}\left( B\left( t+n\right) +Y\left( t+n\right) -\frac{B\left(
t+n+1\right) }{\left( 1+r\right) }-\bar{C}\right) ^{2}\right) \\
&=&\int \dprod\limits_{n=2}^{T}\left( 1+r\right) ^{n}dB\left( t+n\right)
^{\prime }\exp \left( -\left( B\left( t\right) +Y\left( t\right) -\frac{%
B\left( t+1\right) }{\left( 1+r\right) }-\bar{C}\right) ^{2}\right. \\
&&\hspace{3cm}\left. -\sum_{n>0}\left( 1+r\right) ^{n}\beta ^{n}\left(
B^{\prime }\left( t+n\right) -B^{\prime }\left( t+n+1\right) +\frac{Y\left(
t+n\right) -\bar{C}}{\left( 1+r\right) ^{n}}\right) ^{2}\right) \\
&=&\left( \dprod\limits_{n=2}^{T}\left( 1+r\right) ^{n}\right) \exp \left(
-\left( B\left( t\right) +Y\left( t\right) -\frac{B\left( t+1\right) }{%
\left( 1+r\right) }-\bar{C}\right) ^{2}\right. \\
&&\hspace{3cm}\left. -\frac{1}{\sum_{n>0}\left( \beta \left( 1+r\right)
\right) ^{-n}}\left( \frac{B\left( t+1\right) }{\left( 1+r\right) }%
+\sum_{n>0}\frac{Y\left( t+n\right) -\bar{C}}{\left( 1+r\right) ^{n}}\right)
^{2}\right) \\
&=&\left( \dprod\limits_{n=2}^{T}\left( 1+r\right) ^{n}\right) \exp \left(
-\left( B\left( t\right) +Y\left( t\right) -\frac{B\left( t+1\right) }{%
\left( 1+r\right) }-\bar{C}\right) ^{2}-s\left( T\right)\!\!\left( \frac{%
B\left( t+1\right) }{\left( 1+r\right) }+\sum_{n>0}\frac{Y\left( t+n\right) -%
\bar{C}}{\left( 1+r\right) ^{n}}\right) ^{2}\right)
\end{eqnarray*}

where the sum has been performed up to $T$ where $T$ is the time horizon
defined previously and $T\gg 1$, and $s^{-1}\left( T\right) =\sum_{n>0}\left(
\beta \left( 1+r\right) ^{2}\right) ^{-n}$. Since $\beta <1$:%
\begin{equation*}
s^{-1}\left( T\right) >\sum_{n>0}\left( \left( 1+r\right) \right)
^{-n}\simeq \frac{1}{r}
\end{equation*}%
for $r\ll 1$. As a consequence, $s\left( T\right) <r$. For $\beta =0$, $%
s^{-1}\left( T\right) =\frac{1}{r}$.

The factor\bigskip\ $\dprod\limits_{n=2}^{T}\left( 1+r\right) ^{n}$ can be
included in the normalization factor, as explained before, and then we are
left with:%
\begin{eqnarray}
P\left( B_{i}\left( t\right) ,B_{i}\left( t+1\right) \right) &=&\int
\dprod\limits_{n=2}^{T}dB\left( t+n\right) \exp \left( U\left( C\left(
t\right) \right) +\sum_{n>0}U\left( C\left( t+n\right) \right) \right)
\label{pathscndordrbsbt} \\
&=&\exp \left( -\left( B\left( t\right) +Y\left( t\right) -\frac{B\left(
t+1\right) }{\left( 1+r\right) }-\bar{C}\right) ^{2}-s\left( T\right)\!\!\left( 
\frac{B\left( t+1\right) }{\left( 1+r\right) }+\sum_{n>0}\frac{Y\left(
t+n\right) -\bar{C}}{\left( 1+r\right) ^{n}}\right) ^{2}\right)  \notag
\end{eqnarray}%
which is similar to (\ref{gblppdx}), except the $\frac{1}{\left( 1+r\right) }
$ factor in front of $B\left( t+1\right) $ and the $\left( 1+r\right) ^{n}$
multiplying $\left( Y\left( t+n\right) -\bar{C}\right) $. One also replaces $%
T$ by $\frac{1}{s\left( T\right) }$. Then the previous analysis following (%
\ref{gblppdx}) applies, except that, writing $B\left( t+1\right) $ as a
function of the past is now: 
\begin{equation}
\frac{B\left( t+1\right) }{1+r}=\sum_{n\leqslant 0}\frac{Y\left( t+n\right) 
}{\left( 1+r\right) ^{n}}-\sum_{n\leqslant 0}\frac{C\left( t+n\right) }{%
\left( 1+r\right) ^{n}}  \label{pstcrstnn}
\end{equation}%
with $B\left( t\right) \rightarrow 0$, $t\rightarrow T$ to impose the
transversality condition. The number of periods, $T$, is itself unknown, but
as said before $T$ is the expected mean process duration.

If $Y\left( t+n\right) $ is{\LARGE \ }centered on $\bar{Y}$ with variance $%
\left( 1+r\right) ^{2n}\sigma ^{2}$ (we assume that the discounted variable $%
\frac{Y\left( t+n\right) }{\left( 1+r\right) ^{n}}$ has a constant variance $%
\sigma ^{2}$), $\sum_{n>0}Y\left( t+n\right) $ centered on $\bar{Y}$ with
variance $T\sigma ^{2}$, integration over $Y\left( t+n\right) $ yields$:$%
\begin{eqnarray*}
&&\int \dprod dY\left( t+n\right) \exp \left( -s\left( T\right)\!\!\left( \frac{%
B\left( t+1\right) }{\left( 1+r\right) }+\sum_{n>0}\frac{Y\left( t+n\right) -%
\bar{C}}{\left( 1+r\right) ^{n}}\right) ^{2}-\frac{1}{\sigma ^{2}}%
\sum_{n=1}^{T}\left( Y\left( t+n\right) -\bar{Y}\right) ^{2}\right) \\
&=&\int \dprod dY^{\prime }\left( t+n\right) \exp \left( -s\left( T\right)
\left( \frac{B\left( t+1\right) }{\left( 1+r\right) }+\sum_{n>0}\frac{%
Y^{\prime }\left( t+n\right) +\bar{Y}-\bar{C}}{\left( 1+r\right) ^{n}}%
\right) ^{2}-\frac{1}{\sigma ^{2}}\sum_{n=1}^{T}\left( Y\left( t+n\right) -%
\bar{Y}\right) ^{2}\right)
\end{eqnarray*}%
with $Y^{\prime }\left( t+n\right) =Y\left( t+n\right) -\bar{Y}$. Neglecting
the terms $Y^{\prime }\left( t+n\right) Y^{\prime }\left( t+m\right) $ for $%
n\neq m$, since they are null in expectations, the exponential rewrites (for
a time horizon $T\gg 1$):%
\begin{eqnarray*}
&&\exp \left( -s\left( T\right)\!\!\left( \frac{B\left( t+1\right) }{\left(
1+r\right) }+\sum_{n>0}\frac{Y^{\prime }\left( t+n\right) +\bar{Y}-\bar{C}}{%
\left( 1+r\right) ^{n}}\right) ^{2}-\frac{1}{\sigma ^{2}}\sum_{n=1}^{T}%
\left( Y\left( t+n\right) -\bar{Y}\right) ^{2}\right) \\
&=&\exp \left( \left( -s\left( T\right)\!\!\left( \frac{B\left( t+1\right) }{%
\left( 1+r\right) }+\frac{1}{r}\left( \bar{Y}-\bar{C}\right) \right)
^{2}\right. \right) \\
&&\left. -2s\left( T\right)\!\!\left( \frac{B\left( t+1\right) }{\left(
1+r\right) }+\frac{1}{r}\left( \bar{Y}-\bar{C}\right) \right) \sum_{n=1}^{T}%
\frac{Y^{\prime }\left( t+n\right) }{\left( 1+r\right) ^{n}}%
-\sum_{n=1}^{T}\left( \frac{1}{\sigma ^{2}}+\frac{s\left( T\right) }{\left(
1+r\right) ^{2n}}\right)\!\!\left( Y\left( t+n\right) -\bar{Y}\right)
^{2}\right) \\
&\simeq &\exp \left( -s\left( T\right)\!\!\left( \frac{B\left( t+1\right) }{%
\left( 1+r\right) }+\frac{1}{r}\left( \bar{Y}-\bar{C}\right) \right)
^{2}\right. \\
&&\left. -2s\left( T\right) \sum_{n=1}^{T}\left( \frac{B\left( t+1\right) }{%
\left( 1+r\right) }+\frac{1}{r}\left( \bar{Y}-\bar{C}\right) \right) \frac{%
Y^{\prime }\left( t+n\right) }{\left( 1+r\right) ^{n}}-\frac{1}{\sigma ^{2}}%
\sum_{n=1}^{T}\left( Y\left( t+n\right) -\bar{Y}\right) ^{2}\right)
\end{eqnarray*}

for $\sigma ^{2}<1$ and since $s\left( T\right) <r\ll 1$.

and the integration over the $Y^{\prime }\left( t+n\right) $ leads to a
weight: 
\begin{eqnarray*}
&&\exp \left( -s\left( T\right)\!\!\left( \frac{B\left( t+1\right) }{\left(
1+r\right) }+\frac{1}{r}\left( \bar{Y}-\bar{C}\right) \right) ^{2}+\frac{%
\sigma ^{2}s^{2}\left( T\right)\!\!\left( 1+r\right) }{r}\left( \frac{B\left(
t+1\right) }{\left( 1+r\right) }+\frac{1}{r}\left( \bar{Y}-\bar{C}\right)
\right) ^{2}\right) \\
&\simeq &\exp \left( -s\left( T\right)\!\!\left( \frac{B\left( t+1\right) }{%
\left( 1+r\right) }+\frac{1}{r}\left( \bar{Y}-\bar{C}\right) \right)
^{2}\right)
\end{eqnarray*}%
since $s\left( T\right) \ll r$ and thus $\frac{\sigma ^{2}s^{2}\left( T\right)
\left( 1+r\right) }{r}\ll s\left( T\right) \sigma ^{2}\left( 1+r\right)
\ll s\left( T\right) $. Using that: 
\begin{equation*}
\frac{B\left( t+1\right) }{1+r}=\sum_{n\leqslant 0}\frac{Y\left( t+n\right) 
}{\left( 1+r\right) ^{n}}-\sum_{n\leqslant 0}\frac{C\left( t+n\right) }{%
\left( 1+r\right) ^{n}}
\end{equation*}%
the weight can be written:%
\begin{eqnarray*}
&&\exp \left( -\left( B\left( t\right) +Y\left( t\right) -\frac{B\left(
t+1\right) }{\left( 1+r\right) }-\bar{C}\right) ^{2}-s\left( T\right)\!\!\left(
\sum_{n\leqslant 0}\frac{Y\left( t+n\right) }{\left( 1+r\right) ^{n}}%
-\sum_{n\leqslant 0}\frac{C\left( t+n\right) }{\left( 1+r\right) ^{n}}+\frac{%
1}{r}\left( \bar{Y}-\bar{C}\right) \right) ^{2}\right) \\
&=&\exp \left( -\left( C\left( t\right) -\bar{C}\right) ^{2}-s\left(
T\right)\!\!\left( \sum_{n\leqslant 0}\frac{Y\left( t+n\right) }{\left(
1+r\right) ^{n}}-\sum_{n\leqslant 0}\frac{C\left( t+n\right) }{\left(
1+r\right) ^{n}}+\frac{1}{r}\left( \bar{Y}-\bar{C}\right) \right) ^{2}\right)
\\
&=&\exp \left( 
\begin{array}{c}
-\left( 1+s\left( T\right) \right)\!\!\left( C\left( t\right) -\bar{C}\right)
^{2}-s\left( T\right)\!\!\left( \sum_{n\leqslant 0}\frac{Y\left( t+n\right) }{%
\left( 1+r\right) ^{n}}-\sum_{n<0}\frac{C\left( t+n\right) }{\left(
1+r\right) ^{n}}-\bar{C}+\frac{1}{r}\left( \bar{Y}-\bar{C}\right) \right)
^{2} \\ 
+2s\left( T\right)\!\!\left( \sum_{n\leqslant 0}\frac{Y\left( t+n\right) }{%
\left( 1+r\right) ^{n}}-\sum_{n<0}\frac{C\left( t+n\right) }{\left(
1+r\right) ^{n}}-\bar{C}-\frac{1}{r}\left( \bar{C}-\bar{Y}\right) \right)
\left( C\left( t\right) -\bar{C}\right)%
\end{array}%
\right) \\
&\simeq &\exp \left( 
\begin{array}{c}
-\left( C\left( t\right) -\bar{C}\right) ^{2}-s\left( T\right)\!\!\left(
\sum_{n\leqslant 0}\frac{Y\left( t+n\right) }{\left( 1+r\right) ^{n}}%
-\sum_{n<0}\frac{C\left( t+n\right) }{\left( 1+r\right) ^{n}}-\frac{\bar{C}}{%
\left( 1+r\right) }+\frac{1}{r}\left( \bar{Y}-\bar{C}\right) \right) ^{2} \\ 
+2s\left( T\right)\!\!\left( \sum_{n\leqslant 0}\frac{Y\left( t+n\right) }{%
\left( 1+r\right) ^{n}}-\sum_{n<0}\frac{C\left( t+n\right) }{\left(
1+r\right) ^{n}}-\bar{C}-\frac{1}{r}\left( \bar{C}-\bar{Y}\right) \right)
\left( C\left( t\right) -\bar{C}\right)%
\end{array}%
\right)
\end{eqnarray*}%
as in the text, the terms in the exponential depending only of past and
predetermined variables are irrelevant to the statistical weight, so that
this one can be written:%
\begin{eqnarray*}
&&\exp \left( -\left( C\left( t\right) -\bar{C}-s\left( T\right)\!\!\left(
\sum_{n\leqslant 0}\frac{Y\left( t+n\right) }{\left( 1+r\right) ^{n}}%
-\sum_{n<0}\frac{C\left( t+n\right) }{\left( 1+r\right) ^{n}}-\bar{C}+\frac{1%
}{r}\left( \bar{Y}-\bar{C}\right) \right) \right) ^{2}\right) \\
&=&\exp \left( -\left( \left( C\left( t\right) -\bar{C}+s\left( T\right)
\left( \bar{C}-\frac{1}{r}\left( \bar{Y}-\bar{C}\right) \right) \right)
-s\left( T\right)\!\!\left( \sum_{n\leqslant 0}\frac{Y\left( t+n\right) }{%
\left( 1+r\right) ^{n}}-\sum_{n<0}\frac{C\left( t+n\right) }{\left(
1+r\right) ^{n}}\right) \right) ^{2}\right) \\
&\simeq &\exp \left( -\left( \hat{C}\left( t\right) -s\left( T\right)\!\!\left(
\sum_{n\leqslant 0}\frac{\hat{Y}\left( t+n\right) }{\left( 1+r\right) ^{n}}%
-\sum_{n<0}\frac{\hat{C}\left( t+n\right) }{\left( 1+r\right) ^{n}}\right)
\right) ^{2}\right)
\end{eqnarray*}%
for $s\left( T\right) \ll r$, with $\hat{Y}\left( t+n\right) =Y\left(
t+n\right) -\bar{C}+s\left( T\right)\!\!\left( \bar{C}-\frac{1}{r}\left( \bar{Y}%
-\bar{C}\right) \right) $ for $n\leqslant 0$ and $\hat{C}\left( t+n\right)
=C\left( t+n\right) +s\left( T\right)\!\!\left( \bar{C}-\frac{1}{r}\left( \bar{Y%
}-\bar{C}\right) \right) $ for $n\leqslant 0$. For $\beta =1$, $s\left(
T\right) \simeq r$, and then $\hat{Y}\left( t+n\right) =Y\left( t+n\right)
-\left( \bar{Y}-r\bar{C}\right) $, $\hat{C}\left( t+n\right) =C\left(
t+n\right) -r\left( \bar{Y}-r\bar{C}\right) $. In first approximation, $\hat{%
Y}\left( t+n\right) =Y\left( t+n\right) $, $\hat{C}\left( t+n\right)
=C\left( t+n\right) $.

We can now proceed as in the derivation of (\ref{glblwght}), and switch the
representation to express the probabilies for the variables. This weight can
be written differently. Actually, it describes the variables%
\begin{equation*}
X\left( t\right) =\hat{C}\left( t\right) -\frac{s\left( T\right) }{\left(
1+r\right) }\left( \sum_{n\leqslant 0}\frac{\hat{Y}\left( t+n\right) }{%
\left( 1+r\right) ^{n}}-\sum_{n<0}\frac{\hat{C}\left( t+n\right) }{\left(
1+r\right) ^{n}}\right)
\end{equation*}%
as gaussian and independent. Now, remark that at the first order in $r$:%
\begin{eqnarray*}
&&\left( 1+r\right) X\left( t\right) -X\left( t+1\right) \\
&=&\left( 1+r\right) \hat{C}\left( t\right) -\frac{s\left( T\right) }{\left(
1+r\right) }\left( \sum_{n\leqslant 0}\frac{\hat{Y}\left( t+n\right) }{%
\left( 1+r\right) ^{n}}-\sum_{n<0}\frac{\hat{C}\left( t+n\right) }{\left(
1+r\right) ^{n}}\right) \\
&&-\left( \hat{C}\left( t+1\right) -\frac{s\left( T\right) }{\left(
1+r\right) }\left( \sum_{n\leqslant 0}\frac{\hat{Y}\left( t+1+n\right) }{%
\left( 1+r\right) ^{n}}-\sum_{n<0}\frac{\hat{C}\left( t+1+n\right) }{\left(
1+r\right) ^{n}}\right) \right) \\
&\simeq &\left( 1+r\right) \hat{C}\left( t\right) -\hat{C}\left( t+1\right)
+s\left( T\right)\!\!\left( \frac{\hat{Y}\left( t+1\right) }{\left( 1+r\right) }%
-\hat{C}\left( t\right) \right) \\
&=&\left( 1+r-s\left( T\right) \right) \hat{C}\left( t\right) -\hat{C}\left(
t+1\right) +\frac{s\left( T\right) }{\left( 1+r\right) }\hat{Y}\left(
t+1\right) \\
&\simeq &\left( 1+r\right) \hat{C}\left( t\right) -\hat{C}\left( t+1\right) +%
\frac{s\left( T\right) }{\left( 1+r\right) }\hat{Y}\left( t+1\right)
\end{eqnarray*}%
It then allows to compute the density probability of: 
\begin{equation*}
\left( 1+r\right) \hat{C}\left( t\right) -\hat{C}\left( t+1\right) +\frac{%
s\left( T\right) }{\left( 1+r\right) }\hat{Y}\left( t+1\right)
\end{equation*}%
by writing:%
\begin{eqnarray*}
&&\int \exp \left( -X^{2}\left( t\right) -X^{2}\left( t+1\right) \right) \\
&&\times \delta \left( \left( 1+r\right) X\left( t\right) -X\left(
t+1\right) -\left( \left( 1+r\right) \hat{C}\left( t\right) -\hat{C}\left(
t+1\right) +\frac{s\left( T\right) }{\left( 1+r\right) }\hat{Y}\left(
t+1\right) \right) \right) dX\left( t\right) dX\left( t+1\right) \\
&=&\int \exp \left( -X^{2}\left( t\right) -\left( \left( 1+r\right) X\left(
t\right) -\left( \left( 1+r\right) \hat{C}\left( t\right) -\hat{C}\left(
t+1\right) +\frac{s\left( T\right) }{\left( 1+r\right) }\hat{Y}\left(
t+1\right) \right) \right) ^{2}\right) dX\left( t\right) \\
&=&\exp \left( -\frac{\left( \hat{C}\left( t\right) -\frac{\hat{C}\left(
t+1\right) }{\left( 1+r\right) }+\frac{s\left( T\right) }{\left( 1+r\right) }%
\hat{Y}\left( t+1\right) \right) ^{2}}{1+\left( 1+r\right) ^{2}}\right) \\
&\simeq &\exp \left( -\frac{\left( \hat{C}\left( t\right) -\hat{C}\left(
t+1\right)\!\!\left( 1-r\right) +s\left( T\right) \hat{Y}\left( t+1\right)
\right) ^{2}}{2}\right)
\end{eqnarray*}%
This stochastic process is constrained to $X_{T}=0$ through the constraint:%
\begin{equation*}
X\left( T\right) =\hat{C}\left( T\right) -\left( \sum_{0\leqslant n\leqslant
T}\frac{\hat{Y}\left( n\right) }{\left( 1+r\right) ^{n}}-\sum_{0\leqslant
n\leqslant T}\frac{\hat{C}\left( n\right) }{\left( 1+r\right) ^{n}}\right)
=\sum_{0\leqslant n\leqslant T}\frac{\hat{Y}\left( n\right) }{\left(
1+r\right) ^{n}}-\sum_{0\leqslant i\leqslant T}\frac{\hat{C}\left( n\right) 
}{\left( 1+r\right) ^{n}}=0
\end{equation*}%
Given that $s\left( T\right) \hat{Y}\left( t+1\right) $ has variance $%
s^{2}\left( T\right) \sigma ^{2}\ll 1$, including the global constraint yields
the statistical weight over all periods:%
\begin{eqnarray*}
&&\exp \left( -\sum_{t}\frac{\left( \hat{C}\left( t\right) -\hat{C}\left(
t+1\right)\!\!\left( 1-r\right) \right) ^{2}}{2\left( 1+s^{2}\left( T\right)
\sigma ^{2}\right) }\right) \delta \left( \sum_{0\leqslant n\leqslant T}%
\frac{\hat{Y}\left( n\right) }{\left( 1+r\right) ^{n}}-\sum_{0\leqslant
n\leqslant T}\frac{\hat{C}\left( n\right) }{\left( 1+r\right) ^{n}}\right) \\
&\simeq &\exp \left( -\sum_{t}\frac{\left( \hat{C}\left( t+1\right) -\hat{C}%
\left( t\right) -r\hat{C}\left( t+1\right) \right) ^{2}}{2}-\frac{\left(
\sum_{0\leqslant n\leqslant T}\frac{\hat{Y}\left( n\right) }{\left(
1+r\right) ^{n}}-\sum_{0\leqslant n\leqslant T}\frac{\hat{C}\left( n\right) 
}{\left( 1+r\right) ^{n}}\right) ^{2}}{T\sigma ^{2}}\right)
\end{eqnarray*}

\section*{\protect\bigskip Appendix 4}

The results of the first section can be summed up as follows. We described a
set of several individual economic agents by a stochastic process defined in
a space whose dimension depends of the number of degrees of freedom, that is
number of state variables, of the system. We now eplain how this description
can e replaced by a field formalism that will facilitate the computations
for a large number of agents. For the sake of the exposition, we choose a
simplified version of the model developed previously, in its continuous time
version. We start first by considering a single agent discarding the global
part of the constraint. We will then introduce an arbitrary number of agents
without interactions, and include the interactions ultimately. The behavior
of this agent can be represented during a time span $T$\ by a probability
weight for each possible path of actions.\ For a path $X\left( t\right) $\
of actions - such as consumption, production, signals - for $t\in \left[ 0,T%
\right] $\ , the weight (\ref{Crtnnqdr}) is:%
\begin{equation*}
P\left( X\right) =\exp \left( -\left( \frac{1}{\sigma ^{2}}%
\int_{0}^{T}\left( \left( \frac{d}{dt}X\left( t\right) \right)
^{2}+V_{1}\left( X\left( t\right) \right) \right) dt\right) \right)
\end{equation*}%
If we impose some initial and final conditions, $\underline{X}$ and $%
\overline{X}$\ on the path, we can also write, as in the first section, the
probability of transition from $\underline{X}$ to $\overline{X}$: 
\begin{equation}
P\left( \underline{X},\overline{X},T\right) =\exp \left( -\int_{X\left(
0\right) =\underline{X}}^{X\left( T\right) =\overline{X}}\left( \frac{1}{2}%
\left( \frac{d}{dt}X\left( t\right) \right) ^{2}+V_{1}\left( X\left(
t\right) \right) \right) dt\right) \mathcal{D}X\left( t\right)  \label{PRBT}
\end{equation}%
where $K\left( X\left( t\right) \right) $\ is a "potential term" whose form
depends explicitly on the agent's utility function, or any other
intertemporal function the agent optimizes. It represents the probability
for an agent to reach $\overline{X}$\ starting from $\underline{X}$\ during
the time span $T$. It is the probability of social mobility - moving from
point $\underline{X}$ to $\overline{X}$ - for an agent in the social space.
Written under this form, the probability transition (\ref{PRBT}) is given by
a path integral: The weight in the exponential includes a random, brownian
motion, plus a potential $V_{1}\left( X\left( t\right) \right) $\ describing
the individual goals as well as social/economical influences. It can be seen
as an intertemporal utility whose optimization would yield the usual
brownian noise plus some external determinants. As explained before, we have
to compute the Laplace transform (\ref{gblw3bb}) of $P\left( \underline{X},%
\overline{X},T\right) $: 
\begin{equation}
G_{\alpha }\left( \underline{X},\overline{X}\right) =\int_{0}^{\infty }\exp
\left( -\alpha t\right) P_{t}\left( \underline{X},\overline{X}\right) dt
\label{Grnn}
\end{equation}%
To do so, it is straightforward to check that (see Kleinert 1989 for
example) $P_{t}\left( \underline{X},\overline{X}\right) $ satisfies a
partial differential equation: 
\begin{equation*}
\frac{\partial }{\partial t}P_{t}\left( \underline{X},\overline{X}\right)
=\left( \frac{1}{2}\nabla _{\underline{X}}^{2}-V_{1}\left( \underline{X}%
\right) \right) P\left( \underline{X},\overline{X},s\right)
\end{equation*}%
and that, as a consequence, its Laplace transform $G_{\alpha }\left( 
\underline{X},\overline{X}\right) $ satisfies:%
\begin{equation}
\left( -\frac{1}{2}\nabla ^{2}+\alpha +V_{1}\left( \underline{X}\right)
\right) G_{\alpha }\left( \underline{X},\overline{X}\right) =\delta \left( 
\underline{X}-\overline{X}\right)  \label{eqGbs}
\end{equation}%
where $\delta \left( \underline{X}-\overline{X}\right) $ denotes the Dirac
function. The solution of equation (\ref{eqGbs}) is the resolvent, or the
kernel, of the operator: 
\begin{equation*}
L=-\frac{1}{2}\nabla ^{2}+\alpha +V_{1}\left( \underline{X}\right)
\end{equation*}%
Introducing $N$ identical agents without interaction is straightforward.
Without interaction, the agents are indpendent and as a consequence, the
probability transition is a the product of individual probability
transitions:%
\begin{equation*}
P_{t_{1},...,t_{n}}\left( \underline{X},\overline{X},T\right)
=P_{t_{1}}\left( \underline{X_{1}},\overline{X_{1}}\right)
...P_{t_{N}}\left( \underline{X}_{N},\overline{X}_{N}\right)
\end{equation*}%
and the same applies for their Laplace transform:%
\begin{eqnarray}
G_{\alpha }\left( \underline{X},\overline{X}\right) &=&\int_{0}^{\infty
}\exp \left( -\alpha \left( t_{1}+...+t_{N}\right) \right) P_{t_{1}}\left( 
\underline{X_{1}},\overline{X_{1}}\right) ...P_{t_{N}}\left( \underline{X}%
_{N},\overline{X}_{N}\right) dt_{1}...dt_{n}  \label{prdgrnn} \\
&=&G_{\alpha }\left( \underline{X_{1}},\overline{X_{1}}\right) ...G_{\alpha
}\left( \underline{X}_{N},\overline{X}_{N}\right)  \notag
\end{eqnarray}%
From now on, in order to alleviate the notation, we will denote $x_{i}$ and $%
y_{i}$ the inital and final state for agent $i$ (in the case of a single
agent, we will simply use $x$ and $y$). The transition probability for $N$
agents will thus be written: 
\begin{equation*}
\dprod\limits_{i=1}^{N}G_{\alpha }\left( x_{i},y_{i}\right)
\end{equation*}%
We will write $G_{\alpha }\left( \left( x_{i},y_{i}\right) _{N}\right) $ the
probability of transition for $N$ agents.

As explained in the first section, the field theory allows to express the
product in (\ref{prdgrnn}) as derivatives of a single function. To do so we
consider (\ref{eqGbs}). We can actually infer from (\ref{eqGbs}) that the
determinant of the integral operator $G_{\alpha }$\ whose kernel is $%
G_{\alpha }\left( \underline{X},\overline{X}\right) $\ can be expressed as
an infinite dimensional integral:%
\begin{equation}
\left( \det \left( G_{\alpha }\right) \right) ^{-1}=\int \exp \left( -\int
\Psi \left( x\right)\!\!\left( -\frac{1}{2}\nabla ^{2}+\alpha +K\left( x\right)
\right) \Psi ^{\dag }\left( x\right) dx\right) \mathcal{D}\Psi \mathcal{D}%
\Psi ^{\dag }  \label{dtgrnnn}
\end{equation}%
where the integrals over $\Psi \left( x\right) $\ and $\Psi ^{\dag }\left(
x\right) $\ are performed over the space of complex-valued functions of one
variable $x$ living in the same space as $\underline{X}$ and $\overline{X}$,
the initial and final states of one single agent. The function $\Psi ^{\dag
}\left( x\right) $\ is the complex conjugate of $\Psi \left( x\right) $.
These functions are the fields introduced in section 1.

The formula (\ref{dtgrnnn}) is simply the generalization in infinite
dimension of the gaussian integral formula:%
\begin{equation*}
\left( \det \left( M\right) \right) ^{-1}=\int \exp \left( -X\left( M\right)
X^{\dag }\right) \mathcal{D}X\mathcal{D}X^{\dag }
\end{equation*}%
for $X$ a vector of $%
\mathbb{C}
^{N}$, $X^{\dag }$ it's complex conjugate, and $M$ an invertible linear
operator on $%
\mathbb{C}
^{N}$, i.e. an invertible matrix. To recover (\ref{prdgrnn}) from (\ref%
{dtgrnnn}), one introduces the source term $J\left( x\right) \Psi ^{\dag
}\left( x\right) +J^{\dag }\left( x\right) \Psi \left( x\right) $ defined in
section 2. We claim that:%
\begin{eqnarray}
&&\frac{\int \exp \left( -\int \!\!\left( \Psi \left( x\right)\!\!\left( -\frac{1}{2%
}\nabla ^{2}+\alpha +K\left( x\right) \right) \Psi ^{\dag }\left( x\right)
+J\left( x\right) \Psi ^{\dag }\left( x\right) +J^{\dag }\left( x\right)
\Psi \left( x\right) \right) dx\right) \mathcal{D}\Psi \mathcal{D}\Psi
^{\dag }}{\int \exp \left( \left( -\int \!\!\left( \Psi \left( x\right)\!\!\left( -%
\frac{1}{2}\nabla ^{2}+\alpha +K\left( x\right) \right) \Psi ^{\dag }\left(
x\right) \right) dx\right) \right) \mathcal{D}\Psi \mathcal{D}\Psi ^{\dag }}%
\hspace{0.96in}  \label{generatfct} \\
&=&\exp \left( \int J\left( x\right)\!\!\left( -\frac{1}{2}\nabla ^{2}+\alpha
+K\left( x\right) \right) ^{-1}J^{\dag }\left( x\right) dx\right)  \notag \\
&=&\exp \left( \int J\left( x\right) G_{\alpha }\left( x,y\right) J^{\dag
}\left( x\right) dx\right)  \notag
\end{eqnarray}%
This results directly from (\ref{generatfct}) by changing the variable $\Psi
\left( x\right) \rightarrow \Psi \left( x\right) +J\left( x\right) $ in the
numerator and using (\ref{dtgrnnn}). As a consequence, the terms in (\ref%
{prdgrnn}) can be recovered from (\ref{generatfct}). Actually, the
transition function for $N$\ agents (\ref{generatfct}):%
\begin{equation}
\dprod\limits_{i=1}^{N}G_{\alpha }\left( x_{i},y_{i}\right)  \label{Nagentsc}
\end{equation}%
{\LARGE \ } can directly be written as:%
\begin{equation*}
\dprod\limits_{i=1}^{N}G_{\alpha }\left( x_{i},y_{i}\right) =\left[ \left( 
\frac{\delta }{\delta J\left( x_{i_{1}}\right) }\frac{\delta }{\delta
J^{\dag }\left( y_{i_{1}}\right) }\right) ...\left( \frac{\delta }{\delta
J\left( x_{i_{N}}\right) }\frac{\delta }{\delta J^{\dag }\left(
y_{i_{N}}\right) }\right) \exp \left( J\left( x\right) G_{\alpha }\left(
x,y\right) J^{\dag }\left( x\right) \right) \right] _{J=J^{\dag }=0}
\end{equation*}%
Consequently, we now have an infinite dimensional integral representation
for the transition functions for $N$\ agents:%
\begin{eqnarray}
&&\dprod\limits_{i=1}^{N}G_{\alpha }\left( x_{i},y_{i},\alpha \right) =\frac{%
1}{\int \exp \left( \left( -\int \!\!\left( \Psi \left( x\right)\!\!\left( -\frac{1%
}{2}\nabla ^{2}+\alpha +K\left( x\right) \right) \Psi ^{\dag }\left(
x\right) \right) dx\right) \right) \mathcal{D}\Psi \mathcal{D}\Psi ^{\dag }}
\label{grnnfn} \\
&&\times \left[ \left( \frac{\delta }{\delta J\left( x_{i_{1}}\right) }\frac{%
\delta }{\delta J^{\dag }\left( x_{i_{1}}\right) }\right) ...\left( \frac{%
\delta }{\delta J\left( x_{i_{N}}\right) }\frac{\delta }{\delta J^{\dag
}\left( x_{i_{N}}\right) }\right) \right] _{J=J^{\dag }=0}  \notag \\
&&.\int \exp \left( -\int \!\!\left( \Psi \left( x\right)\!\!\left( -\frac{1}{2}%
\nabla ^{2}+\alpha +K\left( x\right) \right) \Psi ^{\dag }\left( x\right)
+J\left( x\right) \Psi ^{\dag }\left( x\right) +J^{\dag }\left( x\right)
\Psi \left( x\right) \right) dx\right) \mathcal{D}\Psi \mathcal{D}\Psi
^{\dag }  \notag
\end{eqnarray}%
The normalization factor{\LARGE \ }%
\begin{equation*}
\frac{1}{\int \exp \left( \left( -\int \!\!\left( \Psi \left( x\right)\!\!\left( -%
\frac{1}{2}\nabla ^{2}+\alpha +K\left( x\right) \right) \Psi ^{\dag }\left(
x\right) \right) dx\right) \right) \mathcal{D}\Psi \mathcal{D}\Psi ^{\dag }}%
{\LARGE \ }=\det \left( -\frac{1}{2}\nabla ^{2}+\alpha +K\left( x\right)
\right) ^{-1}
\end{equation*}%
is constant and will thus be - whenever possible - omitted in the formula.
Thus, the transition functions associated to (\ref{PRBT}) are computed by
taking the derivatives with respect to $J\left( x\right) $\ and $J^{\dag
}\left( x\right) $\ of 
\begin{equation*}
\int \exp \left( -\int \!\!\left( \Psi \left( x\right)\!\!\left( -\frac{1}{2}\nabla
^{2}+\alpha +K\left( x\right) \right) \Psi ^{\dag }\left( x\right) +J\left(
x\right) \Psi ^{\dag }\left( x\right) +J^{\dag }\left( x\right) \Psi \left(
x\right) \right) dx\right) \mathcal{D}\Psi \mathcal{D}\Psi ^{\dag }
\end{equation*}%
The source term is usually implied and only reintroduced ultimately, at the
end of the computations.\ As a consequence, 
\begin{equation}
\int \exp \left( -\int \!\!\left( \Psi \left( x\right)\!\!\left( -\frac{1}{2}\nabla
^{2}+\alpha +K\left( x\right) \right) \Psi ^{\dag }\left( x\right) \right)
dx\right) \mathcal{D}\Psi \mathcal{D}\Psi ^{\dag }  \label{pathintfrm}
\end{equation}%
will describe the same system of identical non interacting agents whose
probabilistic description is (\ref{PRBT}).

We can now consider the case of interacting agents, which means 
including a
potential term involving $k$ agents as in (\ref{gblw3}): 
\begin{equation*}
\sum_{i_{1},...,i_{k}}\int_{0}^{T_{1}}...\int_{0}^{T_{n}}V_{k}\left(
X_{s_{1}}^{\left( i_{1}\right) },...,X_{s_{k}}^{\left( i_{k}\right) }\right)
ds_{1}...ds_{k}
\end{equation*}%
Where we set $\theta =1$ and each agent having its own lifespan $T_{i}$. We
explain how to translate our probabilistic formalism in a field description
similar to (\ref{grnnfn}), and including the interactions.

To do so, we introduce the so called Grand Partition Function for an
infinite set of interacting individual agents associated to the partition
function (\ref{gblw3}): 
\begin{eqnarray}
&&\dsum\limits_{N}\frac{1}{N!}\dprod\limits_{i=1}^{N}\int \exp \left(
-\alpha t_{i}\right) \int \mathcal{D}X_{i}\left( t\right) \exp \left(
-\sum_{i}\int_{X_{i}\left( 0\right) =x_{i}}^{X_{i}\left( t_{i}\right)
=y_{i}}\left( \left( \frac{1}{2}\left( \frac{d}{dt}X_{i}\left( t\right)
\right) ^{2}+V_{1}\left( X_{i}\left( t\right) \right) \right) dt\right)
\right.  \label{gdpr} \\
&&\hspace{0.48in}\hspace{0.48in}\hspace{0.48in}\hspace{0.48in}-\left.
\sum_{i_{1},...i_{k}}\int_{X_{i}\left( 0\right) =x_{i}}^{X_{i}\left(
s\right) =y_{i}}V_{k}\left( X_{i_{1}}\left( t_{1}\right) ...X_{i_{k}}\left(
t_{k}\right) \right) dt_{1}...dt_{k}\right)  \notag
\end{eqnarray}%
Compared to (\ref{gblw3}), two differences arise.

First, the number of agents is variale. This is in line with our description
of the field formalism in section 2. We do not focus on a fixed number of
agents, but rather on the interaction of several agents among a set of an
infinite number of agents. The number $N$ of agents involved in the
interaction process can be variable, eventhough very large, and this is why
we sum over $N$ expressions similar to (\ref{gblw3}). The $N!$ reflects the
fact that agents are identical in that context and is here to avoid
redundancies in the sum over agents.

Second, the lifespan of the agents is different from one agent to another.
As explained in section one we assume this lifespan to be a random Poisson
process of average $\frac{1}{\alpha }$ and we take the average over this
process for all agents. This the reason for the Laplace transform.

\bigskip

As explained above, without interactions, the transition probabilities
associated to (\ref{gdpr}) can be computed with the formalism defined by (%
\ref{pathintfrm}). To include the interaction part, we will now consider the
potential $V_{1}\left( X_{i}\left( t\right) \right) $ as a source term. To
do so, we follow the presentation of (Kleinert 1989). Starting with the
simplest case of no interaction, i.e. $V_{k}\left( X_{1}\left( t_{1}\right)
...X_{k}\left( t_{k}\right) \right) =0$, the function of interest to us is:%
\begin{equation}
\dsum\limits_{N}\frac{1}{N!}\dprod\limits_{i=1}^{N}\int \exp \left( -\alpha
t_{i}\right) \int \mathcal{D}X_{i}\left( t\right) \exp \left(
-\sum_{i}\int_{X_{i}\left( 0\right) =x_{i}}^{X_{i}\left( t_{i}\right)
=y_{i}}\left( \left( \frac{1}{2}\left( \frac{d}{dt}X_{i}\left( t\right)
\right) ^{2}+V_{1}\left( X_{i}\left( t\right) \right) \right) dt\right)
\right)  \label{Nagentsb}
\end{equation}%
Each of these integrals being independent from each others, the results for (%
\ref{Nagentsb}) is: 
\begin{equation}
\dsum\limits_{N}\frac{1}{N!}\dprod\limits_{i=1}^{N}\int \exp \left( -\alpha
t_{i}\right) \int \mathcal{D}X_{i}\left( t\right) \exp \left(
-\sum_{i}\int_{X_{i}\left( 0\right) =x_{i}}^{X_{i}\left( t_{i}\right)
=y_{i}}\left( \left( \frac{1}{2}\left( \frac{d}{dt}X_{i}\left( t\right)
\right) ^{2}+V_{1}\left( X_{i}\left( t\right) \right) \right) dt\right)
\right) =\dsum\limits_{N}\frac{1}{N!}\dprod\limits_{i=1}^{N}G_{\alpha
}\left( x_{i},y_{i}\right)  \label{Nagentscbbr}
\end{equation}%
which is a mixed sum over $N$\ of transition functions for $N$\ agents.\
Each product $\frac{1}{N!}\dprod\limits_{i=1}^{N}G_{K}\left(
x_{i},y_{i},\alpha \right) $\ computes, as needed, the transition
probability from $\left\{ x_{i}\right\} _{i=1...N}$\ to $\left\{
y_{i}\right\} _{i=1...N}$\ for $N$\ ordered agents during a process of mean
duration $\frac{1}{\alpha }$. Thus the sum can be seen as a generating
series for these probabilities with $N$\ agents. However, between identical
agents, order is irrelevant, so that the probability of transition of the
system from $\left\{ x_{i}\right\} _{i=1...N}$\ to $\left\{ y_{i}\right\}
_{i=1...N}$\ is the sum over the permutations with $N$\ elements of the
terms\ on (\ref{Nagentscbbr}) rhs. Since these terms are equal, the "true"
probability of transition is $\dprod\limits_{i=1}^{N}G_{K}\left(
x_{i},y_{i},\alpha \right) $. The whole problem at stake is to recover the
case with interaction (\ref{gdpr}) from the "free" case (\ref{Nagentsb}).
This is done using the following method. Using the functional derivative
with respect to $V_{1}\left( x_{i_{1}}\right) $\ we write:%
\begin{eqnarray*}
&&\frac{\delta }{\delta V_{1}\left( x_{i_{1}}\right) }\dsum\limits_{N}\frac{1%
}{N!}\dprod\limits_{i=1}^{N}\int \exp \left( -\alpha t_{i}\right) \int 
\mathcal{D}X_{i}\left( t\right) \exp \left( -\sum_{i}\int_{X_{i}\left(
0\right) =x_{i}}^{X_{i}\left( t_{i}\right) =y_{i}}\left( \left( \frac{1}{2}%
\left( \frac{d}{dt}X_{i}\left( t\right) \right) ^{2}+V_{1}\left( X_{i}\left(
t\right) \right) \right) dt\right) \right) \\
&=&\dsum\limits_{N}\frac{1}{N!}\dprod\limits_{i=1}^{N}\int \exp \left(
-\alpha t_{i}\right) \int \mathcal{D}X_{i}\left( t\right) \exp \left(
-\sum_{i}\int_{X_{i}\left( 0\right) =x_{i}}^{X_{i}\left( t_{i}\right)
=y_{i}}\left( \left( \frac{1}{2}\left( \frac{d}{dt}X_{i}\left( t\right)
\right) ^{2}+V_{1}\left( X_{i}\left( t\right) \right) \right) dt\right)
\right) \\
&&\times \left\{ -\sum_{i}\int_{X_{i_{1}}\left( 0\right)
=x_{i_{1}}}^{X_{i_{1}}\left( t_{i_{1}}\right) =y_{i_{1}}}dt\delta \left(
X_{i_{1}}\left( t\right) -x_{i_{1}}\right) \right\}
\end{eqnarray*}%
where $\delta \left( x_{i_{1}}\left( t\right) -x_{i_{1}}\right) $\ is the
delta of Dirac function. By extension, this generalizes for any function $%
V\left( x_{i_{1}}\right) $, to yield:%
\begin{eqnarray*}
&&\int dx_{i_{1}}V\left( x_{i_{1}}\right) \frac{\delta }{\delta V_{1}\left(
x_{i_{1}}\right) }\dsum\limits_{N}\frac{1}{N!}\dprod\limits_{i=1}^{N}\int
\exp \left( -\alpha t_{i}\right) \\
&&\hspace{3cm}\times \int \mathcal{D}X_{i}\left( t\right) \exp \left(
-\sum_{i}\int_{X_{i}\left( 0\right) =x_{i}}^{X_{i}\left( t_{i}\right)
=y_{i}}\left( \left( \frac{1}{2}\left( \frac{d}{dt}X_{i}\left( t\right)
\right) ^{2}+V_{1}\left( X_{i}\left( t\right) \right) \right) dt\right)
\right) \\
&=&=\dsum\limits_{N}\frac{1}{N!}\dprod\limits_{i=1}^{N}\int \exp \left(
-\alpha t_{i}\right) \int \mathcal{D}X_{i}\left( t\right) \exp \left(
-\sum_{i}\int_{X_{i}\left( 0\right) =x_{i}}^{X_{i}\left( t_{i}\right)
=y_{i}}\left( \left( \frac{1}{2}\left( \frac{d}{dt}X_{i}\left( t\right)
\right) ^{2}+V_{1}\left( X_{i}\left( t\right) \right) \right) dt\right)
\right) \\
&&\times \left\{ -\sum_{i}\int_{X_{i_{1}}\left( 0\right)
=x_{i_{1}}}^{X_{i_{1}}\left( t_{i_{1}}\right) =y_{i_{1}}}dtV\left(
X_{i_{1}}\left( t\right) \right) \right\}
\end{eqnarray*}%
and for any function of several variables, one has similarly: 
\begin{eqnarray}
&&\dsum\limits_{N}\frac{1}{N!}\dprod\limits_{i=1}^{N}\int \exp \left(
-\alpha t_{i}\right) \int \mathcal{D}X_{i}\left( t\right) \exp \left(
-\sum_{i}\int_{X_{i}\left( 0\right) =x_{i}}^{X_{i}\left( t_{i}\right)
=y_{i}}\left( \left( \frac{1}{2}\left( \frac{d}{dt}X_{i}\left( t\right)
\right) ^{2}+V_{1}\left( X_{i}\left( t\right) \right) \right) dt\right)
\right)  \label{ntrctnnnxpn} \\
&&\times \sum_{i_{1},...i_{k}}\int_{X_{i_{1}}\left( 0\right)
=x_{i_{1}}}^{X_{i_{1}}\left( t_{i_{1}}\right)
=y_{i_{1}}}...\int_{X_{i_{k}}\left( 0\right) =x_{i_{k}}}^{X_{i_{k}}\left(
t_{i_{k}}\right) =y_{i_{k}}}V_{k}\left( X_{i_{1}}\left( t_{k}\right)
...X_{i_{k}}\left( t_{k}\right) \right) dt_{1}...dt_{k}  \notag \\
&=&\sum_{i_{1},...i_{k}}\left\{ \left( -1\right) ^{k}\int
dx_{i_{1}}...dx_{i_{k}}V_{k}\left( x_{i_{1}}...x_{i_{k}}\right) \frac{\delta 
}{\delta V_{1}\left( x_{i_{1}}\right) }...\frac{\delta }{\delta V_{1}\left(
x_{i_{k}}\right) }\right\}  \notag \\
&&\times \dsum\limits_{N}\frac{1}{N!}\dprod\limits_{i=1}^{N}\int \exp \left(
-\alpha t_{i}\right) \int \mathcal{D}X_{i}\left( t\right) \exp \left(
-\sum_{i}\int_{X_{i}\left( 0\right) =x_{i}}^{X_{i}\left( t_{i}\right)
=y_{i}}\left( \left( \frac{1}{2}\left( \frac{d}{dt}X_{i}\left( t\right)
\right) ^{2}+V_{1}\left( X_{i}\left( t\right) \right) \right) dt\right)
\right)  \notag
\end{eqnarray}%
To find (\ref{gdpr}) from (\ref{Nagentsb}), the next step is to exponentiate
(\ref{ntrctnnnxpn}) to express (\ref{gdpr}) as:%
\begin{eqnarray*}
&&\dsum\limits_{N}\frac{1}{N!}\dprod\limits_{i=1}^{N}\int \exp \left(
-\alpha t_{i}\right) \int \mathcal{D}X_{i}\left( t\right) \exp \left(
-\sum_{i}\int_{X_{i}\left( 0\right) =x_{i}}^{X_{i}\left( t_{i}\right)
=y_{i}}\left( \left( \frac{1}{2}\left( \frac{d}{dt}X_{i}\left( t\right)
\right) ^{2}+V_{1}\left( X_{i}\left( t\right) \right) \right) dt\right)
\right. \\
&&\hspace{0.48in}\hspace{0.48in}\hspace{0.48in}\hspace{0.48in}-\left.
\sum_{i_{1},...i_{k}}\int_{X_{i}\left( 0\right) =x_{i}}^{X_{i}\left(
s\right) =y_{i}}V_{k}\left( X_{i_{1}}\left( t_{1}\right) ...X_{i_{k}}\left(
t_{k}\right) \right) dt_{1}...dt_{k}\right) \\
&=&\exp \left( -\int dx_{i_{1}}...dx_{i_{k}}V_{k}\left(
x_{i_{1}}...x_{i_{k}}\right) \frac{\delta }{\delta V_{1}\left(
x_{i_{1}}\right) }...\frac{\delta }{\delta V_{1}\left( x_{i_{k}}\right) }%
\right) \\
&&\times \dsum\limits_{N}\frac{1}{N!}\dprod\limits_{i=1}^{N}\int \exp \left(
-\alpha t_{i}\right) \int \mathcal{D}X_{i}\left( t\right) \exp \left(
-\sum_{i}\int_{X_{i}\left( 0\right) =x_{i}}^{X_{i}\left( t_{i}\right)
=y_{i}}\left( \left( \frac{1}{2}\left( \frac{d}{dt}X_{i}\left( t\right)
\right) ^{2}+V_{1}\left( X_{i}\left( t\right) \right) \right) dt\right)
\right.
\end{eqnarray*}%
In other words, using (\ref{Nagentscbbr}) one finds the partition function
for the system of agents in interaction:%
\begin{eqnarray}
&&\dsum\limits_{N}\frac{1}{N!}\dprod\limits_{i=1}^{N}\int \exp \left(
-\alpha t_{i}\right) \int \mathcal{D}X_{i}\left( t\right) \exp \left(
-\sum_{i}\int_{X_{i}\left( 0\right) =x_{i}}^{X_{i}\left( t_{i}\right)
=y_{i}}\left( \left( \frac{1}{2}\left( \frac{d}{dt}X_{i}\left( t\right)
\right) ^{2}+V_{1}\left( X_{i}\left( t\right) \right) \right) dt\right)
\right.  \label{gdpt} \\
&&\hspace{0.48in}\hspace{0.48in}\hspace{0.48in}\hspace{0.48in}-\left.
\sum_{i_{1},...i_{k}}\int_{X_{i}\left( 0\right) =x_{i}}^{X_{i}\left(
s\right) =y_{i}}V_{k}\left( X_{i_{1}}\left( t_{1}\right) ...X_{i_{k}}\left(
t_{k}\right) \right) dt_{1}...dt_{k}\right)  \notag \\
&=&\exp \left( -\int dx_{i_{1}}...dx_{i_{k}}V_{k}\left(
x_{i_{1}}...x_{i_{k}}\right) \frac{\delta }{\delta V_{1}\left(
x_{i_{1}}\right) }...\frac{\delta }{\delta K\left( x_{i_{k}}\right) }\right)
_{K\equiv 0}\times \dsum\limits_{N}\frac{1}{N!}\dprod\limits_{i=1}^{N}G_{%
\alpha }\left( x_{i},y_{i}\right)  \notag
\end{eqnarray}

\bigskip We can now find the field formulation associated to (\ref{gdpt}).
We have seen that (see (\ref{grnnfn})):%
\begin{eqnarray*}
&&\dprod\limits_{i=1}^{N}G_{\alpha }\left( x_{i},y_{i},\alpha \right) =\det
\left( -\frac{1}{2}\nabla ^{2}+\alpha +K\left( x\right) \right) ^{-1}\times %
\left[ \left( \frac{\delta }{\delta J\left( x_{i_{1}}\right) }\frac{\delta }{%
\delta J^{\dag }\left( x_{i_{1}}\right) }\right) ...\left( \frac{\delta }{%
\delta J\left( x_{i_{N}}\right) }\frac{\delta }{\delta J^{\dag }\left(
x_{i_{N}}\right) }\right) \right] _{J=J^{\dag }=0} \\
&&.\times \int \exp \left( -\int \!\!\left( \Psi \left( x\right)\!\!\left( -\frac{1%
}{2}\nabla ^{2}+\alpha +K\left( x\right) \right) \Psi ^{\dag }\left(
x\right) +J\left( x\right) \Psi ^{\dag }\left( x\right) +J^{\dag }\left(
x\right) \Psi \left( x\right) \right) dx\right) \mathcal{D}\Psi \mathcal{D}%
\Psi ^{\dag }
\end{eqnarray*}%
so that, discarding the constant factor $\det \left( -\frac{1}{2}\nabla
^{2}+\alpha +K\left( x\right) \right) ^{-1}$, one has:%
\begin{eqnarray*}
&&\dsum\limits_{N}\frac{1}{N!}\dprod\limits_{i=1}^{N}\int \exp \left(
-\alpha t_{i}\right) \int \mathcal{D}X_{i}\left( t\right) \exp \left(
-\sum_{i}\int_{X_{i}\left( 0\right) =x_{i}}^{X_{i}\left( t_{i}\right)
=y_{i}}\left( \left( \frac{1}{2}\left( \frac{d}{dt}X_{i}\left( t\right)
\right) ^{2}+V_{1}\left( X_{i}\left( t\right) \right) \right) dt\right)
\right. \\
&&\hspace{0.48in}\hspace{0.48in}\hspace{0.48in}\hspace{0.48in}-\left.
\sum_{i_{1},...i_{k}}\int_{X_{i}\left( 0\right) =x_{i}}^{X_{i}\left(
s\right) =y_{i}}V_{k}\left( X_{i_{1}}\left( t_{1}\right) ...X_{i_{k}}\left(
t_{k}\right) \right) dt_{1}...dt_{k}\right) \\
&=&\exp \left( -\int dx_{i_{1}}...dx_{i_{k}}V_{k}\left(
x_{i_{1}}...x_{i_{k}}\right) \frac{\delta }{\delta V_{1}\left(
x_{i_{1}}\right) }...\frac{\delta }{\delta K\left( x_{i_{k}}\right) }\right)
_{K\equiv 0} \\
&&\times \left[ \left( \frac{\delta }{\delta J\left( x_{i_{1}}\right) }\frac{%
\delta }{\delta J^{\dag }\left( x_{i_{1}}\right) }\right) ...\left( \frac{%
\delta }{\delta J\left( x_{i_{N}}\right) }\frac{\delta }{\delta J^{\dag
}\left( x_{i_{N}}\right) }\right) \right] _{J=J^{\dag }=0} \\
&&\times \int \exp \left( -\int \!\!\left( \Psi \left( x\right)\!\!\left( -\frac{1}{%
2}\nabla ^{2}+\alpha +K\left( x\right) \right) \Psi ^{\dag }\left( x\right)
+J\left( x\right) \Psi ^{\dag }\left( x\right) +J^{\dag }\left( x\right)
\Psi \left( x\right) \right) dx\right) \mathcal{D}\Psi \mathcal{D}\Psi
^{\dag }
\end{eqnarray*}%
and this quantity is equal to:%
\begin{eqnarray*}
&&\left[ \left( \frac{\delta }{\delta J\left( x_{i_{1}}\right) }\frac{\delta 
}{\delta J^{\dag }\left( x_{i_{1}}\right) }\right) ...\left( \frac{\delta }{%
\delta J\left( x_{i_{N}}\right) }\frac{\delta }{\delta J^{\dag }\left(
x_{i_{N}}\right) }\right) \right] _{J=J^{\dag }=0} \\
&&\times \int \exp \left( -\int \Psi \left( x\right)\!\!\left( -\frac{1}{2}%
\nabla ^{2}+\alpha +V_{1}\left( x\right) \right) \Psi ^{\dag }\left(
x\right) dx-\int J\left( x\right) \Psi ^{\dag }\left( x\right) +J^{\dag
}\left( x\right) \Psi \left( x\right) dx\right. \\
&&\left. -\sum_{i_{1},...i_{k}}\int \Psi \left( x_{i_{1}}\right) ...\Psi
\left( x_{i_{k}}\right) V_{k}\left( x_{i_{1}}...x_{i_{k}}\right) \Psi ^{\dag
}\left( x_{i_{1}}\right) ...\Psi ^{\dag }\left( x_{i_{k}}\right)
dx_{i_{1}}...x_{i_{k}}\right) \mathcal{D}\Psi \mathcal{D}\Psi ^{\dag }
\end{eqnarray*}%
In other word, the probabilitic description (\ref{gdpr}) of a large number
of interacting agents is encompassed in the path integral:%
\begin{eqnarray*}
&&\int \exp \left( -\int \Psi \left( x\right)\!\!\left( -\frac{1}{2}\nabla
^{2}+\alpha +V_{1}\left( x\right) \right) \Psi ^{\dag }\left( x\right)
dx-\int J\left( x\right) \Psi ^{\dag }\left( x\right) +J^{\dag }\left(
x\right) \Psi \left( x\right) dx\right. \\
&&\left. -\sum_{i_{1},...i_{k}}\int \Psi \left( x_{i_{1}}\right) ...\Psi
\left( x_{i_{k}}\right) V_{k}\left( x_{i_{1}}...x_{i_{k}}\right) \Psi ^{\dag
}\left( x_{i_{1}}\right) ...\Psi ^{\dag }\left( x_{i_{k}}\right)
dx_{i_{1}}...x_{i_{k}}\right) \mathcal{D}\Psi \mathcal{D}\Psi ^{\dag }
\end{eqnarray*}%
The functional:%
\begin{eqnarray*}
S\left( \Psi ,J\right) &=&\int \Psi \left( x\right)\!\!\left( -\frac{1}{2}%
\nabla ^{2}+\alpha +V_{1}\left( x\right) \right) \Psi ^{\dag }\left(
x\right) dx \\
&&+\sum_{i_{1},...i_{k}}\int \Psi \left( x_{i_{1}}\right) ...\Psi \left(
x_{i_{k}}\right) V_{k}\left( x_{i_{1}}...x_{i_{k}}\right) \Psi ^{\dag
}\left( x_{i_{1}}\right) ...\Psi ^{\dag }\left( x_{i_{k}}\right)
dx_{i_{1}}...x_{i_{k}}+\int J\left( x\right) \Psi ^{\dag }\left( x\right)
+J^{\dag }\left( x\right) \Psi \left( x\right) dx
\end{eqnarray*}%
is a particular case of a field action functional with source as defined in
section 2. It is straightforward to generalize this formula for any type of
potential involving an arbitrary number of agents by introducing over $k$
yielding an action:%
\begin{eqnarray*}
S\left( \Psi ,J\right) &=&\int \Psi \left( x\right)\!\!\left( -\frac{1}{2}%
\nabla ^{2}+\alpha +V_{1}\left( x\right) \right) \Psi ^{\dag }\left(
x\right) dx \\
&&+\sum_{k\geqslant 2}\sum_{i_{1},...i_{k}}\int \Psi \left( x_{i_{1}}\right)
...\Psi \left( x_{i_{k}}\right) V_{k}\left( x_{i_{1}}...x_{i_{k}}\right)
\Psi ^{\dag }\left( x_{i_{1}}\right) ...\Psi ^{\dag }\left( x_{i_{k}}\right)
dx_{i_{1}}...x_{i_{k}}+\int J\left( x\right) \Psi ^{\dag }\left( x\right)
+J^{\dag }\left( x\right) \Psi \left( x\right) dx
\end{eqnarray*}

As a consequence:%
\begin{eqnarray}
&&\int \exp \left( -\int \Psi \left( x\right)\!\!\left( -\frac{1}{2}\nabla
^{2}+\alpha +V_{1}\left( x\right) \right) \Psi ^{\dag }\left( x\right)
dx\right.  \label{Phtntgrl} \\
&&\left. -\sum_{k\geqslant 2}\sum_{i_{1},...i_{k}}\int \Psi \left(
x_{i_{1}}\right) ...\Psi \left( x_{i_{k}}\right) V_{k}\left(
x_{i_{1}}...x_{i_{k}}\right) \Psi ^{\dag }\left( x_{i_{1}}\right) ...\Psi
^{\dag }\left( x_{i_{k}}\right) dx_{i_{1}}...x_{i_{k}}+\int J\left( x\right)
\Psi ^{\dag }\left( x\right) +J^{\dag }\left( x\right) \Psi \left( x\right)
dx\right) \mathcal{D}\Psi \mathcal{D}\Psi ^{\dag }  \notag
\end{eqnarray}%
computes, by successive derivatives with respect to $J\left( x\right) $\ and 
$J^{\dag }\left( x\right) $, the transition functions of a system of
infinite number of identical agents, with arbitrary, non local in time,
interactions $V_{k}\left( X_{i_{1}}\left( t_{1}\right) ...X_{i_{k}}\left(
t_{k}\right) \right) $\ involving $k$\ agents, with $k$\ arbitrary. The
constant $\alpha $\ is the characteristic scale of the interaction process,
and $\frac{1}{\alpha }$\ the mean duration of the interaction process, or
alternately the mean lifespan of the agents. The transition functions are
given by:%
\begin{eqnarray}
&&G_{K}\left( \left\{ x_{i}\right\} ,\left\{ y_{i}\right\} ,\alpha \right)
\label{trnsgr} \\
&=&\left[ \left( \frac{\delta }{\delta J\left( x_{i_{1}}\right) }\frac{%
\delta }{\delta J^{\dag }\left( y_{i_{1}}\right) }\right) ...\left( \frac{%
\delta }{\delta J\left( x_{i_{N}}\right) }\frac{\delta }{\delta J^{\dag
}\left( y_{i_{N}}\right) }\right) \int \exp \left( -\Psi \left( x\right)
\left( -\frac{1}{2}\nabla ^{2}+\alpha +K\left( x\right) \right) \Psi ^{\dag
}\left( x\right) \right. \right.  \notag \\
&&\left. \left. -\sum_{k\geqslant 2}\sum_{i_{1},...i_{k}}\Psi \left(
x_{i_{1}}\right) ...\Psi \left( x_{i_{k}}\right) V_{k}\left(
x_{i_{1}}...x_{i_{k}}\right) \Psi ^{\dag }\left( x_{i_{1}}\right) ...\Psi
^{\dag }\left( x_{i_{k}}\right) +J\left( x\right) \Psi ^{\dag }\left(
x\right) +J^{\dag }\left( x\right) \Psi \left( x\right) \right) \mathcal{D}%
\Psi \mathcal{D}\Psi ^{\dag }\right] _{J=J^{\dag }=0}  \notag
\end{eqnarray}%
and $G_{K}\left( \left\{ x_{i}\right\} ,\left\{ y_{i}\right\} ,\alpha
\right) $\ is the probability of transition for $N$\ agents from a state $%
\left\{ x_{i}\right\} $\ to a state $\left\{ y_{i}\right\} $. Remark that
this formulation realizes what was announced before. The switch in
formulation induces that the transition of the agents, i.e. their dynamical
and stochastic properties, takes place in a surrounding. Instead of
computing directly the dynamic of the system, we derive this behavior from
the global properties of a substratum, the global action for the field $\Psi
\left( x\right) $. By global action we denote the functional, or action:%
\begin{equation*}
S\left( \Psi \right) =\int dx\left( \Psi \left( x\right)\!\!\left( -\frac{1}{2}%
\nabla ^{2}+\alpha +K\left( x\right) \right) \Psi ^{\dag }\left( x\right)
+\sum_{k\geqslant 2}^{A}\sum_{i_{1},...i_{k}}\Psi \left( x_{i_{1}}\right)
...\Psi \left( x_{i_{k}}\right) V_{k}\left( x_{i_{1}}...x_{i_{k}}\right)
\Psi ^{\dag }\left( x_{i_{1}}\right) ...\Psi ^{\dag }\left( x_{i_{k}}\right)
\right)
\end{equation*}

\section*{Appendix 5}

\bigskip We start with the statistical weight associated to the
intertemporal budget constraint:%
\begin{eqnarray*}
&&\exp \left( -\sum_{s}\frac{1}{\varpi ^{2}}\left( C_{s}-\bar{C}-\frac{%
C_{s+1}}{\left( 1+r\right) }\right) ^{2}+\sum_{s}C_{0}\right) \exp \left( -%
\frac{\left( \int_{0}^{T}\left( Y_{i}\left( s\right) -C_{i}\left( s\right)
\right) \exp \left( -\int r_{i}\left( s\right) ds\right) ds\right) ^{2}}{%
\theta ^{2}}\right) \\
&\simeq &\exp \left( -\sum_{s}\frac{\left( C_{s+1}-C_{s}-rC_{s+1}\right) ^{2}%
}{\varpi ^{2}}+\sum_{s}C_{0}\right) \exp \left( -\frac{\left(
\int_{0}^{T}\left( Y_{i}\left( s\right) -C_{i}\left( s\right) \right) \exp
\left( -\int r_{i}\left( s\right) ds\right) ds\right) ^{2}}{\theta ^{2}}%
\right)
\end{eqnarray*}%
with $C_{0}\equiv \frac{1}{2\sigma \varpi ^{2}}$. Now, remark that the budgt
constraint:%
\begin{equation*}
\int \!\!\left( Y_{i}\left( s\right) -C_{s}\right) \exp \left( -rs\right) ds=0
\end{equation*}%
can be expressed as:%
\begin{equation*}
\int \!\!\left( \dot{K}_{i}\left( s\right) +\varepsilon \left( s\right) \right)
\exp \left( -rs\right) ds=0
\end{equation*}%
or as: 
\begin{equation*}
\int \dot{K}_{i}\left( s\right) \exp \left( -rs\right) ds=-\int \varepsilon
\left( s\right) \exp \left( -rs\right) ds
\end{equation*}%
The last term has variance $\frac{\nu ^{2}}{2r}$ which implies that the
overall constraint can be included in the global weight through a term:%
\begin{equation*}
\frac{2\bar{r}}{\nu ^{2}}\left( \int \dot{K}_{i}\left( s\right) \exp \left(
-rs\right) ds\right) ^{2}
\end{equation*}%
with: 
\begin{equation*}
\bar{r}=\frac{1}{\int \exp \left( -rs\right) ds}
\end{equation*}%
\begin{eqnarray*}
\int \dot{K}_{i}\left( s\right) \exp \left( -rs\right) ds &=&\left[
K_{i}\left( s\right) \exp \left( -rs\right) \right] _{0}^{T}+\int
rK_{i}\left( s\right) \exp \left( -rs\right) ds \\
&=&-K_{i}\left( 0\right) +\int rK_{i}\left( s\right) \exp \left( -rs\right)
ds
\end{eqnarray*}%
if the transversality condition is satisfied. At the lowest order in $\bar{r}
$ or $r$, the contribution is approximated by $K_{i}\left( 0\right) $ and
can be neglected. We end up with a contribution:%
\begin{equation*}
\exp \left( -\sum_{s}\frac{1}{\varpi ^{2}}\left( C_{s}-\bar{C}-\frac{\left(
C_{s+1}\right) }{\left( 1+r\right) }\right) ^{2}+\sum_{s}C_{0}\right)
\end{equation*}%
or in continuous time:%
\begin{equation*}
\exp \left( -\int ds\frac{\left( \dot{C}_{s}-rC_{s}+\bar{C}\right) ^{2}}{%
\varpi ^{2}}+C_{0}\int ds\right)
\end{equation*}

\section*{Appendix 6}

\subsection*{Business cycle model, field theoretic representation}

The field theoretic equivalent of (\ref{ndsttwgth}) is obtained by the same
methods we used previously. One obtains the following action for the field:

\begin{eqnarray*}
&&S\left( \Psi \right) =\Psi ^{\dag }\left( K,C,A\right)\!\!\left( -\nabla
.\left( \left( 
\begin{array}{ccc}
\nu ^{2} & 0 & 0 \\ 
0 & \varpi ^{2} & 0 \\ 
0 & 0 & \frac{1}{\lambda ^{2}}%
\end{array}%
\right) \nabla +2\left( 
\begin{array}{c}
-AF\left( K\right) +C+\delta \left( K\right) \\ 
-\left( AF^{\prime }\left( K\right) +r_{c}\right) C+\bar{C} \\ 
0%
\end{array}%
\right) \right) \right. \\
&&\hspace{3cm}\left. +\varsigma ^{2}\left( C-\bar{C}\right) ^{2}+\left(
A_{i}-\bar{A}\right) ^{2}+\left( -\left( AF\left( K\right) -\delta \left(
K\right) \right) ^{\prime }-AF^{\prime }\left( K\right) -r_{c}\right)
+\alpha -C_{0}\right) \Psi \left( K,C,A\right) \\
&&\hspace{3cm}+\gamma \Psi ^{\dag }\left( K_{1},C_{1},A_{1}\right) \Psi
^{\dag }\left( K_{2},C_{2},A_{2}\right) \left\{ A_{2}H\left(
K_{1},K_{2}\right) K_{1}\right\} \Psi \left( K_{1},C_{1},A_{1}\right) \Psi
\left( K_{2},C_{2},A_{2}\right) \\
&=&\int \Psi ^{\dag }\left( K,C,A\right) \left\{ -\varpi ^{2}\frac{\partial
^{2}}{\partial C^{2}}-\frac{1}{\lambda ^{2}}\frac{\partial ^{2}}{\partial
A^{2}}+\varsigma ^{2}\left( C-\bar{C}\right) ^{2}+\left( A_{i}-\bar{A}%
\right) ^{2}-\nu ^{2}\frac{\partial ^{2}}{\partial K^{2}}\right. \\
&&+\left( -2\left( C-AF\left( K\right) +\delta \left( K\right) \right) \frac{%
\partial }{\partial K}+2\left( AF^{\prime }\left( K\right) +r_{c}\right)
\left( C-\bar{C}\right) \frac{\partial }{\partial C}+2\frac{\partial }{%
\partial K}\left( AF\left( K\right) -\delta \left( K\right) \right) \right)
\\
&&\hspace{3cm}\hspace{3cm}\left. +2\left( AF^{\prime }\left( K\right)
+r_{c}\right) \right\} \Psi \left( K,C,A\right) \\
&&+\int \Psi ^{\dag }\left( K,C,A\right)\!\!\left( -\left( \frac{\partial }{%
\partial K}\left( AF\left( K\right) -\delta \left( K\right) \right) \right)
-\left( AF^{\prime }\left( K\right) +r_{c}\right) +\alpha -C_{0}\right) \Psi
\left( K,C,A\right) \\
&&+\gamma \frac{1}{2}\int \Psi ^{\dag }\left( K_{1},C_{1},A_{1}\right) \Psi
^{\dag }\left( K_{2},C_{2},A_{2}\right) \left\{ A_{2}H\left(
K_{1},K_{2}\right) K_{1}+A_{1}H\left( K_{2},K_{1}\right) K_{2}\right\} \Psi
\left( K_{1},C_{1},A_{1}\right) \Psi \left( K_{2},C_{2},A_{2}\right)
\end{eqnarray*}%
Let $\delta \left( K\right) =\delta K$ as usually assumed. The previous
expression simplifies as:%
\begin{eqnarray*}
&&S\left( \Psi \right) =\int \Psi ^{\dag }\left( K,C,A\right) \left\{
-\varpi ^{2}\frac{\partial ^{2}}{\partial C^{2}}-\frac{1}{\lambda ^{2}}\frac{%
\partial ^{2}}{\partial A^{2}}-\nu ^{2}\frac{\partial ^{2}}{\partial K^{2}}%
+\left( A-\bar{A}\right) ^{2}-2\left( C-AF\left( K\right) +\delta K\right) 
\frac{\partial }{\partial K}\right. \\
&&\hspace{3cm}\left. +2\left( AF^{\prime }\left( K\right) +r_{c}\right)
\left( C-\bar{C}\right) \frac{\partial }{\partial C}+\varsigma ^{2}\left( C-%
\bar{C}\right) ^{2}\right\} \Psi \left( K,C,A\right) \\
&&+\int \Psi ^{\dag }\left( K,C,A\right)\!\!\left( \alpha +2AF^{\prime }\left(
K\right) +\left( r_{c}-\delta \right) -C_{0}\right) \Psi \left( K,C,A\right)
\\
&&+\gamma \frac{1}{2}\int \Psi ^{\dag }\left( K_{1},C_{1},A_{1}\right) \Psi
^{\dag }\left( K_{2},C_{2},A_{2}\right) \left\{ A_{2}H\left(
K_{1},K_{2}\right) K_{1}+A_{1}H\left( K_{2},K_{1}\right) K_{2}\right\} \Psi
\left( K_{1},C_{1},A_{1}\right) \Psi \left( K_{2},C_{2},A_{2}\right)
\end{eqnarray*}%
If we consider that the rate $\left( AF^{\prime }\left( K\right)
+r_{c}\right) $ is slowly varying, as an interest rate, we can perform a
change of variable:

\begin{eqnarray*}
\Psi \left( K,C,A\right) &=&\exp \left( \frac{1}{2\varpi ^{2}}\left(
AF^{\prime }\left( K\right) +r_{c}\right)\!\!\left( C-\bar{C}\right)
^{2}\right) \hat{\Psi}\left( K,C,A\right) \\
\hat{\Psi}^{\dag }\left( K,C,A\right) &=&\exp \left( -\frac{1}{2\varpi ^{2}}%
\left( AF^{\prime }\left( K\right) +r_{c}\right)\!\!\left( C-\bar{C}\right)
^{2}\right) \hat{\Psi}\left( K,C,A\right)
\end{eqnarray*}%
and rewrite the action as a function of $\hat{\Psi}$:%
\begin{eqnarray*}
&&S\left( \hat{\Psi}\right) =\int \hat{\Psi}^{\dag }\left( K,C,A\right)
\left\{ -\varpi ^{2}\frac{\partial ^{2}}{\partial C^{2}}-\frac{1}{\lambda
^{2}}\frac{\partial ^{2}}{\partial A^{2}}-\nu ^{2}\frac{\partial ^{2}}{%
\partial K^{2}}+\left( A-\bar{A}\right) ^{2}-2\left( C-AF\left( K\right)
+\delta K\right) \frac{\partial }{\partial K}\right. \\
&&\hspace{3cm}\left. +\left( \varsigma ^{2}+\frac{\left( AF^{\prime }\left(
K\right) +r_{c}\right) ^{2}}{\varpi ^{2}}\right)\!\!\left( C-\bar{C}\right)
^{2}+\alpha -C_{0}+AF^{\prime }\left( K\right) -\delta \right\} \hat{\Psi}%
\left( K,C,A\right) \\
&&+\gamma \frac{1}{2}\int \hat{\Psi}^{\dag }\left( K_{1},C_{1},A_{1}\right) 
\hat{\Psi}^{\dag }\left( K_{2},C_{2},A_{2}\right) \left\{ A_{2}H\left(
K_{1},K_{2}\right) K_{1}+A_{1}H\left( K_{2},K_{1}\right) K_{2}\right\} \hat{%
\Psi}\left( K_{1},C_{1},A_{1}\right) \hat{\Psi}\left(
K_{2},C_{2},A_{2}\right)
\end{eqnarray*}%
Then, a change of variable

\begin{eqnarray*}
K^{\prime } &=&C-AF\left( K\right) +\delta K \\
\frac{\partial }{\partial K} &=&\left( \delta -AF^{\prime }\left( K\right)
\right) \frac{\partial }{\partial K^{\prime }}
\end{eqnarray*}%
associated also with the assumption that the rate $\delta -AF^{\prime
}\left( K\right) $ slowly varying, as well as a rescaling $\frac{\bar{C}}{%
\left( AF^{\prime }\left( K\right) +r_{c}\right) }\rightarrow \bar{C}$ leads
to:%
\begin{eqnarray*}
&&S\left( \hat{\Psi}\right) =\int \hat{\Psi}^{\dag }\left( K^{\prime
},C,A\right) \left\{ -\varpi ^{2}\frac{\partial ^{2}}{\partial C^{2}}-\frac{1%
}{\lambda ^{2}}\frac{\partial ^{2}}{\partial A^{2}}-\nu ^{2}\left( \delta
-AF^{\prime }\left( K\right) \right) ^{2}\frac{\partial ^{2}}{\partial
K^{\prime 2}}+\left( A-\bar{A}\right) ^{2}-2\left( \delta -AF^{\prime
}\left( K\right) \right) K^{\prime }\frac{\partial }{\partial K^{\prime }}%
\right. \\
&&\hspace{3cm}\left. +\left( \varsigma ^{2}+\frac{\left( AF^{\prime }\left(
K\right) +r_{c}\right) ^{2}}{\varpi ^{2}}\right)\!\!\left( C-\bar{C}\right)
^{2}+\alpha -C_{0}+AF^{\prime }\left( K\right) -\delta \right\} \hat{\Psi}%
\left( K^{\prime },C,A\right) \\
&&+\gamma \frac{1}{2}\int \hat{\Psi}^{\dag }\left( K_{1}^{\prime
},C_{1},A_{1}\right) \hat{\Psi}^{\dag }\left( K_{2}^{\prime
},C_{2},A_{2}\right) \left\{ A_{2}H\left( K_{1},K_{2}\right)
K_{1}+A_{1}H\left( K_{2},K_{1}\right) K_{2}\right\} \hat{\Psi}\left(
K_{1}^{\prime },C_{1},A_{1}\right) \hat{\Psi}\left( K_{2}^{\prime
},C_{2},A_{2}\right)
\end{eqnarray*}%
Ultimately, one can recast the action in a tractable form through a second
rescaling of the field:%
\begin{eqnarray*}
\hat{\Psi}\left( K^{\prime },C,A\right) &=&\exp \left( -\frac{\left(
K^{\prime }\right) ^{2}}{2\nu ^{2}\left( \delta -AF^{\prime }\left( K\right)
\right) }\right) \bar{\Psi}\left( K^{\prime },C,A\right) \\
\hat{\Psi}^{\dag }\left( K^{\prime },C,A\right) &=&\exp \left( \frac{\left(
K^{\prime }\right) ^{2}}{2\nu ^{2}\left( \delta -AF^{\prime }\left( K\right)
\right) }\right) \bar{\Psi}\left( K^{\prime },C,A\right)
\end{eqnarray*}%
and thus:%
\begin{eqnarray*}
&&S\left( \bar{\Psi}\right) =\int \bar{\Psi}^{\dag }\left( K^{\prime
},C,A\right) \left\{ -\varpi ^{2}\frac{\partial ^{2}}{\partial C^{2}}-\frac{1%
}{\lambda ^{2}}\frac{\partial ^{2}}{\partial A^{2}}-\nu ^{2}\left( \delta
-AF^{\prime }\left( K\right) \right) ^{2}\frac{\partial ^{2}}{\partial
K^{\prime 2}}+\left( A-\bar{A}\right) ^{2}\right. \\
&&\hspace{3cm}\left. +\left( \varsigma ^{2}+\frac{\left( AF^{\prime }\left(
K\right) +r_{c}\right) ^{2}}{\varpi ^{2}}\right)\!\!\left( C-\bar{C}\right)
^{2}+\frac{\left( K^{\prime }\right) ^{2}}{\nu ^{2}}+\alpha -C_{0}\right\} 
\bar{\Psi}\left( K^{\prime },C,A\right) \\
&&+\gamma \frac{1}{2}\int \bar{\Psi}^{\dag }\left( K_{1}^{\prime
},C_{1},A_{1}\right) \bar{\Psi}^{\dag }\left( K_{2}^{\prime
},C_{2},A_{2}\right) \left\{ A_{2}H\left( K_{1},K_{2}\right)
K_{1}+A_{1}H\left( K_{2},K_{1}\right) K_{2}\right\} \bar{\Psi}\left(
K_{1}^{\prime },C_{1},A_{1}\right) \bar{\Psi}\left( K_{2}^{\prime
},C_{2},A_{2}\right)
\end{eqnarray*}%
The relation between $\Psi \left( K,C,A\right) $ and $\bar{\Psi}\left(
K^{\prime },C,A\right) $ 
\begin{eqnarray*}
\Psi \left( K,C,A\right) &=&\exp \left( \frac{1}{2\varpi ^{2}}\left(
AF^{\prime }\left( K\right) +r_{c}\right)\!\!\left( C-\bar{C}\right)
^{2}\right) \exp \left( -\frac{\left( K^{\prime }\right) ^{2}}{2\nu
^{2}\left( \delta -AF^{\prime }\left( K\right) \right) }\right) \bar{\Psi}%
\left( K^{\prime },C,A\right) \\
\Psi ^{\dag }\left( K,C,A\right) &=&\exp \left( -\frac{1}{2\varpi ^{2}}%
\left( AF^{\prime }\left( K\right) +r_{c}\right)\!\!\left( C-\bar{C}\right)
^{2}\right) \exp \left( \frac{\left( K^{\prime }\right) ^{2}}{2\nu
^{2}\left( \delta -AF^{\prime }\left( K\right) \right) }\right) \bar{\Psi}%
^{\dag }\left( K^{\prime },C,A\right)
\end{eqnarray*}%
implies at first sight that $\bar{\Psi}\left( K^{\prime },C,A\right) $ and $%
\bar{\Psi}^{\dag }\left( K^{\prime },C,A\right) $ are not complex conjugate.
This is the consequence from the fact that the operator involved in the
definition of $S\left( \Psi \right) $ is not hermitian, or self adjoint in
the real interpretation. This non hermiticity is itself the consequence of
an asymmetry in the transition functions: due to a drift term, the
transition probability between two points is not symmetric. However, one can
make sense of the partition function:%
\begin{equation}
\int \exp \left( -S\left( \bar{\Psi}\right) \right) \mathcal{D\bar{\Psi}D}%
\bar{\Psi}^{\dag }  \label{prtf1}
\end{equation}%
and show that it computes the same partition function as:%
\begin{equation}
\int \exp \left( -S\left( \Psi \right) \right) \mathcal{D\Psi D}\Psi ^{\dag }
\label{prtf2}
\end{equation}%
To do so, we first define:%
\begin{eqnarray*}
L &=&-\varpi ^{2}\frac{\partial ^{2}}{\partial C^{2}}-\frac{1}{\lambda ^{2}}%
\frac{\partial ^{2}}{\partial A^{2}}-\nu ^{2}\frac{\partial ^{2}}{\partial
K^{2}}+\left( A-\bar{A}\right) ^{2}-2\left( C-AF\left( K\right) +\delta
K\right) \frac{\partial }{\partial K}+2\left( AF^{\prime }\left( K\right)
+r_{c}\right)\!\!\left( C-\bar{C}\right) \frac{\partial }{\partial C} \\
&&+\varsigma ^{2}\left( C-\bar{C}\right) ^{2}+\alpha -C_{0} \\
L^{\prime } &=&-\varpi ^{2}\frac{\partial ^{2}}{\partial C^{2}}-\frac{1}{%
\lambda ^{2}}\frac{\partial ^{2}}{\partial A^{2}}-\nu ^{2}\left( \delta
-AF^{\prime }\left( K\right) \right) ^{2}\frac{\partial ^{2}}{\partial
K^{\prime 2}}+\left( A-\bar{A}\right) ^{2}+\left( \varsigma ^{2}+\frac{%
\left( AF^{\prime }\left( K\right) +r_{c}\right) ^{2}}{\varpi ^{2}}\right)
\left( C-\bar{C}\right) ^{2} \\
&&+\frac{\left( K^{\prime }\right) ^{2}}{\nu ^{2}}+\alpha -C_{0}
\end{eqnarray*}%
and then:%
\begin{equation*}
\Psi ^{\dag }\left( K,C,A\right) L\Psi \left( K,C,A\right) =\bar{\Psi}^{\dag
}\left( K^{\prime },C,A\right) L^{\prime }\bar{\Psi}\left( K^{\prime
},C,A\right)
\end{equation*}%
so that:%
\begin{eqnarray*}
L^{\prime } &=&\exp \left( -\frac{1}{2\varpi ^{2}}\left( AF^{\prime }\left(
K\right) +r_{c}\right)\!\!\left( C-\bar{C}\right) ^{2}+\frac{\left( K^{\prime
}\right) ^{2}}{2\nu ^{2}\left( \delta -AF^{\prime }\left( K\right) \right) }%
\right) \\
&&\times L\exp \left( \frac{1}{2\varpi ^{2}}\left( AF^{\prime }\left(
K\right) +r_{c}\right)\!\!\left( C-\bar{C}\right) ^{2}-\frac{\left( K^{\prime
}\right) ^{2}}{2\nu ^{2}\left( \delta -AF^{\prime }\left( K\right) \right) }%
\right)
\end{eqnarray*}%
Then, to make sense of the partition function (\ref{prtf1}), we will compare
it to the computation of (\ref{prtf2}). To do so, recall that the partition
function for $\gamma =0$, that is (\ref{prtf2}), is defined as $\det \left(
L^{-1}\right) $ and that this quantities is computed via the eigenvalues of $%
L$:%
\begin{equation*}
\det L^{-1}=\dprod\limits_{n}dx_{n}dy_{n}\exp \left( -x_{n}\lambda
_{n}x_{n}-y_{n}\lambda _{n}y_{n}\right)
\end{equation*}%
This expression makes sense since the eigenvalues of operator $L$ have
positive real part. Let $\alpha _{n}=x_{n}+iy_{n}$ and $\alpha _{n}^{\dag
}=x_{n}-iy_{n}$, $\det L^{-1}$ rewrites: 
\begin{equation*}
\det L^{-1}=\dprod\limits_{n}d\alpha _{n}d\alpha _{n}^{\dag }\exp \left(
\alpha _{n}^{\dag }\lambda _{n}\alpha _{n}\right)
\end{equation*}%
This expression is real, since $L$ is a real operator, and if $\lambda _{n}$
is an eigenvalue of $L$, so is $\bar{\lambda}_{n}$. Consider the expansion
of $\Psi \left( K,C,A\right) $: 
\begin{equation*}
\Psi \left( K,C,A\right) =\sum \alpha _{n}\Psi _{n}\left( K,C,A\right)
\end{equation*}%
where $\Psi _{n}\left( K,C,A\right) $ are eigenfunctions for $\lambda _{n}$
of $L$, then, define $\Psi ^{\dag }\left( K,C,A\right) $ as: 
\begin{equation}
\Psi ^{\dag }\left( K,C,A\right) =\sum \alpha _{n}^{\dag }\Psi _{n}^{\dag
}\left( K,C,A\right)  \label{xpnbr}
\end{equation}%
where $\Psi _{n}^{\dag }\left( K,C,A\right) $, eigenfunctions for $\lambda
_{n}$ of the adjoint $L^{+}$, and $\left\langle \Psi _{m}^{\dag }\left(
K,C,A\right) ,\Psi _{n}\left( K,C,A\right) \right\rangle =\delta _{m,n}$.

As a consequence, the partition function rewrites:%
\begin{equation*}
\det L^{-1}=\exp \left( -S\left( \Psi \right) \right) \mathcal{D\Psi D}\Psi
^{\dag }
\end{equation*}%
This is (\ref{prtf2}), but the field $\Psi ^{\dag }$ is not the complex
conjugate of $\mathcal{\Psi }$, and has rather to be understood as given by
the expansion (\ref{xpnbr}). Now, focusing on (\ref{prtf1}), consider the
transformed eigenfunctions: 
\begin{equation*}
\bar{\Psi}_{n}\left( K^{\prime },C,A\right) =\exp \left( -\frac{1}{2\varpi
^{2}}\left( AF^{\prime }\left( K\right) +r_{c}\right)\!\!\left( C-\bar{C}%
\right) ^{2}+\frac{\left( K^{\prime }\right) ^{2}}{2\nu ^{2}\left( \delta
-AF^{\prime }\left( K\right) \right) }\right) \Psi _{n}\left( K,C,A\right)
\end{equation*}%
that are eigenfunctions of $L^{\prime }$ for eigenvalues $\lambda _{n}$.
Actually:%
\begin{eqnarray}
L^{\prime }\bar{\Psi}_{n}\left( K^{\prime },C,A\right) &=&\exp \left( -\frac{%
1}{2\varpi ^{2}}\left( AF^{\prime }\left( K\right) +r_{c}\right)\!\!\left( C-%
\bar{C}\right) ^{2}+\frac{\left( K^{\prime }\right) ^{2}}{2\nu ^{2}\left(
\delta -AF^{\prime }\left( K\right) \right) }\right) L\Psi _{n}\left(
K,C,A\right)  \label{gnvl} \\
&=&\lambda _{n}\bar{\Psi}_{n}\left( K^{\prime },C,A\right)  \notag
\end{eqnarray}%
Moreover, we define: 
\begin{equation*}
\bar{\Psi}_{n}^{\dag }\left( K^{\prime },C,A\right) =\exp \left( \frac{1}{%
2\varpi ^{2}}\left( AF^{\prime }\left( K\right) +r_{c}\right)\!\!\left( C-\bar{C%
}\right) ^{2}-\frac{\left( K^{\prime }\right) ^{2}}{2\nu ^{2}\left( \delta
-AF^{\prime }\left( K\right) \right) }\right) \Psi _{n}^{\dag }\left(
K,C,A\right)
\end{equation*}%
the eigenfunction of $\left( L^{\prime }\right) ^{+}$ for $\lambda _{n}$.
The functions $\bar{\Psi}_{n}\left( K^{\prime },C,A\right) $ and $\bar{\Psi}%
_{n}\left( K^{\prime },C,A\right) $ are orthogonal :%
\begin{equation}
\left\langle \bar{\Psi}_{m}^{\dag }\left( K,C,A\right) ,\bar{\Psi}_{n}\left(
K,C,A\right) \right\rangle =\left\langle \Psi _{m}^{\dag }\left(
K,C,A\right) ,\Psi _{n}\left( K,C,A\right) \right\rangle =\delta _{m,n}
\label{rthg}
\end{equation}%
as a direct consequence of:%
\begin{eqnarray*}
\left( L^{\prime }\right) ^{+} &=&\exp \left( \frac{1}{2\varpi ^{2}}\left(
AF^{\prime }\left( K\right) +r_{c}\right)\!\!\left( C-\bar{C}\right) ^{2}-\frac{%
\left( K^{\prime }\right) ^{2}}{2\nu ^{2}\left( \delta -AF^{\prime }\left(
K\right) \right) }\right) L^{+} \\
&&\times \exp \left( -\frac{1}{2\varpi ^{2}}\left( AF^{\prime }\left(
K\right) +r_{c}\right)\!\!\left( C-\bar{C}\right) ^{2}+\frac{\left( K^{\prime
}\right) ^{2}}{2\nu ^{2}\left( \delta -AF^{\prime }\left( K\right) \right) }%
\right)
\end{eqnarray*}%
and%
\begin{eqnarray*}
\left( L^{\prime }\right) ^{+}\bar{\Psi}_{n}^{\dag }\left( K^{\prime
},C,A\right) &=&\exp \left( \frac{1}{2\varpi ^{2}}\left( AF^{\prime }\left(
K\right) +r_{c}\right)\!\!\left( C-\bar{C}\right) ^{2}-\frac{\left( K^{\prime
}\right) ^{2}}{2\nu ^{2}\left( \delta -AF^{\prime }\left( K\right) \right) }%
\right) L^{+}\Psi _{n}^{\dag }\left( K,C,A\right) \\
&=&\lambda _{n}\bar{\Psi}_{n}^{\dag }\left( K^{\prime },C,A\right)
\end{eqnarray*}%
since these two relations imply:%
\begin{eqnarray*}
&&\int \bar{\Psi}^{\dag }\left( K^{\prime },C,A\right) \left\{ -\varpi ^{2}%
\frac{\partial ^{2}}{\partial C^{2}}-\frac{1}{\lambda ^{2}}\frac{\partial
^{2}}{\partial A^{2}}-\nu ^{2}\left( \delta -AF^{\prime }\left( K\right)
\right) ^{2}\frac{\partial ^{2}}{\partial K^{\prime 2}}+\left( A-\bar{A}%
\right) ^{2}\right. \\
&&\hspace{3cm}\left. +\left( \varsigma ^{2}+\frac{\left( AF^{\prime }\left(
K\right) +r_{c}\right) ^{2}}{\varpi ^{2}}\right)\!\!\left( C-\bar{C}\right)
^{2}+\frac{\left( K^{\prime }\right) ^{2}}{\nu ^{2}}+\alpha -C_{0}\right\} 
\bar{\Psi}\left( K^{\prime },C,A\right) \\
&=&\int \bar{\Psi}^{\dag }\left( K^{\prime },C,A\right) L^{\prime }\bar{\Psi}%
\left( K^{\prime },C,A\right) =\int \sum \alpha _{m}^{\dag }\bar{\Psi}%
_{m}^{\dag }\left( K^{\prime },C,A\right) \lambda _{n}\alpha _{n}\bar{\Psi}%
_{n}\left( K,C,A\right) \\
&=&\int \sum \alpha _{m}\lambda _{n}\alpha _{n}^{\dag } \\
&=&\int \sum \alpha _{m}^{\dag }\Psi _{m}^{\dag }\left( K,C,A\right) \lambda
_{n}\alpha _{n}\Psi _{n}\left( K,C,A\right) \\
&=&\int \Psi ^{\dag }\left( K,C,A\right) L\Psi \left( K,C,A\right)
\end{eqnarray*}%
As a consequence of (\ref{gnvl}) and (\ref{rthg}), 
\begin{equation*}
\int \mathcal{D\Psi ^{\dag }D}\Psi \exp \left( -\int \Psi ^{\dag }\left(
K,C,A\right) L\Psi \left( K,C,A\right) \right)
\end{equation*}%
and 
\begin{equation*}
\int \mathcal{D\bar{\Psi}D}\bar{\Psi}\exp \left( -\int \bar{\Psi}^{\dag
}\left( K^{\prime },C,A\right) L^{\prime }\bar{\Psi}\left( K^{\prime
},C,A\right) \right)
\end{equation*}%
compute the same partition function. We can thus consider the following
action:%
\begin{eqnarray}
&&S\left( \bar{\Psi}\right) =\int \bar{\Psi}^{\dag }\left( K^{\prime
},C,A\right) \left\{ -\varpi ^{2}\frac{\partial ^{2}}{\partial C^{2}}-\frac{1%
}{\lambda ^{2}}\frac{\partial ^{2}}{\partial A^{2}}-\nu ^{2}\left( \delta
-AF^{\prime }\left( K\right) \right) ^{2}\frac{\partial ^{2}}{\partial
K^{\prime 2}}+\left( A-\bar{A}\right) ^{2}\right.  \label{Sprvr} \\
&&\hspace{3cm}\left. +\left( \varsigma ^{2}+\frac{\left( AF^{\prime }\left(
K\right) +r_{c}\right) ^{2}}{\varpi ^{2}}\right)\!\!\left( C-\bar{C}\right)
^{2}+\frac{\left( K^{\prime }\right) ^{2}}{\nu ^{2}}+\alpha -C_{0}\right\} 
\bar{\Psi}\left( K^{\prime },C,A\right)  \notag \\
&&+\gamma \frac{1}{2}\int \bar{\Psi}^{\dag }\left( K_{1}^{\prime
},C_{1},A_{1}\right) \bar{\Psi}^{\dag }\left( K_{2}^{\prime
},C_{2},A_{2}\right) \left\{ A_{2}H\left( K_{1},K_{2}\right)
K_{1}+A_{1}H\left( K_{2},K_{1}\right) K_{2}\right\} \bar{\Psi}\left(
K_{1}^{\prime },C_{1},A_{1}\right) \bar{\Psi}\left( K_{2}^{\prime
},C_{2},A_{2}\right)  \notag
\end{eqnarray}%
as stated in the text.

\subsection*{Existence of a saddle point}

We first set $H\left( K_{2},K\right) =1$, and $\bar{A}=A_{0}+\varkappa
\left\langle A\right\rangle $ with $\varkappa <1$\ to simplify the
computations, but any function $H\left( K_{2},K\right) $ could be
considered. Before considering the saddle point equation, we can note that
for $\gamma <0$, no minimum can exist for $S\left( \bar{\Psi}\right) $.
Actually, for $\gamma <0$, the quartic term:%
\begin{equation*}
\gamma \frac{1}{2}\int \bar{\Psi}^{\dag }\left( K_{1}^{\prime
},C_{1},A_{1}\right) \bar{\Psi}^{\dag }\left( K_{2}^{\prime
},C_{2},A_{2}\right) \left\{ A_{2}K_{1}+A_{1}K_{2}\right\} \bar{\Psi}\left(
K_{1}^{\prime },C_{1},A_{1}\right) \bar{\Psi}\left( K_{2}^{\prime
},C_{2},A_{2}\right)
\end{equation*}%
is negative, and if we let $\left\Vert \bar{\Psi}\left( K_{1}^{\prime
},C_{1},A_{1}\right) \right\Vert \rightarrow \infty $, then this term
dominates, so that $S\left( \bar{\Psi}\right) \rightarrow -\infty $. Thus,
to inspect the possiblity of a minimum for $S\left( \bar{\Psi}\right) $ we
have to consider $\gamma >0$. The saddle point equation is: 
\begin{eqnarray*}
&&\left\{ -\varpi ^{2}\frac{\partial ^{2}}{\partial C^{2}}-\frac{1}{\lambda
^{2}}\frac{\partial ^{2}}{\partial A^{2}}-\nu ^{2}\left( \delta -AF^{\prime
}\left( K\right) \right) ^{2}\frac{\partial ^{2}}{\partial K^{\prime 2}}%
+\left( A-\bar{A}\right) ^{2}+2\left( \bar{A}-\Gamma _{3}\right) \varkappa
A\right. \\
&&\left. +\left( \varsigma ^{2}+\frac{\left( AF^{\prime }\left( K\right)
+r_{c}\right) ^{2}}{\varpi ^{2}}\right)\!\!\left( C-\bar{C}\right) ^{2}+\frac{%
\left( K^{\prime }\right) ^{2}}{\nu ^{2}}+\alpha -C_{0}\right\} \Psi
_{1}\left( K,C,A\right) \\
&&+\gamma \eta \left( \Gamma _{2}^{\prime }A+\Gamma _{3}K\right) \Psi
_{1}\left( K,C,A\right) =0
\end{eqnarray*}%
where:%
\begin{eqnarray*}
\int \Psi _{1}^{\dag }\left( K_{2},C_{2},A_{2}\right) K_{2}\Psi _{1}\left(
K_{2},C_{2},A_{2}\right) &=&\Gamma _{2}^{\prime } \\
\int \Psi _{1}^{\dag }\left( K_{2},C_{2},A_{2}\right) A_{2}\Psi _{1}\left(
K_{2},C_{2},A_{2}\right) &=&\Gamma _{3} \\
g^{-1}\left( K\right) &=&K-\frac{A}{\delta }F\left( K\right)
\end{eqnarray*}%
The term $2\left( \bar{A}-\Gamma _{3}\right) \varkappa A$ comes from
variation of $\bar{A}=A_{0}+\varkappa \int \bar{\Psi}^{\dag }\left(
K^{\prime },C,A\right) A\bar{\Psi}\left( K^{\prime },C,A\right) $. We impose 
$\left\Vert \Psi _{1}\left( K,C,A\right) \right\Vert =1$, so that $\Psi
\left( K,C,A\right) =\sqrt{\eta }\Psi _{1}\left( K,C,A\right) $. We can also
replace $K$ with $K^{\prime }$ through the relation:%
\begin{equation*}
K^{\prime }=C-AF\left( K\right) +\delta K\equiv \delta g^{-1}\left( K\right)
+C
\end{equation*}%
\begin{eqnarray}
&&\left\{ -\varpi ^{2}\frac{\partial ^{2}}{\partial C^{2}}-\frac{1}{\lambda
^{2}}\frac{\partial ^{2}}{\partial A^{2}}-\nu ^{2}\left( \delta -AF^{\prime
}\left( K\right) \right) ^{2}\frac{\partial ^{2}}{\partial K^{\prime 2}}%
+\left( A-\bar{A}\right) ^{2}+2\left( \bar{A}-\Gamma _{3}\right) \varkappa
A\right.  \label{sddptqnnbs} \\
&&\left. +\left( \varsigma ^{2}+\frac{\left( AF^{\prime }\left( K\right)
+r_{c}\right) ^{2}}{\varpi ^{2}}\right)\!\!\left( C-\bar{C}\right) ^{2}+\frac{%
\left( K^{\prime }\right) ^{2}}{\nu ^{2}}+\alpha -C_{0}\right\} \Psi
_{1}\left( K,C,A\right)  \notag \\
&&+\gamma \eta \left( \Gamma _{2}^{\prime }A+\Gamma _{3}\left( g\left( \frac{%
K^{\prime }-C}{\delta }\right) \right) \right) \Psi _{1}\left( K,C,A\right)
=0  \notag
\end{eqnarray}%
For the usual form $F\left( K\right) =K^{\varepsilon }$, with $\varepsilon
<1 $, 
\begin{equation*}
g^{-1}\left( K\right) =K\left( 1-\frac{A}{\delta K^{1-\varepsilon }}\right)
\end{equation*}%
and above a minimal level $\bar{K}$:%
\begin{eqnarray*}
\frac{AK}{\delta K^{1-\varepsilon }} &\simeq &\frac{A\bar{K}^{\varepsilon }}{%
\delta }\left( 1+\varepsilon \left( \frac{K-\bar{K}}{\bar{K}}\right) -\frac{%
\varepsilon \left( 1-\varepsilon \right) }{2}\left( \frac{K-\bar{K}}{\bar{K}}%
\right) ^{2}\right) \\
&\simeq &\frac{A\bar{K}^{\varepsilon }}{\delta }\left( 1+\varepsilon \left( 
\frac{K-\bar{K}}{\bar{K}}\right) \right)
\end{eqnarray*}%
and:%
\begin{equation*}
g^{-1}\left( K\right) \simeq K-\frac{A\bar{K}^{\varepsilon }}{\delta }\left(
1+\varepsilon \left( \frac{K-\bar{K}}{\bar{K}}\right) \right)
\end{equation*}%
$\allowbreak $so that:%
\begin{equation*}
K=g\left( \frac{K^{\prime }-C}{\delta }\right) =\frac{\left( K^{\prime
}-C\right) +A\bar{K}^{\varepsilon }\left( 1-\varepsilon \right) -\frac{A}{2}%
\varepsilon \left( 1-\varepsilon \right) \bar{K}^{\varepsilon -2}}{\delta
-A\varepsilon \bar{K}^{\varepsilon -1}}
\end{equation*}%
\begin{equation*}
C+\left( \delta -A\varepsilon \bar{K}^{\varepsilon -1}\right)\!\!\left( \frac{%
\varepsilon A}{\delta }\right) ^{\frac{1}{1-\varepsilon }}-A\bar{K}%
^{\varepsilon }\left( 1-\varepsilon \right) <K^{\prime }C-A\bar{K}%
^{\varepsilon }\left( 1-\varepsilon \right)
\end{equation*}%
\begin{equation*}
A>\frac{\delta \bar{K}^{1-\varepsilon }}{\varepsilon }
\end{equation*}%
For this particular form of production function, (\ref{sddptqnnbs}) is then:

\begin{eqnarray}
&&\left\{ -\varpi ^{2}\frac{\partial ^{2}}{\partial C^{2}}-\frac{1}{\lambda
^{2}}\frac{\partial ^{2}}{\partial A^{2}}-\nu ^{2}\left( \delta -AF^{\prime
}\left( K\right) \right) ^{2}\frac{\partial ^{2}}{\partial K^{\prime 2}}%
+\left( A-\bar{A}\right) ^{2}+2\left( \bar{A}-\Gamma _{3}\right) \varkappa
A\right.  \label{sddptqnnbsbb} \\
&&\left. +\left( \varsigma ^{2}+\frac{\left( AF^{\prime }\left( K\right)
+r_{c}\right) ^{2}}{\varpi ^{2}}\right)\!\!\left( C-\bar{C}\right) ^{2}+\frac{%
\left( K^{\prime }\right) ^{2}}{\nu ^{2}}+\alpha -C_{0}\right\} \Psi
_{1}\left( K,C,A\right)  \notag \\
&&+\gamma \eta \left( \Gamma _{2}^{\prime }A+\Gamma _{3}\frac{\left(
K^{\prime }-C\right) +A\bar{K}^{\varepsilon }\left( 1-\varepsilon \right) }{%
\delta -A\varepsilon \bar{K}^{\varepsilon -1}}\right) \Psi _{1}\left(
K,C,A\right) =0  \notag
\end{eqnarray}%
where:%
\begin{eqnarray*}
\Gamma _{1} &=&\left\langle C\right\rangle \text{, }\Gamma _{2}=\left\langle
K^{\prime }\right\rangle \text{, }\Gamma _{3}=\left\langle A\right\rangle \\
\Gamma _{2}^{\prime } &=&\left\langle \frac{\left( K^{\prime }-C\right) +A%
\bar{K}^{\varepsilon }\left( 1-\varepsilon \right) }{\delta -A\varepsilon 
\bar{K}^{\varepsilon -1}}\right\rangle
\end{eqnarray*}%
and the brackets denotes the expectation of the quantities in the state $%
\Psi _{1}$. The potential terms in (\ref{sddptqnnbsbb}):%
\begin{eqnarray*}
&&\left( \varsigma ^{2}+\frac{\left( AF^{\prime }\left( K\right)
+r_{c}\right) ^{2}}{\varpi ^{2}}\right)\!\!\left( C-\bar{C}\right) ^{2}+\frac{%
\left( K^{\prime }\right) ^{2}}{\nu ^{2}}+\left( A-\bar{A}\right)
^{2}+2\left( \bar{A}-\Gamma _{3}\right) \varkappa A+\gamma \eta \left(
\Gamma _{2}^{\prime }A+\Gamma _{3}\frac{\left( K^{\prime }-C\right) +A\bar{K}%
^{\varepsilon }\left( 1-\varepsilon \right) }{\delta -A\varepsilon \bar{K}%
^{\varepsilon -1}}\right) \\
&\simeq &\left( \varsigma ^{2}+\frac{\left( AF^{\prime }\left( K\right)
+r_{c}\right) ^{2}}{\varpi ^{2}}\right)\left( C-\bar{C}\right) ^{2}+\frac{%
\left( K^{\prime }\right) ^{2}}{\nu ^{2}}+\left( A-\bar{A}\right)
^{2}+2\left( \bar{A}-\Gamma _{3}\right) \varkappa A+\gamma \eta \left(
\Gamma _{2}^{\prime }A+\Gamma _{3}\frac{\left( K^{\prime }-C\right) +A\bar{K}%
^{\varepsilon }\left( 1-\varepsilon \right) }{\delta -\Gamma _{3}\varepsilon 
\bar{K}^{\varepsilon -1}}\right)
\end{eqnarray*}%
with%
\begin{equation*}
\Gamma _{2}^{\prime }=\left\langle \frac{\left( K^{\prime }-C\right) +A\bar{K%
}^{\varepsilon }\left( 1-\varepsilon \right) }{\delta -A\varepsilon \bar{K}%
^{\varepsilon -1}}\right\rangle \simeq \frac{\left( \Gamma _{2}-\Gamma
_{1}\right) +\bar{K}^{\varepsilon }\left( 1-\varepsilon \right) \Gamma _{3}}{%
\delta -\Gamma _{3}\varepsilon \bar{K}^{\varepsilon -1}}
\end{equation*}%
and the potential terms are then:

\begin{eqnarray*}
V &=&\left( \varsigma ^{2}+\frac{\left( AF^{\prime }\left( K\right)
+r_{c}\right) ^{2}}{\varpi ^{2}}\right)\!\!\left( C-\bar{C}\right) ^{2}+\frac{%
\left( K^{\prime }\right) ^{2}}{\nu ^{2}}+\left( A-\left( \left( 1-\varkappa
\right) \bar{A}+\varkappa \Gamma _{3}\right) \right) ^{2} \\
&&+\gamma \eta \left( \frac{\left( \Gamma _{2}-\Gamma _{1}\right) +\bar{K}%
^{\varepsilon }\left( 1-\varepsilon \right) \Gamma _{3}}{\delta -\Gamma
_{3}\varepsilon \bar{K}^{\varepsilon -1}}A+\Gamma _{3}\frac{\left( K^{\prime
}-C\right) +A\bar{K}^{\varepsilon }\left( 1-\varepsilon \right) }{\delta
-\Gamma _{3}\varepsilon \bar{K}^{\varepsilon -1}}\right) +\bar{A}^{2}-\left(
\left( 1-\varkappa \right) \bar{A}+\varkappa \Gamma _{3}\right) ^{2}
\end{eqnarray*}%
and can be written in a compact form:%
\begin{eqnarray}
V &=&\left( ^{t}\left( X-\hat{X}\right) \Omega \left( X-\hat{X}\right)
+\left( ^{t}\Gamma \right) MX\right)  \label{cmpct} \\
&=&\left( ^{t}\left( X-\hat{X}+\frac{1}{2}\Omega ^{-1}M\Gamma \right)
\right) \Omega \left( X-\hat{X}+\frac{1}{2}\Omega ^{-1}M\Gamma \right) -%
\frac{1}{4}\left( ^{t}\Gamma M\Omega ^{-1}M\Gamma \right) +\left( ^{t}\Gamma
\right) M\hat{X}  \notag \\
&&+\bar{A}^{2}-\left( \left( 1-\varkappa \right) \bar{A}+\varkappa \Gamma
_{3}\right) ^{2}  \notag
\end{eqnarray}%
with:%
\begin{eqnarray*}
X &=&\left( 
\begin{array}{c}
C \\ 
K^{\prime } \\ 
A%
\end{array}%
\right) ,\hat{X}=\left( 
\begin{array}{c}
\bar{C} \\ 
0 \\ 
\left( 1-\varkappa \right) \bar{A}+\varkappa \Gamma _{3}%
\end{array}%
\right) ,\Gamma =\left( 
\begin{array}{c}
\Gamma _{1} \\ 
\Gamma _{2} \\ 
\Gamma _{3}%
\end{array}%
\right) ,\Omega =\left( 
\begin{array}{ccc}
\left( \varsigma ^{2}+\frac{\left( \Gamma _{3}F^{\prime }\left( K\right)
+r_{c}\right) ^{2}}{\varpi ^{2}}\right) & 0 & 0 \\ 
0 & \frac{1}{\nu ^{2}} & 0 \\ 
0 & 0 & 1%
\end{array}%
\right) \\
M &=&\frac{\gamma \eta }{\delta -\Gamma _{3}\varepsilon \bar{K}^{\varepsilon
-1}}\left( 
\begin{array}{ccc}
0 & 0 & -1 \\ 
0 & 0 & 1 \\ 
-1 & 1 & \bar{K}^{\varepsilon }\left( 1-\varepsilon \right)%
\end{array}%
\right)
\end{eqnarray*}%
$\allowbreak $ From these equations we can identify the mean values of the
variables. Given that we anticipate a gaussian form for $\Psi _{1}\left(
K,C,A\right) $, we have to implement several constraint. Actully, we assume
that $C>0$, $A>0$, so that the distribution for these variables has to be
cut off for the egative values. The variable $K^{\prime }$ on its side is
also constrained by the model and our assumptions. First, given that:%
\begin{equation*}
K=\frac{\left( K^{\prime }-C\right) +A\bar{K}^{\varepsilon }\left(
1-\varepsilon \right) }{\delta -A\varepsilon \bar{K}^{\varepsilon -1}}
\end{equation*}%
The variable $K$ has to be positive, so that:%
\begin{equation*}
K^{\prime }<C-A\bar{K}^{\varepsilon }\left( 1-\varepsilon \right)
\end{equation*}%
Moreover, we have assumed that:%
\begin{equation*}
\delta -\varepsilon A\left( K\right) ^{\varepsilon -1}<0
\end{equation*}%
which translates in: 
\begin{equation*}
\delta -\varepsilon A\left( \frac{\left( K^{\prime }-C\right) +A\bar{K}%
^{\varepsilon }\left( 1-\varepsilon \right) }{\delta -A\varepsilon \bar{K}%
^{\varepsilon -1}}\right) ^{\varepsilon -1}<0
\end{equation*}%
or equivalently: 
\begin{equation*}
C+\left( \delta -A\varepsilon \bar{K}^{\varepsilon -1}\right)\!\!\left( \frac{%
\varepsilon A}{\delta }\right) ^{\frac{1}{1-\varepsilon }}-A\bar{K}%
^{\varepsilon }\left( 1-\varepsilon \right) <K^{\prime }
\end{equation*}%
As a consequence, the domain for is bounded $K^{\prime }$:%
\begin{equation*}
C+\left( \delta -A\varepsilon \bar{K}^{\varepsilon -1}\right)\!\!\left( \frac{%
\varepsilon A}{\delta }\right) ^{\frac{1}{1-\varepsilon }}-A\bar{K}%
^{\varepsilon }\left( 1-\varepsilon \right) <K^{\prime }<C-A\bar{K}%
^{\varepsilon }\left( 1-\varepsilon \right)
\end{equation*}%
We will see later that the existence a non trivial phase is possible if $%
A\gg 0 $. As a consequence, we can assume that $C-A\bar{K}^{\varepsilon
}\left( 1-\varepsilon \right) <0$ and 
\begin{equation*}
\left\vert \delta -A\varepsilon \bar{K}^{\varepsilon -1}\right\vert \left( 
\frac{\varepsilon A}{\delta }\right) ^{\frac{1}{1-\varepsilon }}
\end{equation*}%
wich means that a good approximation for the domain in question is:%
\begin{equation*}
K^{\prime }<C-A\bar{K}^{\varepsilon }\left( 1-\varepsilon \right)
\end{equation*}%
Given these conditions on the variables, the equations for the average
values can thus be written as: 
\begin{eqnarray}
\Gamma &=&\hat{X}-\frac{1}{2}\Omega ^{-1}M\Gamma +\hat{X}_{1}  \label{gmqn}
\\
\Gamma &=&\left( 1+\frac{1}{2}\Omega ^{-1}M\right) ^{-1}\left( \hat{X}+\hat{X%
}_{1}\right)  \notag
\end{eqnarray}%
with $\hat{X}_{1}=\left( 
\begin{array}{c}
C_{1} \\ 
K_{1}^{\prime } \\ 
A_{1}%
\end{array}%
\right) $.The variables $C_{1}$ and $A_{1}$ express the shift due to the cut
off for the negative values of $C$ and $A$.%
\begin{eqnarray}
C_{1} &=&\frac{\int_{0}^{\infty }X_{1}\exp \left( -\frac{\left( X-\hat{X}+%
\frac{1}{2}\Omega ^{-1}M\Gamma \right) _{1}^{2}}{2\varpi ^{2}}\right) dX_{1}%
}{\int_{0}^{\infty }\exp \left( -\frac{\left( X-\hat{X}+\frac{1}{2}\Omega
^{-1}M\Gamma \right) _{1}^{2}}{2\varpi ^{2}}\right) dX_{1}}=\sqrt{\frac{2}{%
\pi }}\varpi \frac{\exp \left( -\frac{\left( \hat{X}-\frac{1}{2}\Omega
^{-1}M\Gamma \right) _{1}^{2}}{2\varpi ^{2}}\right) }{1-\erf\left( 
\frac{\left( \hat{X}-\frac{1}{2}\Omega ^{-1}M\Gamma \right) _{1}}{\sqrt{2}%
\varpi }\right) }  \label{prmt1} \\
A_{1} &=&\frac{\int_{0}^{\infty }X_{2}\exp \left( -\frac{\lambda \left( X-%
\hat{X}+\frac{1}{2}\Omega ^{-1}M\Gamma \right) _{2}^{2}}{2}\right) dX_{2}}{%
\int_{0}^{\infty }\exp \left( -\frac{\lambda \left( X-\hat{X}+\frac{1}{2}%
\Omega ^{-1}M\Gamma \right) _{2}^{2}}{2}\right) dX_{2}}=\frac{2}{\sqrt{\pi
\lambda }}\frac{\exp \left( -\frac{\lambda \left( \hat{X}-\frac{1}{2}\Omega
^{-1}M\Gamma \right) _{3}^{2}}{2}\right) }{1-\erf\left( \frac{\sqrt{%
\lambda }\left( \hat{X}-\frac{1}{2}\Omega ^{-1}M\Gamma \right) _{3}}{\sqrt{2}%
}\right) }  \label{prmt12}
\end{eqnarray}%
whereas $K_{1}^{\prime }$ expresses the shift due to the superior bound for $%
K^{\prime }$:%
\begin{eqnarray*}
K_{1}^{\prime } &=&\frac{\int_{-\infty }^{C-A\bar{K}^{\varepsilon }\left(
1-\varepsilon \right) }X_{3}\exp \left( -\frac{\left( X-\hat{X}+\frac{1}{2}%
\Omega ^{-1}M\Gamma \right) _{2}^{2}}{2\left\vert \delta -\Gamma
_{3}\varepsilon \bar{K}^{\varepsilon -1}\right\vert \nu ^{2}}\right) dX_{3}}{%
\int_{-\infty }^{C-A\bar{K}^{\varepsilon }\left( 1-\varepsilon \right) }\exp
\left( -\frac{\left( X-\hat{X}+\frac{1}{2}\Omega ^{-1}M\Gamma \right)
_{2}^{2}}{2\left\vert \delta -\Gamma _{3}\varepsilon \bar{K}^{\varepsilon
-1}\right\vert \nu ^{2}}\right) dX_{3}} \\
&=&-\frac{\sqrt{\frac{2}{\pi }}\sqrt{\left\vert \delta -\Gamma
_{3}\varepsilon \bar{K}^{\varepsilon -1}\right\vert }\nu \exp \left( -\frac{%
\left( C-A\bar{K}^{\varepsilon }\left( 1-\varepsilon \right) -\left( \hat{X}-%
\frac{1}{2}\Omega ^{-1}M\Gamma \right) _{2}\right) ^{2}}{2\left\vert \delta
-\Gamma _{3}\varepsilon \bar{K}^{\varepsilon -1}\right\vert \nu ^{2}}\right) 
}{\left( \erf\left( \frac{\left( C-A\bar{K}^{\varepsilon }\left(
1-\varepsilon \right) -\left( \hat{X}-\frac{1}{2}\Omega ^{-1}M\Gamma \right)
_{2}\right) }{\sqrt{2}}\right) +1\right) }
\end{eqnarray*}%
and this last expression can be approximated by:%
\begin{equation}
K_{1}^{\prime }\simeq -2\nu ^{2}\left\vert \delta -A\varepsilon \bar{K}%
^{\varepsilon -1}\right\vert \frac{\exp \left( -\frac{\left( C-A\bar{K}%
^{\varepsilon }\left( 1-\varepsilon \right) -\left( \hat{X}-\frac{1}{2}%
\Omega ^{-1}M\Gamma \right) _{2}\right) ^{2}}{2\left\vert \delta -\Gamma
_{3}\varepsilon \bar{K}^{\varepsilon -1}\right\vert \nu ^{2}}\right) }{%
2-\exp \left( -1.9\left( \frac{\left\vert C-A\bar{K}^{\varepsilon }\left(
1-\varepsilon \right) \right\vert }{\sqrt{2\nu ^{2}\left\vert \delta
-A\varepsilon \bar{K}^{\varepsilon -1}\right\vert }}\right) ^{1.3}\right) }
\label{prmt13}
\end{equation}%
For a saddle point equation written as:%
\begin{eqnarray}
&&\hskip-30pt0=\left\{ -\varpi ^{2}\frac{\partial ^{2}}{\partial C^{2}}-\frac{1}{%
\lambda ^{2}}\frac{\partial ^{2}}{\partial A^{2}}-\nu ^{2}\left( \delta
-AF^{\prime }\left( K\right) \right) ^{2}\frac{\partial ^{2}}{\partial
K^{\prime 2}}+{}^{t}\!\!\left( X-\hat{X}+\frac{1}{2}\Omega ^{-1}M\Gamma
\right)  \Omega\! \left( X-\hat{X}+\frac{1}{2}\Omega ^{-1}M\Gamma
\right) \right.  \label{sddptqnnbstt} \\
&&\left. -\frac{1}{4}\left( ^{t}\Gamma M\Omega ^{-1}M\Gamma \right) +\left(
^{t}\Gamma \right) M\hat{X}+\left( \bar{A}^{2}-\left( \left( 1-\varkappa
\right) \bar{A}+\varkappa \Gamma _{3}\right) ^{2}\right) +\alpha
-C_{0}\right\} \Psi _{1}\left( K,C,A\right)  \notag
\end{eqnarray}%
Given that the $\sigma _{i}$ have been considered relatively small for
capital and technology, we can neglect the term proportional to the
exponential in (\ref{gmqn}) for these two variables. We will keep this
additional contribution for $C$ only which shift $\hat{C}$ by $\frac{2}{%
\sqrt{2\pi }}\varpi $ for small $\hat{C}$.

The constant term in (\ref{cmpct}) is thus given by: 
\begin{eqnarray*}
-\frac{1}{4}\left( ^{t}\Gamma M\Omega ^{-1}M\Gamma \right) +\left(
^{t}\Gamma \right) M\hat{X} &=&-\frac{1}{4}\left( ^{t}\Gamma M\Omega
^{-1}M\Gamma \right) +\left( ^{t}\Gamma \right) M\left( 1+\frac{1}{2}\Omega
^{-1}M\right) \Gamma \\
&=&\left( ^{t}\Gamma \right)\!\!\left( M+\frac{1}{4}M\Omega ^{-1}M\right) \Gamma
\\
&=&\left( ^{t}\hat{X}+\hat{X}_{1}\right)\!\!\left( 1+\frac{1}{2}M\Omega
^{-1}\right) ^{-1}\left( M+\frac{1}{4}M\Omega ^{-1}M\right)\!\!\left( 1+\frac{1%
}{2}\Omega ^{-1}M\right) ^{-1}\left( \hat{X}+\hat{X}_{1}\right)
\end{eqnarray*}%
Equation (\ref{gmqn}) can be expressed as an equation for $\Gamma _{3}$:%
\begin{eqnarray}
\Gamma _{3} &=&2\frac{2\left( \left( 1-\varkappa \right) \bar{A}+\varkappa
\Gamma _{3}+A_{1}\right) +\left( \left( \bar{C}+C_{1}\right) -K_{1}^{\prime
}\right) +\alpha \gamma \eta }{4-\left( a+b\right) \alpha ^{2}\left( \gamma
\eta \right) ^{2}+2\alpha \beta \gamma \eta }  \label{gmqnth} \\
&=&2\frac{2\left( \left( 1-\varkappa \right) \bar{A}+\varkappa \Gamma
_{3}+A_{1}\right)\!\!\left( \delta -\Gamma _{3}\varepsilon \bar{K}^{\varepsilon
-1}\right) ^{2}+\left( \left( \bar{C}+C_{1}\right) -K_{1}^{\prime }\right)
\gamma \eta \left( \delta -\Gamma _{3}\varepsilon \bar{K}^{\varepsilon
-1}\right) }{4\left( \delta -\Gamma _{3}\varepsilon \bar{K}^{\varepsilon
-1}\right) ^{2}-\left( a+b\right)\!\!\left( \gamma \eta \right) ^{2}\left(
\delta -\Gamma _{3}\varepsilon \bar{K}^{\varepsilon -1}\right) +2\gamma \eta 
\bar{K}^{\varepsilon }\left( 1-\varepsilon \right) }  \notag \\
&=&2\frac{2\left( \left( 1-\varkappa \right) A_{0}+\left( 2-\varkappa
\right) \varkappa \Gamma _{3}+A_{1}\right)\!\!\left( \delta -\Gamma
_{3}\varepsilon \bar{K}^{\varepsilon -1}\right) ^{2}+\left( \left( \bar{C}%
+C_{1}\right) -K_{1}^{\prime }\right) \gamma \eta \left( \delta -\Gamma
_{3}\varepsilon \bar{K}^{\varepsilon -1}\right) }{4\left( \delta -\Gamma
_{3}\varepsilon \bar{K}^{\varepsilon -1}\right) ^{2}-\left( a+b\right)
\left( \gamma \eta \right) ^{2}\left( \delta -\Gamma _{3}\varepsilon \bar{K}%
^{\varepsilon -1}\right) +2\gamma \eta \bar{K}^{\varepsilon }\left(
1-\varepsilon \right) }  \notag
\end{eqnarray}%
$\allowbreak $where:%
\begin{equation*}
a+b=\frac{\varpi ^{2}}{\varsigma ^{2}\varpi ^{2}+\left( AF^{\prime }\left(
K\right) +r_{c}\right) ^{2}}+\nu ^{2}\ll 1
\end{equation*}%
and:%
\begin{equation*}
\alpha =\frac{1}{\delta -\Gamma _{3}\varepsilon \bar{K}^{\varepsilon -1}}%
,\beta =\frac{\bar{K}^{\varepsilon }\left( 1-\varepsilon \right) }{\delta
-\Gamma _{3}\varepsilon \bar{K}^{\varepsilon -1}}
\end{equation*}%
At the first order in $\gamma \eta $, the equation rewrites:%
\begin{equation*}
\Gamma _{3}=2\frac{2\left( \left( 1-\varkappa \right) A_{0}+\left(
2-\varkappa \right) \varkappa \Gamma _{3}+A_{1}\right)\!\!\left( \delta -\Gamma
_{3}\varepsilon \bar{K}^{\varepsilon -1}\right) ^{2}+\left( \left( \bar{C}%
+C_{1}\right) -K_{1}^{\prime }\right) \gamma \eta \left( \delta -\Gamma
_{3}\varepsilon \bar{K}^{\varepsilon -1}\right) }{4\left( \delta -\Gamma
_{3}\varepsilon \bar{K}^{\varepsilon -1}\right) ^{2}+2\gamma \eta \bar{K}%
^{\varepsilon }\left( 1-\varepsilon \right) }
\end{equation*}%
For $\gamma \eta \ll 1$, we check that $\Gamma _{3}\simeq \frac{A_{0}}{%
1-\varkappa }$, in that case, using (\ref{prmt12}): 
\begin{equation*}
A_{1}=\frac{2}{\sqrt{\lambda }}\frac{\exp \left( -\frac{\lambda \left( \hat{X%
}-\frac{1}{2}\Omega ^{-1}M\Gamma \right) _{3}^{2}}{2}\right) }{2-\erf%
\left( \frac{\sqrt{\lambda }\left( \hat{X}-\frac{1}{2}\Omega ^{-1}M\Gamma
\right) _{3}}{\sqrt{2}}\right) }\simeq \frac{2}{\sqrt{\lambda }}\exp \left( -%
\frac{\lambda \left( \frac{A_{0}}{1-\varkappa }\right) ^{2}}{2}\right) \ll 1
\end{equation*}%
and $A_{1}$ can be neglected, whivh leads to:%
\begin{equation*}
\Gamma _{3}=2\frac{2\left( \left( 1-\varkappa \right) A_{0}+\left(
2-\varkappa \right) \varkappa \Gamma _{3}\right)\!\!\left( \delta -\Gamma
_{3}\varepsilon \bar{K}^{\varepsilon -1}\right) ^{2}+\left( \left( \bar{C}%
+C_{1}\right) -K_{1}^{\prime }\right) \gamma \eta \left( \delta -\Gamma
_{3}\varepsilon \bar{K}^{\varepsilon -1}\right) }{4\left( \delta -\Gamma
_{3}\varepsilon \bar{K}^{\varepsilon -1}\right) ^{2}+2\gamma \eta \bar{K}%
^{\varepsilon }\left( 1-\varepsilon \right) }
\end{equation*}%
We assume that $\left( \delta -\Gamma _{3}\varepsilon \bar{K}^{\varepsilon
-1}\right) <0$, so that the marginal productivity exceeds the depreciation
rate of capital, we see below that the solution for $\gamma \eta $ is of
first order in $\delta -\frac{A_{0}}{\left( 1-\varkappa \right) }\varepsilon 
\bar{K}^{\varepsilon -1}$, which allows for a first order solution in $%
\gamma \eta $ to be considered: 
\begin{equation}
\Gamma _{3}=\frac{A_{0}}{\left( 1-\varkappa \right) }-\frac{1}{2}\frac{%
K^{\varepsilon }A_{0}\left( 1-\varepsilon \right) -\left( \left( \bar{C}%
+C_{1}\right) -K_{1}^{\prime }\right)\!\!\left( \delta -\frac{A_{0}}{\left(
1-\varkappa \right) }\varepsilon \bar{K}^{\varepsilon -1}\right)\!\!\left(
1-\varkappa \right) }{\left( \delta -\frac{A_{0}}{\left( 1-\varkappa \right) 
}\varepsilon \bar{K}^{\varepsilon -1}\right) ^{2}\left( 1-\varkappa \right)
^{3}}\gamma \eta  \label{gmm}
\end{equation}%
$\allowbreak $The quantities $\Gamma _{1}$ and $\Gamma _{2}$ can be
expressed in terms of $\Gamma _{3}$:

\begin{eqnarray*}
\Gamma _{1} &\simeq &\frac{4\left( \bar{C}+C_{1}\right) -\left( \bar{C}%
+C_{1}\right) b\alpha ^{2}\left( \gamma \eta \right) ^{2}+2a\left( \left(
1-\varkappa \right) \bar{A}+\varkappa \Gamma _{3}\right) \alpha \gamma \eta
+2\left( \bar{C}+C_{1}\right) \alpha \beta \gamma \eta +\frac{\alpha
^{2}\varpi ^{2}\left( \gamma \eta \right) ^{2}K_{1}^{\prime }}{\left( \varpi
^{2}\varsigma ^{2}+\left( \Gamma _{3}F^{\prime }\left( K\right)
+r_{c}\right) ^{2}\right) }}{4-\left( a+b\right) \alpha ^{2}\left( \gamma
\eta \right) ^{2}+2\alpha \beta \gamma \eta } \\
&=&\frac{4\left( \bar{C}+C_{1}\right) -\left( \bar{C}+C_{1}\right) b\alpha
^{2}\left( \gamma \eta \right) ^{2}+2a\left( \left( 1-\varkappa \right) \bar{%
A}+\varkappa \Gamma _{3}\right) \alpha \gamma \eta +2\left( \bar{C}%
+C_{1}\right) \alpha \beta \gamma \eta +\frac{\alpha ^{2}\varpi ^{2}\left(
\gamma \eta \right) ^{2}K_{1}^{\prime }}{\left( \varpi ^{2}\varsigma
^{2}+\left( \Gamma _{3}F^{\prime }\left( K\right) +r_{c}\right) ^{2}\right) }%
}{2\left( 2\left( \left( 1-\varkappa \right) \bar{A}+\varkappa \Gamma
_{3}\right) +\left( \bar{C}+C_{1}\right) \alpha \gamma \eta \right) }\Gamma
_{3} \\
\Gamma _{2} &=&-\frac{\left( \gamma \eta \right) ^{2}b\alpha ^{2}\left( \bar{%
C}+C_{1}\right) +2b\alpha \gamma \eta \left( \left( 1-\varkappa \right) \bar{%
A}+\varkappa \Gamma _{3}\right) -\left( 4+2K^{\varepsilon }\left(
1-\varepsilon \right) \alpha \gamma \eta -\frac{\left( \alpha \gamma \eta
\right) ^{2}}{\left( \varsigma ^{2}+\frac{\left( \Gamma _{3}F^{\prime
}\left( K\right) +r_{c}\right) ^{2}}{\varpi ^{2}}\right) }\right)
K_{1}^{\prime }}{4-\left( a+b\right) \alpha ^{2}\left( \gamma \eta \right)
^{2}+2\alpha \beta \gamma \eta } \\
&=&-\frac{\left( \gamma \eta \right) ^{2}b\alpha ^{2}\left( \bar{C}%
+C_{1}\right) +2b\alpha \gamma \eta \left( \left( 1-\varkappa \right) \bar{A}%
+\varkappa \Gamma _{3}\right) -\left( 4+2K^{\varepsilon }\left(
1-\varepsilon \right) \alpha \gamma \eta -\frac{\left( \alpha \gamma \eta
\right) ^{2}}{\left( \varsigma ^{2}+\frac{\left( \Gamma _{3}F^{\prime
}\left( K\right) +r_{c}\right) ^{2}}{\varpi ^{2}}\right) }\right)
K_{1}^{\prime }}{2\left( 2\left( \left( 1-\varkappa \right) \bar{A}%
+\varkappa \Gamma _{3}\right) +\left( \bar{C}+C_{1}\right) \alpha \gamma
\eta \right) }\Gamma _{3}
\end{eqnarray*}%
For $\gamma \eta \neq 0$, one has at the first order in $\gamma \eta $:$%
\allowbreak $%
\begin{equation*}
\Gamma _{1}=\bar{C}+C_{1}-\frac{\varpi ^{2}\left( \frac{A_{0}}{1-\chi }%
\right) }{2\left( \varsigma ^{2}\varpi ^{2}+\left( AF^{\prime }\left(
K\right) +r_{c}\right) ^{2}\right) \left\vert \delta -\Gamma _{3}\varepsilon 
\bar{K}^{\varepsilon -1}\right\vert }\gamma \eta
\end{equation*}%
$C_{1}$ can be found by writing (\ref{prmt1}) as:%
\begin{eqnarray*}
C_{1} &=&\sqrt{\frac{2}{\pi }}\varpi \frac{\exp \left( -\frac{\left( \hat{X}-%
\frac{1}{2}\Omega ^{-1}M\Gamma \right) _{1}^{2}}{2\varpi ^{2}}\right) }{1-%
\erf\left( \frac{\left( \hat{X}-\frac{1}{2}\Omega ^{-1}M\Gamma \right)
_{1}}{\sqrt{2}\varpi }\right) }=\sqrt{\frac{2}{\pi }}\varpi \frac{\exp
\left( -\frac{\left( \Gamma _{1}-C_{1}\right) _{1}^{2}}{2\varpi ^{2}}\right) 
}{1-\erf\left( \frac{\left( \Gamma _{1}-C_{1}\right) _{1}}{\sqrt{2}%
\varpi }\right) } \\
&=&\sqrt{\frac{2}{\pi }}\varpi \frac{\exp \left( -\frac{\left( \bar{C}%
\right) _{1}^{2}}{2\varpi ^{2}}\right) }{1-\erf\left( \frac{\left( 
\bar{C}\right) _{1}}{\sqrt{2}\varpi }\right) }
\end{eqnarray*}%
where we used (\ref{gmqn}). For $\bar{C}\ll 1$, one has:%
\begin{equation*}
C_{1}\simeq \sqrt{\frac{2}{\pi }}\varpi
\end{equation*}%
The value of $\Gamma _{2}$\ is computed in the same way:%
\begin{equation*}
\Gamma _{2}=-\frac{2b\alpha \gamma \eta \left( \left( 1-\varkappa \right) 
\bar{A}+\varkappa \Gamma _{3}\right) -\left( 4+2K^{\varepsilon }\left(
1-\varepsilon \right) \alpha \gamma \eta \right) K_{1}^{\prime }}{4+2\alpha
\beta \gamma \eta }
\end{equation*}%
At the zeroth order in $\gamma \eta $, it reduces to:%
\begin{equation*}
\Gamma _{2}=K_{1}^{\prime }
\end{equation*}%
and the parameter $K_{1}^{\prime }$\ is found by considering (\ref{prmt13}): 
\begin{eqnarray*}
K_{1}^{\prime } &=&-\frac{\sqrt{\frac{2}{\pi }}\sqrt{\left\vert \delta
-\Gamma _{3}\varepsilon \bar{K}^{\varepsilon -1}\right\vert }\nu \exp \left(
-\frac{\left( C-A\bar{K}^{\varepsilon }\left( 1-\varepsilon \right) -\left( 
\hat{X}-\frac{1}{2}\Omega ^{-1}M\Gamma \right) _{2}\right) ^{2}}{2\left\vert
\delta -\Gamma _{3}\varepsilon \bar{K}^{\varepsilon -1}\right\vert \nu ^{2}}%
\right) }{\left( \erf\left( \frac{\left( C-A\bar{K}^{\varepsilon
}\left( 1-\varepsilon \right) -\left( \hat{X}-\frac{1}{2}\Omega ^{-1}M\Gamma
\right) _{2}\right) }{\sqrt{2\left\vert \delta -\Gamma _{3}\varepsilon \bar{K%
}^{\varepsilon -1}\right\vert \nu ^{2}}}\right) +1\right) } \\
&\simeq &-\frac{\sqrt{\frac{2}{\pi }}\sqrt{\left\vert \delta -\Gamma
_{3}\varepsilon \bar{K}^{\varepsilon -1}\right\vert }\nu \exp \left( -\frac{%
\left( C-A\bar{K}^{\varepsilon }\left( 1-\varepsilon \right) \right) ^{2}}{%
2\left\vert \delta -\Gamma _{3}\varepsilon \bar{K}^{\varepsilon
-1}\right\vert \nu ^{2}}\right) }{\left( \erf\left( \frac{\left( C-A%
\bar{K}^{\varepsilon }\left( 1-\varepsilon \right) -\left( \hat{X}-\frac{1}{2%
}\Omega ^{-1}M\Gamma \right) _{2}\right) }{\sqrt{2}}\right) +1\right) } \\
&\simeq &-\frac{2}{\pi }\allowbreak \left\vert \bar{C}+\sqrt{\frac{2}{\pi }}%
\varpi -A\bar{K}^{\varepsilon }\left( 1-\varepsilon \right) \right\vert -%
\frac{\sqrt{2\left\vert \delta -\Gamma _{3}\varepsilon \bar{K}^{\varepsilon
-1}\right\vert }}{\sqrt{\pi }}\nu
\end{eqnarray*}%
for $\frac{\allowbreak \left\vert C-A\bar{K}^{\varepsilon }\left(
1-\varepsilon \right) \right\vert }{\sqrt{\left\vert \delta -\Gamma
_{3}\varepsilon \bar{K}^{\varepsilon -1}\right\vert }\nu }<1$ and using (\ref%
{gmqn}).\bigskip

At the first order in $\gamma \eta $, one finds:%
\begin{equation*}
\Gamma _{2}=K_{1}^{\prime }-\frac{\nu ^{2}\frac{A_{0}}{1-\chi }\left( \delta
-\Gamma _{3}\varepsilon \bar{K}^{\varepsilon -1}\right) +\bar{K}%
^{\varepsilon }\left( 1-\varepsilon \right)\!\!\left( 1-\left( \delta -\Gamma
_{3}\varepsilon \bar{K}^{\varepsilon -1}\right) \right) K_{1}^{\prime }}{%
2\left( \delta -\Gamma _{3}\varepsilon \bar{K}^{\varepsilon -1}\right) ^{2}}%
\gamma \eta
\end{equation*}%
These expressions allow to find the expectation value of $K$ in the state $%
\Psi _{1}$: 
\begin{eqnarray*}
\left\langle K\right\rangle &=&\left\langle \frac{\left( K^{\prime
}-C\right) +A\bar{K}^{\varepsilon }\left( 1-\varepsilon \right) }{\delta
-A\varepsilon \bar{K}^{\varepsilon -1}}\right\rangle \\
&\simeq &\frac{\left( \Gamma _{2}-\Gamma _{1}\right) +\Gamma _{3}\bar{K}%
^{\varepsilon }\left( 1-\varepsilon \right) }{\delta -\Gamma _{3}\varepsilon 
\bar{K}^{\varepsilon -1}} \\
&=&\frac{\left( \Gamma _{2}-\left( \bar{C}+2\varpi \right) \right) +\Gamma
_{3}\bar{K}^{\varepsilon }\left( 1-\varepsilon \right) }{\delta -\Gamma
_{3}\varepsilon \bar{K}^{\varepsilon -1}}
\end{eqnarray*}%
which expands, at first order in $\gamma \eta $, as:%
\begin{eqnarray*}
\left\langle K\right\rangle =- &&\frac{x-K^{\varepsilon }\frac{A_{0}}{%
1-\varkappa }\left( 1-\varepsilon \right) }{Y}-\frac{1}{2}K^{\varepsilon
}\left( K^{\varepsilon }A_{0}\left( 1-\varepsilon \right) -Y\left(
1-\varkappa \right) x\right) \frac{\allowbreak \allowbreak \allowbreak
\varepsilon x-K\delta \left( 1-\varepsilon \right) }{Y^{4}K\left(
1-\varkappa \right) ^{3}}z\gamma \eta \\
&&-\frac{\bar{K}^{\varepsilon }\left( 1-\varepsilon \right)\!\!\left(
1-Y\right) K_{1}^{\prime }}{2Y^{3}}\gamma \eta -\frac{\nu ^{2}\frac{A_{0}}{%
1-\chi }}{2Y^{2}}\gamma \eta -\frac{\varpi ^{2}\left( \frac{A_{0}}{1-\chi }%
\right) }{2\left( \varsigma ^{2}\varpi ^{2}+\left( AF^{\prime }\left(
K\right) +r_{c}\right) ^{2}\right) Y^{2}}\gamma \eta
\end{eqnarray*}%
where:%
\begin{eqnarray*}
Y &=&\left( \delta -\Gamma _{3}\varepsilon \bar{K}^{\varepsilon -1}\right) \\
x &=&\bar{C}+\sqrt{\frac{2}{\pi }}\varpi -K_{1}^{\prime }
\end{eqnarray*}%
One can also compare the average production $\left\langle Y\right\rangle
=\left\langle A\right\rangle \left\langle K\right\rangle $ in both phases.
For $\gamma \eta =0$:%
\begin{equation*}
\left\langle Y\right\rangle _{0}=\frac{A_{0}}{1-\varkappa }\left( \frac{%
x-K^{\varepsilon }\frac{A_{0}}{1-\varkappa }\left( 1-\varepsilon \right) }{%
\left\vert \delta -K^{\varepsilon -1}\varepsilon \frac{A_{0}}{1-\varkappa }%
\right\vert }\right) ^{\varepsilon }
\end{equation*}%
For $\gamma \eta =0$, neglecting the terms proportional to $\nu ^{2}$ and $%
\varpi ^{2}$, one finds: 
\begin{eqnarray*}
\left\langle Y\right\rangle _{1} &<&\frac{A_{0}}{1-\varkappa }\left( \frac{%
x-K^{\varepsilon }\frac{A_{0}}{1-\varkappa }\left( 1-\varepsilon \right) }{%
\left\vert \delta -K^{\varepsilon -1}\varepsilon \frac{A_{0}}{1-\varkappa }%
\right\vert }\right) ^{\varepsilon } \\
&&-\frac{\left( K^{\varepsilon }A_{0}\left( 1-\varepsilon \right) -Yx\left(
1-\varkappa \right) \right) }{2Y^{2}\left( 1-\varkappa \right) ^{3}}\left( 1-%
\frac{\bar{r}}{\bar{r}-\delta }\left( \varepsilon +\frac{\left(
1-\varepsilon \right) \bar{K}}{\left\langle K\right\rangle }\right)
\allowbreak \right)\!\!\left( \frac{x-K^{\varepsilon }\frac{A_{0}}{1-\varkappa }%
\left( 1-\varepsilon \right) }{\left\vert \delta -K^{\varepsilon
-1}\varepsilon \frac{A_{0}}{1-\varkappa }\right\vert }\right) ^{\varepsilon
}\gamma \eta
\end{eqnarray*}%
with:%
\begin{equation*}
\bar{r}=\frac{\varepsilon K^{\varepsilon }A_{0}}{1-\varkappa }
\end{equation*}%
For a minimal stock of capital lower than the average one, one has $\frac{%
\bar{K}}{\left\langle K\right\rangle }\ll 1$ and $\bar{r}\gg \delta $ since for
the minimal capital stock, the marginal productivity exceeds the
depreciation rate to allow accumulation. As a consequence:%
\begin{equation*}
1-\frac{\bar{r}}{\bar{r}-\delta }\left( \varepsilon +\frac{\left(
1-\varepsilon \right) \bar{K}}{\left\langle K\right\rangle }\right) \simeq
1-\varepsilon \left( 1+\frac{\delta }{\bar{r}}\right) >0
\end{equation*}%
for usual values of $\varepsilon $, $\varepsilon \precsim 0.3$. As a
consequence $\left\langle Y\right\rangle _{1}<\left\langle Y\right\rangle
_{0}$ in most cases.

For the rest of the section, we redefine $\bar{C}+\sqrt{\frac{2}{\pi }}%
\varpi -K_{1}^{\prime }\rightarrow \bar{C}$. The previous results lead then
to the quadratic term in (\ref{sddptqnnbsbb}): 
\begin{eqnarray*}
&&\left( ^{t}\hat{X}\right)\!\!\left( 1+\frac{1}{2}M\Omega ^{-1}\right)
^{-1}\left( M+\frac{1}{4}M\Omega ^{-1}M\right)\!\!\left( 1+\frac{1}{2}\Omega
^{-1}M\right) ^{-1}\hat{X} \\
&=&-\frac{4\gamma \eta \left( \delta -\Gamma _{3}\varepsilon \bar{K}%
^{\varepsilon -1}\right)\!\!\left( 2\left( \frac{\gamma \eta \bar{K}%
^{\varepsilon }\left( 1-\varepsilon \right) }{\delta -\Gamma _{3}\varepsilon 
\bar{K}^{\varepsilon -1}}+4\right)\!\!\left( \left( 1-\varkappa \right) \bar{A}%
+\varkappa \Gamma _{3}\right) \bar{C}+\left( \frac{\gamma \eta \bar{K}%
^{\varepsilon }\left( 1-\varepsilon \right) }{\delta -\Gamma _{3}\varepsilon 
\bar{K}^{\varepsilon -1}}+3\right) \bar{C}^{2}\right) }{4\left( \gamma \eta 
\bar{K}^{\varepsilon }\left( 1-\varepsilon \right) +2\left( \delta -\Gamma
_{3}\varepsilon \bar{K}^{\varepsilon -1}\right) \right) ^{2}} \\
&&+\frac{4\gamma \eta \left( \delta -\Gamma _{3}\varepsilon \bar{K}%
^{\varepsilon -1}\right)\!\!\left( \frac{\gamma \eta \bar{K}^{\varepsilon
}\left( 1-\varepsilon \right) }{\delta -\Gamma _{3}\varepsilon \bar{K}%
^{\varepsilon -1}}+4\right)\!\!\left( \left( 1-\varkappa \right) \bar{A}%
+\varkappa \Gamma _{3}\right) ^{2}}{4\left( \gamma \eta \bar{K}^{\varepsilon
}\left( 1-\varepsilon \right) +2\left( \delta -\Gamma _{3}\varepsilon \bar{K}%
^{\varepsilon -1}\right) \right) ^{2}}
\end{eqnarray*}%
Gathering the equation for $\Gamma _{3}$, and the condition for the
cancellation of the denominator leads to the system:%
\begin{eqnarray*}
\gamma \eta \bar{K}^{\varepsilon }\left( 1-\varepsilon \right) +2\left(
\delta -\Gamma _{3}\varepsilon \bar{K}^{\varepsilon -1}\right) &=&0 \\
\Gamma _{3} &=&\frac{2\left( \left( 1-\varkappa \right) A_{0}+\left(
2-\varkappa \right) \varkappa \Gamma _{3}\right)\!\!\left( \delta -\Gamma
_{3}\varepsilon \bar{K}^{\varepsilon -1}\right) ^{2}+\bar{C}\gamma \eta
\left( \delta -\Gamma _{3}\varepsilon \bar{K}^{\varepsilon -1}\right) }{%
2\left( \delta -\Gamma _{3}\varepsilon \bar{K}^{\varepsilon -1}\right)
^{2}-\left( a+b\right)\!\!\left( \gamma \eta \right) ^{2}\left( \delta -\Gamma
_{3}\varepsilon \bar{K}^{\varepsilon -1}\right) +\gamma \eta \bar{K}%
^{\varepsilon }\left( 1-\varepsilon \right) }
\end{eqnarray*}%
\begin{equation*}
\frac{2\delta +x\bar{K}^{\varepsilon }\left( 1-\varepsilon \right) }{%
2\varepsilon \bar{K}^{\varepsilon -1}}=2\frac{2\left( \left( 1-\varkappa
\right) A_{0}+\left( 2-\varkappa \right) \varkappa \frac{2\delta +x\bar{K}%
^{\varepsilon }\left( 1-\varepsilon \right) }{2\varepsilon \bar{K}%
^{\varepsilon -1}}\right)\!\!\left( x\frac{\bar{K}^{\varepsilon }\left(
1-\varepsilon \right) }{2}\right) ^{2}-\bar{C}x^{2}\frac{\bar{K}%
^{\varepsilon }\left( 1-\varepsilon \right) }{2}}{4\left( x\frac{\bar{K}%
^{\varepsilon }\left( 1-\varepsilon \right) }{2}\right) ^{2}+\left(
a+b\right) x^{3}\frac{\bar{K}^{\varepsilon }\left( 1-\varepsilon \right) }{2}%
+2x\bar{K}^{\varepsilon }\left( 1-\varepsilon \right) }
\end{equation*}%
$\allowbreak \allowbreak $which has no solution, and thus the potential is
defined for all $\gamma \eta $. The solution $\Gamma _{3}$ for $\gamma \eta
\rightarrow \infty $ has the asymptotic form:%
\begin{equation*}
\Gamma _{3}=c\gamma \eta
\end{equation*}%
with constant $c$ satisfying:%
\begin{equation*}
c=\frac{\left( 2-\varkappa \right) \varkappa \varepsilon \bar{K}%
^{\varepsilon -1}c^{2}}{\left( a+b\right) }
\end{equation*}%
so that:%
\begin{equation*}
\Gamma _{3}=\frac{\left( a+b\right) \bar{K}^{1-\varepsilon }}{\left(
2-\varkappa \right) \varkappa \varepsilon }\gamma \eta
\end{equation*}%
The term $\delta -\Gamma _{3}\varepsilon \bar{K}^{\varepsilon -1}$ cancels
at $\gamma \eta \gg 1$, for:%
\begin{equation*}
\gamma \eta \rightarrow \frac{\delta \left( 2-\varkappa \right) \varkappa }{%
a+b}\gg 1
\end{equation*}%
\begin{equation*}
\Gamma _{3}-\rightarrow \frac{\delta }{\varepsilon \bar{K}^{\varepsilon -1}}
\end{equation*}%
and our approximations are no more valid above these values. For $\gamma
\eta \rightarrow \frac{\delta \left( 2-\varkappa \right) \varkappa }{a+b}$,
one finds:%
\begin{eqnarray*}
&&-\frac{4\gamma \eta \left( 2\left( \gamma \eta \bar{K}^{\varepsilon
}\left( 1-\varepsilon \right) \right)\!\!\left( \left( 1-\varkappa \right) \bar{%
A}+\varkappa \Gamma _{3}\right) \bar{C}+\left( \gamma \eta \bar{K}%
^{\varepsilon }\left( 1-\varepsilon \right) \right) \bar{C}^{2}-\left(
\gamma \eta \bar{K}^{\varepsilon }\left( 1-\varepsilon \right) \right)
\left( \left( 1-\varkappa \right) \bar{A}+\varkappa \Gamma _{3}\right)
^{2}\right) }{4\left( \gamma \eta \bar{K}^{\varepsilon }\left( 1-\varepsilon
\right) \right) ^{2}} \\
&=&\frac{\left( \left( \left( 1-\varkappa \right) \bar{A}+\varkappa \Gamma
_{3}\right) ^{2}-2\bar{C}\left( \left( 1-\varkappa \right) \bar{A}+\varkappa
\Gamma _{3}\right) -\bar{C}^{2}\right) }{\left( 1-\varepsilon \right)
K^{\varepsilon }}
\end{eqnarray*}%
If $\left( \left( 1-\varkappa \right) \bar{A}+\varkappa \Gamma _{3}\right)
^{2}-2C\left( \left( 1-\varkappa \right) \bar{A}+\varkappa \Gamma
_{3}\right) -C^{2}>0$, i.e. for:%
\begin{equation*}
\left( 1-\varkappa \right) A_{0}+\left( 2-\varkappa \right) \varkappa \Gamma
_{3}>\left( 1+\sqrt{2}\right) \bar{C}
\end{equation*}%
that is:%
\begin{equation*}
A_{0}>\frac{\left( 1+\sqrt{2}\right) \bar{C}-\frac{\left( 2-\varkappa
\right) \varkappa \delta }{\varepsilon }\bar{K}^{1-\varepsilon }}{%
1-\varkappa }
\end{equation*}%
then:%
\begin{eqnarray*}
0 &=&\left( ^{t}\hat{X}\right)\!\!\left( 1+\frac{1}{2}M\Omega ^{-1}\right)
^{-1}\left( M+\frac{1}{4}M\Omega ^{-1}M\right)\!\!\left( 1+\frac{1}{2}\Omega
^{-1}M\right) ^{-1}\hat{X}\left( 0\right) \\
&<&\left( ^{t}\hat{X}\right)\!\!\left( 1+\frac{1}{2}M\Omega ^{-1}\right)
^{-1}\left( M+\frac{1}{4}M\Omega ^{-1}M\right)\!\!\left( 1+\frac{1}{2}\Omega
^{-1}M\right) ^{-1}\hat{X}\left( \frac{\delta \left( 2-\varkappa \right)
\varkappa }{a+b}\right)
\end{eqnarray*}%
Equation (\ref{sddptqnntt}) has solutions of the type:%
\begin{eqnarray*}
\Psi _{n_{1},n_{2},n_{3}} &=&H_{n_{1}}\left( \left( C-\Gamma _{1}\right)
\right) \exp \left( -\frac{\left( \sqrt{\frac{\varsigma ^{2}}{\varpi ^{2}}+%
\frac{\left( \hat{r}+r_{c}\right) ^{2}}{\varpi ^{4}}}\right)\!\!\left( C-\Gamma
_{1}\right) ^{2}}{2}\right) \\
&&\times H_{n_{2}}\left( A-\Gamma _{3}\right) \exp \left( -\frac{\lambda
\left( A-\Gamma _{3}\right) ^{2}}{2}\right) H_{n_{3}}\left( K^{\prime
}-\Gamma _{2}\right) \exp \left( -\frac{\left( K^{\prime }-\Gamma
_{2}\right) ^{2}}{2\left\vert \delta -\Gamma _{3}\varepsilon \bar{K}%
^{\varepsilon -1}\right\vert \nu ^{2}}\right)
\end{eqnarray*}%
with $\hat{r}$ defined by:%
\begin{equation*}
\hat{r}+r_{c}\simeq \left\langle \varepsilon A\left( \frac{\left( K^{\prime
}-C\right) +A\bar{K}^{\varepsilon }\left( 1-\varepsilon \right) }{\delta
-A\varepsilon \bar{K}^{\varepsilon -1}}\right) ^{\varepsilon
-1}\right\rangle +r_{c}\simeq \varepsilon \Gamma _{3}\left( \frac{\left(
\Gamma _{2}-\Gamma _{1}\right) +\bar{K}^{\varepsilon }\left( 1-\varepsilon
\right) \Gamma _{3}}{\delta -\Gamma _{3}\varepsilon \bar{K}^{\varepsilon -1}}%
\right) ^{\varepsilon -1}+r_{c}
\end{equation*}%
if the integers $n_{1},n_{2},n_{3}$ and $C_{0}$ satisfy the compatibility
condition:%
\begin{eqnarray*}
&&\alpha -C_{0}+\left( ^{t}\Gamma \right)\!\!\left( M+\frac{1}{4}M\Omega
^{-1}M\right) \Gamma +\bar{A}^{2}-\left( \left( 1-\varkappa \right) \bar{A}%
+\varkappa \Gamma _{3}\right) ^{2} \\
&=&-\left( 2n_{1}+1\right) \sqrt{\varsigma ^{2}\varpi ^{2}+\left( \hat{r}%
+r_{c}\right) ^{2}}-\frac{2n_{2}+1}{\lambda }-\left( 2n_{3}+1\right)
\left\vert \delta -\Gamma _{3}\varepsilon \bar{K}^{\varepsilon -1}\right\vert
\end{eqnarray*}%
A minimum for the action may thus exist for 
\begin{equation*}
C_{0}>\alpha +Min_{\eta }\left( \left( ^{t}\Gamma \right)\!\!\left( M+\frac{1}{4%
}M\Omega ^{-1}M\right) \Gamma \right) +\sqrt{\varsigma ^{2}\varpi
^{2}+\left( \hat{r}+r_{c}\right) ^{2}}+\frac{1}{\lambda }+\left\vert \delta
-\Gamma _{3}\varepsilon \bar{K}^{\varepsilon -1}\right\vert +\left( \bar{A}%
^{2}-\left( \left( 1-\varkappa \right) \bar{A}+\varkappa \Gamma _{3}\right)
^{2}\right)
\end{equation*}%
Since the minimum for $\left( \left( ^{t}\Gamma \right)\!\!\left( M+\frac{1}{4}%
M\Omega ^{-1}M\right) \Gamma \right) $ is $0$ for $\gamma \eta =0$, and
since for $\gamma \eta =0$, \ $\Gamma _{3}=\frac{A_{0}}{\left( 1-\varkappa
\right) }$, $\bar{A}=\frac{A_{0}}{\left( 1-\varkappa \right) }$, so that:%
\begin{equation*}
\bar{A}^{2}-\left( \left( 1-\varkappa \right) \bar{A}+\varkappa \Gamma
_{3}\right) ^{2}=0
\end{equation*}%
and the condition reduces to: 
\begin{equation}
C_{0}>\alpha +\sqrt{\varsigma ^{2}\varpi ^{2}+\left( \hat{r}\left( 0\right)
+r_{c}\right) ^{2}}+\frac{1}{\lambda }+\left\vert \delta -\Gamma
_{3}\varepsilon \bar{K}^{\varepsilon -1}\right\vert  \label{cmpbtcndt}
\end{equation}%
In that case, the compatibility fixes the value for $\gamma \eta $. For $%
n_{1}=n_{2}=n_{3}=0$ it is:%
\begin{eqnarray}
0 &=&\hskip-8pt\alpha -C_{0}-\frac{4\gamma \eta \left( \delta -\Gamma _{3}\varepsilon 
\bar{K}^{\varepsilon -1}\right)\!\!\left( 2\left( \frac{\gamma \eta \bar{K}%
^{\varepsilon }\left( 1-\varepsilon \right) }{\delta -\Gamma _{3}\varepsilon 
\bar{K}^{\varepsilon -1}}+4\right) \bar{A}\bar{C}+\left( \frac{\gamma \eta 
\bar{K}^{\varepsilon }\left( 1-\varepsilon \right) }{\delta -\Gamma
_{3}\varepsilon \bar{K}^{\varepsilon -1}}+3\right) \bar{C}^{2}-\left( \frac{%
\gamma \eta \bar{K}^{\varepsilon }\left( 1-\varepsilon \right) }{\delta
-\Gamma _{3}\varepsilon \bar{K}^{\varepsilon -1}}+4\right) \bar{A}%
^{2}\right) }{4\left( \gamma \eta \bar{K}^{\varepsilon }\left( 1-\varepsilon
\right) +2\left( \delta -\Gamma _{3}\varepsilon \bar{K}^{\varepsilon
-1}\right) \right) ^{2}}  \label{cmpblt} \\
&&+\sqrt{\varsigma ^{2}\varpi ^{2}+\left( \hat{r}+r_{c}\right) ^{2}}+\frac{1%
}{\lambda }+\left\vert \delta -\Gamma _{3}\varepsilon \bar{K}^{\varepsilon
-1}\right\vert +\left( \bar{A}^{2}-\left( \left( 1-\varkappa \right) \bar{A}%
+\varkappa \Gamma _{3}\right) ^{2}\right)  \notag \\
&\equiv &g\left( \eta \right)  \notag
\end{eqnarray}%
for $\left( a+b\right) \ll 1$. If we find a solution to (\ref{cmpblt}) with $%
\eta \neq 0$, then (\ref{sddptqnntt}) will have a solution with $\gamma \eta
\neq 0$. \ To inspect (\ref{cmpblt}), we compare the case $\gamma \eta =0$
and the case $\gamma \eta \rightarrow \frac{\delta \left( 2-\varkappa
\right) \varkappa }{a+b}$. For $\gamma \eta =0$: $\allowbreak $%
\begin{equation*}
g\left( 0\right) =\alpha -C_{0}+\sqrt{\varsigma ^{2}\varpi ^{2}+\left(
\varepsilon \left( \frac{A_{0}}{\left( 1-\varkappa \right) }\right)
^{\varepsilon }\left( \frac{\bar{K}^{\varepsilon }\left( 1-\varepsilon
\right) -\bar{C}}{\delta -\frac{A_{0}}{\left( 1-\varkappa \right) }%
\varepsilon \bar{K}^{\varepsilon -1}}\right) ^{\varepsilon -1}+r_{c}\right)
^{2}\varpi ^{2}}+\frac{1}{\lambda }+\left\vert \delta -\frac{%
A_{0}\varepsilon \bar{K}^{\varepsilon -1}}{\left( 1-\varkappa \right) }%
\right\vert
\end{equation*}%
If one has:%
\begin{equation*}
\left\vert \delta -\frac{A_{0}\varepsilon \bar{K}^{\varepsilon -1}}{\left(
1-\varkappa \right) }\right\vert \ll 1
\end{equation*}%
then:%
\begin{equation*}
\frac{\bar{K}^{\varepsilon }\left( 1-\varepsilon \right) -\bar{C}}{\delta -%
\frac{A_{0}}{\left( 1-\varkappa \right) }\varepsilon \bar{K}^{\varepsilon -1}%
}\gg 1
\end{equation*}%
which implies:%
\begin{equation*}
\varepsilon \left( \frac{A_{0}}{\left( 1-\varkappa \right) }\right)
^{\varepsilon }\left( \frac{\bar{K}^{\varepsilon }\left( 1-\varepsilon
\right) -\bar{C}}{\delta -\frac{A_{0}}{\left( 1-\varkappa \right) }%
\varepsilon \bar{K}^{\varepsilon -1}}\right) ^{\varepsilon -1}\ll 1
\end{equation*}%
so that:%
\begin{equation*}
g\left( 0\right) =\alpha -C_{0}+\sqrt{\varsigma ^{2}\varpi ^{2}+\left(
r_{c}\right) ^{2}\varpi ^{2}}+\frac{1}{\lambda }
\end{equation*}%
Now consider $\gamma \eta \rightarrow \frac{\delta \left( 2-\varkappa
\right) \varkappa }{a+b}$. In that case:

\begin{eqnarray*}
\bar{A}^{2}-\left( \left( 1-\varkappa \right) \bar{A}+\varkappa \Gamma
_{3}\right) ^{2} &=&\left( A_{0}+\varkappa \Gamma _{3}\right) ^{2}-\left(
\left( 1-\varkappa \right) A_{0}+\left( 2-\varkappa \right) \varkappa \Gamma
_{3}\right) ^{2} \\
&=&\left( A_{0}+\varkappa \frac{\delta }{\varepsilon \bar{K}^{\varepsilon -1}%
}\right) ^{2}-\left( \left( 1-\varkappa \right) A_{0}+\left( 2-\varkappa
\right) \varkappa \frac{\delta }{\varepsilon \bar{K}^{\varepsilon -1}}%
\right) ^{2} \\
&=&-\allowbreak \left( 1-\varkappa \right) K\frac{\varkappa \left( \delta
-K^{\varepsilon -1}\varepsilon \frac{A_{0}}{\left( 1-\varkappa \right) }%
\right)\!\!\left( \left( 2-\varkappa \right) A_{0}+\left( 3-\varkappa \right)
\varkappa \frac{\delta }{\varepsilon \bar{K}^{\varepsilon -1}}\right) }{%
\varepsilon K^{\varepsilon }}
\end{eqnarray*}%
and this is positive given our assumptions. Moreover, given the same
assumptions:%
\begin{eqnarray*}
&&\frac{\left( \left( \left( 1-\varkappa \right) \bar{A}+\varkappa \Gamma
_{3}\right) ^{2}-2\bar{C}\left( \left( 1-\varkappa \right) \bar{A}+\varkappa
\Gamma _{3}\right) -\bar{C}^{2}\right) }{\left( 1-\varepsilon \right)
K^{\varepsilon }} \\
&=&\frac{\left( \left( \left( 1-\varkappa \right)\!\!\left( A_{0}+\varkappa 
\frac{\delta \bar{K}^{1-\varepsilon }}{\varepsilon }\right) +\varkappa \frac{%
\delta \bar{K}^{1-\varepsilon }}{\varepsilon }\right) ^{2}-2\bar{C}\left(
\left( 1-\varkappa \right)\!\!\left( A_{0}+\varkappa \frac{\delta \bar{K}%
^{1-\varepsilon }}{\varepsilon }\right) +\varkappa \frac{\delta \bar{K}%
^{1-\varepsilon }}{\varepsilon }\right) -\bar{C}^{2}\right) }{\left(
1-\varepsilon \right) K^{\varepsilon }}>0
\end{eqnarray*}%
and as a consequence:%
\begin{equation*}
U=\bar{A}^{2}-\left( \left( 1-\varkappa \right) \bar{A}+\varkappa \Gamma
_{3}\right) ^{2}+\frac{\left( \left( \left( 1-\varkappa \right) \bar{A}%
+\varkappa \Gamma _{3}\right) ^{2}-2\bar{C}\left( \left( 1-\varkappa \right) 
\bar{A}+\varkappa \Gamma _{3}\right) -\bar{C}^{2}\right) }{\left(
1-\varepsilon \right) K^{\varepsilon }}>0
\end{equation*}%
$\allowbreak $ $\allowbreak $ $\allowbreak $%
\begin{eqnarray*}
\hat{r}+r_{c} &\simeq &\left\langle \varepsilon A\left( \frac{\left(
K^{\prime }-C\right) +A\bar{K}^{\varepsilon }\left( 1-\varepsilon \right) }{%
\delta -A\varepsilon \bar{K}^{\varepsilon -1}}\right) ^{\varepsilon
-1}\right\rangle +r_{c} \\
&\simeq &\varepsilon \Gamma _{3}\left( \frac{\left( \Gamma _{2}-\Gamma
_{1}\right) +\bar{K}^{\varepsilon }\left( 1-\varepsilon \right) \Gamma _{3}}{%
\delta -\Gamma _{3}\varepsilon \bar{K}^{\varepsilon -1}}\right)
^{\varepsilon -1}+r_{c}\rightarrow r_{c}
\end{eqnarray*}%
since $\delta -\Gamma _{3}\varepsilon \bar{K}^{\varepsilon -1}\rightarrow 0$
and $\varepsilon -1<0$. Then:%
\begin{equation*}
g\left( \frac{\delta \left( 2-\varkappa \right) \varkappa }{a+b}\right)
=\alpha -C_{0}+\sqrt{\varsigma ^{2}\varpi ^{2}+\left( r_{c}\right)
^{2}\varpi ^{2}}+\frac{1}{\lambda }+U
\end{equation*}%
with $U>0$. As a consequence, $g\left( 0\right) <g\left( \frac{\delta \left(
2-\varkappa \right) \varkappa }{a+b}\right) $, and for all the parameters
satisfying our assumptions:%
\begin{eqnarray*}
A_{0} &>&\frac{\left( 1+\sqrt{2}\right) \bar{C}-\frac{\left( 2-\varkappa
\right) \varkappa \delta }{\varepsilon }\bar{K}^{1-\varepsilon }}{%
1-\varkappa } \\
\left\vert \delta -\frac{A_{0}\varepsilon \bar{K}^{\varepsilon -1}}{\left(
1-\varkappa \right) }\right\vert &<&<1
\end{eqnarray*}%
and all $C_{0}$ such that $g\left( \frac{\delta \left( 2-\varkappa \right)
\varkappa }{a+b}\right) >0>g\left( 0\right) $, that is:%
\begin{equation*}
C_{0}\in \left] \alpha +\sqrt{\varsigma ^{2}\varpi ^{2}+\left( r_{c}\right)
^{2}\varpi ^{2}}+\frac{1}{\lambda },\alpha +\sqrt{\varsigma ^{2}\varpi
^{2}+\left( r_{c}\right) ^{2}\varpi ^{2}}+\frac{1}{\lambda }+U\right[
\end{equation*}%
there is $\left( \gamma \eta \right) _{0}$ such that the equation $g\left(
\left( \gamma \eta \right) _{0}\right) =0$. An estimation for $\gamma \eta $
can be obtained by rewriting the compatibility condition (\ref{cmpblt}):

\begin{eqnarray*}
0 &=&\alpha -C_{0}-\frac{4\gamma \eta \left( 2\left( \frac{\gamma \eta \bar{K%
}^{\varepsilon }\left( 1-\varepsilon \right) }{\delta -\Gamma
_{3}\varepsilon \bar{K}^{\varepsilon -1}}+4\right) \bar{A}\bar{C}+\left( 
\frac{\gamma \eta \bar{K}^{\varepsilon }\left( 1-\varepsilon \right) }{%
\delta -\Gamma _{3}\varepsilon \bar{K}^{\varepsilon -1}}+3\right) \bar{C}%
^{2}-\left( \frac{\gamma \eta \bar{K}^{\varepsilon }\left( 1-\varepsilon
\right) }{\delta -\Gamma _{3}\varepsilon \bar{K}^{\varepsilon -1}}+4\right) 
\bar{A}^{2}\right) }{4\left( \delta -\Gamma _{3}\varepsilon \bar{K}%
^{\varepsilon -1}\right)\!\!\left( \frac{\gamma \eta \bar{K}^{\varepsilon
}\left( 1-\varepsilon \right) }{\left( \delta -\Gamma _{3}\varepsilon \bar{K}%
^{\varepsilon -1}\right) }+2\right) ^{2}} \\
&&+\sqrt{\varsigma ^{2}\varpi ^{2}+\left( \hat{r}+r_{c}\right) ^{2}}+\frac{1%
}{\lambda }+\left\vert \delta -\Gamma _{3}\varepsilon \bar{K}^{\varepsilon
-1}\right\vert +\bar{A}^{2}-\left( \left( 1-\varkappa \right) \bar{A}%
+\varkappa \Gamma _{3}\right) ^{2}
\end{eqnarray*}%
as:%
\begin{equation*}
\frac{x\left( 2\left( x+4\right) \bar{A}\bar{C}+\left( x+3\right) \bar{C}%
^{2}-\left( x+4\right) \bar{A}^{2}\right) }{\left( x+2\right) ^{2}}=D
\end{equation*}%
with:%
\begin{eqnarray*}
x &=&\frac{\gamma \eta \bar{K}^{\varepsilon }\left( 1-\varepsilon \right) }{%
\left( \delta -\Gamma _{3}\varepsilon \bar{K}^{\varepsilon -1}\right) } \\
D &=&\bar{K}^{\varepsilon }\left( 1-\varepsilon \right)\!\!\left( \alpha -C_{0}+%
\sqrt{\varsigma ^{2}\varpi ^{2}+\left( \hat{r}+r_{c}\right) ^{2}}+\frac{1}{%
\lambda }+\left\vert \delta -\Gamma _{3}\varepsilon \bar{K}^{\varepsilon
-1}\right\vert +\bar{A}^{2}-\left( \left( 1-\varkappa \right) \bar{A}%
+\varkappa \Gamma _{3}\right) ^{2}\right)
\end{eqnarray*}%
for $\varkappa \ll 1$, we can approximate $\left( \bar{A}^{2}-\left( \left(
1-\varkappa \right) \bar{A}+\varkappa \Gamma _{3}\right) ^{2}\right) $ by
its value for $\gamma \eta =0$:%
\begin{equation*}
\bar{A}^{2}-\left( \left( 1-\varkappa \right) \bar{A}+\varkappa \Gamma
_{3}\right) ^{2}=0
\end{equation*}%
as well as $\hat{r}$ by $0$. Note that $D>0$, due to (\ref{cmpbtcndt}). The
compatibility equation can be be expanded as:%
\begin{eqnarray*}
&&x\left( 2\left( x+4\right) \bar{A}\bar{C}+\left( x+3\right) \bar{C}%
^{2}-\left( x+4\right) \bar{A}^{2}\right) -\left( x+2\right) ^{2}D \\
&=&\allowbreak \left( C^{2}+2C\bar{A}-\bar{A}^{2}-D\right) x^{2}+\left(
3C^{2}+8C\bar{A}-4\bar{A}^{2}-4D\right) \allowbreak x-4D
\end{eqnarray*}%
Given our assumptions $C^{2}+2C\bar{A}-\bar{A}^{2}<0$ and $3C^{2}+8C\bar{A}-4%
\bar{A}^{2}<4\left( C^{2}+2C\bar{A}-\bar{A}^{2}\right) <0$, so that:%
\begin{equation*}
x=\frac{-\left( 3C^{2}+8C\bar{A}-4\bar{A}^{2}-4D\right) -\sqrt{\left(
3C^{2}+8C\bar{A}-4\bar{A}^{2}-4D\right) ^{2}+16D\allowbreak \left( C^{2}+2C%
\bar{A}-\bar{A}^{2}-D\right) }}{2\allowbreak \left( C^{2}+2C\bar{A}-\bar{A}%
^{2}-D\right) }
\end{equation*}

\bigskip For $A_{0}\gg 1$ and $D<A_{0}$ and thus:%
\begin{eqnarray*}
\left\vert x\right\vert &\simeq &\frac{4D}{\left\vert 3C^{2}+8C\bar{A}-4\bar{%
A}^{2}-4D\right\vert }\ll 1 \\
\gamma \eta &<&\frac{\left\vert \delta -\frac{A_{0}}{1-\chi }\varepsilon 
\bar{K}^{\varepsilon -1}\right\vert }{\bar{K}^{\varepsilon }\left(
1-\varepsilon \right) }\frac{4D}{\left\vert 3C^{2}+8C\bar{A}-4\bar{A}%
^{2}-4D\right\vert }\ll 1
\end{eqnarray*}

Since:%
\begin{equation*}
C_{0}\in \left] \alpha +\sqrt{\varsigma ^{2}\varpi ^{2}+\left( r_{c}\right)
^{2}\varpi ^{2}}+\frac{1}{\lambda },\alpha +\sqrt{\varsigma ^{2}\varpi
^{2}+\left( r_{c}\right) ^{2}\varpi ^{2}}+\frac{1}{\lambda }+U\right[
\end{equation*}%
with $U$ defined by:%
\begin{eqnarray}
U &=&-\allowbreak \left( 1-\varkappa \right) K\frac{\varkappa \left( \delta
-K^{\varepsilon -1}\varepsilon \frac{A_{0}}{\left( 1-\varkappa \right) }%
\right)\!\!\left( \left( 2-\varkappa \right) A_{0}+\left( 3-\varkappa \right)
\varkappa \frac{\delta }{\varepsilon \bar{K}^{\varepsilon -1}}\right) }{%
\varepsilon K^{\varepsilon }}  \label{stmnbb} \\
&&+\frac{\left( \left( \left( 1-\varkappa \right)\!\!\left( A_{0}+\varkappa 
\frac{\delta \bar{K}^{1-\varepsilon }}{\varepsilon }\right) +\varkappa \frac{%
\delta \bar{K}^{1-\varepsilon }}{\varepsilon }\right) ^{2}-2\bar{C}\left(
\left( 1-\varkappa \right)\!\!\left( A_{0}+\varkappa \frac{\delta \bar{K}%
^{1-\varepsilon }}{\varepsilon }\right) +\varkappa \frac{\delta \bar{K}%
^{1-\varepsilon }}{\varepsilon }\right) -\bar{C}^{2}\right) }{\left(
1-\varepsilon \right) K^{\varepsilon }}  \notag
\end{eqnarray}%
then%
\begin{equation*}
\alpha -C_{0}+\sqrt{\varsigma ^{2}\varpi ^{2}+\left( r_{c}\right) ^{2}}+%
\frac{1}{\lambda }\in \left] 0,U\right[
\end{equation*}%
and:%
\begin{equation}
\gamma \eta \in \left] 0,8\frac{\left\vert \delta -\frac{A_{0}}{1-\chi }%
\varepsilon \bar{K}^{\varepsilon -1}\right\vert }{\bar{K}^{\varepsilon
}\left( 1-\varepsilon \right) }U\right[  \label{stmn}
\end{equation}

\subsection*{Saddle point stability}

The solution of (\ref{sddptqnn}) may thus present a non trivial minimum, as
asserted before. To prove this point, we have to show that among the set of
possible solutions of (\ref{sddptqnn}), the action $S\left( \Psi \right) $
is bounded from below. \ Moreover, the second order variation of $S\left(
\Psi \right) $ around the solution with the lowest value of $S\left( \Psi
\right) $ has to be positive. We write this second order variation $\delta
^{2}S\left( \Psi \right) $. We decompose the variation $\varphi \left(
K,C,A\right) $ in three parts. The first part, $\varphi \left( K,C,A\right) $
is orthogonal to the fundamental $\Psi _{1}\left( K,C,A\right) $. We compute
below its contribution to $\delta ^{2}S\left( \Psi \right) $. The second
part is proportional to $\Psi _{1}\left( K,C,A\right) $ and corresponds to a
variation of the norm $\eta $ of $\sqrt{\eta }\Psi _{1}\left( K,C,A\right) $%
, and we write this variation $\delta \sqrt{\eta }\Psi _{1}\left(
K,C,A\right) $. The variation of the action with respect to $\eta $ is thus:

\begin{eqnarray*}
\frac{1}{2}\delta _{\sqrt{\eta }}S\left( \Psi \right) &=&\left( \delta \sqrt{%
\eta }\right) ^{2}\int \Psi _{1}^{\dag }\left( K,C,A\right) \left\{ -\varpi
^{2}\frac{\partial ^{2}}{\partial C^{2}}-\frac{1}{\lambda ^{2}}\frac{%
\partial ^{2}}{\partial A^{2}}-\nu ^{2}\left( \delta -AF^{\prime }\left(
K\right) \right) ^{2}\frac{\partial ^{2}}{\partial K^{\prime 2}}\right. \\
&&\left. +\left( A-\bar{A}\right) ^{2}+2\left( \bar{A}-\Gamma _{3}\right)
\varkappa A+\left( \varsigma ^{2}+\frac{\left( AF^{\prime }\left( K\right)
+r_{c}\right) ^{2}}{\varpi ^{2}}\right)\!\!\left( C-\bar{C}\right) ^{2}+\frac{%
\left( K^{\prime }\right) ^{2}}{\nu ^{2}}+\alpha -C_{0}\right\} \Psi
_{1}\left( K,C,A\right) \\
&&+3\gamma \left( \delta \sqrt{\eta }\right) ^{2}\eta \int \Psi _{1}^{\dag
}\left( K,C,A\right) K\Psi _{1}\left( K,C,A\right) \int \Psi _{1}^{\dag
}\left( K,C,A\right) A\Psi _{1}\left( K,C,A\right)
\end{eqnarray*}%
Given the saddle point equation (\ref{sddptqnnbsbb}), it reduces to:%
\begin{equation}
\frac{1}{2}\delta _{\sqrt{\eta }}S\left( \Psi \right) =2\gamma \left( \delta 
\sqrt{\eta }\right) ^{2}\eta \int \Psi _{1}^{\dag }\left( K,C,A\right) K\Psi
_{1}\left( K,C,A\right) \int \Psi _{1}^{\dag }\left( K,C,A\right) A\Psi
_{1}\left( K,C,A\right) >0  \label{delsqrt}
\end{equation}%
Since this the average value of $A$ multiplied by the average value of $K$
in the state $\Psi _{1}\left( K,C,A\right) $. The third part is a
combination of the variation in the direction of $\Psi _{1}\left(
K,C,A\right) $ and of the $\varphi \left( K,C,A\right) $ variation
orthogonal to $\Psi _{1}\left( K,C,A\right) $. The corresponding second
order variation is:%
\begin{eqnarray}
\frac{1}{2}\delta _{\sqrt{\eta },\varphi }S\left( \Psi \right) &=&\delta 
\sqrt{\eta }\int \!\!\left( \varphi ^{\dag }\left( K,C,A\right) +\varphi \left(
K,C,A\right) \right)\times    \label{delsqrtbb} \ \\ &&\qquad\qquad
\left\{ -\varpi ^{2}\frac{\partial ^{2}}{\partial C^{2}}%
-\frac{1}{\lambda ^{2}}\frac{\partial ^{2}}{\partial A^{2}}-\nu ^{2}\left(
\delta -AF^{\prime }\left( K\right) \right) ^{2}\frac{\partial ^{2}}{%
\partial K^{\prime 2}}+\left( A-\bar{A}\right) ^{2}+2\left( \bar{A}-\Gamma
_{3}\right) \varkappa A\right.  \notag \\
&&\left. +\left( \varsigma ^{2}+\frac{\left( AF^{\prime }\left( K\right)
+r_{c}\right) ^{2}}{\varpi ^{2}}\right)\!\!\left( C-\bar{C}\right) ^{2}+\frac{%
\left( K^{\prime }\right) ^{2}}{\nu ^{2}}+\alpha -C_{0}\right\} \Psi
_{1}\left( K,C,A\right)  \notag \\
&&+2\gamma \left( \delta \sqrt{\eta }\right) ^{2}\eta \int \!\!\left( \varphi
^{\dag }\left( K,C,A\right) +\varphi \left( K,C,A\right) \right) K\Psi
_{1}\left( K,C,A\right) \int \Psi _{1}^{\dag }\left( K,C,A\right) A\Psi
_{1}\left( K,C,A\right)  \notag \\
&&+2\gamma \left( \delta \sqrt{\eta }\right) ^{2}\eta \int \Psi _{1}^{\dag
}\left( K,C,A\right) K\Psi _{1}\left( K,C,A\right) \int \!\!\left( \varphi
^{\dag }\left( K,C,A\right) +\varphi \left( K,C,A\right) \right) A\Psi
_{1}\left( K,C,A\right)  \notag \\
&=&\gamma \left( \delta \sqrt{\eta }\right) ^{2}\eta \int \!\!\left( \varphi
^{\dag }\left( K,C,A\right) +\varphi \left( K,C,A\right) \right) K\Psi
_{1}\left( K,C,A\right) \int \Psi _{1}^{\dag }\left( K,C,A\right) A\Psi
_{1}\left( K,C,A\right)  \notag \\
&&+\gamma \left( \delta \sqrt{\eta }\right) ^{2}\eta \int \Psi _{1}^{\dag
}\left( K,C,A\right) K\Psi _{1}\left( K,C,A\right) \int \!\!\left( \varphi
^{\dag }\left( K,C,A\right) +\varphi \left( K,C,A\right) \right) A\Psi
_{1}\left( K,C,A\right)  \notag
\end{eqnarray}%
where the saddle point equation (\ref{sddptqnnbsbb}) has been used in the
last equation.We will show below that such a contibution is neglible with
respect to (\ref{delsqrt}) or with respect variations involving $\varphi
\left( K,C,A\right) $ only. As a consequence, the second order variation
involving a variation in the direction of $\Psi _{1}\left( K,C,A\right) $\
is positive. We can now turn to the part involving only variations $\varphi
\left( K,C,A\right) $ orthogonal to $\Psi _{1}\left( K,C,A\right) $.\ 
\begin{eqnarray}
\frac{1}{2}\delta ^{2}S\left( \Psi \right) &=&\int \varphi ^{\dag }\left(
K,C,A\right) \left\{ -\varpi ^{2}\frac{\partial ^{2}}{\partial C^{2}}-\frac{1%
}{\lambda ^{2}}\frac{\partial ^{2}}{\partial A^{2}}-\nu ^{2}\left( \delta
-AF^{\prime }\left( K\right) \right) ^{2}\frac{\partial ^{2}}{\partial
K^{\prime 2}}+\left( A-\Gamma _{3}\right) ^{2}\right.  \label{delta2Sbb} \\
&&\left. +\left( \varsigma ^{2}+\frac{\left( \hat{r}+r_{c}\right) ^{2}}{%
\varpi ^{2}}\right)\!\!\left( C-\Gamma _{1}\right) ^{2}+\frac{\left( K^{\prime
}-\Gamma _{2}\right) ^{2}}{\nu ^{2}}\right.  \notag \\
&&\left. +\alpha +\left( ^{t}\Gamma \right)\!\!\left( M+\frac{1}{4}M\Omega
^{-1}M\right) \Gamma -C_{0}+\varkappa \left( \left( 2-\varkappa \right)
A_{0}+\varkappa \left( 3-\varkappa \right) \Gamma _{3}\right)\!\!\left(
A_{0}-\Gamma _{3}\left( 1-\varkappa \right) \right) \right\} \varphi \left(
K,C,A\right)  \notag \\
&&-2\left( 1-\varkappa \right) \varkappa \left( \int \!\!\left( \varphi \left(
K,C,A\right) +\varphi ^{\dag }\left( K,C,A\right) \right) A\Psi _{1}\left(
K,C,A\right) \right) ^{2}  \notag \\
&&+\gamma \eta \int \!\!\left( \varphi \left( K,C,A\right) +\varphi ^{\dag
}\left( K,C,A\right) \right) K\Psi _{1}\left( K,C,A\right) \int \!\!\left(
\varphi \left( K,C,A\right) +\varphi ^{\dag }\left( K,C,A\right) \right)
A\Psi _{1}\left( K,C,A\right)  \notag
\end{eqnarray}

\bigskip Where $\sqrt{\eta }\Psi _{1}\left( K,C,A\right) $ is the
fundamental previously computed for $n_{1}=n_{2}=n_{3}=0$, $\left(
^{t}\Gamma \right) $, and $\left( ^{t}\Gamma \right)\!\!\left( M+\frac{1}{4}%
M\Omega ^{-1}M\right) \Gamma $ is evaluated for this state. The perturbation 
$\varphi \left( K,C,A\right) $ orthogonal to this fundamental state $%
n_{1}=n_{2}=n_{3}=0$, and normalized to $1$.

Given the compatibility condition,%
\begin{equation*}
0=\alpha -C_{0}+\left( ^{t}\Gamma \right)\!\!\left( M+\frac{1}{4}M\Omega
^{-1}M\right) \Gamma +\sqrt{\varsigma ^{2}\varpi ^{2}+\left( \hat{r}%
+r_{c}\right) ^{2}}+\frac{1}{\sqrt{\lambda }}+\left\vert \delta
-A\varepsilon \bar{K}^{\varepsilon -1}\right\vert +\left( \bar{A}^{2}-\left(
\left( 1-\varkappa \right) \bar{A}+\varkappa \Gamma _{3}\right) ^{2}\right)
\end{equation*}%
and the variation becomes:%
\begin{eqnarray*}
\frac{1}{2}\delta ^{2}S\left( \Psi \right) &>&\int \varphi ^{\dag }\left(
K,C,A\right) \left\{ -\varpi ^{2}\frac{\partial ^{2}}{\partial C^{2}}-\frac{1%
}{\lambda ^{2}}\frac{\partial ^{2}}{\partial A^{2}}-\nu ^{2}\left( \delta
-AF^{\prime }\left( K\right) \right) ^{2}\frac{\partial ^{2}}{\partial
K^{\prime 2}}+\left( A-\Gamma _{3}\right) ^{2}\right. \\
&&\left. +\left( \varsigma ^{2}+\frac{\left( \hat{r}+r_{c}\right) ^{2}}{%
\varpi ^{2}}\right)\!\!\left( C-\Gamma _{1}\right) ^{2}+\frac{\left( K^{\prime
}-\Gamma _{2}\right) ^{2}}{\nu ^{2}}-\left( \sqrt{\varsigma ^{2}\varpi
^{2}+\left( \hat{r}+r_{c}\right) ^{2}}+\frac{1}{\sqrt{\lambda }}+\left\vert
\delta -A\varepsilon \bar{K}^{\varepsilon -1}\right\vert \right) \right\}
\varphi \left( K,C,A\right) \\
&&+\gamma \eta \int \!\!\left( \varphi \left( K,C,A\right) +\varphi ^{\dag
}\left( K,C,A\right) \right) K\Psi _{1}\left( K,C,A\right) \int \!\!\left(
\varphi \left( K,C,A\right) +\varphi ^{\dag }\left( K,C,A\right) \right)
A\Psi _{1}\left( K,C,A\right)
\end{eqnarray*}%
The first part of $\frac{1}{2}\delta ^{2}S\left( \Psi \right) $ is positive
given the definition of the operator, only the last part can be negative.
Given that:%
\begin{eqnarray*}
&&\gamma \eta \int \!\!\left( \varphi \left( K,C,A\right) +\varphi ^{\dag
}\left( K,C,A\right) \right) K\Psi _{1}\left( K,C,A\right) \int \!\!\left(
\varphi \left( K,C,A\right) +\varphi ^{\dag }\left( K,C,A\right) \right)
A\Psi _{1}\left( K,C,A\right) \\
&\simeq &\gamma \eta \int \!\!\left( \varphi \left( K,C,A\right) +\varphi ^{\dag
}\left( K,C,A\right) \right) \frac{\left( K^{\prime }-C\right) +A\bar{K}%
^{\varepsilon }\left( 1-\varepsilon \right) }{\delta -\Gamma _{3}\varepsilon 
\bar{K}^{\varepsilon -1}}\Psi _{1}\left( K,C,A\right) \int \!\!\left( \varphi
\left( K,C,A\right) +\varphi ^{\dag }\left( K,C,A\right) \right) A\Psi
_{1}\left( K,C,A\right)
\end{eqnarray*}%
in first order approximation in $\Gamma _{3}\varepsilon \bar{K}^{\varepsilon
-1}$. This expression is non null for the components $\left(
n_{1},n_{2},n_{3}\right) =\left( 1,0,0\right) $, $\left( 0,1,0\right) $ or $%
\left( 0,0,1\right) $ of $\varphi \left( K,C,A\right) $, to inspect the sign
of $\frac{1}{2}\delta ^{2}S\left( \Psi \right) $, we can restrict to:%
\begin{equation*}
\varphi =a_{1}\Psi _{\left( 1,0,0\right) }+a_{2}\Psi _{\left( 0,1,0\right)
}+a_{3}\Psi _{\left( 0,0,1\right) }\equiv \sum_{i=1}^{3}a_{i}\Psi _{i}
\end{equation*}%
with 
\begin{equation*}
\left\vert a_{1}\right\vert ^{2}+\left\vert a_{2}\right\vert ^{2}+\left\vert
a_{3}\right\vert ^{2}=1
\end{equation*}%
For each variable $X_{i}$, 
\begin{eqnarray*}
&&\int \!\!\left( \varphi \left( K,C,A\right) +\varphi ^{\dag }\left(
K,C,A\right) \right) X_{i}\Psi _{1}\left( K,C,A\right) \\
&=&\int \!\!\left( \varphi \left( K,C,A\right) +\varphi ^{\dag }\left(
K,C,A\right) \right)\!\!\left( X_{i}-\Gamma _{i}\right) \Psi _{1}\left(
K,C,A\right) +\Gamma _{i}\int \!\!\left( \varphi \left( K,C,A\right) +\varphi
^{\dag }\left( K,C,A\right) \right) \Psi _{1}\left( K,C,A\right) \\
&=&\left( a_{i}+a_{i}^{\ast }\right) \int \Psi _{i}\left( K,C,A\right)
\left( X_{i}-\Gamma _{i}\right) \Psi _{1}\left( K,C,A\right)
\end{eqnarray*}%
Now, 
\begin{equation*}
\left( X_{i}-\Gamma _{i}\right) =\frac{A_{i}^{\dag }+A_{i}}{\sqrt{2\omega
_{i}}}
\end{equation*}%
where are the annihilation/creation operators, and with:%
\begin{eqnarray*}
\omega _{1} &=&\sqrt{\frac{\varsigma ^{2}}{\varpi ^{2}}+\frac{\left( \hat{r}%
+r_{c}\right) ^{2}}{\varpi ^{4}}} \\
\omega _{2} &=&\frac{1}{\left\vert \delta -AF^{\prime }\left( K\right)
\right\vert \nu } \\
\omega _{3} &=&\lambda
\end{eqnarray*}%
As a consequence:%
\begin{equation}
\int \!\!\left( \varphi \left( K,C,A\right) +\varphi ^{\dag }\left( K,C,A\right)
\right) X_{i}\Psi _{1}\left( K,C,A\right) =\frac{\left( a_{i}+a_{i}^{\ast
}\right) }{\sqrt{2\omega _{i}}}  \label{dtls}
\end{equation}%
and this allows to estimate the various contributions in (\ref{delta2Sbb}).
The term proportional to $\varkappa $ can be computed as:$\allowbreak $ 
\begin{equation*}
-2\left( 1-\varkappa \right) \varkappa \left( \int \!\!\left( \varphi \left(
K,C,A\right) +\varphi ^{\dag }\left( K,C,A\right) \right) A\Psi _{1}\left(
K,C,A\right) \right) ^{2}=-2\left( 1-\varkappa \right) \varkappa \left( 
\frac{a_{3}+a_{3}^{\ast }}{\sqrt{2\lambda }}\right) ^{2}
\end{equation*}

and 
\begin{equation*}
8\frac{\left\vert \delta -\frac{A_{0}}{1-\chi }\varepsilon \bar{K}%
^{\varepsilon -1}\right\vert }{\bar{K}^{\varepsilon }\left( 1-\varepsilon
\right) }\left\vert \alpha -C_{0}+\sqrt{\varsigma ^{2}\varpi ^{2}+\left(
r_{c}\right) ^{2}}+\frac{1}{\lambda }\right\vert
\end{equation*}%
\begin{equation*}
-\allowbreak \frac{4\left( K^{\varepsilon }A_{0}\left( 1-\varepsilon \right)
-CY\left( 1-\varkappa \right) \right) }{\left( 1-\varkappa \right)
\left\vert \delta -\frac{A_{0}}{1-\chi }\varepsilon \bar{K}^{\varepsilon
-1}\right\vert \bar{K}^{\varepsilon }\left( 1-\varepsilon \right) }%
\left\vert \alpha -C_{0}+\sqrt{\varsigma ^{2}\varpi ^{2}+\left( r_{c}\right)
^{2}}+\frac{1}{\sqrt{\lambda }}\right\vert
\end{equation*}%
Other terms can be estimated in the same way:

\begin{eqnarray*}
&&\gamma \eta \int \!\!\left( \varphi \left( K,C,A\right) +\varphi ^{\dag
}\left( K,C,A\right) \right) K\Psi _{1}\left( K,C,A\right) \int \!\!\left(
\varphi \left( K,C,A\right) +\varphi ^{\dag }\left( K,C,A\right) \right)
A\Psi _{1}\left( K,C,A\right) \\
&=&\gamma \eta \int \!\!\left( \varphi \left( K,C,A\right) +\varphi ^{\dag
}\left( K,C,A\right) \right) \frac{\left( K^{\prime }-C\right) +A\bar{K}%
^{\varepsilon }\left( 1-\varepsilon \right) }{\delta -\Gamma _{3}\varepsilon 
\bar{K}^{\varepsilon -1}}\Psi _{1}\left( K,C,A\right) \int \!\!\left( \varphi
\left( K,C,A\right) +\varphi ^{\dag }\left( K,C,A\right) \right) A\Psi
_{1}\left( K,C,A\right) \\
&=&\gamma \eta \frac{\left( \frac{\left( a_{2}+a_{2}^{\ast }\right) }{\sqrt{%
2\omega _{2}}}-\frac{\left( a_{1}+a_{1}^{\ast }\right) }{\sqrt{2\omega _{1}}}%
\right) +\frac{\left( a_{3}+a_{3}^{\ast }\right) }{\sqrt{2\omega _{3}}}\bar{K%
}^{\varepsilon }\left( 1-\varepsilon \right) }{\delta -\Gamma
_{3}\varepsilon \bar{K}^{\varepsilon -1}}\frac{\left( a_{3}+a_{3}^{\ast
}\right) }{\sqrt{2\omega _{3}}} \\
&=&\gamma \eta \frac{\left( \frac{\left( a_{2}+a_{2}^{\ast }\right) \sqrt{%
\left\vert \delta -AF^{\prime }\left( K\right) \right\vert }\nu }{2\sqrt{%
\lambda }}-\frac{\left( a_{1}+a_{1}^{\ast }\right) \varpi }{2\sqrt{\lambda 
\sqrt{\varsigma ^{2}\varpi ^{2}+\left( \hat{r}+r_{c}\right) ^{2}}}}\right)
\left( a_{3}+a_{3}^{\ast }\right) +\frac{\left( a_{3}+a_{3}^{\ast }\right)
^{2}}{2\lambda }\bar{K}^{\varepsilon }\left( 1-\varepsilon \right) }{\delta
-\Gamma _{3}\varepsilon \bar{K}^{\varepsilon -1}}
\end{eqnarray*}%
Moreover, for:%
\begin{equation*}
\varphi =a_{1}\Psi _{\left( 1,0,0\right) }+a_{2}\Psi _{\left( 0,1,0\right)
}+a_{3}\Psi _{\left( 0,0,1\right) }\equiv \sum_{i=1}^{3}a_{i}\Psi _{i}
\end{equation*}%
the first part of $\frac{1}{2}\delta ^{2}S\left( \Psi \right) $ is equal to: 
\begin{eqnarray*}
&&\int \varphi ^{\dag }\left( K,C,A\right) \left\{ -\varpi ^{2}\frac{%
\partial ^{2}}{\partial C^{2}}-\frac{1}{\lambda ^{2}}\frac{\partial ^{2}}{%
\partial A^{2}}-\nu ^{2}\left( \delta -AF^{\prime }\left( K\right) \right)
^{2}\frac{\partial ^{2}}{\partial K^{\prime 2}}+\left( A-\Gamma _{3}\right)
^{2}\right. \\
&&\left. +\left( \varsigma ^{2}+\frac{\left( \hat{r}+r_{c}\right) ^{2}}{%
\varpi ^{2}}\right)\!\!\left( C-\Gamma _{1}\right) ^{2}+\frac{\left( K^{\prime
}-\Gamma _{2}\right) ^{2}}{\nu ^{2}}-\left( \sqrt{\varsigma ^{2}\varpi
^{2}+\left( \hat{r}+r_{c}\right) ^{2}}+\frac{1}{\lambda }+\left\vert \delta
-A\varepsilon \bar{K}^{\varepsilon -1}\right\vert \right) \right\} \varphi
\left( K,C,A\right) \\
&=&2\left\vert a_{1}\right\vert ^{2}\sqrt{\varsigma ^{2}\varpi ^{2}+\left( 
\hat{r}+r_{c}\right) ^{2}\varpi ^{2}}+\frac{2\left\vert a_{3}\right\vert ^{2}%
}{\lambda }+2\left\vert \delta -A\varepsilon \bar{K}^{\varepsilon
-1}\right\vert \left\vert a_{2}\right\vert ^{2}
\end{eqnarray*}%
and then:%
\begin{eqnarray*}
\frac{1}{2}\delta ^{2}S\left( \Psi \right) &>&2\left\vert a_{1}\right\vert
^{2}\sqrt{\varsigma ^{2}\varpi ^{2}+\left( \hat{r}+r_{c}\right) ^{2}}+\frac{%
2\left\vert a_{3}\right\vert ^{2}}{\lambda }+2\left\vert \delta
-A\varepsilon \bar{K}^{\varepsilon -1}\right\vert \left\vert
a_{2}\right\vert ^{2}+\gamma \eta \frac{\frac{\left( a_{3}+a_{3}^{\ast
}\right) ^{2}}{2\lambda }\bar{K}^{\varepsilon }\left( 1-\varepsilon \right) 
}{\delta -\Gamma _{3}\varepsilon \bar{K}^{\varepsilon -1}} \\
&&+\gamma \eta \frac{\left( \frac{\left( a_{2}+a_{2}^{\ast }\right) \sqrt{%
\left\vert \delta -\Gamma _{3}\varepsilon \bar{K}^{\varepsilon
-1}\right\vert }\nu }{2\sqrt{\lambda }}-\frac{\left( a_{1}+a_{1}^{\ast
}\right) \varpi }{2\sqrt{\lambda \sqrt{\varsigma ^{2}\varpi ^{2}+\left( \hat{%
r}+r_{c}\right) ^{2}}}}\right)\!\!\left( a_{3}+a_{3}^{\ast }\right) }{\delta
-\Gamma _{3}\varepsilon \bar{K}^{\varepsilon -1}}-\left( 1-\varkappa \right)
\varkappa \frac{\left( a_{3}+a_{3}^{\ast }\right) ^{2}}{\lambda }
\end{eqnarray*}%
Given that:%
\begin{eqnarray*}
\left\vert \left( a_{2}+a_{2}^{\ast }\right)\!\!\left( a_{3}+a_{3}^{\ast
}\right) \right\vert &<&4\left\vert a_{2}\right\vert \left\vert
a_{3}\right\vert <2\left( \left\vert a_{3}\right\vert ^{2}+\left\vert
a_{2}\right\vert ^{2}\right) \\
\left\vert \left( a_{1}+a_{1}^{\ast }\right)\!\!\left( a_{3}+a_{3}^{\ast
}\right) \right\vert &<&2\left( \left\vert a_{3}\right\vert ^{2}+\left\vert
a_{1}\right\vert ^{2}\right)
\end{eqnarray*}%
\begin{eqnarray*}
\left\vert \left( a_{2}+a_{2}^{\ast }\right)\!\!\left( a_{3}+a_{3}^{\ast
}\right) \right\vert &<&4\left\vert a_{2}\right\vert \left\vert
a_{3}\right\vert <2\left( \left\vert a_{3}\right\vert ^{2}+\left\vert
a_{2}\right\vert ^{2}\right) \\
\left\vert \left( a_{1}+a_{1}^{\ast }\right)\!\!\left( a_{3}+a_{3}^{\ast
}\right) \right\vert &<&2\left( \left\vert a_{3}\right\vert ^{2}+\left\vert
a_{1}\right\vert ^{2}\right)
\end{eqnarray*}%
one has:%
\begin{eqnarray*}
&&\gamma \eta \left\vert \left( \frac{\left( a_{2}+a_{2}^{\ast }\right) 
\sqrt{\left\vert \delta -\Gamma _{3}\varepsilon \bar{K}^{\varepsilon
-1}\right\vert }\nu }{2\sqrt{\lambda }}-\frac{\left( a_{1}+a_{1}^{\ast
}\right) \varpi }{2\sqrt{\lambda \sqrt{\varsigma ^{2}\varpi ^{2}+\left( \hat{%
r}+r_{c}\right) ^{2}}}}\right)\!\!\left( a_{3}+a_{3}^{\ast }\right) \right. \\
&&\left. +\gamma \eta \frac{\frac{\left( a_{3}+a_{3}^{\ast }\right) ^{2}}{%
2\lambda }\bar{K}^{\varepsilon }\left( 1-\varepsilon \right) }{\delta
-\Gamma _{3}\varepsilon \bar{K}^{\varepsilon -1}}-\left( 1-\varkappa \right)
\varkappa \frac{\left( a_{3}+a_{3}^{\ast }\right) ^{2}}{\lambda }\right\vert
\\
&<&\frac{\varpi \left( \gamma \eta \left( \left\vert a_{3}\right\vert
^{2}+\left\vert a_{2}\right\vert ^{2}\right) \frac{\sqrt{\left\vert \delta
-\Gamma _{3}\varepsilon \bar{K}^{\varepsilon -1}\right\vert }\nu }{\sqrt{%
\lambda }}+\gamma \eta \left( \left\vert a_{3}\right\vert ^{2}+\left\vert
a_{1}\right\vert ^{2}\right) \right) }{\sqrt{\lambda \sqrt{\varsigma
^{2}\varpi ^{2}+\left( \hat{r}+r_{c}\right) ^{2}}}}+\gamma \eta \frac{%
2\left\vert a_{3}\right\vert ^{2}\bar{K}^{\varepsilon }\left( 1-\varepsilon
\right) }{\lambda \left\vert \delta -\Gamma _{3}\varepsilon \bar{K}%
^{\varepsilon -1}\right\vert }+4\left( 1-\varkappa \right) \varkappa \frac{%
\left\vert a_{3}\right\vert ^{2}}{\lambda } \\
&<&\gamma \eta \left\vert a_{1}\right\vert ^{2}\frac{\varpi }{\sqrt{\lambda 
\sqrt{\varsigma ^{2}\varpi ^{2}+\left( \hat{r}+r_{c}\right) ^{2}}}}+\gamma
\eta \left\vert a_{2}\right\vert ^{2}\frac{\sqrt{\left\vert \delta -\Gamma
_{3}\varepsilon \bar{K}^{\varepsilon -1}\right\vert }\nu }{\sqrt{\lambda }}
\\
&&+\left( \gamma \eta \left( \frac{\sqrt{\left\vert \delta -\Gamma
_{3}\varepsilon \bar{K}^{\varepsilon -1}\right\vert }\nu }{2\sqrt{\lambda }}+%
\frac{\varpi }{2\sqrt{\lambda \sqrt{\varsigma ^{2}\varpi ^{2}+\left( \hat{r}%
+r_{c}\right) ^{2}}}}+\frac{2\bar{K}^{\varepsilon }\left( 1-\varepsilon
\right) }{\lambda \left\vert \delta -\Gamma _{3}\varepsilon \bar{K}%
^{\varepsilon -1}\right\vert }\right) +\frac{1}{\lambda }\right) \left\vert
a_{3}\right\vert ^{2}
\end{eqnarray*}

\bigskip As a consequence:%
\begin{eqnarray*}
\frac{1}{2}\delta ^{2}S\left( \Psi \right) &>&\left( 2\sqrt{\varsigma
^{2}\varpi ^{2}+\left( \hat{r}+r_{c}\right) ^{2}}-\gamma \eta \frac{\varpi }{%
\sqrt{\lambda \sqrt{\varsigma ^{2}\varpi ^{2}+\left( \hat{r}+r_{c}\right)
^{2}}}}\right) \left\vert a_{1}\right\vert ^{2} \\
&&+\left( 2\left\vert \delta -A\varepsilon \bar{K}^{\varepsilon
-1}\right\vert -\gamma \eta \left\vert a_{2}\right\vert ^{2}\frac{\sqrt{%
\left\vert \delta -\Gamma _{3}\varepsilon \bar{K}^{\varepsilon
-1}\right\vert }\nu }{\sqrt{\lambda }}\right) \left\vert a_{2}\right\vert
^{2} \\
&&+\left( \frac{1}{\lambda }-\gamma \eta \left( \frac{\sqrt{\left\vert
\delta -\Gamma _{3}\varepsilon \bar{K}^{\varepsilon -1}\right\vert }\nu }{2%
\sqrt{\lambda }}+\frac{\varpi }{2\sqrt{\lambda \sqrt{\varsigma ^{2}\varpi
^{2}+\left( \hat{r}+r_{c}\right) ^{2}}}}+\frac{2\bar{K}^{\varepsilon }\left(
1-\varepsilon \right) }{\lambda \left\vert \delta -\Gamma _{3}\varepsilon 
\bar{K}^{\varepsilon -1}\right\vert }\right) \right) \left\vert
a_{3}\right\vert ^{2}
\end{eqnarray*}%
Since we found that:%
\begin{equation*}
\gamma \eta <\frac{\left\vert \delta -\frac{A_{0}}{1-\chi }\varepsilon \bar{K%
}^{\varepsilon -1}\right\vert }{\bar{K}^{\varepsilon }\left( 1-\varepsilon
\right) }\frac{4D}{\left\vert 3C^{2}+8C\bar{A}-4\bar{A}^{2}-4D\right\vert }
\end{equation*}%
For $A_{0}\gg 1$, and for $C_{0}$ such that $D<A_{0}$, there is a large range
of parameters such that the contributions proportional to $\left\vert
a_{i}\right\vert ^{2}$, $i=1,2,3$ are positive, and thus $\frac{1}{2}\delta
^{2}S\left( \Psi \right) >0$.

We conclude by noting that for similar arguments, the sum of (\ref{delsqrtbb}%
) and (\ref{delsqrt}) is positive. Actually (\ref{delsqrtbb}) has the same
form as (\ref{dtls}) and is equal to:%
\begin{eqnarray*}
&&2\gamma \left( \delta \sqrt{\eta }\right) ^{2}\eta \int \Psi _{1}^{\dag
}\left( K,C,A\right) K\Psi _{1}\left( K,C,A\right) \int \Psi _{1}^{\dag
}\left( K,C,A\right) A\Psi _{1}\left( K,C,A\right) \\
&&+\gamma \left( \delta \sqrt{\eta }\right) ^{2}\eta \int \Psi _{1}^{\dag
}\left( K,C,A\right) K\Psi _{1}\left( K,C,A\right) \int \!\!\left( \varphi
^{\dag }\left( K,C,A\right) +\varphi \left( K,C,A\right) \right) A\Psi
_{1}\left( K,C,A\right) \\
&&+\gamma \left( \delta \sqrt{\eta }\right) ^{2}\eta \int \!\!\left( \varphi
^{\dag }\left( K,C,A\right) +\varphi \left( K,C,A\right) \right) K\Psi
_{1}\left( K,C,A\right) \int \Psi _{1}^{\dag }\left( K,C,A\right) A\Psi
_{1}\left( K,C,A\right) \\
&=&2\gamma \left( \delta \sqrt{\eta }\right) ^{2}\eta \left\langle
K\right\rangle \left\langle A\right\rangle \\
&&+\gamma \left( \delta \sqrt{\eta }\right) ^{2}\eta \left( \left\langle
K\right\rangle \frac{\left( a_{3}+a_{3}^{\ast }\right) }{\sqrt{\lambda }}+%
\frac{\left( \frac{\left( a_{2}+a_{2}^{\ast }\right) \sqrt{\left\vert \delta
-\Gamma _{3}\varepsilon \bar{K}^{\varepsilon -1}\right\vert }\nu }{2}-\frac{%
\left( a_{1}+a_{1}^{\ast }\right) \varpi }{2\sqrt{\sqrt{\varsigma ^{2}\varpi
^{2}+\left( \hat{r}+r_{c}\right) ^{2}}}}\right)\!\!\left( a_{3}+a_{3}^{\ast
}\right) }{\delta -\Gamma _{3}\varepsilon \bar{K}^{\varepsilon -1}}%
\left\langle A\right\rangle \right)
\end{eqnarray*}%
and 
\begin{eqnarray*}
&&\left\vert \left\langle K\right\rangle \frac{\left( a_{3}+a_{3}^{\ast
}\right) }{\sqrt{\lambda }}+\frac{\left( \frac{\left( a_{2}+a_{2}^{\ast
}\right) \sqrt{\left\vert \delta -\Gamma _{3}\varepsilon \bar{K}%
^{\varepsilon -1}\right\vert }\nu }{2}-\frac{\left( a_{1}+a_{1}^{\ast
}\right) \varpi }{2\sqrt{\sqrt{\varsigma ^{2}\varpi ^{2}+\left( \hat{r}%
+r_{c}\right) ^{2}}}}\right)\!\!\left( a_{3}+a_{3}^{\ast }\right) }{\delta
-\Gamma _{3}\varepsilon \bar{K}^{\varepsilon -1}}\left\langle A\right\rangle
\right\vert \\
&<&2\left( \frac{\left\langle K\right\rangle }{\sqrt{\lambda }}+\frac{\left( 
\sqrt{\left\vert \delta -\Gamma _{3}\varepsilon \bar{K}^{\varepsilon
-1}\right\vert }\nu +\frac{\varpi }{\sqrt{\sqrt{\varsigma ^{2}\varpi
^{2}+\left( \hat{r}+r_{c}\right) ^{2}}}}\right) }{\left\vert \delta -\Gamma
_{3}\varepsilon \bar{K}^{\varepsilon -1}\right\vert }\left\langle
A\right\rangle \right) \\
&<&2\gamma \left( \delta \sqrt{\eta }\right) ^{2}\eta \left\langle
K\right\rangle \left\langle A\right\rangle
\end{eqnarray*}

\bigskip For values of $A_{0}$ and $\bar{K}$ that are large enough.

\subsection*{Computation of the Green functions in both Phases}

As explained in the text, to inspect the transition functions in the various
phases of the system, one has to come back to the initial set of variables $%
\left( K,C,A\right) $. We consider each phase separately.

In the phase $\Psi _{1}\left( K,C,A\right) =0$, one can directly come back
to the initial action for $\Psi \left( K,C,A\right) $, and write: 
\begin{eqnarray}
&&\int \Psi ^{\dag }\left( K,C,A\right) \left\{ -\varpi ^{2}\frac{\partial
^{2}}{\partial C^{2}}-\frac{1}{\lambda ^{2}}\frac{\partial ^{2}}{\partial
A^{2}}-\nu ^{2}\frac{\partial ^{2}}{\partial K^{2}}+\left( A-\bar{A}\right)
^{2}-2\frac{\partial }{\partial K}\left( C-AF\left( K\right) +\delta
K\right) \right.  \label{Cmptgrn} \\
&&\hspace{3cm}\left. -2\frac{\partial }{\partial C}\left( AF^{\prime }\left(
K\right) +r_{c}\right)\!\!\left( C-\bar{C}\right) +\left( \alpha -C_{0}\right)
\right\} \Psi \left( K,C,A\right)  \notag \\
&&+\frac{\gamma }{2}\int \!\!\left( \Psi ^{\dag }\left( K_{1},C_{1},A_{1}\right)
A_{1}\Psi ^{\dag }\left( K_{1},C_{1},A_{1}\right) \right)\!\!\left( \Psi ^{\dag
}\left( K_{2},C_{2},A_{2}\right) K_{2}\Psi ^{\dag }\left(
K_{2},C_{2},A_{2}\right) \right)  \notag
\end{eqnarray}%
We have set $\varsigma =0$, since we assumed \bigskip $\varsigma \ll 1$. We
first neglect the interaction term and compute the transition function
associated to:%
\begin{eqnarray}
&&\int \Psi ^{\dag }\left( K,C,A\right) \left\{ -\varpi ^{2}\frac{\partial
^{2}}{\partial C^{2}}-\frac{1}{\lambda ^{2}}\frac{\partial ^{2}}{\partial
A^{2}}-\nu ^{2}\frac{\partial ^{2}}{\partial K^{2}}+\left( A-\bar{A}\right)
^{2}-2\frac{\partial }{\partial K}\left( C-AF\left( K\right) +\delta
K\right) \right.  \label{Phs1} \\
&&\hspace{3cm}\left. +2\frac{\partial }{\partial C}\left( AF^{\prime }\left(
K\right) +r_{c}\right)\!\!\left( C-\bar{C}\right) +\left( \alpha -C_{0}\right)
\right\} \Psi \left( K,C,A\right)  \notag
\end{eqnarray}%
with $\bar{A}=\frac{A_{0}}{1-\chi }$ as computed in the previous sections.
The corrections due to the interactions will be inspected in the next
section. The transition function is equal to the Green function for the
operator:%
\begin{eqnarray*}
L &=&-\varpi ^{2}\frac{\partial ^{2}}{\partial C^{2}}-\frac{1}{\lambda ^{2}}%
\frac{\partial ^{2}}{\partial A^{2}}-\nu ^{2}\frac{\partial ^{2}}{\partial
K^{2}}+\left( A-\bar{A}\right) ^{2}-2\left( C-AF\left( K\right) +\delta
K\right) \frac{\partial }{\partial K}-2\left( AF^{\prime }\left( K\right)
+r_{c}\right)\!\!\left( C-\bar{C}\right) \frac{\partial }{\partial C} \\
&&+\alpha -C_{0}
\end{eqnarray*}%
In the phase where $\Psi _{1}\left( K,C,A\right) \neq 0$, one has to proceed
indirectly. Starting with action (\ref{Sprvr}) (with $H\left(
K_{1},K_{2}\right) =1$) whose saddle point equation is (\ref{sddptqnnbstt}):%
\begin{eqnarray}
&&S\left( \bar{\Psi}\right) =\int \bar{\Psi}^{\dag }\left( K^{\prime
},C,A\right) \left\{ -\varpi ^{2}\frac{\partial ^{2}}{\partial C^{2}}-\frac{1%
}{\lambda ^{2}}\frac{\partial ^{2}}{\partial A^{2}}-\nu ^{2}\left( \delta
-AF^{\prime }\left( K\right) \right) ^{2}\frac{\partial ^{2}}{\partial
K^{\prime 2}}+\left( A-\bar{A}\right) ^{2}\right.  \label{Sprvr2} \\
&&\hspace{3cm}\left. +\left( \varsigma ^{2}+\frac{\left( AF^{\prime }\left(
K\right) +r_{c}\right) ^{2}}{\varpi ^{2}}\right)\!\!\left( C-\bar{C}\right)
^{2}+\frac{\left( K^{\prime }\right) ^{2}}{\nu ^{2}}+\alpha -C_{0}\right\} 
\bar{\Psi}\left( K^{\prime },C,A\right)  \notag \\
&&+\gamma \frac{1}{2}\int \bar{\Psi}^{\dag }\left( K_{1}^{\prime
},C_{1},A_{1}\right) \bar{\Psi}^{\dag }\left( K_{2}^{\prime
},C_{2},A_{2}\right) \left\{ A_{2}K_{1}+A_{1}K_{2}\right\} \bar{\Psi}\left(
K_{1}^{\prime },C_{1},A_{1}\right) \bar{\Psi}\left( K_{2}^{\prime
},C_{2},A_{2}\right)  \notag
\end{eqnarray}%
One has to shift the field by letting: $\bar{\Psi}\left( K,C,A\right) =\Psi
\left( K^{\prime },C,A\right) +\Psi _{1}\left( K^{\prime },C,A\right) $.
Using that $\Psi _{1}\left( K,C,A\right) $ is a saddle point, it yields the
following expansion in the shifted field $\Psi \left( K^{\prime },C,A\right) 
$: 
\begin{eqnarray}
\hat{S}\left( \Psi \right) &=&\Psi \left( K,C,A\right) \left\{ -\varpi ^{2}%
\frac{\partial ^{2}}{\partial C^{2}}-\frac{1}{\lambda ^{2}}\frac{\partial
^{2}}{\partial A^{2}}-\nu ^{2}\left( \delta -AF^{\prime }\left( K\right)
\right) ^{2}\frac{\partial ^{2}}{\partial K^{\prime 2}}+\left( A-\bar{A}%
\right) ^{2}\right.  \label{Cmptgrnbb} \\
&&\left. +2\left( \bar{A}-\Gamma _{3}\right) \varkappa A+\left( \varsigma
^{2}+\frac{\left( AF^{\prime }\left( K\right) +r_{c}\right) ^{2}}{\varpi ^{2}%
}\right)\!\!\left( C-\bar{C}\right) ^{2}+\frac{\left( K^{\prime }\right) ^{2}}{%
\nu ^{2}}+\alpha -C_{0}\right\} \Psi \left( K,C,A\right)  \notag \\
&&+\gamma \eta \int \!\!\left( \Psi \left( K,C,A\right) +\Psi ^{\dag }\left(
K,C,A\right) \right) K\Psi _{1}\left( K,C,A\right) \int \!\!\left( \Psi \left(
K,C,A\right) +\Psi ^{\dag }\left( K,C,A\right) \right) A\Psi _{1}\left(
K,C,A\right)  \notag \\
&&+\frac{\gamma }{2}\int \!\!\left( \Psi _{1}^{\dag }\left( K,C,A\right) A\Psi
_{1}^{\dag }\left( K,C,A\right) \right) \int \!\!\left( \Psi ^{\dag }\left(
K,C,A\right) K\Psi ^{\dag }\left( K,C,A\right) \right)  \notag \\
&&+\frac{\gamma }{2}\int \!\!\left( \Psi ^{\dag }\left( K,C,A\right) A\Psi
^{\dag }\left( K,C,A\right) \right) \int \!\!\left( \Psi _{1}^{\dag }\left(
K,C,A\right) K\Psi _{1}^{\dag }\left( K,C,A\right) \right)  \notag \\
&&+\frac{\gamma }{2}\int \!\!\left( \Psi ^{\dag }\left( K,C,A\right) A\Psi
\left( K,C,A\right) \right) \int \!\!\left( \Psi ^{\dag }\left( K,C,A\right)
K\Psi \left( K,C,A\right) \right)  \notag
\end{eqnarray}%
The term $2\left( \bar{A}-\Gamma _{3}\right) \varkappa A$ comes from the
second order expansion of $\bar{A}=\int \bar{\Psi}^{\dag }\left( K^{\prime
},C,A\right) A\bar{\Psi}\left( K^{\prime },C,A\right) $, using that, as
shown in the previous section (stability analysis), the product of the
projections of $\left( \Psi \left( K,C,A\right) +\Psi ^{\dag }\left(
K,C,A\right) \right) K$ and $\left( \Psi \left( K,C,A\right) +\Psi ^{\dag
}\left( K,C,A\right) \right) A$ on $\Psi _{1}$ can be neglected. For the
same reasons, the term: 
\begin{equation*}
\gamma \eta \int \!\!\left( \Psi \left( K,C,A\right) +\Psi ^{\dag }\left(
K,C,A\right) \right) K\Psi _{1}\left( K,C,A\right) \int \!\!\left( \Psi \left(
K,C,A\right) +\Psi ^{\dag }\left( K,C,A\right) \right) A\Psi _{1}\left(
K,C,A\right)
\end{equation*}%
can also be neglected. The last term:%
\begin{equation*}
\frac{\gamma }{2}\int \!\!\left( \Psi ^{\dag }\left( K,C,A\right) A\Psi \left(
K,C,A\right) \right) \int \!\!\left( \Psi ^{\dag }\left( K,C,A\right) K\Psi
\left( K,C,A\right) \right)
\end{equation*}%
is a quartic interaction term in $\Psi \left( K,C,A\right) $. As for the
other phase, we discard this term that will be considered perturbatively.
Let:%
\begin{eqnarray*}
\Gamma _{3} &=&\int \Psi _{1}^{\dag }\left( K^{\prime },C,A\right) A\Psi
_{1}\left( K^{\prime },C,A\right) \\
\Gamma _{1} &=&\int \Psi _{1}^{\dag }\left( K^{\prime },C,A\right) C\Psi
_{1}\left( K,C,A\right) \\
\Gamma _{2} &=&\int \Psi _{1}^{\dag }\left( K^{\prime },C,A\right) K^{\prime
}\Psi _{1}\left( K,C,A\right)
\end{eqnarray*}%
Moreover:%
\begin{equation*}
\int \Psi _{1}^{\dag }\left( K,C,A\right) K\Psi _{1}\left( K,C,A\right) =%
\frac{\left( \Gamma _{2}-\Gamma _{1}\right) +\Gamma _{3}\bar{K}^{\varepsilon
}\left( 1-\varepsilon \right) }{\delta -\Gamma _{3}\varepsilon \bar{K}%
^{\varepsilon -1}}
\end{equation*}%
We are thus left with the following quadratic action:%
\begin{eqnarray*}
\hat{S}\left( \Psi \right) &=&\Psi \left( K,C,A\right) \left\{ -\varpi ^{2}%
\frac{\partial ^{2}}{\partial C^{2}}-\frac{1}{\lambda ^{2}}\frac{\partial
^{2}}{\partial A^{2}}-\nu ^{2}\left( \delta -AF^{\prime }\left( K\right)
\right) ^{2}\frac{\partial ^{2}}{\partial K^{\prime 2}}+\left( A-\bar{A}%
\right) ^{2}+2\left( \bar{A}-\Gamma _{3}\right) \varkappa A\right. \\
&&\left. +\left( \varsigma ^{2}+\frac{\left( AF^{\prime }\left( K\right)
+r_{c}\right) ^{2}}{\varpi ^{2}}\right)\!\!\left( C-\bar{C}\right) ^{2}+\frac{%
\left( K^{\prime }\right) ^{2}}{\nu ^{2}}+\alpha -C_{0}+\gamma \eta \Gamma
_{3}K+\gamma \eta A\frac{\left( \Gamma _{2}-\Gamma _{1}\right) +\Gamma _{3}%
\bar{K}^{\varepsilon }\left( 1-\varepsilon \right) }{\delta -\Gamma
_{3}\varepsilon \bar{K}^{\varepsilon -1}}\right\} \Psi \left( K,C,A\right)
\end{eqnarray*}%
It is shown in the previous sections that under our assumptions $\Gamma
_{2}=0$, and that $\Gamma _{1}$ and $\Gamma _{3}$ are defined such that the
previous action rewrites:%
\begin{eqnarray*}
\hat{S}\left( \Psi \right) &=&\Psi \left( K,C,A\right) \left\{ -\varpi ^{2}%
\frac{\partial ^{2}}{\partial C^{2}}-\frac{1}{\lambda ^{2}}\frac{\partial
^{2}}{\partial A^{2}}-\nu ^{2}\left( \delta -AF^{\prime }\left( K\right)
\right) ^{2}\frac{\partial ^{2}}{\partial K^{\prime 2}}+\left( A-\bar{A}%
_{1}\right) ^{2}\right. \\
&&\left. +\left( \varsigma ^{2}+\frac{\left( AF^{\prime }\left( K\right)
+r_{c}\right) ^{2}}{\varpi ^{2}}\right)\!\!\left( C-\bar{C}_{1}\right) ^{2}+%
\frac{\left( K^{\prime }\right) ^{2}}{\nu ^{2}}+\left( ^{t}\Gamma \right)
\left( M+\frac{1}{4}M\Omega ^{-1}M\right) \Gamma +m_{1}+\alpha
-C_{0}\right\} \Psi \left( K,C,A\right)
\end{eqnarray*}%
where:%
\begin{eqnarray*}
\bar{A}_{1} &=&A_{0}+\chi \Gamma _{3} \\
\bar{C}_{1} &=&\Gamma _{1}
\end{eqnarray*}%
have been computed previously, and:%
\begin{equation*}
m_{1}=\left( ^{t}\Gamma \right)\!\!\left( M+\frac{1}{4}M\Omega ^{-1}M\right)
\Gamma +\left( \bar{A}^{2}-\left( \left( 1-\varkappa \right) \bar{A}%
+\varkappa \Gamma _{3}\right) ^{2}\right) +\alpha -C_{0}
\end{equation*}%
\begin{eqnarray*}
\Gamma &=&\left( 
\begin{array}{c}
\Gamma _{1} \\ 
\Gamma _{2} \\ 
\Gamma _{3}%
\end{array}%
\right) ,\Omega =\left( 
\begin{array}{ccc}
\left( \varsigma ^{2}+\frac{\left( \Gamma _{3}F^{\prime }\left( K\right)
+r_{c}\right) ^{2}}{\varpi ^{2}}\right) & 0 & 0 \\ 
0 & \frac{1}{\nu ^{2}} & 0 \\ 
0 & 0 & 1%
\end{array}%
\right) \\
M &=&\frac{\gamma \eta }{\delta -\Gamma _{3}\varepsilon \bar{K}^{\varepsilon
-1}}\left( 
\begin{array}{ccc}
0 & 0 & -1 \\ 
0 & 0 & 1 \\ 
-1 & 1 & \bar{K}^{\varepsilon }\left( 1-\varepsilon \right)%
\end{array}%
\right)
\end{eqnarray*}%
Remark that $m_{1}$ has been computed in the previous section, and $m_{1}>0$%
. More precisely:%
\begin{equation*}
m_{1}=\bar{A}_{1}^{2}-\left( \left( 1-\varkappa \right) \bar{A}%
_{1}+\varkappa \Gamma _{3}\right) ^{2}+\frac{\left( \left( \left(
1-\varkappa \right) \bar{A}_{1}+\varkappa \Gamma _{3}\right) ^{2}-2\bar{C}%
\left( \left( 1-\varkappa \right) \bar{A}_{1}+\varkappa \bar{A}_{1}\right) -%
\bar{C}_{1}^{2}\right) }{\left( 1-\varepsilon \right) \bar{K}^{\varepsilon }}
\end{equation*}

Now, it is possible to come back to the initial variables, through a
transformation similar to the one performed in the dirst section of this
Appendix (and in the reverse direction):%
\begin{equation*}
\Psi _{n}\left( K,C,A\right) =\exp \left( \frac{1}{2\varpi ^{2}}\left(
\left( AF^{\prime }\left( K\right) +r_{c}\right)\!\!\left( C-\bar{C}_{1}\right)
^{2}\right) -\frac{\left( K^{\prime }\right) ^{2}}{2\nu ^{2}\left( \delta
-AF^{\prime }\left( K\right) \right) }\right) \bar{\Psi}_{n}\left( K^{\prime
},C,A\right)
\end{equation*}%
\begin{equation*}
\Psi _{n}^{\dag }\left( K,C,A\right) =\exp \left( -\frac{1}{2\varpi ^{2}}%
\left( \left( AF^{\prime }\left( K\right) +r_{c}\right)\!\!\left( C-\bar{C}%
_{1}\right) ^{2}\right) +\frac{\left( K^{\prime }\right) ^{2}}{2\nu
^{2}\left( \delta -AF^{\prime }\left( K\right) \right) }\right) \bar{\Psi}%
_{n}\left( K^{\prime },C,A\right)
\end{equation*}%
This leads to the following action:%
\begin{eqnarray}
&&\int \Psi ^{\dag }\left( K,C,A\right) \left\{ -\varpi ^{2}\frac{\partial
^{2}}{\partial C^{2}}-\frac{1}{\lambda ^{2}}\frac{\partial ^{2}}{\partial
A^{2}}-\nu ^{2}\frac{\partial ^{2}}{\partial K^{2}}+\left( A-\bar{A}%
_{1}\right) ^{2}-2\frac{\partial }{\partial K}\left( C-AF\left( K\right)
+\delta K\right) \right.  \label{Phs2} \\
&&\hspace{3cm}\left. +2\frac{\partial }{\partial C}\left( AF^{\prime }\left(
K\right) +r_{c}\right)\!\!\left( C-\bar{C}_{1}\right) +m_{1}+\left( \alpha
-C_{0}\right) \right\} \Psi \left( K,C,A\right)  \notag
\end{eqnarray}

Then, the two phases (\ref{Phs1}) and (\ref{Phs2}) can be put on the same
footing. Both actions have the form:%
\begin{eqnarray}
&&\int \Psi ^{\dag }\left( K,C,A\right) \left\{ -\varpi ^{2}\frac{\partial
^{2}}{\partial C^{2}}-\frac{1}{\lambda ^{2}}\frac{\partial ^{2}}{\partial
A^{2}}-\nu ^{2}\frac{\partial ^{2}}{\partial K^{2}}+\left( A-\bar{A}%
_{i}\right) ^{2}-2\frac{\partial }{\partial K}\left( C-AF\left( K\right)
+\delta K\right) \right.  \label{Phs} \\
&&\hspace{3cm}\left. +2\frac{\partial }{\partial C}\left( \left( AF^{\prime
}\left( K\right) +r_{c}\right)\!\!\left( C-\bar{C}_{i}\right) \right)
+m_{i}+\left( \alpha -C_{0}\right) \right\} \Psi \left( K,C,A\right)  \notag
\end{eqnarray}%
for $i=0$ (phase with $\Psi _{1}=0$) or $i=0$ (phase with $\Psi _{1}\neq 0$%
). We have defined:%
\begin{eqnarray*}
\bar{A}_{0} &=&\frac{A_{0}}{1-\chi } \\
\bar{C}_{0} &=&\bar{C} \\
m_{0} &=&0
\end{eqnarray*}%
The difference between manifests thus both through the different values of $%
\bar{A}$ and $\bar{A}_{1}$ and by the "mass terms" $m_{i}$ In the phase (\ref%
{Phs2}), characterized by a non zero fundamental $\Psi _{1}$, the mass term $%
m_{1}$ is greater than $0$, which implies reduced transitions probabilities
compared to the other phases. The duration of interaction is lower than
phase "$0$", which means a more static system. To solve explicitly for the
transition function in both phases, one can use, as before, the expansion
above a minimal level of capital to express the production function: 
\begin{equation*}
AF\left( K\right) =K^{\varepsilon }\simeq A\bar{K}^{\varepsilon
}+\varepsilon \bar{A}_{i}\frac{K-\bar{K}}{\bar{K}^{1-\varepsilon }}
\end{equation*}%
with:%
\begin{equation*}
\bar{A}_{0}=\bar{A}=\frac{A_{0}}{1-\chi },\bar{A}_{1}=A_{0}+\chi \Gamma _{3}
\end{equation*}%
However, we will see below how to avoid this approximations. Neglecting the
interaction term, and rescaling:%
\begin{equation*}
G\left( K,C,A,K^{\prime },C^{\prime },A^{\prime }\right) \rightarrow \exp
\left( -m_{i}s\right) G\left( K,C,A,K^{\prime },C^{\prime },A^{\prime
}\right)
\end{equation*}%
the Green functions for phase $i=0,1$ satisfy a similar equation:%
\begin{eqnarray}
&&\left\{ -\varpi ^{2}\frac{\partial ^{2}}{\partial C^{2}}-\frac{1}{\lambda
^{2}}\frac{\partial ^{2}}{\partial A^{2}}-\nu ^{2}\frac{\partial ^{2}}{%
\partial K^{2}}+\left( A-\bar{A}_{i}\right) ^{2}-2\frac{\partial }{\partial K%
}\left( C-\bar{C}-\left( A\bar{K}^{\varepsilon }+\varepsilon \bar{A}_{i}%
\frac{K-\bar{K}}{\bar{K}^{1-\varepsilon }}\right) +\delta \left( K-\bar{K}%
\right) +\left( \delta \bar{K}+\bar{C}\right) \right) \right.
\qquad\label{smlrgrn} \\
&&\hspace{3cm}\left. +2\frac{\partial }{\partial C}\left( \bar{A}%
_{i}\varepsilon \bar{K}^{\varepsilon -1}+r_{c}\right)\!\!\left( C-\bar{C}%
\right) +m_{i}\right\} G\left( K,C,A,K^{\prime },C^{\prime },A^{\prime
}\right) =\delta \left( \left( K,C,A\right) -\left( K^{\prime },C^{\prime
},A^{\prime }\right) \right)  \notag
\end{eqnarray}%
Since $\lambda ^{2}\gg 1$, we first neglect the term $\left( A-\bar{A}%
_{i}\right) ^{2}$, to reintroduce it later.

in Fourier components:%
\begin{equation*}
\hat{G}=\int \exp \left( il_{C}\left( C-\bar{C}\right) +il_{K}\left( K-\bar{K%
}\right) +il_{A}A\right) G
\end{equation*}%
and (\ref{smlrgrn}) becomes:%
\begin{eqnarray*}
&&\left\{ \varpi ^{2}l_{C}^{2}+\frac{1}{\lambda ^{2}}l_{A}^{2}+\nu
^{2}l_{K}^{2}-2il_{K}\left( \delta \bar{K}+\bar{C}\right) -2l_{K}\left( 
\frac{\partial }{\partial l_{C}}-\bar{K}^{\varepsilon }\frac{\partial }{%
\partial l_{A}}+\left( \delta -\frac{\varepsilon \bar{A}_{i}}{\bar{K}%
^{1-\varepsilon }}\right) \frac{\partial }{\partial l_{K}}\right) \right. \\
&&\hspace{3cm}\left. +2l_{C}\left( \bar{A}_{i}\varepsilon \bar{K}%
^{\varepsilon -1}+r_{c}\right) \frac{\partial }{\partial l_{C}}%
+m_{i}\right\} \hat{G}\left( K,C,A,K^{\prime },C^{\prime },A^{\prime
}\right) =\exp \left( -i\sum_{i}l_{i}.x_{i}\right)
\end{eqnarray*}%
We also define $\bar{G}$ as:%
\begin{equation*}
\hat{G}=\exp \left( -i\frac{\delta \bar{K}+\bar{C}}{\delta -\frac{%
\varepsilon \bar{A}_{i}}{\bar{K}^{1-\varepsilon }}}l_{K}\right) \bar{G}
\end{equation*}%
and this function satisfies the equation:%
\begin{eqnarray}
&&\left\{ \varpi ^{2}l_{C}^{2}+\frac{1}{\lambda ^{2}}l_{A}^{2}+\nu
^{2}l_{K}^{2}-2l_{K}\left( \frac{\partial }{\partial l_{C}}-\bar{K}%
^{\varepsilon }\frac{\partial }{\partial l_{A}}+\left( \delta -\frac{%
\varepsilon \bar{A}_{i}}{\bar{K}^{1-\varepsilon }}\right) \frac{\partial }{%
\partial l_{K}}\right) \right.  \label{smlrgrnfr} \\
&&\hspace{3cm}\left. +2l_{C}\left( \bar{A}_{i}\varepsilon \bar{K}%
^{\varepsilon -1}+r_{c}\right) \frac{\partial }{\partial l_{C}}%
+m_{i}\right\} \bar{G}\left( K,C,A,K^{\prime },C^{\prime },A^{\prime
}\right) =\exp \left( -i\sum_{i}l_{i}.x_{i}\right)  \notag
\end{eqnarray}%
whose solution is:%
\begin{equation*}
\bar{G}=\int \exp \left( -m_{i}s\right) \exp \left( -\frac{1}{2}%
\sum_{i,j}l_{i}H_{i,j}l_{j}-i\sum_{i}J_{i}l_{i}\right) ds
\end{equation*}%
with $H$ and $J$\ satisfying the equations:%
\begin{eqnarray}
2\hat{\Omega}-NH-H\left( ^{t}N\right) &=&2\frac{\partial }{\partial s}H
\label{Grnqn} \\
\frac{\partial }{\partial s}J &=&-\frac{1}{2}NJ  \notag
\end{eqnarray}%
and where:%
\begin{eqnarray*}
\hat{\Omega} &=&\left( 
\begin{array}{ccc}
\varpi ^{2} & 0 & 0 \\ 
0 & \nu ^{2} & 0 \\ 
0 & 0 & \frac{1}{\lambda ^{2}}%
\end{array}%
\right) ,N=\left( 
\begin{array}{ccc}
-2\left( \bar{A}_{i}\varepsilon \bar{K}^{\varepsilon -1}+r_{c}\right) & 0 & 0
\\ 
-2 & -2\left( \delta -\frac{\varepsilon \bar{A}_{i}}{\bar{K}^{1-\varepsilon }%
}\right) & 2\bar{K}^{\varepsilon } \\ 
0 & 0 & 0%
\end{array}%
\right) \\
H &=&\left( 
\begin{array}{ccc}
a & b & c \\ 
b & d & e \\ 
c & e & f%
\end{array}%
\right)
\end{eqnarray*}%
The initial initial conditions to solve (\ref{smlrgrnfr}):%
\begin{eqnarray}
H\left( 0\right) &=&0  \label{ntcnd} \\
J\left( 0\right) &=&\left( 
\begin{array}{c}
C^{\prime }-\bar{C} \\ 
K^{\prime }-\bar{K} \\ 
A^{\prime }%
\end{array}%
\right)  \notag
\end{eqnarray}%
\begin{eqnarray*}
J_{1} &=&\left( C^{\prime }-\bar{C}\right) \exp \left( \left( \bar{A}%
_{i}\varepsilon \bar{K}^{\varepsilon -1}+r_{c}\right) s\right) \\
J_{2} &=&\left( K^{\prime }-\bar{K}-\left[ \frac{\left( C^{\prime }-\bar{C}%
\right) }{2\bar{A}_{i}\varepsilon \bar{K}^{\varepsilon -1}+r_{c}-\delta }+%
\frac{\bar{K}^{\varepsilon }A^{\prime }}{\delta -\frac{\varepsilon \bar{A}%
_{i}}{\bar{K}^{1-\varepsilon }}}\right] \right) \exp \left( -\left( \delta -%
\frac{\varepsilon \bar{A}_{i}}{\bar{K}^{1-\varepsilon }}\right) s\right) \\
&&+\frac{\left( C^{\prime }-\bar{C}\right) \exp \left( \left( \bar{A}%
_{i}\varepsilon \bar{K}^{\varepsilon -1}+r_{c}\right) s\right) }{2\bar{A}%
_{i}\varepsilon \bar{K}^{\varepsilon -1}+r_{c}-\delta }+\frac{\bar{K}%
^{\varepsilon }A^{\prime }}{\delta -\frac{\varepsilon \bar{A}_{i}}{\bar{K}%
^{1-\varepsilon }}} \\
J_{3} &=&\bar{J}_{3}=A^{\prime }
\end{eqnarray*}%
To find $H$, we first compute $NH+H\left( ^{t}N\right) $:$\allowbreak $

\begin{eqnarray*}
&&NH+H\left( ^{t}N\right) \\
&=&\left( 
\begin{array}{ccc}
-2\left( \bar{A}_{i}\varepsilon \bar{K}^{\varepsilon -1}+r_{c}\right) & 0 & 0
\\ 
-2 & -2\left( \delta -\frac{\varepsilon \bar{A}_{i}}{\bar{K}^{1-\varepsilon }%
}\right) & 2\bar{K}^{\varepsilon } \\ 
0 & 0 & 0%
\end{array}%
\right)\!\!\left( 
\begin{array}{ccc}
a & b & c \\ 
b & d & e \\ 
c & e & f%
\end{array}%
\right) \\
&&+\left( 
\begin{array}{ccc}
a & b & c \\ 
b & d & e \\ 
c & e & f%
\end{array}%
\right)\!\!\left( 
\begin{array}{ccc}
-2\left( \bar{A}_{i}\varepsilon \bar{K}^{\varepsilon -1}+r_{c}\right) & -2 & 
0 \\ 
0 & -2\left( \delta -\frac{\varepsilon \bar{A}_{i}}{\bar{K}^{1-\varepsilon }}%
\right) & 0 \\ 
0 & 2\bar{K}^{\varepsilon } & 0%
\end{array}%
\right) \\
&=&\left( 
\begin{array}{ccc}
-2a\left( 2r_{c}+2K^{\varepsilon -1}\bar{A}_{i}\varepsilon \right) & \left( 
\begin{array}{c}
2\bar{K}^{\varepsilon }c-b\left( 2r_{c}+2\bar{K}^{\varepsilon -1}\bar{A}%
_{i}\varepsilon \right) \\ 
-2a-b\left( 2\delta -2\bar{K}^{\varepsilon -1}\bar{A}_{i}\varepsilon \right)%
\end{array}%
\right) & -c\left( 2r_{c}+2\bar{K}^{\varepsilon -1}\bar{A}_{i}\varepsilon
\right) \\ 
\left( 
\begin{array}{c}
2\bar{K}^{\varepsilon }c-b\left( 2r_{c}+2\bar{K}^{\varepsilon -1}\bar{A}%
_{i}\varepsilon \right) \\ 
-2a-b\left( 2\delta -2\bar{K}^{\varepsilon -1}\bar{A}_{i}\varepsilon \right)%
\end{array}%
\right) & 4\bar{K}^{\varepsilon }e-4b-4d\left( \delta -\bar{K}^{\varepsilon
-1}\bar{A}_{i}\varepsilon \right) & 2\bar{K}^{\varepsilon }f-2c-e\left(
2\delta -2\bar{K}^{\varepsilon -1}\bar{A}_{i}\varepsilon \right) \\ 
-c\left( 2r_{c}+2\bar{K}^{\varepsilon -1}\bar{A}_{i}\varepsilon \right) & 2%
\bar{K}^{\varepsilon }f-2c-e\left( 2\delta -2\bar{K}^{\varepsilon -1}\bar{A}%
_{i}\varepsilon \right) & 0%
\end{array}%
\right)
\end{eqnarray*}%
and (\ref{Grnqn}) leads to the set of differential equations:

\begin{eqnarray*}
\dot{a} &=&2a\left( 2r_{c}+2\bar{K}^{\varepsilon -1}\bar{A}_{i}\varepsilon
\right) +2\varpi ^{2} \\
\dot{b} &=&-2\bar{K}^{\varepsilon }c+b\left( 2r_{c}+2\bar{K}^{\varepsilon -1}%
\bar{A}_{i}\varepsilon \right) +2a+b\left( 2\delta -2\bar{K}^{\varepsilon -1}%
\bar{A}_{i}\varepsilon \right) \\
\dot{c} &=&2c\left( r_{c}+\bar{K}^{\varepsilon -1}\bar{A}_{i}\varepsilon
\right) \\
\dot{d} &=&-\left( 4\bar{K}^{\varepsilon }e-4b-4d\left( \delta -\bar{K}%
^{\varepsilon -1}\bar{A}_{i}\varepsilon \right) \right) +2\nu ^{2} \\
\dot{e} &=&-2\bar{K}^{\varepsilon }f+2c+2e\left( \delta -\bar{K}%
^{\varepsilon -1}\bar{A}_{i}\varepsilon \right) \\
\dot{f} &=&\frac{2}{\lambda ^{2}}
\end{eqnarray*}%
whose solution involves $6$ constants of integration, and given the initial
conditions, it yields 
\begin{eqnarray}
a &=&-\frac{\varpi ^{2}}{2\left( r_{c}+\bar{K}^{\varepsilon -1}\bar{A}%
_{i}\varepsilon \right) }+a_{1}\exp \left( 4\left( r_{c}+\bar{K}%
^{\varepsilon -1}\bar{A}_{i}\varepsilon \right) s\right)  \label{cfth} \\
&=&\frac{\varpi ^{2}\left( \exp \left( 4\left( r_{c}+\bar{K}^{\varepsilon -1}%
\bar{A}_{i}\varepsilon \right) s\right) -1\right) }{2\left( r_{c}+\bar{K}%
^{\varepsilon -1}\bar{A}_{i}\varepsilon \right) }  \notag \\
f &=&\frac{2}{\lambda ^{2}}s  \notag \\
c &=&a_{3}\exp \left( 2\left( r_{c}+\bar{K}^{\varepsilon -1}\bar{A}%
_{i}\varepsilon \right) s\right) =0  \notag \\
b &=&a_{2}\exp \left( 2\left( \delta +r_{c}\right) s\right) +\frac{\varpi
^{2}\exp \left( 4\left( r_{c}+\bar{K}^{\varepsilon -1}\bar{A}_{i}\varepsilon
\right) s\right) }{2\left( r_{c}+\bar{K}^{\varepsilon -1}\bar{A}%
_{i}\varepsilon \right)\!\!\left( r_{c}-\delta +2\bar{K}^{\varepsilon -1}\bar{A}%
_{i}\varepsilon \right) }+\frac{\varpi ^{2}}{2\left( \delta +r_{c}\right)
\left( r_{c}+K^{\varepsilon -1}\bar{A}_{i}\varepsilon \right) }  \notag \\
&=&-\frac{\varpi ^{2}\exp \left( 2\left( \delta +r_{c}\right) s\right) }{%
\left( r_{c}-\delta +2\bar{K}^{\varepsilon -1}\bar{A}_{i}\varepsilon \right)
\left( \delta +r_{c}\right) }+\frac{\varpi ^{2}\exp \left( 4\left( r_{c}+%
\bar{K}^{\varepsilon -1}\bar{A}_{i}\varepsilon \right) s\right) }{2\left(
r_{c}+\bar{K}^{\varepsilon -1}\bar{A}_{i}\varepsilon \right)\!\!\left(
r_{c}-\delta +2\bar{K}^{\varepsilon -1}\bar{A}_{i}\varepsilon \right) } 
\notag \\
&&+\frac{\varpi ^{2}}{2\left( \delta +r_{c}\right)\!\!\left( r_{c}+\bar{K}%
^{\varepsilon -1}\bar{A}_{i}\varepsilon \right) }  \notag \\
e &=&a_{5}\exp \left( 2\left( \delta -\bar{K}^{\varepsilon -1}\bar{A}%
_{i}\varepsilon \right) s\right) +\frac{2\bar{K}^{\varepsilon }}{\lambda
^{2}\left( \delta -\bar{K}^{\varepsilon -1}\bar{A}_{i}\varepsilon \right) }s+%
\frac{\bar{K}^{\varepsilon }}{\lambda ^{2}\left( \delta -\bar{K}%
^{\varepsilon -1}\bar{A}_{i}\varepsilon \right) ^{2}}  \notag \\
&=&-\frac{K^{\varepsilon }\left( \exp \left( 2\left( \delta -\bar{K}%
^{\varepsilon -1}\bar{A}_{i}\varepsilon \right) s\right) -1\right) }{\lambda
^{2}\left( \delta -\bar{K}^{\varepsilon -1}\bar{A}_{i}\varepsilon \right)
^{2}}+\frac{2\bar{K}^{\varepsilon }}{\lambda ^{2}\left( \delta -\bar{K}%
^{\varepsilon -1}\bar{A}_{i}\varepsilon \right) }s  \notag \\
d &=&-\frac{\nu ^{2}}{2\left( \delta -\bar{K}^{\varepsilon -1}\bar{A}%
_{i}\varepsilon \right) }+a_{4}\exp \left( 4\left( \delta -\bar{K}%
^{\varepsilon -1}\bar{A}_{i}\varepsilon \right) s\right)  \notag \\
&&-\frac{2\bar{K}^{2\varepsilon }\left( \exp \left( 2\left( \delta -\bar{K}%
^{\varepsilon -1}\bar{A}_{i}\varepsilon \right) s\right) \right) }{\lambda
^{2}\left( \delta -\bar{K}^{\varepsilon -1}\bar{A}_{i}\varepsilon \right)
^{3}}  \notag \\
&&+\frac{2\bar{K}^{2\varepsilon }}{\lambda ^{2}\left( \delta -\bar{K}%
^{\varepsilon -1}\bar{A}_{i}\varepsilon \right) ^{2}}s+\frac{\bar{K}%
^{2\varepsilon }}{2\lambda ^{2}\left( \delta -\bar{K}^{\varepsilon -1}\bar{A}%
_{i}\varepsilon \right) ^{3}}  \notag \\
&&-\frac{\varpi ^{2}}{2\left( \delta -\bar{K}^{\varepsilon -1}\bar{A}%
_{i}\varepsilon \right)\!\!\left( \delta +r_{c}\right)\!\!\left( r_{c}+\bar{K}%
^{\varepsilon -1}\bar{A}_{i}\varepsilon \right) }  \notag \\
&&+\frac{\varpi ^{2}\exp \left( 4\left( r_{c}+\bar{K}^{\varepsilon -1}\bar{A}%
_{i}\varepsilon \right) s\right) }{2\left( r_{c}+\bar{K}^{\varepsilon -1}%
\bar{A}_{i}\varepsilon \right)\!\!\left( r_{c}-\delta +2\bar{K}^{\varepsilon -1}%
\bar{A}_{i}\varepsilon \right) ^{2}}  \notag \\
&&-\frac{\varpi ^{2}\exp \left( 2\left( \delta +r_{c}\right) s\right) }{%
2\left( r_{c}-\delta +2\bar{K}^{\varepsilon -1}\bar{A}_{i}\varepsilon
\right) ^{2}\left( \delta +r_{c}\right) }  \notag
\end{eqnarray}%
which is the result stated in the text. These expressions can be simplified
given our assumptions about the parameters and for $m_{i}$ relatively large, 
$m_{i}>$ $\delta ,r_{c},\bar{K}^{\varepsilon -1}\bar{A}_{i}\varepsilon $.
Equivalently it corresponds to consider $s\ll 1$. In this case, this can be
approximated by: 
\begin{eqnarray*}
a &=&2\varpi ^{2}s \\
f &=&\frac{2}{\lambda ^{2}}s, \\
c &=&0, \\
b &=&0 \\
e &=&0 \\
d &=&2\nu ^{2}s+\frac{4\bar{K}^{2\varepsilon }s}{\lambda ^{2}\left( \delta -%
\bar{K}^{\varepsilon -1}\bar{A}_{i}\varepsilon \right) ^{2}}-\frac{3\varpi
^{2}s}{\left( \delta +r_{c}\right)\!\!\left( \delta -r_{c}-2\bar{K}%
^{\varepsilon -1}\bar{A}_{i}\varepsilon \right) } \\
&\simeq &\frac{4\bar{K}^{2\varepsilon }s}{\lambda ^{2}\left( \delta -\bar{K}%
^{\varepsilon -1}\bar{A}_{i}\varepsilon \right) ^{2}}
\end{eqnarray*}%
so that: 
\begin{equation*}
H=\left( 
\begin{array}{ccc}
2\varpi ^{2} & 0 & 0 \\ 
0 & 2\nu ^{2}+\frac{4\bar{K}^{2\varepsilon }}{\lambda ^{2}\left( \delta -%
\bar{K}^{\varepsilon -1}\bar{A}_{i}\varepsilon \right) ^{2}}-\frac{3\varpi
^{2}}{\left( \delta +r_{c}\right)\!\!\left( \delta -r_{c}-2\bar{K}^{\varepsilon
-1}\bar{A}_{i}\varepsilon \right) } & 0 \\ 
0 & 0 & \frac{2}{\lambda ^{2}}%
\end{array}%
\right) s
\end{equation*}%
For $J$, the formula simplify as: $\allowbreak $%
\begin{eqnarray*}
J_{1} &=&\left( C^{\prime }-\bar{C}\right)\!\!\left( 1+\left( \bar{A}%
_{i}\varepsilon \bar{K}^{\varepsilon -1}+r_{c}\right) s\right) \\
J_{2} &=&\left( K^{\prime }-\bar{K}-\left[ \frac{\left( C^{\prime }-\bar{C}%
\right) }{2\bar{A}_{i}\varepsilon \bar{K}^{\varepsilon -1}+r_{c}-\delta }+%
\frac{A^{\prime }\bar{K}^{\varepsilon }}{\delta -\bar{K}^{\varepsilon -1}%
\bar{A}_{i}\varepsilon }\right] \right)\!\!\left( 1+\left( \delta -\varepsilon 
\bar{A}_{i}\bar{K}^{\varepsilon -1}\right) s\right) \\
&&+\frac{\left( C^{\prime }-\bar{C}\right) }{2\bar{A}_{i}\varepsilon \bar{K}%
^{\varepsilon -1}+r_{c}-\delta }\left( 1+\left( \bar{A}_{i}\varepsilon \bar{K%
}^{\varepsilon -1}+r_{c}\right) s\right) +\frac{A^{\prime }\bar{K}%
^{\varepsilon }}{\delta -\bar{K}^{\varepsilon -1}\bar{A}_{i}\varepsilon } \\
&=&\left( K^{\prime }-\bar{K}\right)\!\!\left( 1+\left( \delta -\varepsilon 
\bar{A}_{i}\bar{K}^{\varepsilon -1}\right) s\right) -\left( C^{\prime }-\bar{%
C}\right) s+2A^{\prime }\bar{K}^{\varepsilon }s \\
J_{3} &=&\bar{J}_{3}=A^{\prime }
\end{eqnarray*}%
The Green function is computed through the inverse Fourier transform:%
\begin{eqnarray*}
&&G\left( C,K,A,C^{\prime },K^{\prime },A^{\prime },s\right) \\
&=&\int \exp \left( -il_{C}\left( C-\bar{C}\right) -il_{K}\left( K-\bar{K}+%
\frac{\delta \bar{K}+\bar{C}}{\delta -\bar{K}^{\varepsilon -1}\bar{A}%
_{i}\varepsilon }\right) -il_{A}A\right) \exp \left( -\frac{1}{2}%
\sum_{i,j}l_{i}H_{i,j}l_{j}-i\sum_{i}J_{i}l_{i}\right) dl_{C}dl_{K}dl_{A} \\
&=&\frac{\exp \left( -\frac{1}{2}\left( ^{t}XH^{-1}X\right) \right) }{\sqrt{%
\det H}}
\end{eqnarray*}%
with:%
\begin{eqnarray*}
X &=&\left( 
\begin{array}{c}
\left( C-\bar{C}\right) +J_{1} \\ 
\left( K-\bar{K}+\frac{\delta \bar{K}+\bar{C}}{\delta -\bar{K}^{\varepsilon
-1}\bar{A}_{i}\varepsilon }\right) +J_{2} \\ 
A+J_{3}%
\end{array}%
\right) \\
H &=&\left( 
\begin{array}{ccc}
\varpi ^{2} & 0 & 0 \\ 
0 & \nu ^{2}+\frac{2\bar{K}^{2\varepsilon }}{\lambda ^{2}\alpha ^{2}}+\frac{%
3\varpi ^{2}}{2\left( 2\alpha +\beta \right) \beta } & 0 \\ 
0 & 0 & \frac{1}{\lambda ^{2}}%
\end{array}%
\right)
\end{eqnarray*}

As a consequence, the green function between two points $\left( C^{\prime
},K^{\prime },A^{\prime }\right) $ and $\left( C,K,A\right) $ is:%
\begin{eqnarray}
&&G\left( C,K,A,C^{\prime },K^{\prime },A^{\prime },s\right) =\frac{\exp
\left( -\frac{1}{2}\left( ^{t}XH^{-1}X\right) \right) }{\sqrt{2\pi \det H}}
\label{cptfr} \\
&=&\frac{\exp \left( -\frac{\left( \left( C-\bar{C}_{i}\right) -\left(
C^{\prime }-\bar{C}_{i}\right)\!\!\left( 1+\left( \alpha +\beta \right)
s\right) \right) ^{2}}{2\varpi ^{2}s}-\frac{\left( \left( K-\bar{K}+\frac{%
\delta \bar{K}+\bar{C}_{i}}{\alpha }\right) -\left( \left( K^{\prime }-\bar{K%
}+\frac{\delta \bar{K}+\bar{C}_{i}}{\alpha }\right)\!\!\left( 1-\alpha s\right)
-\left( C^{\prime }-\bar{C}_{i}\right) s+A^{\prime }\bar{K}^{\varepsilon
}s\right) \right) ^{2}}{2\left( \nu ^{2}+\frac{2\bar{K}^{2\varepsilon }}{%
\lambda ^{2}\alpha ^{2}}+\frac{3\varpi ^{2}}{2\left( 2\alpha +\beta \right)
\beta }\right) s}-\frac{\lambda ^{2}\left( A-A^{\prime }\right) ^{2}}{2s}%
\right) }{2\sqrt{2\pi \frac{\varpi ^{2}}{\lambda ^{2}}\left( \nu ^{2}+\frac{2%
\bar{K}^{2\varepsilon }}{\lambda ^{2}\alpha ^{2}}+\frac{3\varpi ^{2}}{%
2\left( 2\alpha +\beta \right) \beta }\right) s}}  \notag
\end{eqnarray}%
with:%
\begin{eqnarray*}
\alpha &=&\delta -\bar{K}^{\varepsilon -1}\bar{A}_{i}\varepsilon \\
\beta &=&2\bar{A}_{i}\varepsilon \bar{K}^{\varepsilon -1}+r_{c}-\delta \\
\delta +r_{c} &=&2\alpha +\beta \\
\bar{A}_{i}\varepsilon \bar{K}^{\varepsilon -1}+r_{c} &=&\alpha +\beta
\end{eqnarray*}%
The contribution due to the potential term $\left( A-\bar{A}\right) ^{2}$
can be reintroduced. Since the contribution of $A$ in the exponential is
gaussian, that is, $\frac{\lambda ^{2}\left( A-A^{\prime }\right) ^{2}}{4s}$%
, the quadratic contribution can be introduced by adding a contribution $%
\frac{\left( \frac{A+A^{\prime }}{2}-\bar{A}_{i}\right) ^{2}}{2}s$.
Reintroducing also the factor $\exp \left( -m_{i}s\right) $, yields: 
\begin{eqnarray}
&&G\left( C,K,A,C^{\prime },K^{\prime },A^{\prime },s\right) =\frac{\exp
\left( -m_{i}s\right) \exp \left( -\frac{1}{2}\left( ^{t}XH^{-1}X\right)
\right) }{\sqrt{2\pi \det H}}  \label{Trstngn} \\
&=&\frac1{2\sqrt{2\pi \frac{\varpi ^{2}}{\lambda ^{2}}\left( \nu
^{2}+\frac{2\bar{K}^{2\varepsilon }}{\lambda ^{2}\alpha ^{2}}+\frac{3\varpi
^{2}}{2\left( 2\alpha +\beta \right) \beta }\right) s}} \times \exp \left( -\frac{\left( \left( C-\bar{C}_{i}\right) -\left(
C^{\prime }-\bar{C}_{i}\right)\!\!\left( 1+\left( \alpha +\beta \right)
s\right) \right) ^{2}}{2\varpi ^{2}s}\right) \notag
\\
&&
\times {\exp \left(  -\frac{\left( \left( K-\bar{K}+\frac{%
\delta \bar{K}+\bar{C}_{i}}{\alpha }\right) -\left( \left( K^{\prime }-\bar{K%
}+\frac{\delta \bar{K}+\bar{C}_{i}}{\alpha }\right)\!\!\left( 1-\alpha s\right)
-\left( C^{\prime }-\bar{C}_{i}\right) s+A^{\prime }\bar{K}^{\varepsilon
}s\right) \right) ^{2}}{2\left( \nu ^{2}+\frac{2\bar{K}^{2\varepsilon }}{%
\lambda ^{2}\alpha ^{2}}+\frac{3\varpi ^{2}}{2\left( 2\alpha +\beta \right)
\beta }\right) s}\right)}\notag
\\
&&
\times {\exp\left(
-\frac{\lambda ^{2}\left( A-A^{\prime }\right) ^{2}}{2s}-%
\frac{\left( \frac{A+A^{\prime }}{2}-\bar{A}_{i}\right) ^{2}}{2}%
s-m_{i}s\right) }
\notag\\\notag
\end{eqnarray}%
Remark that we could also find an expression for $G\left( C,K,A,C^{\prime
},K^{\prime },A^{\prime },s\right) $ for all $s$, using (\ref{cfth}) and (%
\ref{cptfr}): the inversion of the matrix $H$ would produce an exponential
weight with non quadratic exponents. We will not develop this point here.

As said before we can also avoid our approximations about the production
function. Since we have considered $s$ relatively small, the expansion of
the production function could have been done between the final and initial
point, for any form of production function. It amounts to replace the
coefficients in the previous expression by:%
\begin{eqnarray}
\alpha &=&\delta -\frac{A+A^{\prime }}{2}F^{\prime }\left( \frac{K+K^{\prime
}}{2}\right)  \label{cftdn} \\
\beta &=&2\frac{A+A^{\prime }}{2}F^{\prime }\left( \frac{K+K^{\prime }}{2}%
\right) +r_{c}-\delta  \notag \\
\delta +r_{c} &=&2\alpha +\beta  \notag \\
\frac{A+A^{\prime }}{2}F^{\prime }\left( \frac{K+K^{\prime }}{2}\right)
+r_{c} &=&\alpha +\beta  \notag
\end{eqnarray}%
Equation (\ref{Trstngn}) represent a stochatic motion around an average
path. The equilibrium value can be found by letting:%
\begin{eqnarray*}
K &=&K^{\prime }=K_{e} \\
C &=&C^{\prime }=C_{e} \\
A &=&A^{\prime }=A_{e}
\end{eqnarray*}%
and by setting the exponent equal to $0$. One finds:%
\begin{eqnarray*}
K_{e} &=&\frac{\left( 1-\varepsilon \right) \bar{A}_{i}\bar{K}^{\varepsilon
}-\bar{C}_{i}}{\delta -\bar{K}^{\varepsilon -1}\bar{A}_{i}\varepsilon } \\
C_{e} &=&\bar{C}_{i} \\
A_{e} &=&\bar{A}_{i}
\end{eqnarray*}%
and replacing these values in the exponent, and equating this one with $0$,
yields directly the relations: 
\begin{eqnarray*}
\left( C-\bar{C}_{i}\right) &=&\left( C^{\prime }-\bar{C}_{i}\right) +\left(
C^{\prime }-\bar{C}_{i}\right)\!\!\left( \alpha +\beta \right) s \\
\left( K-K_{e}\right) &=&\left( K^{\prime }-K_{e}\right) -\alpha \left(
K^{\prime }-K_{e}\right) s-\left( C^{\prime }-\bar{C}_{i}\right) s \\
\lambda ^{2}\left( A-A^{\prime }\right) &=&-\frac{\frac{A+A^{\prime }}{2}-%
\bar{A}_{i}}{2}s
\end{eqnarray*}%
In the limit of small $s$, and using (\ref{cftdn}), as well as $\frac{%
K+K^{\prime }}{2}\rightarrow K$ leads to a differential equation for the
average path:%
\begin{eqnarray*}
\frac{d}{dt}\left( C\left( t\right) -\bar{C}_{i}\right) &=&\left( C\left(
t\right) -\bar{C}_{i}\right)\!\!\left( AF^{\prime }\left( K\left( t\right)
\right) +r_{c}-\delta \right) \\
\frac{d}{dt}\left( K\left( t\right) -K_{e}\right) &=&\left( AF^{\prime
}\left( K\left( t\right) \right) -\delta \right)\!\!\left( K\left( t\right)
-K_{e}\right) -\left( C\left( t\right) -\bar{C}_{i}\right) \\
\lambda ^{2}\frac{d\left( A-\bar{A}_{i}\right) }{dt} &=&-\frac{\left( A-\bar{%
A}_{i}\right) }{2}
\end{eqnarray*}%
This describes a simplified model of capital accumulation: the first
equation id the usual Euler equation with interest rate. The second one is
the dynamic for the capital variable. The last equation describes the
dynamic for the technology level. The fixed point $\bar{A}_{i}$ depends on
the whole system and it's interaction as seen in (\ref{gmm}).

We end up this section by computing the Laplace transform of $G\left(
C,K,A,C^{\prime },K^{\prime },A^{\prime },s\right) $, that is, the one agent
propagator of the system:%
\begin{eqnarray}
&&G\left( C,K,A,C^{\prime },K^{\prime },A^{\prime },\alpha \right)
\label{Grnbc} \\
&=&\int \exp \left( -\left( m_{i}+\alpha -C_{0}\right) s\right)  \times \frac1{2\sqrt{2\pi \frac{\varpi ^{2}}{\lambda ^{2}}\left( \nu
^{2}+\frac{2\bar{K}^{2\varepsilon }}{\lambda ^{2}\alpha ^{2}}+\frac{3\varpi
^{2}}{2\left( 2\alpha +\beta \right) \beta }\right) s}}   \notag\\
&&\times  \exp \left( -\frac{\left( \left( C-\bar{C}\right) -\left(
C^{\prime }-\bar{C}_{i}\right)\!\!\left( 1+\left( \alpha +\beta \right)
s\right) \right) ^{2}}{2\varpi ^{2}s}\right) \notag\\
&&\times  \exp \left(-\frac{\left( \left( K-\bar{K}+\frac{%
\delta \bar{K}+\bar{C}_{i}}{\alpha }\right) -\left( \left( K^{\prime }-\bar{K%
}+\frac{\delta \bar{K}+\bar{C}_{i}}{\alpha }\right)\!\!\left( 1-\alpha s\right)
-\left( C^{\prime }-\bar{C}_{i}\right) s+A^{\prime }\bar{K}^{\varepsilon
}s\right) \right) ^{2}}{2\left( \nu ^{2}+\frac{2\bar{K}^{2\varepsilon }}{%
\lambda ^{2}\alpha ^{2}}+\frac{3\varpi ^{2}}{2\left( 2\alpha +\beta \right)
\beta }\right) s}\right) \notag\\
&&\times  \exp \left(-\frac{\lambda ^{2}\left( A-A^{\prime }\right) ^{2}}{2s}-%
\frac{\left( \frac{A+A^{\prime }}{2}-\bar{A}_{i}\right) ^{2}}{2}%
s-m_{i}s\right)ds\notag \\
\notag
\end{eqnarray}%
\begin{equation*}
C=\left( C^{\prime }-\bar{C}_{i}\right)\!\!\left( 1-\left( \alpha +\beta
\right) s\right)
\end{equation*}%
This can be found explicitly as the Laplace transform of a gaussian
expression. Actually:%
\begin{eqnarray*}
&&\int \exp \left( -m_{i}s\right) \frac{\exp \left( -\frac{1}{2s}\left(
^{t}\left( X-sY\right) H^{-1}\left( X-sY\right) \right) \right) }{\sqrt{2\pi
\det Hs}}ds \\
&=&\int \exp \left( -\left( m_{i}+\frac{1}{2}\left( ^{t}Y\right)
H^{-1}Y\right) s\right) \frac{\exp \left( -\frac{1}{2s}\left(
^{t}XH^{-1}X\right) \right) }{\sqrt{2\pi \det Hs}}ds\exp \left( \left(
^{t}X\right) H^{-1}\left( Y\right) \right) \\
&=&\frac{\exp \left( -\sqrt{2m_{i}+\left( ^{t}Y\right) H^{-1}Y}\sqrt{\left(
^{t}XH^{-1}X\right) }+\left( ^{t}X\right) H^{-1}\left( Y\right) \right) }{%
\sqrt{2m_{i}+\left( ^{t}Y\right) H^{-1}Y}}
\end{eqnarray*}%
and as a consequence:

\begin{eqnarray*}
&&G\left( C,K,A,C^{\prime },K^{\prime },A^{\prime },m_{i}\right) \\
&=&\frac{\exp \left( -\sqrt{2m_{i}+\left( ^{t}Y\right) H^{-1}Y}\sqrt{\left(
^{t}XH^{-1}X\right) }+\left( ^{t}X\right) H^{-1}\left( Y\right) \right) }{%
\sqrt{2m_{i}+\left( ^{t}Y\right) H^{-1}Y}} \\
&=&\frac{\exp \left( -\sqrt{2m_{i}+\frac{\left( \alpha +\beta \right)
^{2}\left( C^{\prime }-\bar{C}\right) ^{2}}{\varpi ^{2}}+\frac{\left( \left(
K^{\prime }-\bar{K}\right) \alpha +\delta \bar{K}+C^{\prime }-A^{\prime }%
\bar{K}^{\varepsilon }\right) ^{2}}{\left( \nu ^{2}+\frac{2\bar{K}%
^{2\varepsilon }}{\lambda ^{2}\alpha ^{2}}+\frac{3\varpi ^{2}}{2\left(
2\alpha +\beta \right) \beta }\right) }}\sqrt{\frac{\left( C-C^{\prime
}\right) ^{2}}{4\varpi ^{2}}+\frac{\left( \left( K-K^{\prime }\right)
\right) ^{2}}{4\left( \nu ^{2}+\frac{2\bar{K}^{2\varepsilon }}{\lambda
^{2}\alpha ^{2}}+\frac{3\varpi ^{2}}{2\left( 2\alpha +\beta \right) \beta }%
\right) }+\frac{\lambda ^{2}\left( A-A^{\prime }\right) ^{2}}{4}}\right) }{%
\sqrt{2m_{i}+\frac{\left( \alpha +\beta \right) ^{2}\left( C^{\prime }-\bar{C%
}\right) ^{2}}{\varpi ^{2}}+\frac{\left( \left( K^{\prime }-\bar{K}\right)
\alpha +\left( C^{\prime }-\bar{C}\right) -A^{\prime }\bar{K}^{\varepsilon
}\right) ^{2}}{\left( \nu ^{2}+\frac{2\bar{K}^{2\varepsilon }}{\lambda
^{2}\alpha ^{2}}+\frac{3\varpi ^{2}}{2\left( 2\alpha +\beta \right) \beta }%
\right) }}} \\
&&\times \exp \left( \frac{\left( \alpha +\beta \right)\!\!\left( C-C^{\prime
}\right)\!\!\left( C^{\prime }-\bar{C}\right) }{2\varpi ^{2}}+\frac{\left(
\left( K-K^{\prime }\right) \right)\!\!\left( \left( K^{\prime }-\bar{K}\right)
\alpha +\delta \bar{K}+C^{\prime }-A^{\prime }\bar{K}^{\varepsilon }\right) 
}{2\left( \nu ^{2}+\frac{2\bar{K}^{2\varepsilon }}{\lambda ^{2}\alpha ^{2}}+%
\frac{3\varpi ^{2}}{2\left( 2\alpha +\beta \right) \beta }\right) }\right)
\end{eqnarray*}

$\bigskip $

\subsection*{Correction to the Green function due to the interaction term}

The interaction term%
\begin{equation*}
\frac{\gamma }{2}\int \Psi ^{\dag }\left( K_{1},C_{1},A_{1}\right) \Psi
^{\dag }\left( K_{2},C_{2},A_{2}\right) \left\{
A_{2}K_{1}+A_{1}K_{2}\right\} \Psi \left( K_{1},C_{1},A_{1}\right) \Psi
\left( K_{2},C_{2},A_{2}\right)
\end{equation*}%
modifies the Green functions of individual agents. The correction at first
order in $\gamma $ is obtained by the application of the Wick theorem to the
interaction term. The contractions $\underbrace{\Psi ^{\dag }\left(
K_{i},C_{i},A_{i}\right) \Psi ^{\dag }\left( K_{j},C_{j},A_{j}\right) }$
being replaced by propagators $G\left(
K_{i},C_{i},A_{i},K_{j},C_{j},A_{j},m_{i}\right) $. It leads to the
contribution: 
\begin{eqnarray*}
&&\delta G\left( C,K,A,C^{\prime },K^{\prime },A^{\prime },m_{i}\right) \\
&=&\gamma \int G\left( C,K,A,C_{1},K_{1},A_{1},m_{i}\right)\!\!\left(
A_{2}K_{1}+A_{1}K_{2}\right) G\left(
C_{1},K_{1},A_{1},C_{2},K_{2},A_{2},m_{i}\right) \\
&&\hspace{3cm}\hspace{3cm}\times G\left( C_{2},K_{2},A_{2},C^{\prime
},K^{\prime },A^{\prime },m_{i}\right) d\left( C_{1},K_{1},A_{1}\right)
d\left( C_{2},K_{2},A_{2}\right)
\end{eqnarray*}%
which is given by the following contribution.%
\begin{eqnarray*}
&&\frac{\exp \left( -\sqrt{2m_{i}+\left( ^{t}X_{1}\right) H^{-1}X_{1}}\sqrt{%
\left( ^{t}XH^{-1}X\right) }+\left( ^{t}X\right) H^{-1}\left( X_{1}\right)
\right) }{\sqrt{2m_{i}+\left( ^{t}X_{1}\right) H^{-1}X_{1}}} \\
&&\times \left( ^{t}X_{1}\right) B\left( X_{2}\right) \frac{\exp \left( -%
\sqrt{2m_{i}+\left( ^{t}X_{2}\right) H^{-1}X_{2}}\sqrt{\left(
^{t}X_{1}H^{-1}X_{1}\right) }+\left( ^{t}X_{1}\right) H^{-1}\left(
X_{2}\right) \right) }{\sqrt{2m_{i}+\left( ^{t}X_{2}\right) H^{-1}X_{2}}} \\
\times &&\frac{\exp \left( -\sqrt{2m_{i}+\left( ^{t}X^{\prime }\right)
H^{-1}X^{\prime }}\sqrt{\left( ^{t}X_{2}H^{-1}X_{2}\right) }+\left(
^{t}X_{2}\right) H^{-1}\left( X^{\prime }\right) \right) }{\sqrt{%
2m_{i}+\left( ^{t}X^{\prime }\right) H^{-1}X^{\prime }}}
\end{eqnarray*}%
However, in this case, it is much more convenient to work with the time
representation and to compute rather   $\delta G\left( C,K,A,C^{\prime
},K^{\prime },A^{\prime },s\right) $. The correction at first order in $%
\gamma $ is:%
\begin{eqnarray}
&&\delta G\left( C,K,A,C^{\prime },K^{\prime },A^{\prime },s\right)
\label{crgrn} \\
&=&\gamma \int G\left( C,K,A,C_{1},K_{1},A_{1},s_{1}\right)\!\!\left(
A_{2}K_{1}+A_{1}K_{2}\right) G\left(
C_{1},K_{1},A_{1},C_{2},K_{2},A_{2},s_{2}\right)  \notag \\
&&G\left( C_{2},K_{2},A_{2},C^{\prime },K^{\prime },A^{\prime
},s-s_{1}-s_{2}\right) d\left( C_{1},K_{1},A_{1}\right)\!\!\left(
C_{2},K_{2},A_{2}\right) ds_{1}ds_{2}  \notag \\
&=&\gamma \left\langle A_{2}K_{1}+A_{1}K_{2}\right\rangle  \notag
\end{eqnarray}%
with a mean taken for a stochastic process constrained to start at $%
\allowbreak \left( C^{\prime },K^{\prime },A^{\prime }\right) $ and to end
at $\left( C,K,A\right) $. \ In first approximation, one can approximate $%
\left\langle A_{2}K_{1}+A_{1}K_{2}\right\rangle $ by its value along the
average path. This one is given by the minimization of:%
\begin{equation}
\int \!\!\left( ^{t}\left( \frac{d}{ds}X+MX\right) H^{-1}\left( \frac{d}{ds}%
X+MX\right) \right)  \label{tmnn}
\end{equation}%
for a path starting at $\left( C^{\prime },K^{\prime },A^{\prime },s\right) $%
, and ending at $\left( C,K,A,C^{\prime }\right) $. The matrices $M$ and $H$
are given by the exponential weight (\ref{Grnbc}):%
\begin{eqnarray*}
&&\exp \left( -\frac{\left( \left( C-C^{\prime }\right) +2\left( C^{\prime }-%
\bar{C}\right)\!\!\left( \alpha +\beta \right) s\right) ^{2}}{4\varpi ^{2}s}%
\right. \\
&&\hspace{3cm}\left. -\frac{\left( \left( K-K^{\prime }\right) +2\left(
\left( K^{\prime }-\bar{K}+\frac{\delta \bar{K}+\bar{C}}{\alpha }\right)
\alpha +\left( C^{\prime }-\bar{C}\right) -A^{\prime }\bar{K}^{\varepsilon
}\right) s\right) ^{2}}{4\left( \nu ^{2}+\frac{2\bar{K}^{2\varepsilon }}{%
\lambda ^{2}\alpha ^{2}}+\frac{3\varpi ^{2}}{2\left( 2\alpha +\beta \right)
\beta }\right) s}-\frac{\lambda ^{2}\left( A-A^{\prime }\right) ^{2}}{4s}%
\right)
\end{eqnarray*}%
so that:%
\begin{eqnarray*}
M &=&\left( 
\begin{array}{ccc}
\left( \alpha +\beta \right) & 0 & 0 \\ 
1 & \alpha & -\bar{K}^{\varepsilon } \\ 
0 & 0 & 0%
\end{array}%
\right) ,^{t}M=\left( 
\begin{array}{ccc}
\left( \alpha +\beta \right) & 1 & 0 \\ 
0 & \alpha & 0 \\ 
0 & -K^{\varepsilon } & 0%
\end{array}%
\right) \\
H &=&\left( 
\begin{array}{ccc}
2\varpi ^{2} & 0 & 0 \\ 
0 & 2\left( \nu ^{2}+\frac{2\bar{K}^{2\varepsilon }}{\lambda ^{2}\alpha ^{2}}%
+\frac{3\varpi ^{2}}{2\left( 2\alpha +\beta \right) \beta }\right) & 0 \\ 
0 & 0 & \frac{2}{\lambda ^{2}}%
\end{array}%
\right) =\left( 
\begin{array}{ccc}
a & 0 & 0 \\ 
0 & b & 0 \\ 
0 & 0 & c%
\end{array}%
\right)
\end{eqnarray*}%
\begin{equation*}
\left( 
\begin{array}{ccc}
a & 0 & 0 \\ 
0 & b & 0 \\ 
0 & 0 & c%
\end{array}%
\right)\!\!\left( 
\begin{array}{ccc}
\left( \alpha +\beta \right) & 1 & 0 \\ 
0 & \alpha & 0 \\ 
0 & -K^{\varepsilon } & 0%
\end{array}%
\right)\!\!\left( 
\begin{array}{ccc}
a & 0 & 0 \\ 
0 & b & 0 \\ 
0 & 0 & c%
\end{array}%
\right) ^{-1}\left( 
\begin{array}{ccc}
\left( \alpha +\beta \right) & 0 & 0 \\ 
1 & \alpha & -\bar{K}^{\varepsilon } \\ 
0 & 0 & 0%
\end{array}%
\right)
\end{equation*}%
The paths that minimize (\ref{tmnn}) satisfiy:%
\begin{equation*}
H^{-1}\frac{d^{2}}{ds^{2}}X+H^{-1}M\frac{d}{ds}X-\left( ^{t}\left(
H^{-1}M\right) \right) \frac{d}{ds}X-\left( ^{t}M\right) H^{-1}MX=0
\end{equation*}%
or equivalently:%
\begin{equation}
\frac{d^{2}}{ds^{2}}X+H\left( H^{-1}M-\left( ^{t}\left( H^{-1}M\right)
\right) \right) \frac{d}{ds}X-H\left( ^{t}M\right) H^{-1}MX=0
\label{qnvrgpt}
\end{equation}%
$\allowbreak \allowbreak $ $\allowbreak \allowbreak \allowbreak $We will
solve (\ref{qnvrgpt}), by looking first for a solution $\exp \left(
Ns\right) \allowbreak $ at the first order in $\alpha $ and $\beta $. A
straightforward computation yields:$\allowbreak $ $\allowbreak $

\begin{eqnarray*}
N &=&\left( 
\begin{array}{ccc}
-\left( \alpha +\beta \right) & 0 & 0 \\ 
-1 & -\alpha & K^{\varepsilon } \\ 
0 & 0 & 0%
\end{array}%
\right) +O\left( \alpha ^{2}\right) \\
\exp \left( Nu\right) &=&\left( 
\begin{array}{ccc}
1-u\left( \beta +\alpha \right) & 0 & 0 \\ 
u\left( u\left( \alpha +\frac{\beta }{2}\right) -1\right) & 1-u\alpha & 
K^{\varepsilon }u\left( 1-\frac{\alpha }{2}u\right) \\ 
0 & 0 & 1%
\end{array}%
\right) +O\left( \alpha ^{2}\right)
\end{eqnarray*}%
Then, a factorization $X=\exp \left( Ns\right) Y$ in the equation (\ref%
{qnvrgpt}) leads to:%
\begin{equation*}
\left( \frac{d^{2}}{ds^{2}}Y+\exp \left( -Ns\right) H\left( H^{-1}M-\left(
^{t}\left( H^{-1}M\right) \right) \right) \exp \left( Ns\right) \frac{d}{ds}%
Y+2N\frac{d}{ds}Y\right) =0
\end{equation*}%
$\allowbreak $That is:$\allowbreak $ $\allowbreak \allowbreak $ 
\begin{equation*}
\frac{d^{2}}{ds^{2}}Y+\allowbreak \left( 
\begin{array}{ccc}
-2\left( \alpha +\beta \right) & 0 & 0 \\ 
-s\beta -1 & -2\alpha & \left( 1-s\alpha \right) K^{\varepsilon } \\ 
0 & 0 & 0%
\end{array}%
\right) \frac{d}{ds}Y=0
\end{equation*}%
$\allowbreak $ $\allowbreak $ $\allowbreak $ which is solved as:%
\begin{equation*}
\left( \frac{d}{ds}Y\right) =\exp \left( N^{\prime }\left( s\right) \right) A
\end{equation*}%
$\allowbreak $for $A$ an initial condition and $\allowbreak $ 
\begin{equation*}
N^{\prime }\left( s\right) =\allowbreak \left( 
\begin{array}{ccc}
2s\left( \alpha +\beta \right) & 0 & 0 \\ 
\frac{1}{2}s\left( s\beta +2\right) & 2s\alpha & \frac{1}{2}K^{\varepsilon
}s\left( s\alpha -2\right) \\ 
0 & 0 & 0%
\end{array}%
\right)
\end{equation*}%
One finds given our assumptions of first order approximation: 
\begin{equation*}
\allowbreak \exp \left( N^{\prime }\left( s\right) \right) =\left( 
\begin{array}{ccc}
1+2s\left( \alpha +\beta \right) & 0 & 0 \\ 
s\left( 2s\alpha +3s\frac{\beta }{2}+1\right) & 1+2s\alpha & -K^{\varepsilon
}s\left( s\frac{\alpha }{2}+1\right) \\ 
0 & 0 & 1%
\end{array}%
\right)
\end{equation*}%
$\allowbreak \allowbreak $and the solutions of (\ref{qnvrgpt}) are thus:%
\begin{eqnarray*}
X\left( u\right) &=&\exp \left( Nu\right)\!\!\left( B+\int \exp \left(
N^{\prime }u\right) A\right) \\
&=&\exp \left( Nu\right)\!\!\left( B+P\left( u\right) A\right)
\end{eqnarray*}%
where:%
\begin{equation*}
\allowbreak P\left( u\right) =\int \exp \left( N^{\prime }u\right) =\left( 
\begin{array}{ccc}
u\left( u\alpha +u\beta +1\right) & 0 & 0 \\ 
\frac{1}{6}u^{2}\left( 4u\alpha +3u\beta +3\right) & u\left( u\alpha
+1\right) & -\frac{1}{6}K^{\varepsilon }u^{2}\left( u\alpha +3\right) \\ 
0 & 0 & u%
\end{array}%
\right)
\end{equation*}%
For a path where $X\left( 0\right) $ and $X\left( s\right) $ are fixed, the
constants $A$ and $B$ satisfy: 
\begin{equation*}
B=X\left( 0\right) ,A=\left( P\left( s\right) \right) ^{-1}\left( \exp
\left( -Ns\right) X\left( s\right) -X\left( 0\right) \right)
\end{equation*}%
To find the correction (\ref{crgrn}) in terms of initial and final points,
we define:%
\begin{eqnarray*}
\bar{X} &=&\frac{X\left( 0\right) +X\left( s\right) }{2} \\
\Delta X &=&X\left( s\right) -X\left( 0\right)
\end{eqnarray*}%
and the solution of (\ref{qnvrgpt}) rewrites:$\allowbreak $

\begin{eqnarray}
X\left( u\right) &=&\exp \left( Nu\right)\!\!\left( \left( 1-\left( P\left(
u\right) \right)\!\!\left( P\left( s\right) \right) ^{-1}\right) X\left(
0\right) +\left( P\left( u\right) \right)\!\!\left( P\left( s\right) \right)
^{-1}\exp \left( -Ns\right) X\left( s\right) \right)  \label{slntn} \\
&=&\exp \left( Nu\right)\!\!\left( \left( 1-\left( P\left( u\right) \right)
\left( P\left( s\right) \right) ^{-1}\left( 1-\exp \left( -Ns\right) \right)
\right) \bar{X}\right)  \notag \\
&&+\exp \left( Nu\right)\!\!\left( \left( P\left( u\right) \right)\!\!\left(
P\left( s\right) \right) ^{-1}\left( \exp \left( -Ns\right) +1\right)
-1\right) \frac{\Delta X}{2}  \notag \\
&=&\exp \left( Nu\right)\!\!\left( P\left( u\right) \right)\!\!\left( P\left(
s\right) \right) ^{-1}\exp \left( -Ns\right) \Delta X  \notag \\
&&+\exp \left( Nu\right)\!\!\left( 1-\left( P\left( u\right) \right)\!\!\left(
P\left( s\right) \right) ^{-1}\left( 1-\exp \left( -Ns\right) \right)
\right) X\left( 0\right)  \notag
\end{eqnarray}%
Some computations yield intermediate results:%
\begin{equation*}
\left( P\left( u\right) \right)\!\!\left( P\left( s\right) \right) ^{-1}=\left( 
\begin{array}{ccc}
u\frac{1-\left( s-u\right)\!\!\left( \alpha +\beta \right) }{s} & 0 & 0 \\ 
\allowbreak \frac{1}{6}u\left( s-u\right) \frac{2s\alpha -4u\alpha -3u\beta
-3}{s} & u\frac{1-\left( s-u\right) \alpha }{s} & -\frac{1}{6}\frac{%
K^{\varepsilon }}{s}u\left( s-u\right)\!\!\left( 2s\alpha -u\alpha -3\right) \\ 
0 & 0 & \frac{1}{s}u%
\end{array}%
\right)
\end{equation*}%
\begin{equation*}
\exp \left( Nu\right)\!\!\left( P\left( u\right) \right)\!\!\left( P\left(
s\right) \right) ^{-1}\exp \left( -Ns\right) =\left( 
\begin{array}{ccc}
\frac{u}{s} & 0 & 0 \\ 
-\frac{1}{6}u\left( s-u\right) \frac{s\alpha +u\alpha -3}{s} & \frac{u}{s} & 
\frac{1}{6}uK^{\varepsilon }\left( s-u\right) \frac{s\alpha +u\alpha -3}{s}
\\ 
0 & 0 & \frac{u}{s}%
\end{array}%
\right)
\end{equation*}%
\begin{equation*}
\exp \left( Nu\right)\!\!\left( 1-\left( P\left( u\right) \right)\!\!\left(
P\left( s\right) \right) ^{-1}\left( 1-\exp \left( -Ns\right) \right)
\right) =\allowbreak \left( 
\begin{array}{ccc}
1 & 0 & 0 \\ 
-\frac{1}{2}u\alpha \left( s-u\right) & 1 & \frac{1}{2}K^{\varepsilon
}u\alpha \left( s-u\right) \\ 
0 & 0 & 1%
\end{array}%
\right)
\end{equation*}%
\bigskip

and one finds for $X\left( u\right) $:$\allowbreak $ $\allowbreak $%
\begin{eqnarray*}
X\left( u\right) &=&\left( 
\begin{array}{ccc}
\frac{u}{s} & 0 & 0 \\ 
-\frac{1}{6}u\left( s-u\right) \frac{s\alpha +u\alpha -3}{s} & \frac{u}{s} & 
\frac{1}{6}uK^{\varepsilon }\left( s-u\right) \frac{s\alpha +u\alpha -3}{s}
\\ 
0 & 0 & \frac{u}{s}%
\end{array}%
\right) \Delta X \\
&&+\allowbreak \left( 
\begin{array}{ccc}
1 & 0 & 0 \\ 
-\frac{1}{2}u\alpha \left( s-u\right) & 1 & \frac{1}{2}K^{\varepsilon
}u\alpha \left( s-u\right) \\ 
0 & 0 & 1%
\end{array}%
\right) X\left( 0\right) \\
&=&\left( 
\begin{array}{ccc}
\frac{u}{s} & 0 & 0 \\ 
-\frac{1}{6}u\left( s-u\right) \frac{s\alpha +u\alpha -3}{s} & \frac{u}{s} & 
\frac{1}{6}uK^{\varepsilon }\left( s-u\right) \frac{s\alpha +u\alpha -3}{s}
\\ 
0 & 0 & \frac{u}{s}%
\end{array}%
\right) X\left( s\right) \\
&&+\left( 
\begin{array}{ccc}
\frac{s-u}{s} & 0 & 0 \\ 
-\frac{1}{6}u\left( s-u\right) \frac{2s\alpha -u\alpha +3}{s} & \frac{s-u}{s}
& \frac{1}{6}K^{\varepsilon }u\left( s-u\right) \frac{2s\alpha -u\alpha +3}{s%
} \\ 
0 & 0 & \frac{s-u}{s}%
\end{array}%
\right) X\left( 0\right)
\end{eqnarray*}%
so that the correction to the statistical weight can be found directly. One
has:%
\begin{equation*}
\gamma \int_{0}^{s}X\left( u\right) du=\gamma \left( 
\begin{array}{ccc}
\frac{1}{2}s & 0 & 0 \\ 
\frac{1}{12}s^{2} & \frac{1}{2}s & -\frac{1}{12}K^{\varepsilon }s^{2} \\ 
0 & 0 & \frac{1}{2}s%
\end{array}%
\right) \Delta X+\gamma sX\left( 0\right)
\end{equation*}%
and ultimately:%
\begin{eqnarray*}
&&\gamma \left\langle A_{2}K_{1}+A_{1}K_{2}\right\rangle
=\int_{0}^{s}X\left( u\right) duM\gamma \int_{0}^{s}X\left( u\right) du \\
&=&\left( ^{t}X\left( 0\right) \right)\!\!\left( 
\begin{array}{ccc}
0 & 0 & 0 \\ 
0 & 0 & s^{2} \\ 
0 & s^{2} & 0%
\end{array}%
\right) X\left( 0\right) +\left( ^{t}\Delta X\right)\!\!\left( 
\begin{array}{ccc}
0 & 0 & \frac{1}{24}s^{3} \\ 
0 & 0 & \frac{1}{4}s^{2} \\ 
\frac{1}{24}s^{3} & \frac{1}{4}s^{2} & -\frac{1}{12}K^{\varepsilon }s^{3}%
\end{array}%
\right) \Delta X \\
&&+2\left( ^{t}\Delta X\right)\!\!\left( 
\begin{array}{ccc}
0 & 0 & \frac{1}{6}s^{3} \\ 
0 & 0 & \frac{1}{2}s^{2} \\ 
0 & \frac{1}{2}s^{2} & -\frac{1}{6}K^{\varepsilon }s^{3}%
\end{array}%
\right) X\left( 0\right)
\end{eqnarray*}%
This term modifies the transition functions as:%
\begin{eqnarray}
&&\bar{G}\left( C,K,A,C^{\prime },K^{\prime },A^{\prime },s\right)
\label{GrnMd} \\
&=&\frac{\exp \left( -\frac{\left( \left( C-\bar{C}\right) -\left( C^{\prime
}-\bar{C}\right)\!\!\left( 1-2\left( \alpha +\beta \right) s\right) \right) ^{2}%
}{4\varpi ^{2}s}-\frac{\left( \left( K-\bar{K}+\frac{\delta \bar{K}+\bar{C}}{%
\alpha }\right) -\left( \left( K^{\prime }-\bar{K}+\frac{\delta \bar{K}+\bar{%
C}}{\alpha }\right)\!\!\left( 1-2\alpha s\right) -2\left( C^{\prime }-\bar{C}%
\right) s+2A^{\prime }\bar{K}^{\varepsilon }s\right) \right) ^{2}}{4\left(
\nu ^{2}+\frac{2\bar{K}^{2\varepsilon }}{\lambda ^{2}\alpha ^{2}}+\frac{%
3\varpi ^{2}}{2\left( 2\alpha +\beta \right) \beta }\right) s}-\frac{\lambda
^{2}\left( A-A^{\prime }\right) ^{2}}{4s}\right) }{4\sqrt{\pi \frac{\varpi
^{2}}{\lambda ^{2}}\left( \nu ^{2}+\frac{2\bar{K}^{2\varepsilon }}{\lambda
^{2}\alpha ^{2}}+\frac{3\varpi ^{2}}{2\left( 2\alpha +\beta \right) \beta }%
\right) s}}  \notag \\
&&\times \exp \left( -\gamma \left\langle A_{2}K_{1}+A_{1}K_{2}\right\rangle
\right)  \notag
\end{eqnarray}%
We can write more precisely this correction. To do so, let us first remark
that a weight of the form:%
\begin{equation*}
\left( ^{t}\left( \Delta X+MX\left( 0\right) \right) H^{-1}\left( \Delta
X+MX\left( 0\right) \right) \right) +\gamma \left( ^{t}\Delta X\right)
R_{1}\left( \Delta X\right) +2\gamma \left( ^{t}\Delta X\right) R_{2}X\left(
0\right) +\gamma \left( ^{t}X\left( 0\right) \right) R_{3}X\left( 0\right)
\end{equation*}%
with:%
\begin{equation*}
R_{1}=\left( 
\begin{array}{ccc}
0 & 0 & \frac{1}{24}s^{3} \\ 
0 & 0 & \frac{1}{4}s^{2} \\ 
\frac{1}{24}s^{3} & \frac{1}{4}s^{2} & -\frac{1}{12}K^{\varepsilon }s^{3}%
\end{array}%
\right) ,R_{2}=\left( 
\begin{array}{ccc}
0 & 0 & \frac{1}{6}s^{3} \\ 
0 & 0 & \frac{1}{2}s^{2} \\ 
0 & \frac{1}{2}s^{2} & -\frac{1}{6}K^{\varepsilon }s^{3}%
\end{array}%
\right) ,R_{3}=\left( 
\begin{array}{ccc}
0 & 0 & 0 \\ 
0 & 0 & s^{2} \\ 
0 & s^{2} & 0%
\end{array}%
\right)
\end{equation*}%
and the log of (\ref{GrnMdcan}) be rewritten:%
\begin{eqnarray}
&&\left( ^{t}\left( \Delta X+\bar{M}X\left( 0\right) \right)\!\!\left(
H^{-1}+\gamma A_{1}\right)\!\!\left( \Delta X+\bar{M}X\left( 0\right) \right)
\right)  \label{crtn} \\
&&+\left( ^{t}\left( MX\left( 0\right) \right) \right)\!\!\left( H^{-1}\right)
\left( MX\left( 0\right) \right) +\gamma \left( ^{t}X\left( 0\right) \right)
A_{3}X\left( 0\right)  \notag \\
&&-\left( ^{t}\left( \bar{M}X\left( 0\right) \right) \right)\!\!\left(
H^{-1}+\gamma A_{1}\right)\!\!\left( \bar{M}X\left( 0\right) \right)  \notag
\end{eqnarray}%
where $\bar{M}$ satisfies:%
\begin{equation*}
\left( H^{-1}+\gamma R_{1}\right) \bar{M}=H^{-1}M+\gamma R_{2}
\end{equation*}%
that is:%
\begin{equation*}
\bar{M}=\left( 1-\gamma HR_{1}\right)\!\!\left( M+\gamma HR_{2}\right)
\end{equation*}%
In our case, it leads to:%
\begin{equation*}
\bar{M}=\left( 
\begin{array}{ccc}
\alpha +\beta & 0 & \frac{1}{6}as^{3}\gamma \\ 
1 & \alpha -\frac{1}{8}bcs^{4}\gamma ^{2} & \frac{1}{2}bs^{2}\gamma
-K^{\varepsilon } \\ 
-\frac{1}{4}cs^{2}\gamma & \frac{1}{2}cs^{2}\gamma & \frac{1}{4}%
K^{\varepsilon }cs^{2}\gamma -\frac{1}{6}K^{\varepsilon }cs^{3}\gamma%
\end{array}%
\right)
\end{equation*}%
To complete the computation, we rewrite the two last terms in (\ref{crtn})
as:%
\begin{eqnarray*}
&&\left( ^{t}\left( MX\left( 0\right) \right) \right)\!\!\left( H^{-1}\right)
\left( MX\left( 0\right) \right) +\gamma \left( ^{t}X\left( 0\right) \right)
R_{3}X\left( 0\right) -\left( ^{t}\left( \bar{M}X\left( 0\right) \right)
\right)\!\!\left( H^{-1}+\gamma R_{1}\right)\!\!\left( \bar{M}X\left( 0\right)
\right) \\
&=&\left( ^{t}\left( MX\left( 0\right) \right) \right)\!\!\left( H^{-1}\right)
\left( MX\left( 0\right) \right) +\gamma \left( ^{t}X\left( 0\right) \right)
R_{3}X\left( 0\right) \\
&&-\left( ^{t}X\left( 0\right) \right)\!\!\left( ^{t}M+\gamma \left(
^{t}R_{2}\right) H\right)\!\!\left( 1-\gamma \left( ^{t}R_{1}\right) H\right)
\left( H^{-1}M+\gamma R_{2}\right) X\left( 0\right) \\
&=&\gamma \left( ^{t}X\left( 0\right) \right) R_{3}X\left( 0\right) -\left(
^{t}X\left( 0\right) \right)\!\!\left( \gamma \left( ^{t}R_{2}\right) M-\gamma
\left( ^{t}R_{1}\right) M+\gamma \left( ^{t}M\right) R_{2}\right) X\left(
0\right) \\
&=&\gamma \left( ^{t}X\left( 0\right) \right)\!\!\left( R_{3}-\left( \left(
^{t}M\right)\!\!\left( 2R_{2}-R_{1}\right) \right) \right) X\left( 0\right)
\end{eqnarray*}

\bigskip Defining $\bar{H}^{-1}$ by:%
\begin{eqnarray*}
\bar{H}^{-1} &=&\left( H^{-1}+\gamma R_{1}\right) \\
&=&\left( 
\begin{array}{ccc}
\frac{1}{as} & 0 & \frac{1}{12}s^{3}\gamma \\ 
0 & \frac{1}{bs} & \frac{1}{4}s^{2}\gamma \\ 
\frac{1}{12}s^{3}\gamma & \frac{1}{4}s^{2}\gamma & -\frac{1}{6}\frac{%
K^{\varepsilon }cs^{4}\gamma -6}{cs}%
\end{array}%
\right)
\end{eqnarray*}%
and:%
\begin{eqnarray*}
\bar{H} &=&\left( H^{-1}+\gamma R_{1}\right) ^{-1}=H-H\gamma R_{1}H \\
&=&\left( 
\begin{array}{ccc}
as & 0 & -\frac{1}{24}acs^{5}\gamma \\ 
0 & bs & -\frac{1}{8}bcs^{4}\gamma \\ 
0 & -\frac{1}{8}bcs^{4}\gamma & cs+\frac{1}{24}K^{\varepsilon
}c^{2}s^{5}\gamma%
\end{array}%
\right)
\end{eqnarray*}%
$\allowbreak $ $\allowbreak \allowbreak \allowbreak $the weight including
the correction (\ref{GrnMd}) is:%
\begin{equation*}
\exp \left( -\left( ^{t}\left( \Delta X+\bar{M}X\left( 0\right) \right) \bar{%
H}^{-1}\left( \Delta X+\bar{M}X\left( 0\right) \right) \right) -\gamma
\left( ^{t}X\left( 0\right) \right)\!\!\left( R_{3}-\left( \left( ^{t}M\right)
\left( 2R_{2}-R_{1}\right) \right) \right) X\left( 0\right) \right)
\end{equation*}%
This result can also be studied in terms of trajectories. Actually, in (\ref%
{GrnMd}) a term is added to the initial action. It has the form:%
\begin{eqnarray*}
&&\frac{\gamma }{2}\left( ^{t}\left( \int_{0}^{s}\left( X\left( v\right)
\right) dv\right) \right) M\left( \int_{0}^{s}\left( X\left( u\right)
\right) du\right) \\
&=&\gamma \left( ^{t}\left( \int_{0}^{s}\left( X\left( v\right) \right)
dv\right) \right) M\left( \int_{0}^{v}\left( X\left( u\right) \right)
du\right)
\end{eqnarray*}%
and the correction to the dynamic equations due to the agents interaction is:

\begin{equation}
\frac{d^{2}}{ds^{2}}X+H\left( \left( H^{-1}M-\left( ^{t}\left(
H^{-1}M\right) \right) \right) \right) \frac{d}{ds}X-\left( \left( H\left(
^{t}M\right) H^{-1}M\right) \right) X-\gamma M\left( \int_{0}^{s}\left(
X\left( u\right) \right) du\right) =0  \label{qnprtb}
\end{equation}%
We can approximate $M\left( \int_{0}^{v}\left( X\left( u\right) \right)
du\right) $ with its mean path approximation, so that (\ref{qnprtb})
rewrites:%
\begin{equation}
\frac{d^{2}}{ds^{2}}X+H\left( \left( H^{-1}M-\left( ^{t}\left(
H^{-1}M\right) \right) \right) \right) \frac{d}{ds}X-\left( \left( H\left(
^{t}M\right) H^{-1}M\right) +\gamma HM_{1}\right) X-\gamma HM_{2}X\left(
0\right) =0  \label{qnprtbn}
\end{equation}%
with%
\begin{eqnarray*}
M\left( \int_{0}^{v}\left( X\left( u\right) \right) du\right) &=&\gamma
\left( 
\begin{array}{ccc}
0 & 0 & 0 \\ 
0 & 0 & 1 \\ 
0 & 1 & 0%
\end{array}%
\right)\!\!\left( \left( 
\begin{array}{ccc}
\frac{1}{2}s & 0 & 0 \\ 
\frac{1}{12}s^{2} & \frac{1}{2}s & -\frac{1}{12}K^{\varepsilon }s^{2} \\ 
0 & 0 & \frac{1}{2}s%
\end{array}%
\right) \Delta X+sX\left( 0\right) \right) \\
&=&\gamma \allowbreak \left( 
\begin{array}{ccc}
0 & 0 & 0 \\ 
0 & 0 & \frac{1}{2}s \\ 
\frac{1}{12}s^{2} & \frac{1}{2}s & -\frac{1}{12}K^{\varepsilon }s^{2}%
\end{array}%
\right) X\left( s\right) +\gamma \allowbreak \left( 
\begin{array}{ccc}
0 & 0 & 0 \\ 
0 & 0 & \frac{1}{2}s \\ 
-\frac{1}{12}s^{2} & \frac{1}{2}s & \frac{1}{12}K^{\varepsilon }s^{2}%
\end{array}%
\right) X\left( 0\right) \\
&\equiv &M_{1}X\left( s\right) +M_{2}X\left( 0\right)
\end{eqnarray*}%
To find the solution of (\ref{qnprtbnbs}), we first consider:%
\begin{equation}
\frac{d^{2}}{ds^{2}}X+H\left( \left( H^{-1}M-\left( ^{t}\left(
H^{-1}M\right) \right) \right) \right) \frac{d}{ds}X-\left( \left( H\left(
^{t}M\right) H^{-1}M\right) +\gamma HM_{1}\right) X=0  \label{qnprtbnbs}
\end{equation}%
and proceed as for (\ref{qnvrgpt}). We look for a solution of (\ref%
{qnprtbnbs}) of the form:%
\begin{equation*}
\exp \left( Ns\right)\!\!\left( 1+\gamma \hat{N}\left( s\right) \right)
\end{equation*}%
so that $\hat{N}\left( s\right) $ satisfies: 
\begin{equation*}
\frac{d^{2}}{ds^{2}}\hat{N}\left( s\right) +2N\frac{d}{ds}\hat{N}\left(
s\right) +\exp \left( -Ns\right) H\left( H^{-1}M-\left( ^{t}\left(
H^{-1}M\right) \right) \right) \exp \left( Ns\right) \frac{d}{ds}\hat{N}%
\left( s\right) -\exp \left( -Ns\right) HM_{1}\exp \left( Ns\right) =0
\end{equation*}%
$\allowbreak $whose expanded form in our order of approximation is:$%
\allowbreak \allowbreak \allowbreak \allowbreak $ 
\begin{eqnarray*}
0 &=&\gamma \frac{d^{2}}{ds^{2}}\hat{N}\left( s\right) +\gamma \left( 
\begin{array}{ccc}
0 & 0 & 0 \\ 
-\beta s+1 & 0 & -K^{\varepsilon }-K^{\varepsilon }s\alpha \\ 
0 & 0 & 0%
\end{array}%
\right) \frac{d}{ds}\hat{N}\left( s\right) \\
&&-\gamma \left( 
\begin{array}{ccc}
0 & 0 & 0 \\ 
\frac{5}{12}K^{\varepsilon }cs^{3} & -\frac{1}{2}K^{\varepsilon }cs^{2} & 
\frac{1}{2}bs \\ 
-\frac{5}{12}cs^{2} & \frac{1}{2}cs & \frac{5}{12}K^{\varepsilon }cs^{2}%
\end{array}%
\right)
\end{eqnarray*}%
The solution $\hat{N}\left( s\right) $ is computed at the zeroth order in $%
\alpha $ and as a consequence, it satisfies:%
\begin{equation*}
\frac{d^{2}}{ds^{2}}\hat{N}\left( s\right) +\left( 
\begin{array}{ccc}
0 & 0 & 0 \\ 
-1 & 0 & K^{\varepsilon } \\ 
0 & 0 & 0%
\end{array}%
\right) \frac{d}{ds}\hat{N}\left( s\right) -\left( 
\begin{array}{ccc}
0 & 0 & 0 \\ 
\frac{5}{12}K^{\varepsilon }cs^{3} & -\frac{1}{2}K^{\varepsilon }cs^{2} & 
\frac{1}{2}bs \\ 
-\frac{5}{12}cs^{2} & \frac{1}{2}cs & \frac{5}{12}K^{\varepsilon }cs^{2}%
\end{array}%
\right)
\end{equation*}%
$\allowbreak \allowbreak \allowbreak $ $\allowbreak $ $\allowbreak $ $%
\allowbreak $ $\allowbreak $ $\allowbreak $

\bigskip

\begin{equation*}
\hat{N}\left( s\right) =\left( 
\begin{array}{ccc}
0 & 0 & 0 \\ 
\frac{1}{36}K^{\varepsilon }cs^{5} & -\frac{3}{48}K^{\varepsilon }cs^{4} & -%
\frac{c}{144}K^{2\varepsilon }s^{5}+\frac{1}{12}bs^{3} \\ 
-\frac{5}{144}cs^{4} & \frac{1}{12}cs^{3} & \frac{5}{144}K^{\varepsilon
}cs^{4}%
\end{array}%
\right)
\end{equation*}%
$\allowbreak $The equation (\ref{qnprtbnbs}) can then be solved in the
following way. We first solve%
\begin{equation*}
\frac{d^{2}}{ds^{2}}X+H\left( \left( H^{-1}M-\left( ^{t}\left(
H^{-1}M\right) \right) \right) \right) \frac{d}{ds}X-\left( \left( H\left(
^{t}M\right) H^{-1}M+\gamma HN_{1}\right) +\gamma HM_{1}\right) X=0
\end{equation*}%
as before by setting $X=\exp \left( Ns\right)\!\!\left( 1+\gamma \hat{N}\left(
s\right) \right) Y\left( s\right) $ and $Y\left( s\right) $ satisfies:%
\begin{equation*}
\frac{d^{2}}{ds^{2}}Y+L\frac{d}{ds}Y=0
\end{equation*}%
with:$\allowbreak \allowbreak $ $\allowbreak \allowbreak \allowbreak
\allowbreak \allowbreak \allowbreak $ $\allowbreak $

$\allowbreak \allowbreak $%
\begin{eqnarray*}
L &=&\left( 1-\gamma \hat{N}\left( s\right) \right) \exp \left( -Ns\right)
\left( H\left( H^{-1}M-\left( ^{t}\left( H^{-1}M\right) \right) \right)
\right) \exp \left( Ns\right)\!\!\left( 1+\gamma \hat{N}\left( s\right) \right)
\\
&&+2\left( 1-\gamma \hat{N}\left( s\right) \right) \exp \left( -Ns\right) 
\frac{d}{ds}\left( \exp \left( Ns\right)\!\!\left( 1+\gamma \hat{N}\left(
s\right) \right) \right) \\
&=&\left( 
\begin{array}{ccc}
0 & 0 & 0 \\ 
\frac{7}{72}K^{\varepsilon }cs^{4}\gamma -s\beta +1 & -\frac{1}{12}%
K^{\varepsilon }cs^{3}\gamma & -K^{\varepsilon }-K^{\varepsilon }s\alpha -%
\frac{7}{72}K^{2\varepsilon }cs^{4}\gamma \\ 
-\frac{1}{12}cs^{3}\gamma & 0 & \frac{1}{12}K^{\varepsilon }cs^{3}\gamma%
\end{array}%
\right) \\
&&+2\left( 
\begin{array}{ccc}
-\alpha -\beta & 0 & 0 \\ 
\frac{1}{24}K^{\varepsilon }cs^{4}\gamma -1 & -\frac{1}{6}K^{\varepsilon
}cs^{3}\gamma -\alpha & K^{\varepsilon }+\frac{1}{4}bs^{2}\gamma +\frac{1}{16%
}K^{2\varepsilon }cs^{4}\gamma \\ 
\frac{1}{144}cs^{5}\beta ^{2}\gamma -\frac{1}{18}cs^{3}\gamma & \frac{1}{4}%
cs^{2}\gamma & \frac{1}{18}K^{\varepsilon }cs^{3}\gamma%
\end{array}%
\right) \\
&=&\left( 
\begin{array}{ccc}
-2\alpha -2\beta & 0 & 0 \\ 
\frac{13}{72}K^{\varepsilon }cs^{4}\gamma -s\beta -1 & -2\alpha -\frac{5}{12}%
K^{\varepsilon }cs^{3}\gamma & K^{\varepsilon }+\frac{1}{2}bs^{2}\gamma
-K^{\varepsilon }s\alpha +\frac{1}{36}K^{2\varepsilon }cs^{4}\gamma \\ 
\frac{1}{72}cs^{5}\beta ^{2}\gamma -\frac{7}{36}cs^{3}\gamma & \frac{1}{2}%
cs^{2}\gamma & \frac{7}{36}K^{\varepsilon }cs^{3}\gamma%
\end{array}%
\right)
\end{eqnarray*}%
\begin{eqnarray*}
\int L\left( u\right) du &=&\int_{0}^{u}\allowbreak \left( 
\begin{array}{ccc}
-2\alpha -2\beta & 0 & 0 \\ 
\frac{13}{72}K^{\varepsilon }cs^{4}\gamma -s\beta -1 & -2\alpha -\frac{5}{12}%
K^{\varepsilon }cs^{3}\gamma & K^{\varepsilon }+\frac{1}{2}bs^{2}\gamma
-K^{\varepsilon }s\alpha +\frac{1}{36}K^{2\varepsilon }cs^{4}\gamma \\ 
\frac{1}{72}cs^{5}\beta ^{2}\gamma -\frac{7}{36}cs^{3}\gamma & \frac{1}{2}%
cs^{2}\gamma & \frac{7}{36}K^{\varepsilon }cs^{3}\gamma%
\end{array}%
\right) ds \\
&=&\left( 
\begin{array}{ccc}
-2u\left( \alpha +\beta \right) & 0 & 0 \\ 
\frac{13}{360}K^{\varepsilon }cu^{5}\gamma -\frac{1}{2}u^{2}\beta -u & 
-2u\alpha -\frac{5}{48}K^{\varepsilon }cu^{4}\gamma & K^{\varepsilon }u-%
\frac{1}{2}K^{\varepsilon }u^{2}\alpha +\frac{1}{6}bu^{3}\gamma +\frac{1}{180%
}K^{2\varepsilon }cu^{5}\gamma \\ 
\frac{1}{432}cu^{6}\beta ^{2}\gamma -\frac{7}{144}cu^{4}\gamma & \frac{1}{6}%
cu^{3}\gamma & \frac{7}{144}K^{\varepsilon }cu^{4}\gamma%
\end{array}%
\right)
\end{eqnarray*}%
As a consequence:$\allowbreak $%
\begin{eqnarray*}
&&\exp \left( -\int L\left( u\right) du\right) \\
&=&\left( 
\begin{array}{ccc}
1 & 0 & 0 \\ 
0 & 1 & 0 \\ 
0 & 0 & 1%
\end{array}%
\right) -\left( 
\begin{array}{ccc}
-2u\left( \alpha +\beta \right) & 0 & 0 \\ 
\frac{13K^{\varepsilon }cu^{5}\gamma }{360}-\frac{1}{2}u^{2}\beta -u & 
-2u\alpha -\frac{5K^{\varepsilon }cu^{4}\gamma }{48} & K^{\varepsilon }u-%
\frac{K^{\varepsilon }u^{2}\alpha }{2}+\frac{bu^{3}\gamma }{6}+\frac{%
K^{2\varepsilon }cu^{5}\gamma }{180} \\ 
\frac{cu^{6}\beta ^{2}\gamma }{432}-\frac{7cu^{4}\gamma }{144} & \frac{1}{6}%
cu^{3}\gamma & \frac{7K^{\varepsilon }cu^{4}\gamma }{144}%
\end{array}%
\right) \\
&&+\frac{1}{2}\left( 
\begin{array}{ccc}
-2u\left( \alpha +\beta \right) & 0 & 0 \\ 
\frac{13}{360}K^{\varepsilon }cu^{5}\gamma -\frac{1}{2}u^{2}\beta -u & 
-2u\alpha -\frac{5}{48}K^{\varepsilon }cu^{4}\gamma & K^{\varepsilon }u-%
\frac{1}{2}K^{\varepsilon }u^{2}\alpha +\frac{1}{6}bu^{3}\gamma +\frac{1}{180%
}K^{2\varepsilon }cu^{5}\gamma \\ 
\frac{1}{432}cu^{6}\beta ^{2}\gamma -\frac{7}{144}cu^{4}\gamma & \frac{1}{6}%
cu^{3}\gamma & \frac{7}{144}K^{\varepsilon }cu^{4}\gamma%
\end{array}%
\right) ^{2}
\end{eqnarray*}%
and then$\allowbreak $: 
\begin{eqnarray*}
Y\left( u\right) &=&B+\left( \int \exp \left( -\int L\left( u\right)
du\right) \right) A \\
&=&B+\left( 
\begin{array}{ccc}
s\left( s\alpha +s\beta +1\right) & 0 & 0 \\ 
\frac{s^{2}\left( 480s\alpha +360s\beta -K^{\varepsilon }cs^{4}\gamma
+360\right) }{720} & s+s^{2}\alpha +\frac{3}{80}K^{\varepsilon }cs^{5}\gamma
& -\frac{K^{\varepsilon }s^{2}}{2}-\frac{K^{\varepsilon }s^{3}\alpha }{6}-%
\frac{bs^{4}\gamma }{24}-\frac{K^{2\varepsilon }cs^{6}\gamma }{180} \\ 
-\frac{cs^{5}\gamma }{144} & -\frac{cs^{4}\gamma }{24} & s+\frac{%
K^{\varepsilon }cs^{5}\gamma }{144}%
\end{array}%
\right) A
\end{eqnarray*}%
which leads to the solution of (\ref{qnprtbnbs}):

\begin{equation*}
X\left( u\right) =\exp \left( Nu\right)\!\!\left( 1+\gamma \hat{N}\left(
u\right) \right)\!\!\left( B+\left( \int \exp \left( -\int L\left( u\right)
du\right) \right) A\right)
\end{equation*}%
$\allowbreak \allowbreak $ Given that:$\allowbreak \allowbreak \allowbreak
\allowbreak \allowbreak \allowbreak $ $\allowbreak \allowbreak $

\begin{eqnarray*}
&&\exp \left( Nu\right)\!\!\left( 1+\gamma \hat{N}\left( u\right) \right) \\
&=&\left( 
\begin{array}{ccc}
1-s\left( \beta +\alpha \right) & 0 & 0 \\ 
s\left( s\left( \alpha +\frac{\beta }{2}\right) -1\right) & 1-s\alpha & 
K^{\varepsilon }s\left( 1-\frac{\alpha }{2}s\right) \\ 
0 & 0 & 1%
\end{array}%
\right) \\
&&\times \left( \left( 
\begin{array}{ccc}
1 & 0 & 0 \\ 
0 & 1 & 0 \\ 
0 & 0 & 1%
\end{array}%
\right) +\gamma \left( 
\begin{array}{ccc}
0 & 0 & 0 \\ 
\frac{1}{36}K^{\varepsilon }cs^{5} & -\frac{3}{48}K^{\varepsilon }cs^{4} & -%
\frac{c}{144}K^{2\varepsilon }s^{5}+\frac{1}{12}bs^{3} \\ 
-\frac{5}{144}cs^{4} & \frac{1}{12}cs^{3} & \frac{5}{144}K^{\varepsilon
}cs^{4}%
\end{array}%
\right) \right) \\
&=&\left( 
\begin{array}{ccc}
1-s\beta -s\alpha & 0 & 0 \\ 
s^{2}\alpha -s+\frac{1}{2}s^{2}\beta -\frac{1}{144}K^{\varepsilon
}cs^{5}\gamma & \frac{1}{48}K^{\varepsilon }cs^{4}\gamma -s\alpha +1 & 
K^{\varepsilon }s-\frac{1}{2}K^{\varepsilon }s^{2}\alpha +\frac{1}{12}%
bs^{3}\gamma +\frac{1}{36}K^{2\varepsilon }cs^{5}\gamma \\ 
-\frac{5}{144}cs^{4}\gamma & \frac{1}{12}cs^{3}\gamma & \frac{5}{144}%
K^{\varepsilon }cs^{4}\gamma +1%
\end{array}%
\right)
\end{eqnarray*}%
$\allowbreak $

\begin{eqnarray*}
&&\exp \left( Nu\right)\!\!\left( 1+\gamma \hat{N}\left( u\right) \right)
\left( \int \exp \left( -\int L\left( u\right) du\right) \right) \\
&=&\left( 
\begin{array}{ccc}
s & 0 & 0 \\ 
\frac{1}{6}s^{3}\alpha -\frac{1}{2}s^{2}-\frac{7}{1440}K^{\varepsilon
}cs^{6}\gamma & s+\frac{1}{60}K^{\varepsilon }cs^{5}\gamma & \frac{1}{2}%
K^{\varepsilon }s^{2}-\frac{1}{6}K^{\varepsilon }s^{3}\alpha +\frac{1}{24}%
bs^{4}\gamma +\frac{3}{160}K^{2\varepsilon }cs^{6}\gamma \\ 
0 & \frac{1}{24}cs^{4}\gamma & s%
\end{array}%
\right)
\end{eqnarray*}%
we have ultimately the general solution of (\ref{qnprtbnbs}):%
\begin{eqnarray*}
X\left( u\right) &=&\exp \left( Nu\right)\!\!\left( 1+\gamma \hat{N}\left(
u\right) \right)\!\!\left( B+\left( \int \exp \left( -\int L\left( u\right)
du\right) \right) A\right) \\
&=&\left( 
\begin{array}{ccc}
1-s\beta -s\alpha & 0 & 0 \\ 
s^{2}\alpha -s+\frac{1}{2}s^{2}\beta -\frac{1}{144}K^{\varepsilon
}cs^{5}\gamma & \frac{1}{48}K^{\varepsilon }cs^{4}\gamma -s\alpha +1 & 
K^{\varepsilon }s-\frac{1}{2}K^{\varepsilon }s^{2}\alpha +\frac{1}{12}%
bs^{3}\gamma +\frac{1}{36}K^{2\varepsilon }cs^{5}\gamma \\ 
-\frac{5}{144}cs^{4}\gamma & \frac{1}{12}cs^{3}\gamma & \frac{5}{144}%
K^{\varepsilon }cs^{4}\gamma +1%
\end{array}%
\right) B \\
&&+\left( 
\begin{array}{ccc}
s & 0 & 0 \\ 
\frac{1}{6}s^{3}\alpha -\frac{1}{2}s^{2}-\frac{7}{1440}K^{\varepsilon
}cs^{6}\gamma & s+\frac{1}{60}K^{\varepsilon }cs^{5}\gamma & \frac{1}{2}%
K^{\varepsilon }s^{2}-\frac{1}{6}K^{\varepsilon }s^{3}\alpha +\frac{1}{24}%
bs^{4}\gamma +\frac{3}{160}K^{2\varepsilon }cs^{6}\gamma \\ 
0 & \frac{1}{24}cs^{4}\gamma & s%
\end{array}%
\right) A
\end{eqnarray*}%
Then, adding the particular solution of (\ref{qnprtbn}):%
\begin{equation*}
\frac{d^{2}}{ds^{2}}X+H\left( \left( H^{-1}M-\left( ^{t}\left(
H^{-1}M\right) \right) \right) \right) \frac{d}{ds}X-\left( \left( H\left(
^{t}M\right) H^{-1}M\right) +\gamma HM_{1}\right) X-\gamma HM_{2}X\left(
0\right) =0
\end{equation*}%
which is:%
\begin{equation*}
X\left( s\right) =\gamma \left( 
\begin{array}{ccc}
0 & 0 & 0 \\ 
0 & 0 & \frac{1}{12}bs^{3} \\ 
-\frac{1}{144}cs^{4} & \frac{1}{12}cs^{3} & \frac{K^{\varepsilon }}{144}%
cs^{4}%
\end{array}%
\right) X\left( 0\right)
\end{equation*}%
we obtain the a full solution of (\ref{qnprtbn}):$\allowbreak \allowbreak $%
\begin{eqnarray}
X\left( s\right) &=&\left( 
\begin{array}{ccc}
1-s\beta -s\alpha & 0 & 0 \\ 
s^{2}\alpha -s+\frac{1}{2}s^{2}\beta -\frac{K^{\varepsilon }cs^{5}\gamma }{%
144} & \frac{K^{\varepsilon }cs^{4}\gamma }{48}-s\alpha +1 & K^{\varepsilon
}s-\frac{1}{2}K^{\varepsilon }s^{2}\alpha +\frac{1}{12}bs^{3}\gamma +\frac{%
K^{2\varepsilon }cs^{5}\gamma }{36} \\ 
-\frac{5cs^{4}\gamma }{144} & \frac{1}{12}cs^{3}\gamma & \frac{5}{144}%
K^{\varepsilon }cs^{4}\gamma +1%
\end{array}%
\right) B  \label{dnmcndvp} \\
&&+\left( 
\begin{array}{ccc}
s & 0 & 0 \\ 
\frac{1}{6}s^{3}\alpha -\frac{1}{2}s^{2}-\frac{7K^{\varepsilon }cs^{6}\gamma 
}{1440} & s+\frac{K^{\varepsilon }cs^{5}\gamma }{60} & \frac{K^{\varepsilon
}s^{2}}{2}-\frac{K^{\varepsilon }s^{3}\alpha }{6}+\frac{bs^{4}\gamma }{24}+%
\frac{3K^{2\varepsilon }cs^{6}\gamma }{160} \\ 
0 & \frac{cs^{4}\gamma }{24} & s%
\end{array}%
\right) A  \notag \\
&&+\gamma \left( 
\begin{array}{ccc}
0 & 0 & 0 \\ 
0 & 0 & \frac{1}{12}bs^{3} \\ 
-\frac{1}{144}cs^{4} & \frac{1}{12}cs^{3} & \frac{K^{\varepsilon }}{144}%
cs^{4}%
\end{array}%
\right) X\left( 0\right)  \notag
\end{eqnarray}%
$\allowbreak \allowbreak \allowbreak $ $\allowbreak $ $\allowbreak
\allowbreak $Several types of initial conditions are possible. The
most relevant will be to chose $X\left( 0\right) $ and $\dot{X}\left(
0\right) $ as initial conditions, one finds $A$ by writing: 
\begin{equation*}
A=\dot{X}\left( 0\right) -\left( 
\begin{array}{ccc}
-\alpha -\beta & 0 & 0 \\ 
-1 & -\alpha & K^{\varepsilon } \\ 
0 & 0 & 0%
\end{array}%
\right) X\left( 0\right)
\end{equation*}%
Inserting this result in (\ref{dnmcndvp}) leads ultimately to$\allowbreak $: 
\begin{eqnarray*}
X\left( s\right) &=&\left( 
\begin{array}{ccc}
1 & 0 & 0 \\ 
\frac{1}{2}s^{2}\alpha +\frac{7}{720}K^{\varepsilon }cs^{5}\gamma & \frac{1}{%
48}K^{\varepsilon }cs^{4}\gamma +1 & \frac{1}{6}bs^{3}\gamma -\frac{1}{2}%
K^{\varepsilon }s^{2}\alpha +\frac{1}{90}K^{2\varepsilon }cs^{5}\gamma \\ 
0 & \frac{1}{6}c\gamma s^{3} & 1%
\end{array}%
\right) X\left( 0\right) \\
&&+\left( 
\begin{array}{ccc}
s & 0 & 0 \\ 
\frac{1}{6}s^{3}\alpha -\frac{1}{2}s^{2}-\frac{7}{1440}K^{\varepsilon
}cs^{6}\gamma & s+\frac{1}{60}K^{\varepsilon }cs^{5}\gamma & \frac{1}{2}%
K^{\varepsilon }s^{2}-\frac{1}{6}K^{\varepsilon }s^{3}\alpha +\frac{1}{24}%
bs^{4}\gamma +\frac{3}{160}K^{2\varepsilon }cs^{6}\gamma \\ 
0 & \frac{1}{24}cs^{4}\gamma & s%
\end{array}%
\right) \dot{X}\left( 0\right)
\end{eqnarray*}%
Then, replacing $b$ and $c$ by their values, and $\gamma $ by $\frac{\gamma 
}{\bar{A}^{2}K^{\varepsilon }}$, so that $\gamma $ is dimensionless, and
noting that $X\left( 0\right) =B$, yields the result stated in the text and
the deviation to this trajectory due to the interaction term, with the same
initial conditions, is thus:%
\begin{eqnarray*}
\delta X\left( s\right) &=&\gamma \left( 
\begin{array}{ccc}
0 & 0 & 0 \\ 
\frac{7}{720}K^{\varepsilon }cs^{5} & \frac{1}{48}K^{\varepsilon }cs^{4} & 
\frac{1}{6}bs^{3}+\frac{1}{90}K^{2\varepsilon }cs^{5} \\ 
0 & \frac{1}{6}cs^{3} & 0%
\end{array}%
\right) X\left( 0\right) \\
&&+\gamma \left( 
\begin{array}{ccc}
0 & 0 & 0 \\ 
-\frac{7}{1440}K^{\varepsilon }cs^{6} & \frac{1}{60}K^{\varepsilon }cs^{5} & 
\frac{1}{24}bs^{4}+\frac{3}{160}K^{2\varepsilon }cs^{6} \\ 
0 & \frac{1}{24}cs^{4} & 0%
\end{array}%
\right) \dot{X}\left( 0\right)
\end{eqnarray*}

\section*{\protect\bigskip Case 3: 2 Agents interaction via 4 points Green
function}

The field theoretic context allows also to study the impact of one type of
agent on an other. Consider the transition functions, for two agents,
without interaction. The probability of transition between $\left(
K_{1},C_{1},A_{1}\right) _{i}$ and $\left( K_{1},C_{1},A_{1}\right) _{f}$
for the first one, and $\left( K_{2},C_{2},A_{2}\right) _{i}$ and $\left(
K_{2},C_{2},A_{2}\right) _{f}$ for the second one, is:%
\begin{eqnarray*}
&&G\left( \left( K_{1},C_{1},A_{1}\right) _{i},\left(
K_{2},C_{2},A_{2}\right) _{i},\left( K_{1},C_{1},A_{1}\right) _{f},\left(
K_{2},C_{2},A_{2}\right) _{f}\right) \\
&\equiv &G\left( \left( K_{1},C_{1},A_{1}\right) _{i},\left(
K_{2},C_{2},A_{2}\right) _{i},s\right) G\left( \left(
K_{1},C_{1},A_{1}\right) _{f},\left( K_{2},C_{2},A_{2}\right) _{f},s\right)
\end{eqnarray*}%
An application of the Wick theorem to the field interaction term: 
\begin{equation*}
\frac{\gamma }{2}\int \Psi ^{\dag }\left( K_{1},C_{1},A_{1}\right) \Psi
^{\dag }\left( K_{2},C_{2},A_{2}\right) \left\{
A_{2}K_{1}+A_{1}K_{2}\right\} \Psi \left( K_{1},C_{1},A_{1}\right) \Psi
\left( K_{2},C_{2},A_{2}\right)
\end{equation*}%
leads directly to a correction, on Green function Laplace transform:%
\begin{eqnarray*}
&&\int \gamma G\left( \left( K_{1},C_{1},A_{1}\right) _{i},\left(
K_{1},C_{1},A_{1}\right) ,m_{i}\right) G\left( \left(
K_{1},C_{1},A_{1}\right) ,\left( K_{1},C_{1},A_{1}\right) _{f},m_{i}\right)
\\
&&\times \left\{ A_{2}K_{1}+A_{1}K_{2}\right\} \\
&&\times G\left( \left( K_{2},C_{2},A_{2}\right) _{i},\left(
K_{2},C_{2},A_{2}\right) ,m_{i}\right) G\left( \left(
K_{2},C_{2},A_{2}\right) ,\left( K_{2},C_{2},A_{2}\right) _{f},m_{i}\right)
\\
&\equiv &\int \gamma G\left( \left( X_{1}\right) _{i},X_{1},m_{i}\right)
G\left( X_{1},\left( X_{1}\right) _{f},m_{i}\right) \left\{ \left(
^{t}X_{1}\right) MX_{2}\right\} G\left( \left( X_{2}\right)
_{i},X_{2},m_{i}\right) G\left( X_{2},\left( X_{2}\right) _{f},m_{i}\right)
\end{eqnarray*}%
and in time representation:

\begin{eqnarray*}
&&\gamma \int G\left( \left( X_{1}\right) _{i},X_{1}\left( s_{1}\right)
,s_{1}\right) G\left( X_{1}\left( s_{1}\right) ,\left( X_{1}\right)
_{f},s-s_{1}\right) \\
&&\times \left\{ \left( ^{t}X_{1}\left( s_{1}\right) \right) MX_{2}\left(
s_{2}\right) \right\} G\left( \left( X_{2}\right) _{i},X_{2}\left(
s_{2}\right) ,s_{2}\right) G\left( X_{2}\left( s_{2}\right) ,\left(
X_{2}\right) _{f},s-s_{2}\right) ds_{1}ds_{2} \\
&=&\gamma \int \!\!\left\langle \left( ^{t}X_{1}\left( s_{1}\right) \right)
MX_{2}\left( s_{2}\right) \right\rangle ds_{1}ds_{2}
\end{eqnarray*}%
where the expectation is taken for path $X_{1}\left( s_{1}\right) $ starting
from $\left( K_{1},C_{1},A_{1}\right) _{i}$ and ending at $\left(
K_{1},C_{1},A_{1}\right) _{f}$ and path $X_{2}\left( s_{2}\right) $ starting
from $\left( K_{2},C_{2},A_{2}\right) _{i}$ and ending at $\left(
K_{2},C_{2},A_{2}\right) _{f}$. Given our assumptions about the parameters,
we can, as in the previous paragraph, approximate these paths by their
average values to the zeroth order in the parameters:%
\begin{equation*}
X_{i}\left( u\right) =\left( 
\begin{array}{ccc}
\frac{u}{s} & 0 & 0 \\ 
\frac{u}{2s}\left( s-u\right) & \frac{u}{s} & -\frac{u}{2s}K^{\varepsilon
}\left( s-u\right) \\ 
0 & 0 & \frac{u}{s}%
\end{array}%
\right) X_{i}\left( s\right) +\left( 
\begin{array}{ccc}
\frac{s-u}{s} & 0 & 0 \\ 
-\frac{1}{2}\frac{u}{s}\left( s-u\right) & \frac{s-u}{s} & K^{\varepsilon }%
\frac{u}{2s}\left( s-u\right) \\ 
0 & 0 & \frac{s-u}{s}%
\end{array}%
\right) X_{i}\left( 0\right)
\end{equation*}%
so that:$\allowbreak $%
\begin{eqnarray*}
\int_{0}^{s}X_{i}\left( u\right) du &=&\left( 
\begin{array}{ccc}
\frac{1}{2}s & 0 & 0 \\ 
\frac{1}{12}s^{2} & \frac{1}{2}s & -\frac{1}{12}K^{\varepsilon }s^{2} \\ 
0 & 0 & \frac{1}{2}s%
\end{array}%
\right) X_{i}\left( s\right) +\left( 
\begin{array}{ccc}
\frac{1}{2}s & 0 & 0 \\ 
-\frac{1}{12}s^{2} & \frac{1}{2}s & \frac{1}{12}K^{\varepsilon }s^{2} \\ 
0 & 0 & \frac{1}{2}s%
\end{array}%
\right) X_{i}\left( 0\right) \label{trjcl} \\
&=&\left( 
\begin{array}{ccc}
\frac{1}{2}s & 0 & 0 \\ 
\frac{1}{12}s^{2} & \frac{1}{2}s & -\frac{1}{12}K^{\varepsilon }s^{2} \\ 
0 & 0 & \frac{1}{2}s%
\end{array}%
\right) \Delta X_{i}+sX_{i}\left( 0\right) \\
&=&\left( 
\begin{array}{ccc}
0 & 0 & 0 \\ 
\frac{1}{6}s^{2} & 0 & -\frac{1}{6}K^{\varepsilon }s^{2} \\ 
0 & 0 & 0%
\end{array}%
\right) \frac{\Delta X_{i}}{2}+s\bar{X}_{i}
\end{eqnarray*}%
with $\bar{X}_{i}=\frac{X_{i}\left( 0\right) +X_{i}\left( s\right) }{2}$. 
\begin{eqnarray*}
&&\gamma \int \!\!\left\langle \left( ^{t}X_{1}\left( s_{1}\right) \right)
MX_{2}\left( s_{2}\right) \right\rangle ds_{1}ds_{2} \\
&=&\gamma \left( ^{t}\Delta X_{2}\right)\!\!\left( 
\begin{array}{ccc}
0 & 0 & \frac{1}{24}s^{3} \\ 
0 & 0 & \frac{1}{4}s^{2} \\ 
\frac{1}{24}s^{3} & \frac{1}{4}s^{2} & -\frac{1}{12}K^{\varepsilon }s^{3}%
\end{array}%
\right) \allowbreak \allowbreak \Delta X_{1}+\gamma \left( ^{t}X_{2}\left(
0\right) \right)\!\!\left( 
\begin{array}{ccc}
0 & 0 & 0 \\ 
0 & 0 & s^{2} \\ 
0 & s^{2} & 0%
\end{array}%
\right) \allowbreak \allowbreak X_{1}\left( 0\right) \\
&&+\gamma \left( ^{t}\Delta X_{2}\right)\!\!\left( 
\begin{array}{ccc}
0 & 0 & \frac{1}{12}s^{3} \\ 
0 & 0 & \frac{1}{2}s^{2} \\ 
0 & \frac{1}{2}s^{2} & -\frac{1}{12}K^{\varepsilon }s^{3}%
\end{array}%
\right) X_{1}\left( 0\right) +\gamma \left( ^{t}X_{2}\left( 0\right) \right)
\left( 
\begin{array}{ccc}
0 & 0 & 0 \\ 
0 & 0 & \frac{1}{2}s^{2} \\ 
\frac{1}{12}s^{3} & \frac{1}{2}s^{2} & -\frac{1}{12}K^{\varepsilon }s^{3}%
\end{array}%
\right) \Delta X_{1} \\
&=&\gamma \left( ^{t}\bar{X}_{2}\right)\!\!\left( 
\begin{array}{ccc}
0 & 0 & 0 \\ 
0 & 0 & s^{2} \\ 
0 & s^{2} & 0%
\end{array}%
\right) \allowbreak \allowbreak \bar{X}_{1}+\frac{\gamma \left( ^{t}\Delta
X_{2}\right) }{2}\left( 
\begin{array}{ccc}
0 & 0 & \frac{1}{6}s^{3} \\ 
0 & 0 & 0 \\ 
0 & 0 & -\frac{1}{6}K^{\varepsilon }s^{3}%
\end{array}%
\right) \allowbreak \allowbreak \bar{X}_{1} \\
&&+\gamma \left( ^{t}\bar{X}_{2}\right)\!\!\left( 
\begin{array}{ccc}
0 & 0 & 0 \\ 
0 & 0 & 0 \\ 
\frac{1}{6}s^{3} & 0 & -\frac{1}{6}K^{\varepsilon }s^{3}%
\end{array}%
\right) \frac{\Delta X_{1}}{2}
\end{eqnarray*}%
This term modifies the 4 points Green function to an interaction Green
function $G_{\gamma }$:%
\begin{eqnarray*}
&&G_{\gamma }\left( \left( K_{1},C_{1},A_{1}\right) _{i},\left(
K_{2},C_{2},A_{2}\right) _{i},\left( K_{1},C_{1},A_{1}\right) _{f},\left(
K_{2},C_{2},A_{2}\right) _{f}\right) \\
&\equiv &G\left( \left( K_{1},C_{1},A_{1}\right) _{i},\left(
K_{2},C_{2},A_{2}\right) _{i},\left( K_{1},C_{1},A_{1}\right) _{f},\left(
K_{2},C_{2},A_{2}\right) _{f}\right) \exp \left( -\gamma \int \!\!\left\langle
\left( ^{t}X_{1}\left( s_{1}\right) \right) MX_{2}\left( s_{2}\right)
\right\rangle ds_{1}ds_{2}\right)
\end{eqnarray*}%
In terms of trajectory, this means that the deviation of $X_{1}\left(
s\right) $ due to $X_{2}\left( s_{2}\right) $ is given by $\gamma
\int_{0}^{s}MX_{2}\left( s_{2}\right) ds_{2}$ (and the deviation for $%
X_{2}\left( s\right) $ is $\gamma H\int_{0}^{s}MX_{1}\left( s_{1}\right)
ds_{1}$). Given (\ref{trjcl}), one has:%
\begin{equation*}
\gamma HM\int_{0}^{s}X_{2}\left( u\right) du=\gamma \left( 
\begin{array}{ccc}
0 & 0 & 0 \\ 
0 & 0 & 0 \\ 
\frac{1}{6}cs^{3} & 0 & -\frac{1}{6}cK^{\varepsilon }s^{3}%
\end{array}%
\right) \frac{\Delta X_{2}}{2}+\gamma \left( 
\begin{array}{ccc}
0 & 0 & 0 \\ 
0 & 0 & bs \\ 
0 & cs & 0%
\end{array}%
\right) \bar{X}_{2}
\end{equation*}

\end{document}